%% file: g2modspaces.tex
\documentclass[12pt]{article}
\usepackage{jheppub}
\pdfoutput=1

\usepackage{amsmath,bbm,array,amsfonts,graphicx,wrapfig,lscape,float,mathtools,slashbox,multirow,longtable,rotating,subfigure}

\definecolor{strawberry}{RGB}{255,0,128}

\input{pref}

\title{Double Handled Brane Tilings}

\author{Stefano Cremonesi,}
\author{Amihay Hanany,}
\author{Rak-Kyeong Seong}

\affiliation{
Theoretical Physics Group, The Blackett Laboratory,
Imperial College London, \\
Prince Consort Road, London SW7 2AZ, United Kingdom
}

\emailAdd{s.cremonesi@imperial.ac.uk, a.hanany@imperial.ac.uk, rak-kyeong.seong@imperial.ac.uk}

\preprint{Imperial/TP/13/AH/01}

\abstract{
We classify the first few brane tilings on a genus $2$ Riemann surface and identify their toric Calabi-Yau moduli spaces. These brane tilings are extensions of tilings on the 2-torus, which represent one of the largest known classes of $4d$ $\mathcal{N}=1$ superconformal field theories for D3-branes. The classification consists of $16$ distinct genus $2$ brane tilings with up to $8$ quiver fields and $4$ superpotential terms. The Higgs mechanism is used to relate the different theories.
\\
}

\begin{document}

\maketitle

\section{Introduction \label{s_intro}}

Brane tilings \cite{Hanany:2005ve,Franco:2005rj} provide one of the largest known classes of $4d$ $\mathcal{N}=1$ supersymmetric gauge theories living on D3-branes which probe Calabi-Yau 3-fold singularities. As bipartite periodic graphs on the 2-torus, which encode both field theory information and geometry, brane tilings represent an epitome of the rich interface between algebraic geometry and string theory. Our work attempts to upgrade this active relationship by introducing and classifying brane tilings not confined to the traditional 2-torus. 

Brane tilings have been used to classify $4d$ $\mathcal{N}=1$ toric quiver gauge theories with their mesonic and baryonic moduli spaces \cite{Benvenuti:2006qr,Feng:2007ur,Butti:2007jv,Grant:2007ze,Davey:2009bp,2007arXiv0710.1898I,Forcella:2009vw,Hanany:2007zz,Hanany:2010zz,Zaffaroni:2008zz,Forcella:2009bv,Forcella:2008eh,Forcella:2008bb}, dualities \cite{Feng:2002zw,Feng:2001xr,Feng:2001bn,2001JHEP...12..001B} and symmetries \cite{Butti:2005vn,Franco:2004rt}. With the understanding of 3d $\mathcal{N}=2$ Chern-Simons theories as worldvolume theories of M2-branes \cite{Aharony:2008ug,Bagger:2006sk,Bagger:2007jr,Bagger:2007vi,Gustavsson:2007vu,Gustavsson:2008dy}, this tour de force of research and discovery reached new heights and led to the introduction of Chern-Simons levels on brane tilings \cite{Hanany:2008cd,Hanany:2008fj,Ueda:2008hx,Davey:2009qx,Davey:2011mz,Benini:2009qs,Benini:2011cma,Closset:2012ep}.

The work on brane boxes \cite{Hanany:1997tb} described the construction of a prototypical brane tiling on a surface with boundaries such as a disc or cylinder. This idea recently re-emerged as bipartite graphs on discs in relation to string scattering amplitudes \cite{Franco:2012mm,ArkaniHamed:2012nw}. The connection between supersymmetric gauge theories and brane tilings on surfaces with boundaries was further studied in \cite{Franco:2012wv}.

 In parallel, brane tilings associated to Calabi-Yau geometries whose toric diagrams are reflexive polygons \cite{Hanany:2012hi} were found to have the same combined mesonic and baryonic moduli spaces under a map which is known as specular duality \cite{Hanany:2012vc}. The fascinating properties of specular duality further motivates our work.
 
Specular duality makes use of the untwisting map \cite{Feng:2005gw,Butti:2007jv} which relates theories with the same master space \cite{Forcella:2008ng,Hanany:2010zz,Zaffaroni:2008zz,Forcella:2009bv,Forcella:2008eh,Forcella:2008bb} and generates new brane tilings that are not necessarily confined to the 2-torus. The simplest example of this capability is the $\mathbb{C}^3/\mathbb{Z}_5$ $(1,1,3)$ orbifold theory \cite{Hanany:2010cx,Davey:2010px,Davey:2011dd,Hanany:2011iw,Hanany:2010ne} whose brane tiling can be untwisted to give a dual on a $g=2$ Riemann surface. This is an important example of a brane tiling beyond the 2-torus and sheds light on a new infinite class of unexplored field theories.

\begin{figure}[h]
\begin{center}
\includegraphics[trim=0cm 0cm 0cm 0cm,totalheight=5 cm]{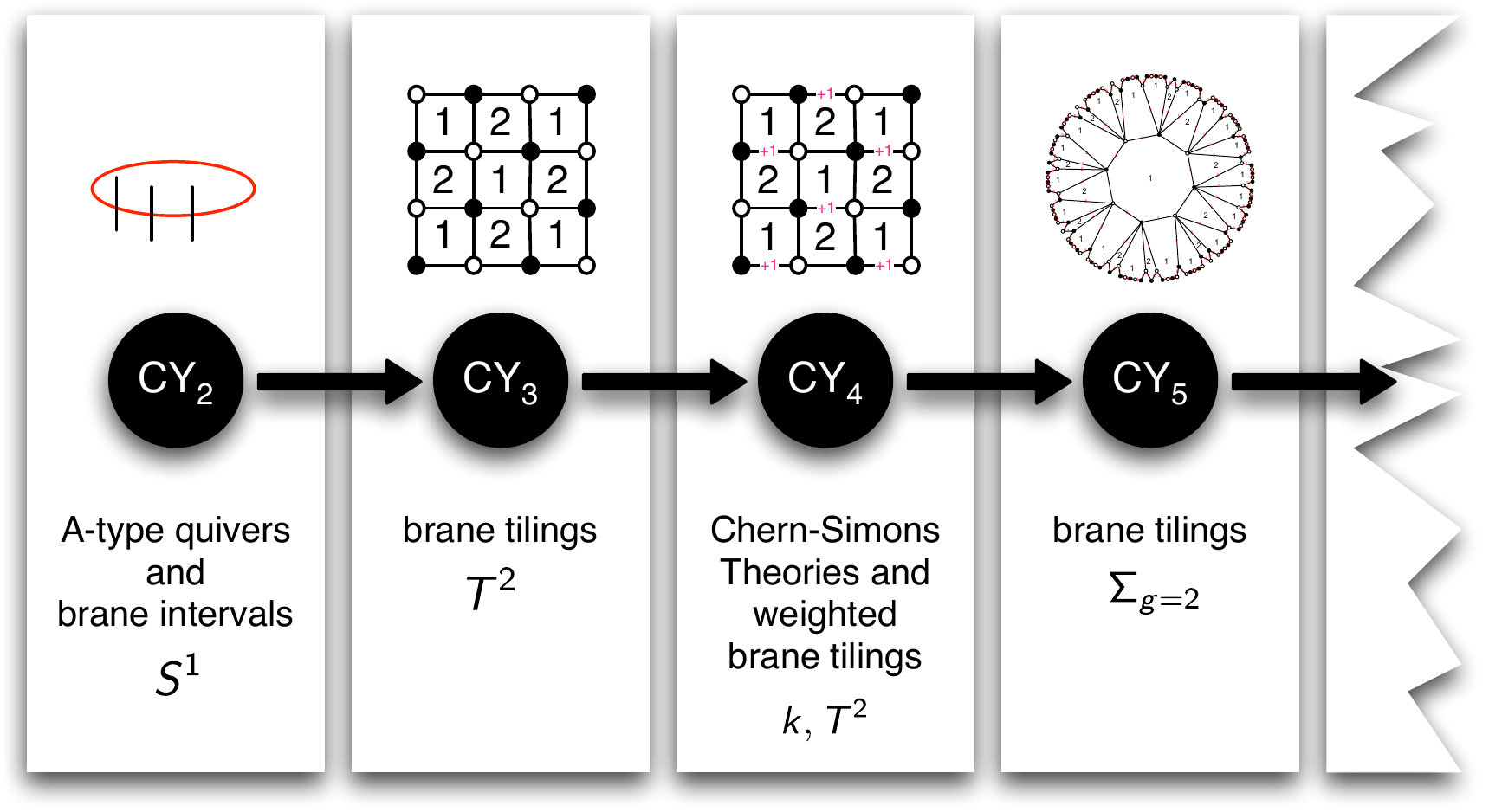}
\caption{\textit{The evolution of brane tilings.} Brane tilings have evolved from  representing A-type quivers to $\mathcal{N}=1$ $4d$ supersymmetric theories and $\mathcal{N}=2$ $3d$ Chern-Simons theories. This paper studies brane tilings on $g=2$ Riemann surfaces associated to Calabi-Yau 5-folds.
\label{ffundd}}
 \end{center}
 \end{figure}

This paper introduces a new procedure of classifying brane tilings on Riemann surfaces. We continue to call the new periodic bipartite graphs on Riemann surfaces as \textit{brane tilings} since they are natural generalisations of the tilings on the 2-torus. Although the brane construction for the generalisation is not yet fully understood, we believe that our classification is an important step towards a better understanding of the problem.

Despite the efficiency of generating brane tilings on $g=2$ or higher genus Riemann surfaces with specular duality, only a subset of these new brane tilings can be identified with this method. Most other brane tilings, often with much smaller number of fields and gauge groups, can only be obtained via a direct construction on the Riemann surface.\footnote{These are in fact under specular duality often related to \textit{inconsistent} brane tilings on the 2-torus. Consistency of brane tilings on the 2-torus has been studied from many perspectives \cite{Franco:2005sm,Hanany:2005ss,Broomhead:2008an,Gulotta:2008ef,2011arXiv1104.1592B}, and the most important properties are reviewed in this work.} The work will give the first classification of brane tilings on a $g=2$ Riemann surface with up to 8 quiver fields and 4 superpotential terms. Our classification identifies precisely 16 distinct $g=2$ brane tilings which can be related by a successive application of the Higgs mechanism.

The mesonic moduli space of each brane tiling in the classification is computed by imposing F-and D-term constraints. These moduli spaces are all toric Calabi-Yau 5-folds. The moduli space dimension is in general $2g+1$ where the number of homology 1-cyles on the genus $g$ Riemann surface is $2g$. By computing the Hilbert series, we specify the explicit algebraic structure of the moduli space and relate new geometries to classical field theories.

For generic ranks of the gauge groups, it is not clear whether the beta functions of all couplings can be set to zero.\footnote{It is well known \cite{Hanany:2005ss} that if the ranks of the gauge groups are all equal and none of the couplings vanish, the beta functions cannot all be zero.} Accordingly, understanding the IR behaviour of the brane tilings may be challenging. For the moment, the classification of $g=2$ brane tilings should be considered as an important step towards a better understanding of recent lines of thought. We believe that such extensions to the field theories classified in this work along with a better understanding of the brane construction will lead to new exciting progress in the near future.

The structure of the paper is as follows. Section \sref{s_riemm} gives a first glimpse of a $g=2$ brane tiling by untwisting the brane tiling for the $\mathbb{C}^3/\mathbb{Z}_5$ $(1,1,3)$ theory and then proceeds to outline an algorithm for classifying all distinct brane tilings on a $g=2$ Riemann surface. The results are summarized in section \sref{s_class}. Section \sref{s_consist} continues with a discussion on consistency of brane tilings that plays an important role in the case of the 2-torus. The section explains that restrictions are set on $g=2$ brane tilings to reduce the number of models in the classification even though the restrictions are not well motivated from a field theory perspective. Section \sref{s_moduli} summarises the basic properties of the mesonic moduli spaces and continues with section \sref{s_higgs} by explaining how the Higgs mechanism relates the theories in the classification and acts as a check of the classification. In the second part of the paper, section \sref{s_classification} summarises the full classification data for $g=2$ brane tilings, including the computation of the Hilbert series. Appendix \sref{app_con} includes a more concise summary of the classification. In addition, $g=2$ brane tilings with self-intersecting zig-zag paths are presented in appendix \sref{app_incon}. Appendix \sref{app_mesonic} gives a short summary of the forward algorithm which is used to identify the mesonic moduli spaces.
\\

\section{Brane Tilings on Riemann Surfaces \label{s_riemm}}

In this section we present the classification scheme which we used for the $g=2$ brane tilings. A brief summary is given for what is meant by a $g>1$ brane tiling, with an overview of their field theoretic and geometric properties. 
\\

\subsection{The Construction \label{s_constr}}

\begin{figure}[ht!!]
\begin{center}
\includegraphics[trim=0cm 0cm 0cm 0cm,totalheight=4.5 cm]{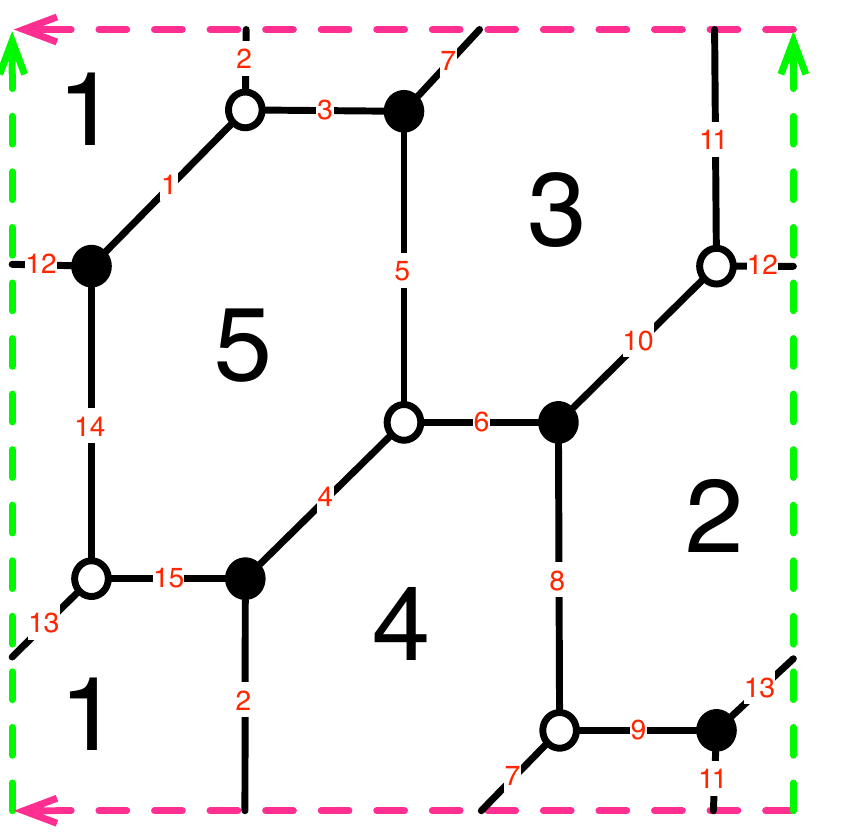}~~
\includegraphics[trim=0cm 0cm 0cm 0cm,totalheight=4.5 cm]{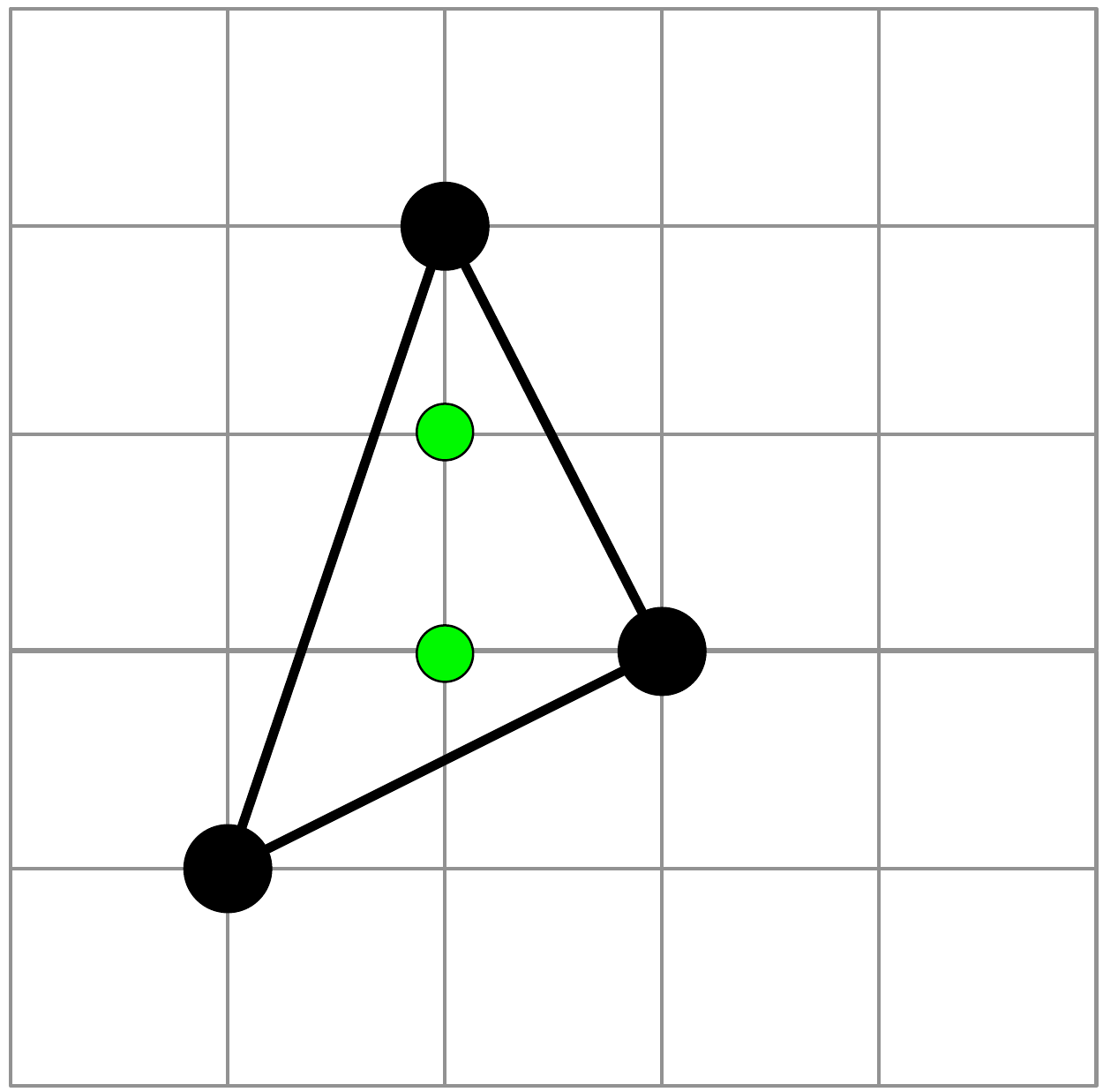}\\
\vspace{0.3cm}
\resizebox{\hsize}{!}{
\begin{tabular}{cccccccccccccccccccc}
{\color[rgb]{1.000000,0.000000,0.000000} 1} &
{\color[rgb]{1.000000,0.000000,0.000000} 2} &
{\color[rgb]{1.000000,0.000000,0.000000} 3} &
{\color[rgb]{1.000000,0.000000,0.000000} 4} &
{\color[rgb]{1.000000,0.000000,0.000000} 5} &
{\color[rgb]{1.000000,0.000000,0.000000} 6} &
{\color[rgb]{1.000000,0.000000,0.000000} 7} &
{\color[rgb]{1.000000,0.000000,0.000000} 8} &
{\color[rgb]{1.000000,0.000000,0.000000} 9} &
{\color[rgb]{1.000000,0.000000,0.000000} 10}&
{\color[rgb]{1.000000,0.000000,0.000000} 11} &
{\color[rgb]{1.000000,0.000000,0.000000} 12} &
{\color[rgb]{1.000000,0.000000,0.000000} 13} &
{\color[rgb]{1.000000,0.000000,0.000000} 14} &
{\color[rgb]{1.000000,0.000000,0.000000} 15} 
\\
$X_{51}^{1}$ & $X_{14}$ & $X_{45}^{1}$ &
$X_{45}^{2}$ & $X_{53}$ & $X_{34}^{1}$ &
$X_{34}^{2}$ & $X_{42}$ & $X_{23}^{1}$ &
$X_{23}^{2}$ & $X_{31}$ & $X_{12}^{1}$ &
$X_{12}^{2}$ & $X_{25}$ & $X_{51}^{2}$
\end{tabular}
}
\caption{Brane tiling and toric diagram of $\mathbb{C}^3/\mathbb{Z}_5$ (1,1,3).
\label{fc3z5}}
 \end{center}
 \end{figure}

\begin{figure}[ht!!]
\begin{center}
\includegraphics[trim=0cm 0cm 0cm 0cm,totalheight=7 cm]{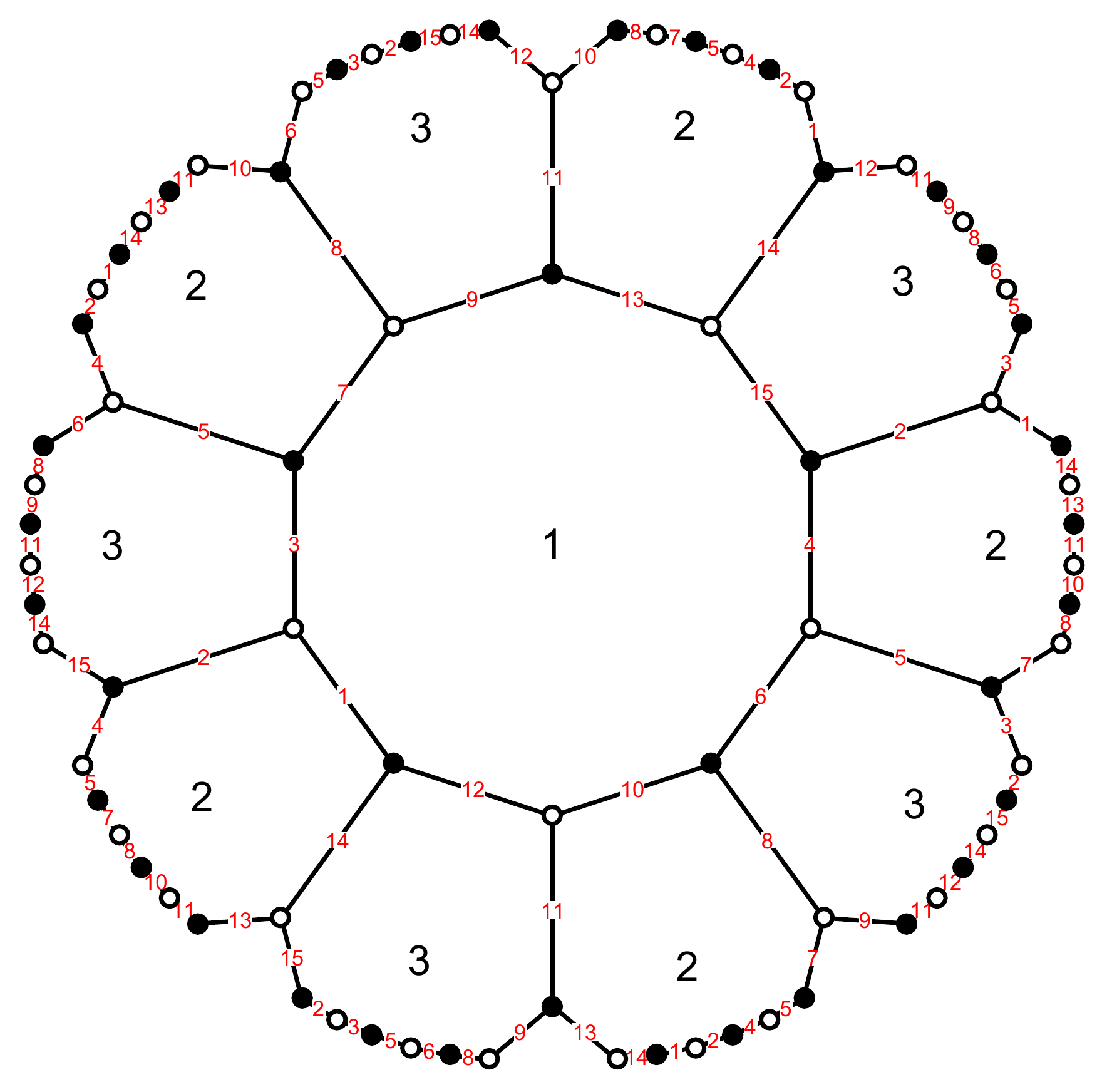}~~
\includegraphics[trim=0cm 0cm 0cm 0cm,totalheight=7 cm]{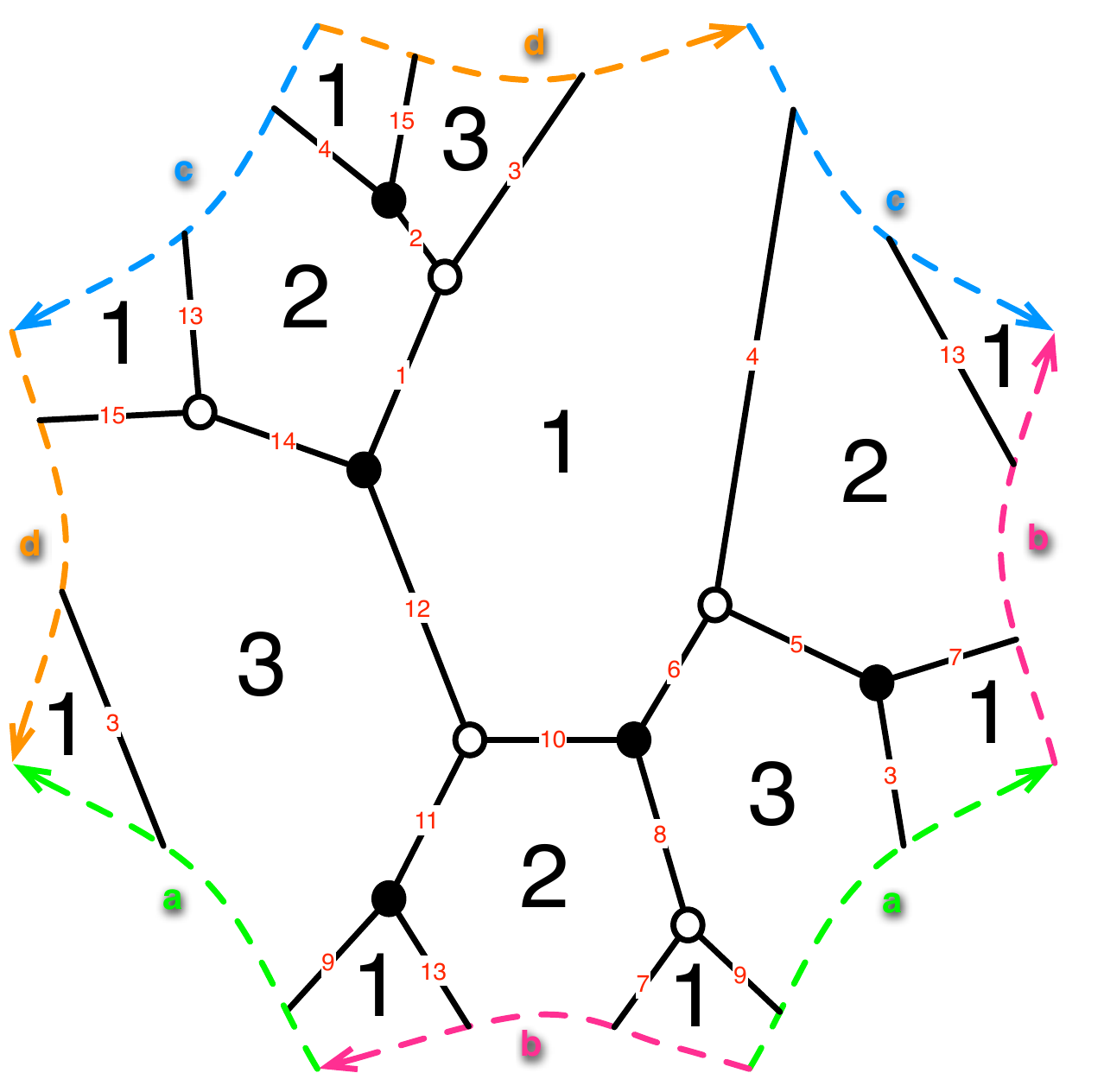}\\
\vspace{0.3cm}
\resizebox{\hsize}{!}{
\begin{tabular}{cccccccccccccccccccc}
{\color[rgb]{1.000000,0.000000,0.000000} 1} &
{\color[rgb]{1.000000,0.000000,0.000000} 2} &
{\color[rgb]{1.000000,0.000000,0.000000} 3} &
{\color[rgb]{1.000000,0.000000,0.000000} 4} &
{\color[rgb]{1.000000,0.000000,0.000000} 5} &
{\color[rgb]{1.000000,0.000000,0.000000} 6} &
{\color[rgb]{1.000000,0.000000,0.000000} 7} &
{\color[rgb]{1.000000,0.000000,0.000000} 8} &
{\color[rgb]{1.000000,0.000000,0.000000} 9} &
{\color[rgb]{1.000000,0.000000,0.000000} 10}&
{\color[rgb]{1.000000,0.000000,0.000000} 11} &
{\color[rgb]{1.000000,0.000000,0.000000} 12} &
{\color[rgb]{1.000000,0.000000,0.000000} 13} &
{\color[rgb]{1.000000,0.000000,0.000000} 14} &
{\color[rgb]{1.000000,0.000000,0.000000} 15} 
\\
$X_{12}^{1}$ & $X_{23}^{1}$ & $X_{31}^{1}$ &
$X_{12}^{2}$ & $X_{23}^{2}$ & $X_{31}^{2}$ &
$X_{12}^{3}$ & $X_{23}^{3}$ & $X_{31}^{3}$ &
$X_{12}^{4}$ & $X_{23}^{4}$ & $X_{31}^{4}$ &
$X_{12}^{5}$ & $X_{23}^{5}$ & $X_{31}^{5}$
\end{tabular}
}
\caption{Specular dual brane tiling of $\mathbb{C}^3/\mathbb{Z}_5$ (1,1,3) on a $g=2$ Riemann surface with its fundamental domain.
\label{fspecc3z5}}
 \end{center}
 \end{figure}

\begin{figure}[ht!!]
\begin{center}
\includegraphics[trim=0cm 0cm 0cm 0cm,totalheight=4 cm]{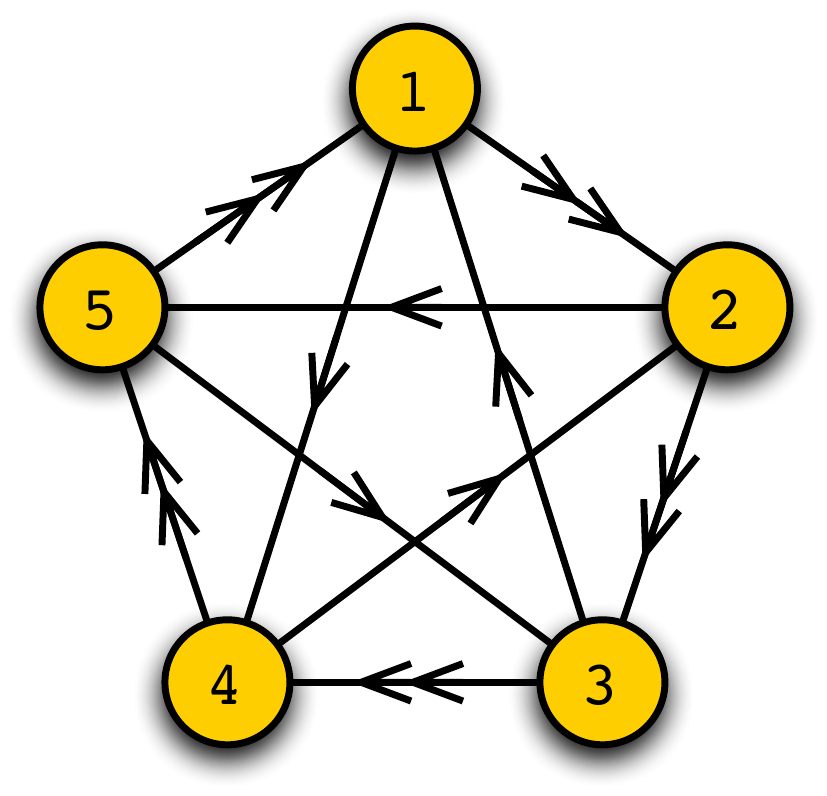}~~
\includegraphics[trim=0cm 0cm 0cm 0cm,totalheight=3 cm]{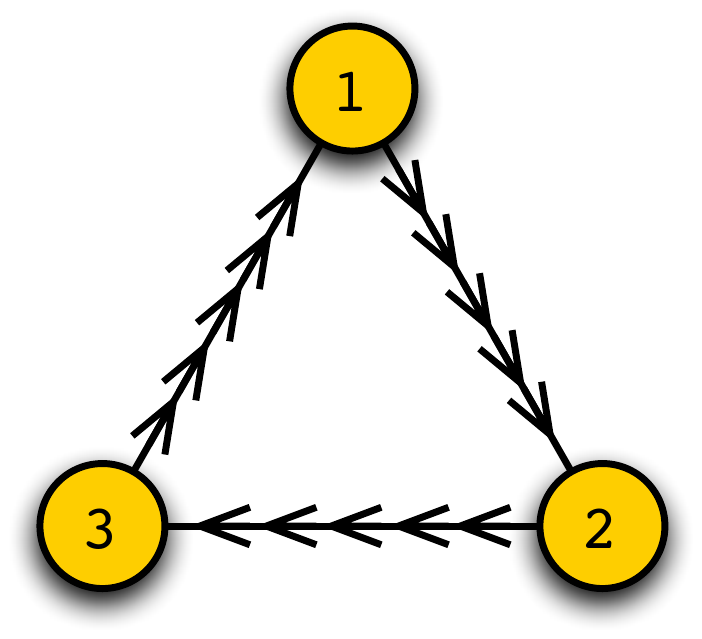}\\
\hspace{0.3cm}(a) \hspace{3.4cm} (b)
\\
\vspace{0.3cm}
\resizebox{\hsize}{!}{
\begin{tabular}{cccccccccccccccccccc}
{\color[rgb]{1.000000,0.000000,0.000000} 1} &
{\color[rgb]{1.000000,0.000000,0.000000} 2} &
{\color[rgb]{1.000000,0.000000,0.000000} 3} &
{\color[rgb]{1.000000,0.000000,0.000000} 4} &
{\color[rgb]{1.000000,0.000000,0.000000} 5} &
{\color[rgb]{1.000000,0.000000,0.000000} 6} &
{\color[rgb]{1.000000,0.000000,0.000000} 7} &
{\color[rgb]{1.000000,0.000000,0.000000} 8} &
{\color[rgb]{1.000000,0.000000,0.000000} 9} &
{\color[rgb]{1.000000,0.000000,0.000000} 10}&
{\color[rgb]{1.000000,0.000000,0.000000} 11} &
{\color[rgb]{1.000000,0.000000,0.000000} 12} &
{\color[rgb]{1.000000,0.000000,0.000000} 13} &
{\color[rgb]{1.000000,0.000000,0.000000} 14} &
{\color[rgb]{1.000000,0.000000,0.000000} 15} 
\\
$X_{51}^{1}$ & $X_{14}$ & $X_{45}^{1}$ &
$X_{45}^{2}$ & $X_{53}$ & $X_{34}^{1}$ &
$X_{34}^{2}$ & $X_{42}$ & $X_{23}^{1}$ &
$X_{23}^{2}$ & $X_{31}$ & $X_{12}^{1}$ &
$X_{12}^{2}$ & $X_{25}$ & $X_{51}^{2}$
\\
\hline
$X_{12}^{1}$ & $X_{23}^{1}$ & $X_{31}^{1}$ &
$X_{12}^{2}$ & $X_{23}^{2}$ & $X_{31}^{2}$ &
$X_{12}^{3}$ & $X_{23}^{3}$ & $X_{31}^{3}$ &
$X_{12}^{4}$ & $X_{23}^{4}$ & $X_{31}^{4}$ &
$X_{12}^{5}$ & $X_{23}^{5}$ & $X_{31}^{5}$
\end{tabular}
}
\caption{The (a) quiver of $\mathbb{C}^3/\mathbb{Z}_5$ (1,1,3) and (b) its specular dual quiver with the field map under the untwisting move.
\label{fquivercomp}}
 \end{center}
 \end{figure}

As seen in \cite{Hanany:2012vc}, specular duality and the untwisting map \cite{Butti:2007jv,Feng:2005gw} can be used to generate brane tilings on Riemann surfaces with genus $g>1$. The simplest example is the brane tiling for $\mathbb{C}^3/\mathbb{Z}_5$ with orbifold action $(1,1,3)$, whose toric diagram is a lattice triangle with exactly two internal points. The toric diagram and the brane tiling are in \fref{fc3z5} with the quiver diagram in \fref{fquivercomp}. The superpotential has the form
\beal{esi1}
W&=&
+ X_{51}^{1} X_{14} X_{45}^{1}
+ X_{45}^{2} X_{53} X_{34}^{1} 
+ X_{34}^{2} X_{42} X_{23}^{1} 
+ X_{23}^{2} X_{31} X_{12}^{1} 
+ X_{12}^{2} X_{25} X_{51}^{2}
\nn\\
&&
- X_{51}^{1} X_{12}^{1} X_{25}
- X_{45}^{2} X_{51}^{2} X_{14}
- X_{34}^{2} X_{45}^{1} X_{53}
- X_{23}^{2} X_{32}^{1} X_{42}
- X_{12}^{2} X_{23}^{1} X_{31}
~~.
\nn\\
\eea
Given that the superpotential has an overall trace, which is omitted for brevity, let us use the notation which replaces terms in the superpotential as a cyclic permutation of integers \cite{Jejjala:2010vb}. The integers themselves label fields with the dictionary given in \fref{fc3z5},
\beal{esi2}
W&=& 
+ (1 ~ 2 ~ 3)
+ (4 ~ 5 ~ 6)
+ (7 ~ 8 ~ 9)
+ (10 ~ 11 ~ 12)
+ (13 ~ 14 ~ 15)
\nn\\
&&
- (1 ~ 12 ~ 14)
- (4 ~ 15 ~ 2)
- (7 ~ 3 ~ 5)
- (10 ~ 6 ~ 8)
- (13 ~ 9 ~ 11)
~~.
\eea

The specular dual tiling is on a $g=2$ Riemann surface and the corresponding supersymmetric field theory has a $5d$ toric Calabi-Yau mesonic moduli space. The brane tiling is shown in \fref{fspecc3z5} with the quiver in \fref{fquivercomp}. The superpotential of the specular dual is easily obtained by reversing the permutations which correspond to the negative (or equivalently the positive) terms in the original superpotential in \eref{esi2}. 

This $g=2$ brane tiling is the one that can be generated via specular duality with the least number of fields. In fact, there are $g=2$ brane tilings with much fewer fields that cannot be obtained via specular duality on 2-torus tilings. In the following section, we illustrate a method of generating such tilings and give a full classification up to 8 quiver fields and 4 superpotential terms. 
\\

\subsection{Classification of $g=2$ Brane Tilings \label{s_class}}

\begin{figure}[H]
\begin{center}
\includegraphics[trim=0cm 0cm 0cm 0cm,totalheight=7 cm]{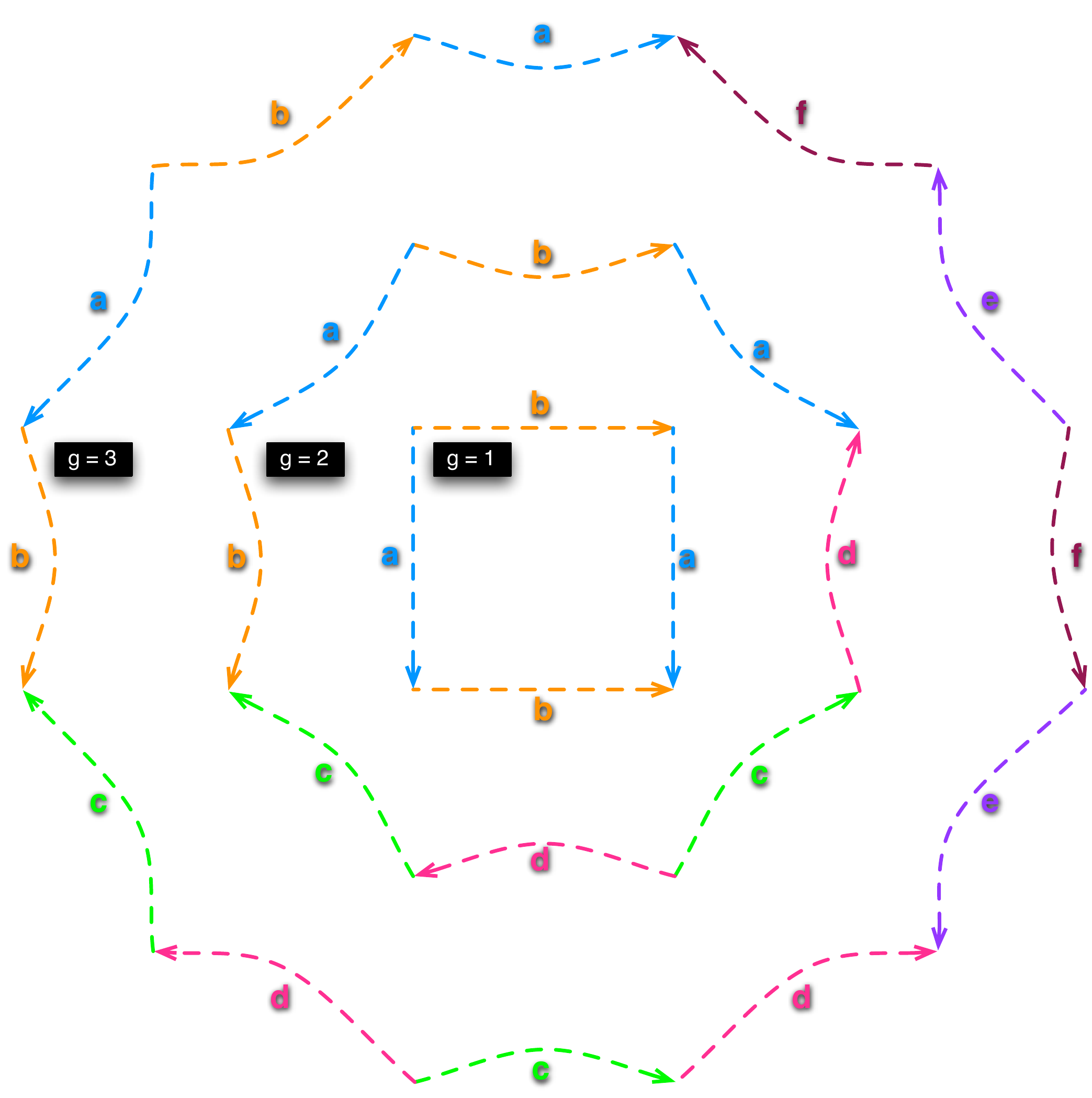}
\caption{\textit{Fundamental domains of higher genus brane tilings.} 
These are choices for fundamental domains for Riemann surfaces of genus $g=1,2,3$.
\label{ffundd}}
 \end{center}
 \end{figure}
 
The brane tiling as a bipartite graph satisfies the \textit{Euler formula},
\beal{esi20}
F-E+V = 2-2g~~,
\eea 
where $E$, $V$ and $F$ are respectively the number of edges, nodes and faces of the brane tiling and $g$ is the genus of the Riemann surface. The fundamental domain of the genus $g$ brane tiling is a $4g$-sided polygon with our identification of sides being the one shown in \fref{ffundd}. Accordingly, there are $2g$ fundamental cycles with every zig-zag path\footnote{A zig-zag path is a closed path along the edges on the brane tiling which alternates between white and black nodes. The path is such that it makes precisely one maximal clockwise turn around a white note and the a maximal anti-clockwise turn around the next black node before reaching the next edge and node in the sequence.} of the brane tiling having $2g$ winding numbers. This leads to rank $2g$ mesonic symmetry in the associated field theory \cite{Hanany:2005ve,Kennaway:2007tq}.

\begin{table}[ht!!]
\begin{center}
\begin{tabular}{|c|c|c||c|}
\hline
$E$ & $V$ & $F$ & \# Models \\
\hline \hline
5 & 2 & 1 & 1 \\
6 & 2 & 2 & 3 \\
7 & 2 & 3 & 1 \\
7 & 4 & 1 & 1 \\
8 & 2 & 4 & 2 \\
8 & 4 & 2 & 8 \\
\hline
\end{tabular}
\end{center}
\caption{
\textit{The Euler formula and the classification.} These are the numbers of distinct brane tilings on a $g=2$ Riemann surface without self-intersecting zig-zag paths and without multi-bonded edges for specific numbers of edges $E$, number of vertices $V$ and faces $F$.
\label{tclass}
}
\end{table}

For $g=2$, the first few values of $E$, $V$ and $F$ satisfying the Euler formula are given in \tref{tclass}. By setting $(E,V,F)$ for $g=2$, we generate all possible permutations of $E$ integers. From this set of permutations, all possible pairings of permutations are taken. For each permutation pair one is marked as positive and the other one as negative. We associate a pairing to a brane tiling if it satisfies the following brane tiling conditions:
\begin{itemize}
\item The number of cycles in the positive permutation is the same as the number of cycles in the negative permutation. This translates to the condition that there are the same number of positive and negative superpotential terms.
\item Every integer precisely appears once in a positive permutation cycle and a negative permutation cycle. This by construction satisfies the \textit{toric condition} of the brane tiling.
\item The associated brane tiling has no self-intersecting zig-zag paths and no multi-bonded edges \cite{Franco:2005sm,Hanany:2005ss,Broomhead:2008an} as discussed in \sref{s_consist}. We adopt these restrictions in the classification for $g=2$ brane tilings to reduce the number of identified models.\end{itemize}

Two brane tilings on any genus Riemann surface are the same if they satisfy the following \textit{equivalence conditions}:
\begin{itemize}
\item The brane tilings are on the same Riemann surface with the same genus $g$.
\item The quiver diagrams are equivalent graphs.
\item The superpotential as a permutation pairing is the same partition of integers.
\item The zig-zag paths \cite{Gulotta:2008ef,Butti:2005ps} are the same partition of integers.
\item The mesonic moduli spaces $\mesonic$ \cite{Butti:2007jv,Hanany:2012hi,Feng:2000mi} are the same.
\end{itemize}

\begin{figure}[H]
\begin{center}
\resizebox{0.98\hsize}{!}{
\includegraphics[trim=0cm 0cm 0cm 0cm,totalheight=18.5 cm]{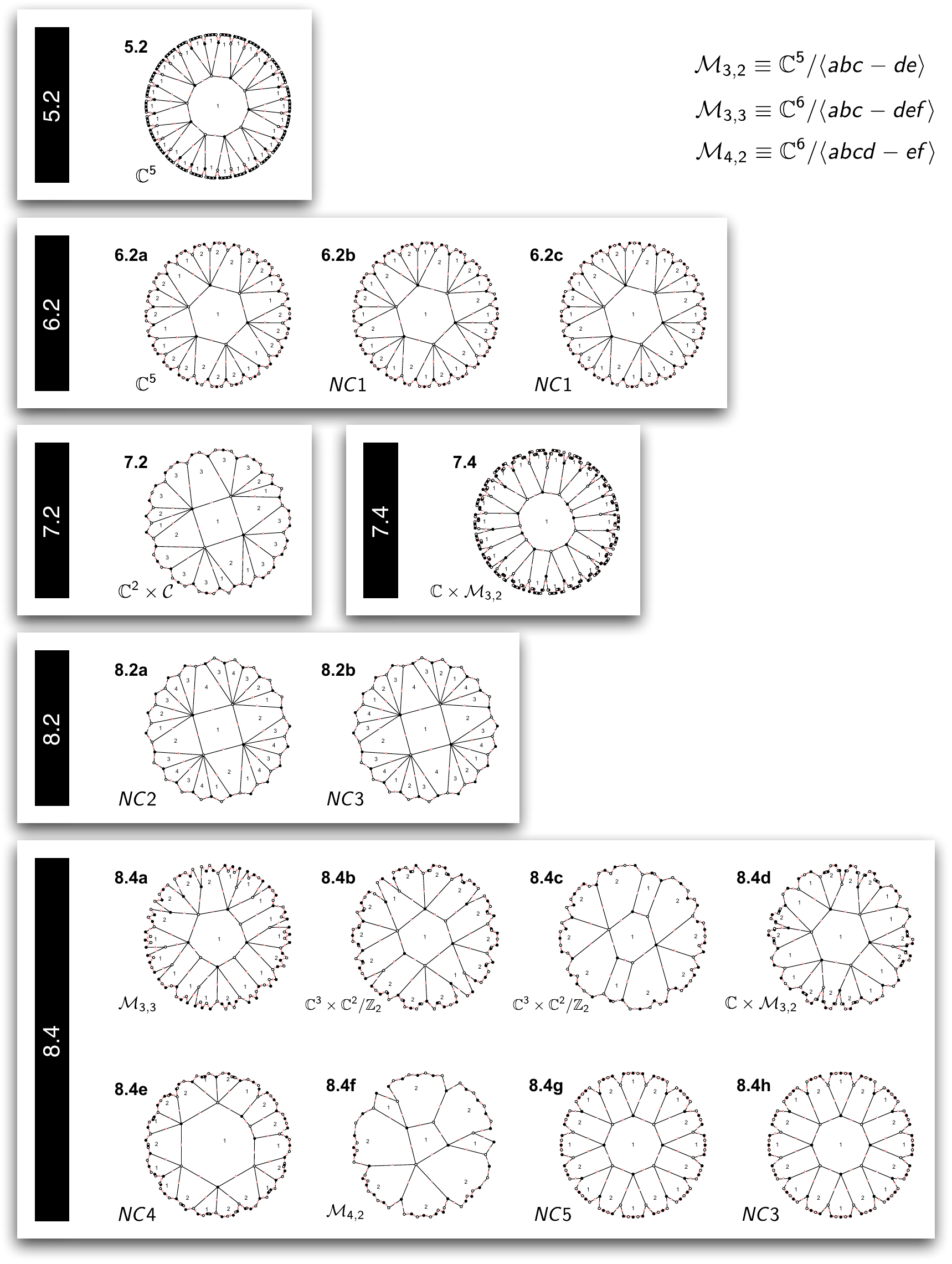}
}
\caption{\textit{Classification of $g=2$ brane tilings with no self-intersecting zig-zag paths and no multi-bonded edges.} These are the first 16 brane tiling on a $g=2$ Riemann surface with up to $E=8$ and $V=4$.
\label{fg2summary}}
 \end{center}
 \end{figure}

\noindent Note that a subset of the conditions above may not be enough to identify brane tiling equivalence. An example is a pair of distinct toric dual brane tilings which are related by the \textit{urban renewal} move. The dual brane tilings have the same mesonic moduli space \cite{Hanany:2005ve}. In fact, for $g>1$ brane tilings, two distinct brane tilings which are not related by the urban renewal move can have the same mesonic moduli space.

Following the procedure which is outlined above, we classify all distinct brane tilings on a $g=2$ Riemann surface with up to $E=8$ edges and $V=4$ superpotential terms. We identify $16$ distinct $g=2$ brane tilings. They are summarized in \fref{fg2summary}, and their mesonic moduli spaces are identified and discussed in Section \sref{s_classification}. We emphasise that the 16 brane tilings are \textit{restricted}, in other words they do not have self-intersecting zig-zag paths and no multi-bonded edges. All other tilings are not discussed in detail in this paper and are subject for future studies.
\\

\subsection{Consistency of Brane Tilings on a 2-torus \label{s_consist}}

\begin{figure}[ht!!]
\begin{center}
\includegraphics[trim=0cm 0cm 0cm 0cm,totalheight=4 cm]{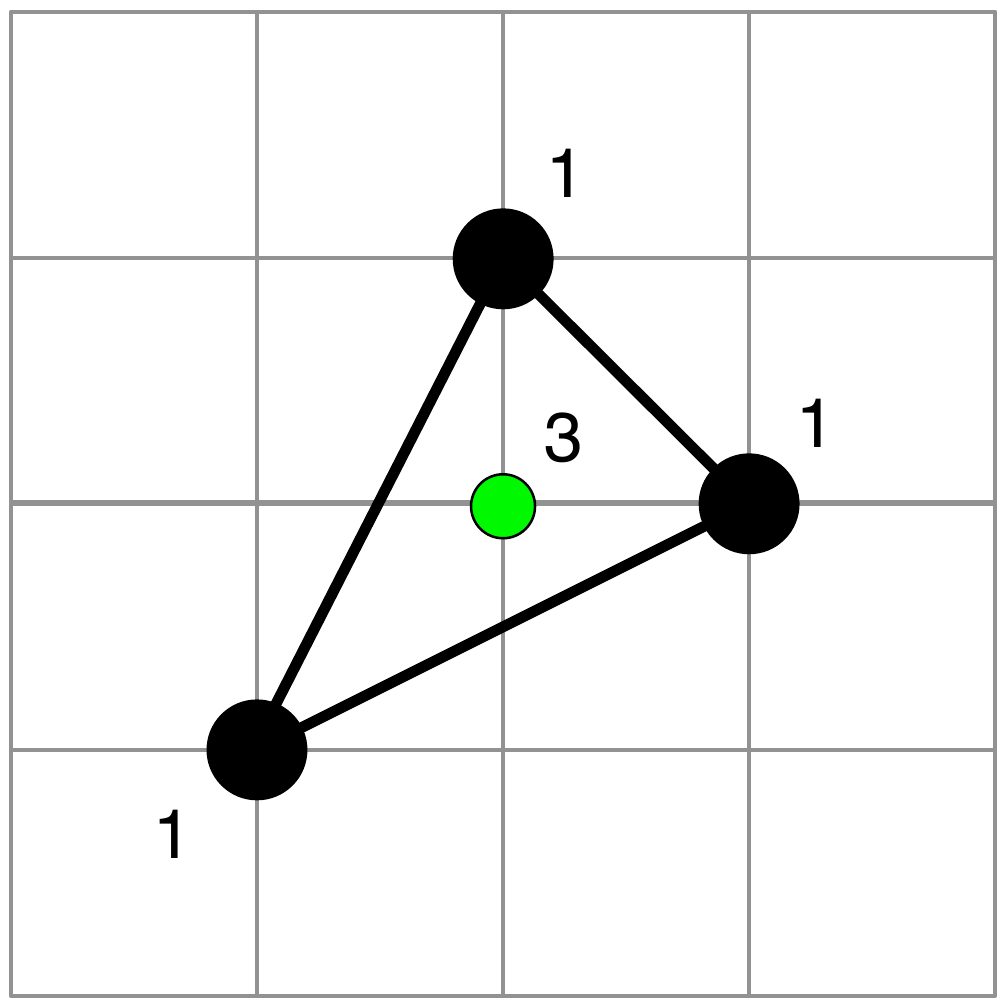}~
\includegraphics[trim=0cm 0cm 0cm 0cm,totalheight=4 cm]{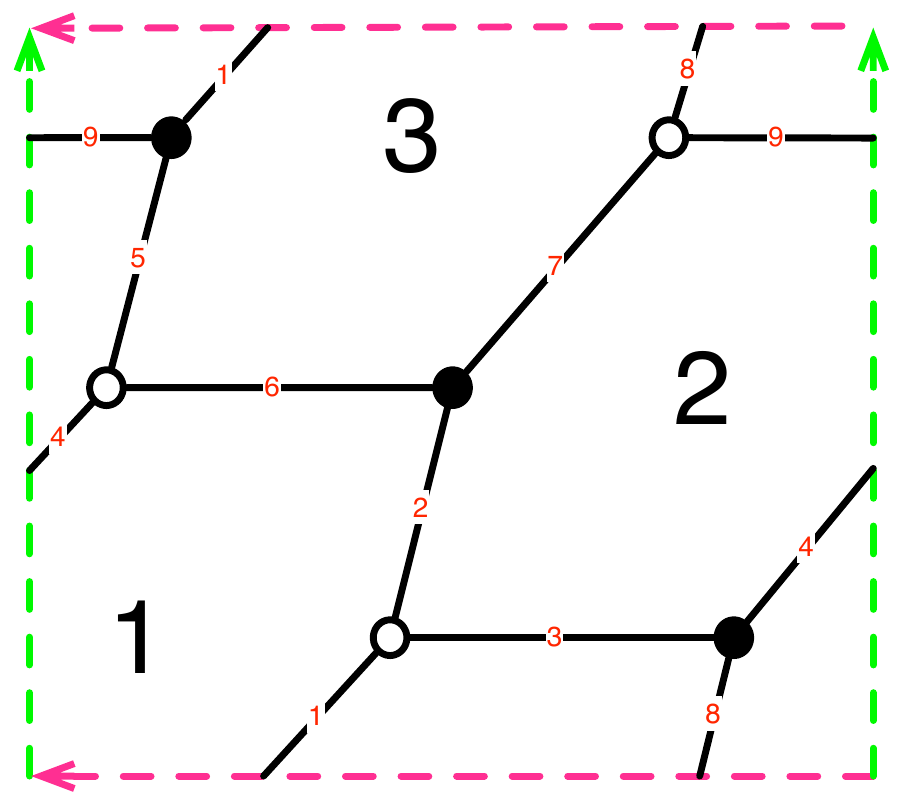}~
\includegraphics[trim=0cm 0cm 0cm 0cm,totalheight=4 cm]{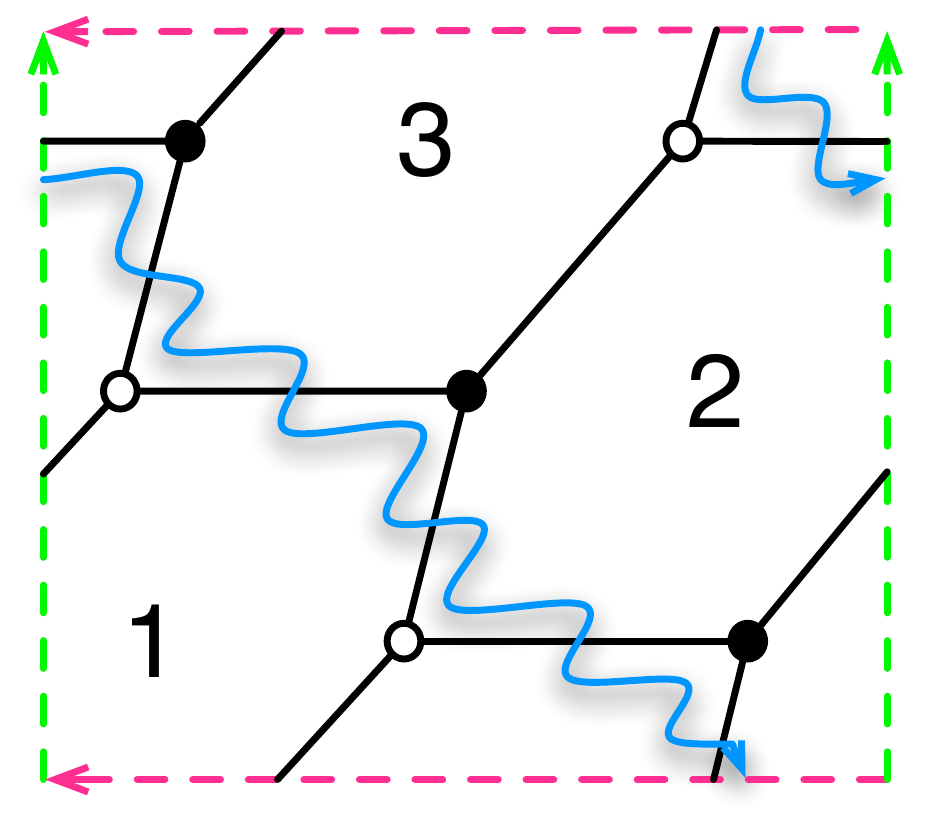}
\\
\vspace{0.8cm}
\includegraphics[trim=0cm 0cm 0cm 0cm,totalheight=4 cm]{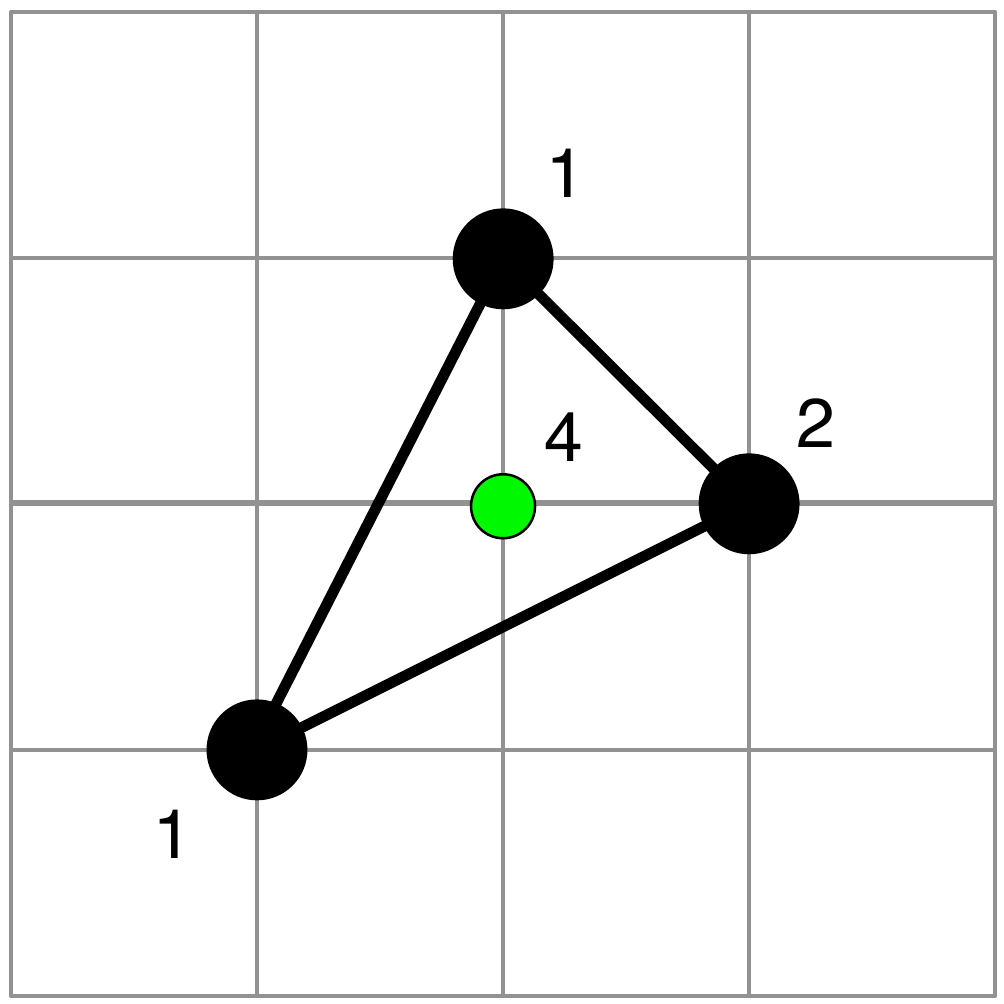}~
\includegraphics[trim=0cm 0cm 0cm 0cm,totalheight=4 cm]{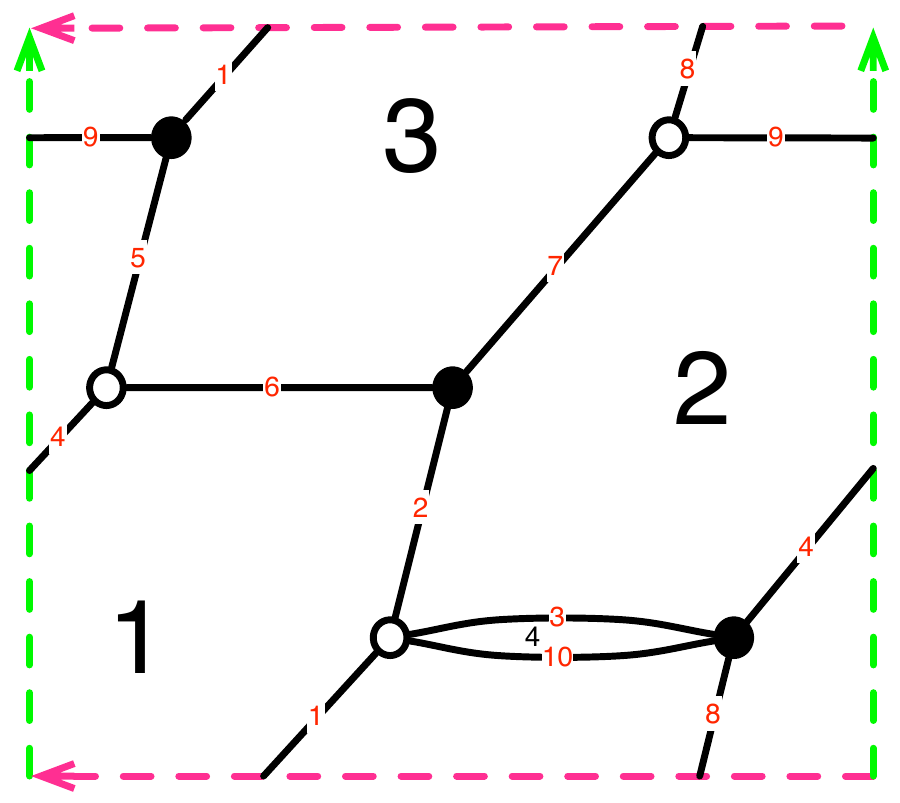}~
\includegraphics[trim=0cm 0cm 0cm 0cm,totalheight=4 cm]{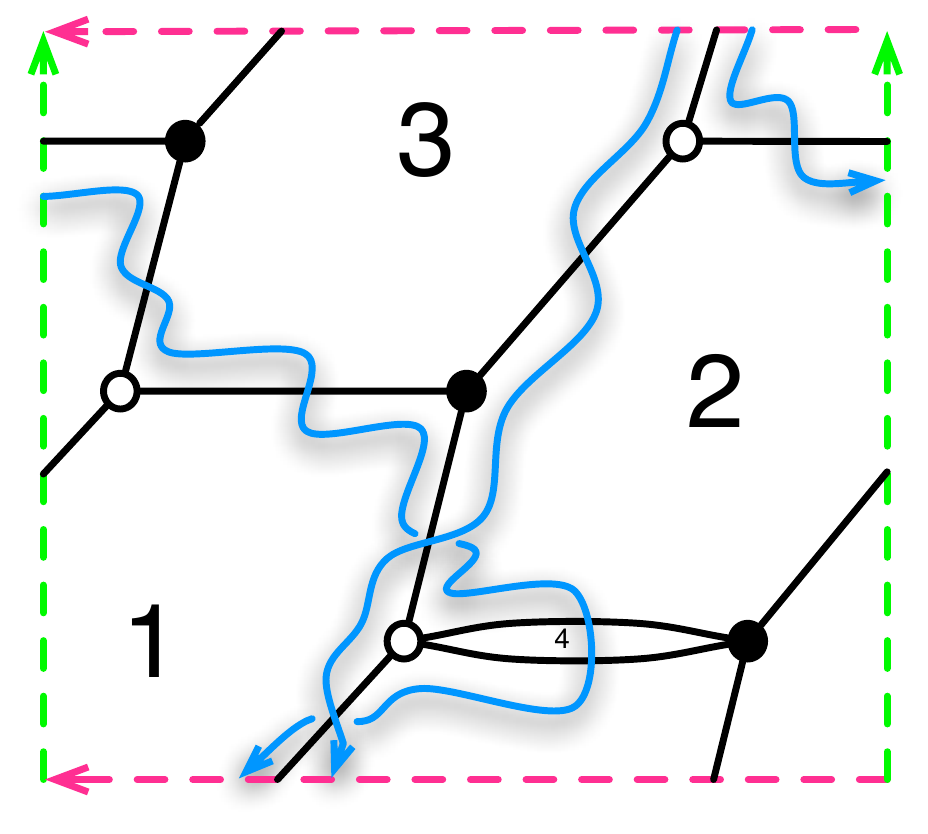}
\caption{\textit{Inconsistent $\text{dP}_0$ Model.} The top row shows the toric diagram of the $\text{dP}_0$ model with the brane tiling and zig-zag path of the brane tiling going around the 2-torus. The bottom row shows an inconsistent toric diagram with an extremal toric point having a multiplicity greater than 1, and its corresponding double-bonded brane tiling with self-intersecting zig-zag path.
\label{fincons}}
 \end{center}
 \end{figure}

The notion of \textit{consistency} of a brane tiling on the 2-torus was first discussed in \cite{Hanany:2005ss}. Consistent torus brane tilings are expected to flow in the IR to a superconformal fixed point with a preferred $U(1)$ R-symmetry which appears in the superconformal algebra and determines the scaling dimension of BPS operators. If the consistency conditions are not satisfied, one normally can expect zero superconformal R-charges to be assigned to bifundamental fields under a-maximisation \cite{Butti:2005ps,Butti:2005vn,Martelli:2005tp}. In this case, some dibaryon operators would violate the unitarity bound on the scaling dimension.

In order to discuss brane tiling consistency from a geometric and combinatorial point of view, we recall that the classical vacuum moduli space of the 1-brane theory\footnote{In the following sections, we call a quiver gauge theory with a superpotential associated to a brane tiling Abelian if it has only $U(1)$ gauge factors.} is a toric Calabi-Yau 3-fold. It is represented by a convex lattice polygon known as the toric diagram. This mesonic moduli space can be expressed as a K\"ahler quotient in terms of a gauge linear sigma model (GLSM) \cite{Witten:1993yc} description of the theory. Perfect matchings of the brane tiling are associated to GLSM fields and are identified as lattice points on the toric diagram. In summary, \textit{inconsistency} can be observed when
\begin{itemize}
\item Twice the area of the toric diagram is \textit{not} the number of gauge groups in the brane tiling.\comment{ as it would be expected for branes wrapping holomorphic cycles in a Calabi-Yau 3-fold.}
\item More than one GLSM field of the brane tiling is associated to a corner (extremal) point of the toric diagram.\comment{ which would indicate a mismatch between the baryon number and the  }
\end{itemize}

From a purely graphical point of view, a brane tiling is \textit{consistent} if it has the following properties:
\begin{itemize}
\item No zig-zag paths self-intersect.
\item No edges are `multi-bonded' and hence no faces are 2-sided.
\end{itemize}
The above consistency conditions are illustrated in \fref{fincons}. 

For the following classification of brane tilings on a $g=2$ Riemann surface, we restrict ourselves to brane tilings with no self-intersecting zig-zag paths and no multi-bonded edges. We call these \textit{restricted $g=2$ brane tilings}. We apply the restriction in order to reduce the number of brane tilings identified in the classification, even though we believe that it is of interest to study unrestricted brane tilings on $g=2$ Riemann surfaces. We leave the study of unrestricted brane tilings for future work.
\\

\subsection{Mesonic Moduli Spaces \label{s_moduli}} 

\begin{table}[ht!!]
\centering
\begin{tabular}{|c|c|c|}
\hline
\# & $\mesonic$ & Global Symmetry \\
\hline \hline
5.2 & $\mathbb{C}^5$ &$SU(5) \times U(1)_R$
\\
6.2a & $\mathbb{C}^5$ & $SU(5) \times U(1)_R$ 
\\
6.2b & $NC1$ & $SU(3)^2 \times U(1)_R$
\\ 
6.2c & $NC1$ & $SU(3)^2 \times U(1)_R$
\\
7.2 & $\mathbb{C}^2\times\mathcal{C}$ & $SU(2)^3 \times U(1) \times U(1)_R$
\\
7.4 & $\mathbb{C}\times\mathcal{M}_{3,2}$ &$U(1)^4 \times U(1)_R$
\\
8.2a & $NC2$ & $SU(2)^2 \times U(1)^2 \times U(1)_R$
\\
8.2b & $NC3$ & $SU(2)^4 \times U(1)_R$
\\
8.4a & $\mathcal{M}_{3,3}$ & $U(1)^4 \times U(1)_R$
\\
8.4b & $\mathbb{C}^3\times\mathbb{C}^2/\mathbb{Z}_2$ & $SU(3) \times SU(2) \times U(1) \times U(1)_R$
\\
8.4c & $\mathbb{C}^3\times\mathbb{C}^2/\mathbb{Z}_2$ & $SU(3) \times SU(2) \times U(1) \times U(1)_R$
\\
8.4d & $\mathbb{C}\times\mathcal{M}_{3,2}$ & $U(1)^4 \times U(1)_R$
\\
8.4e & $NC4$ & $U(1)^4 \times U(1)_R$
\\
8.4f & $\mathcal{M}_{4,2}$ & $U(1)^4 \times U(1)_R$
\\
8.4g & $NC5$ & $U(1)^4 \times U(1)_R$
\\
8.4h & $NC3$ & $SU(2)^4 \times U(1)_R$
\\
\hline
\end{tabular}
\caption{\textit{Mesonic moduli spaces and global symmetries.} These are the  theories in the classification with their mesonic moduli spaces and global symmetries of total rank 5. \label{tglobal}}
\end{table} 

The mesonic moduli space $\mesonic$ of a brane tiling is the vacuum moduli space of the corresponding supersymmetric gauge theory under both F-and D-term constraints. The \textit{forward algorithm} \cite{Feng:2000mi,Feng:2001xr,Feng:2002zw,Feng:2002fv,Hanany:2005ve,Franco:2005rj,Franco:2006gc} has been used extensively in the case for brane tilings on $T^2$ to identify the mesonic moduli space of Abelian gauge theories with only $U(1)$ gauge groups. It is summarized in appendix \sref{app_mesonic}.

The forward algorithm can be used to identify $\mesonic$ for supersymmetric gauge theories represented by brane tilings on Riemann surfaces of arbitrary genus. The mesonic moduli spaces of the Abelian gauge theory is a $(2g+1)$-dimensional toric Calabi-Yau variety.

\comment{
The forward algorithm can be in fact used to identify the mesonic moduli spaces of brane tilings on $g=2$ Riemann surfaces. In general, without modifications, we expect it to be used to compute $\mesonic$ for supersymmetric gauge theories that are represented by brane tilings on Riemann surfaces with arbitrary genus. 

The mesonic moduli spaces have the following properties:
\begin{itemize}
\item The dimension of the mesonic moduli space is $2g+1$.
\item The mesonic moduli space is toric Calabi-Yau.
\end{itemize}
}

In order to compute the structure of the mesonic moduli space, we evaluate the Hilbert series of $\mesonic$. The Hilbert series is refined with fugacities which count charges under the global symmetries. The global symmetry group has total rank $2g+1$ and can have for the case of $g=2$ brane tilings $SU(2)$, $SU(3)$, $SU(4)$ and $SU(5)$ enhancements. \tref{tglobal} summarises the global symmetries which are observed in the classification.

In field theory, the superpotential is conventionally assigned R-charge $2$, when the supercharges have unit R-charge. For simplicity, we rescale the R-symmetry generator: quiver fields are assigned R-charges such that every perfect matching carries a R-charge of 1. This is a notational simplification in the following sections. For the actual R-charges the reader is reminded that the charges for perfect matchings should be rescaled such that the superpotential carries R-charge 2 rather than equal to the number of perfect matchings.

\begin{figure}[ht!!]
\begin{center}
\includegraphics[trim=0cm 0cm 0cm 0cm,totalheight=3.5 cm]{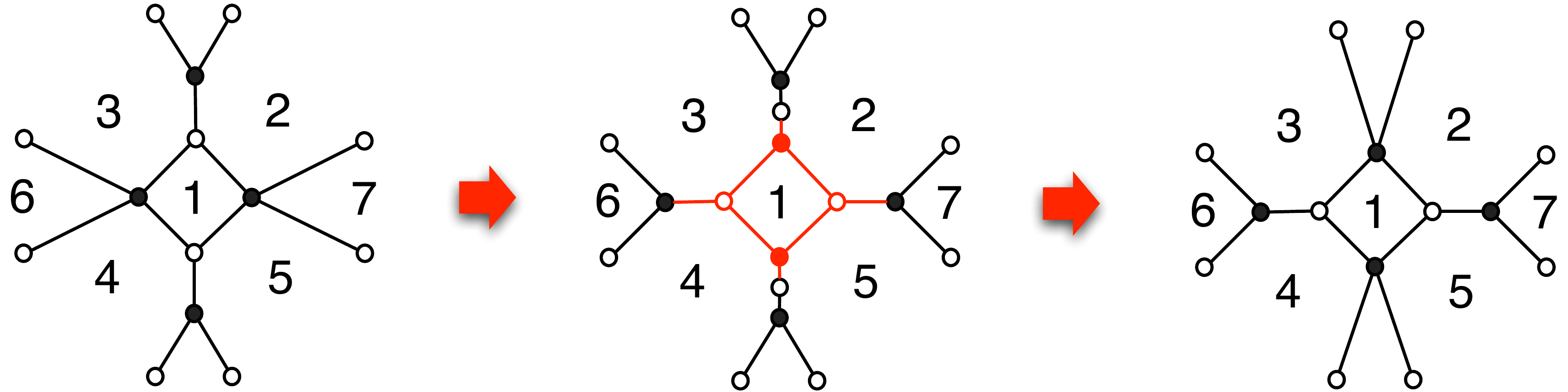}
\caption{
\textit{Urban renewal move of a brane tiling.} The first step shows the urban renewal move which creates bivalent nodes. These correspond to mass terms that are integrated out and removed in the second step.
\label{fseiberg}}
 \end{center}
 \end{figure}

\begin{table}
\begin{center}
\begin{tabular}{|c|c|}
\hline
$\mesonic$ & $\#E.T$\\
\hline \hline
$\mathbb{C}^5$ & 5.2, 6.2a \\
$NC1$ & 6.2b, 6.2c \\
$\mathbb{C}\times\mathcal{M}_{3,2}$ & 7.4, 8.4d \\
$NC3$ & 8.2b, 8.4h \\
$\mathbb{C}^3\times\mathbb{C}^2/\mathbb{Z}_2$ & 8.4b, 8.4c\\
\hline
\end{tabular}
\end{center}
\caption{Brane tilings on $g=2$ which share the same Abelian mesonic moduli space. $NC1$ is the first non-complete intersection mesonic moduli space in the classification. \label{fsamemesonic}}
\end{table}

By analysing the mesonic moduli spaces of the $g=2$ brane tilings in the classification shown in \fref{fg2summary}, we observe interesting new phenomena. In the case of torus brane tilings, the mesonic moduli spaces of two brane tilings are the same if the brane tilings are related by an urban renewal move as depicted in \fref{fseiberg}. Such a move seems to be still a sufficient condition for moduli space equivalence for brane tilings on higher genus Riemann surfaces. However, we observe examples of $g=2$ brane tilings which are not related by urban renewal, but have the same mesonic moduli space. The examples identified in the classification are shown in \tref{fsamemesonic}. 

The above classification of the mesonic moduli spaces are based on the fact that we restrict to Abelian theories with only $U(1)$ gauge groups. Whether as in the case of toric duality the supersymmetric theories share the same mesonic moduli spaces in the non-Abelian extension is unclear. It is of great interest to study this problem in future work. 
\\

\subsection{Higgsing $g=2$ Brane Tilings \label{s_higgs}}

\begin{figure}[ht!!]
\begin{center}
\includegraphics[trim=0cm 0cm 0cm 0cm,totalheight=3.5 cm]{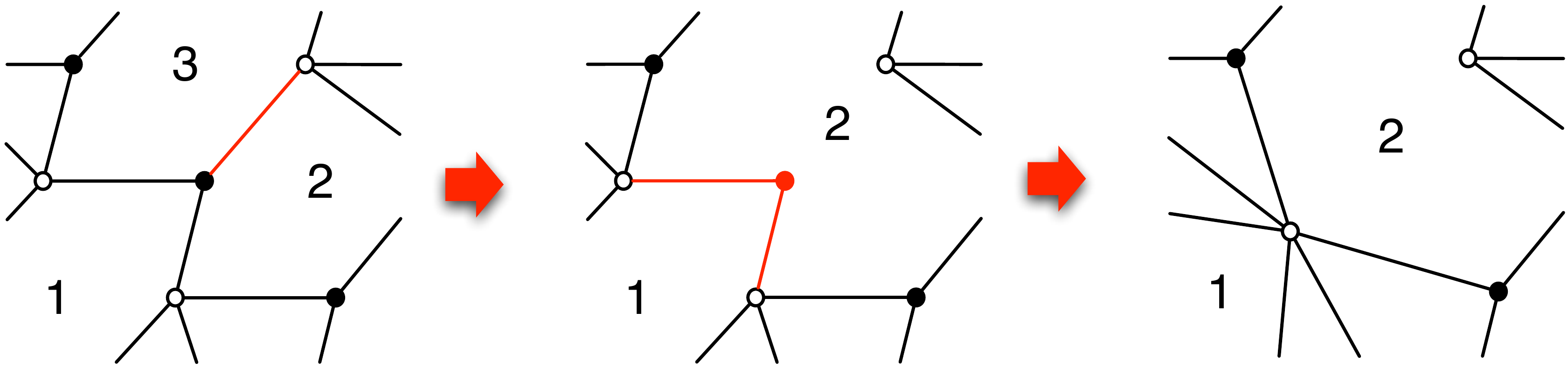}
\caption{
\textit{Higgsing in a brane tiling.} The first step shows the removal of the edge which corresponds to the bifundamental field which is assigned a VEV. The Higgsing results in a bivalent node which corresponds to a mass term. This is integrated out in the second step.
\label{fhiggs}}
 \end{center}
 \end{figure}

Section \sref{s_class} explained the procedure which is followed in this work to identify $g=2$ brane tilings with up to $E=8$ fields and $V=4$ superpotential terms. We expect \textit{Higgsing} \cite{Hanany:2005ve,Feng:2002fv,Hanany:2012hi} to be an exploratory way to relate the discovered brane tilings and at the same time to check the classification for consistency. Higgsing is the procedure of giving VEVs to bifundamental fields in order to solve D-term equations in the presence of FI parameters, and to integrate out mass terms in the resulting superpotential of the theory. It translates to removing edges in the brane tiling and reducing the graph such that there are no bivalent nodes. The procedure is illustrated in \fref{fhiggs}.

\begin{figure}[H]
\begin{center}
\includegraphics[trim=0cm 0cm 0cm 0cm,totalheight=19 cm]{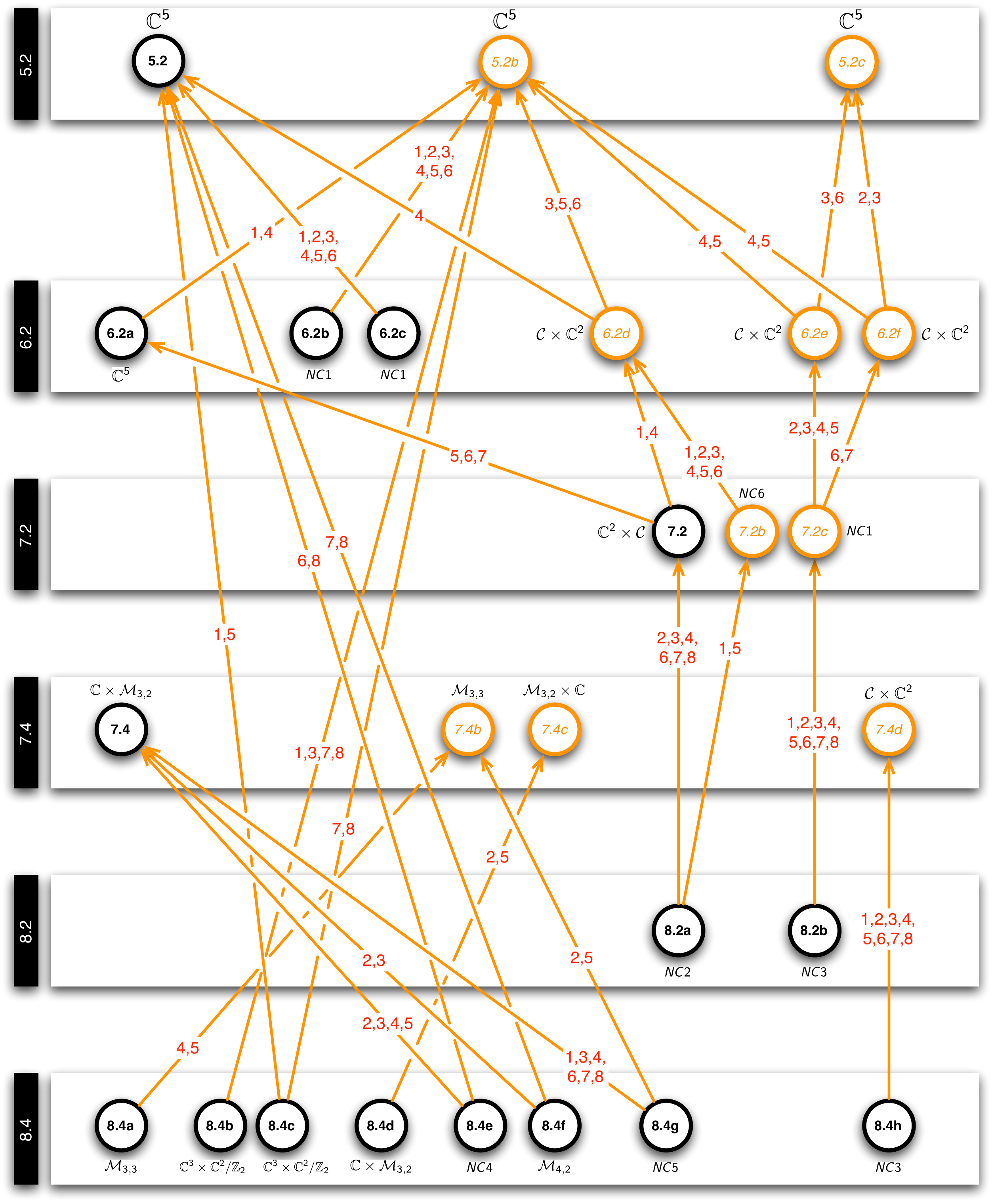}
\caption{\textit{Higgsing tree for $g=2$ brane tilings with up to $8$ quiver fields.} The models labeled with italics correspond to unrestricted brane tilings with self-intersecting zig-zag paths. The arrows correspond to a single field Higgsing, with the field numbers given on the arrows (see \sref{app_con} and \sref{app_incon} for field labels).
\label{fhiggsum}}
 \end{center}
 \end{figure}

Higgsing relates the $16$ restricted brane tilings in \fref{fg2summary} with each other. Intriguingly, Higgsing also relates restricted brane tilings with unrestricted ones which are not part of our classification. In fact, via Higgsing one identifies 10 unrestricted brane tilings with self-intersecting zig-zag paths which are summarized with the corresponding superpotentials and quiver diagrams in appendix \sref{app_incon}. A `Higgsing tree', which illustrates brane tilings as nodes and VEVs as arrows, is shown in \fref{fhiggsum}. It is of great interest to understand the mechanism that relates restricted and unrestricted brane tilings in future studies.
\\

\section{A Classification of $g=2$ Brane Tilings \label{s_classification}}

This section summarises the classification of $g=2$ brane tilings with up to $E=8$ fields and $V=4$ superpotential terms. The mesonic moduli spaces are studied by computing the Hilbert series of the corresponding algebraic variety. We discover several interesting geometries which are related to the new brane tilings.
\\

\subsection{5 Fields, 2 Superpotential Terms, 1 Gauge Group \label{s_m1}}

\subsubsection{Model 5.2: $\mathbb{C}^5$ \label{s_m52a}}

\begin{figure}[ht!!]
\begin{center}
\includegraphics[trim=0cm 0cm 0cm 0cm,totalheight=8 cm]{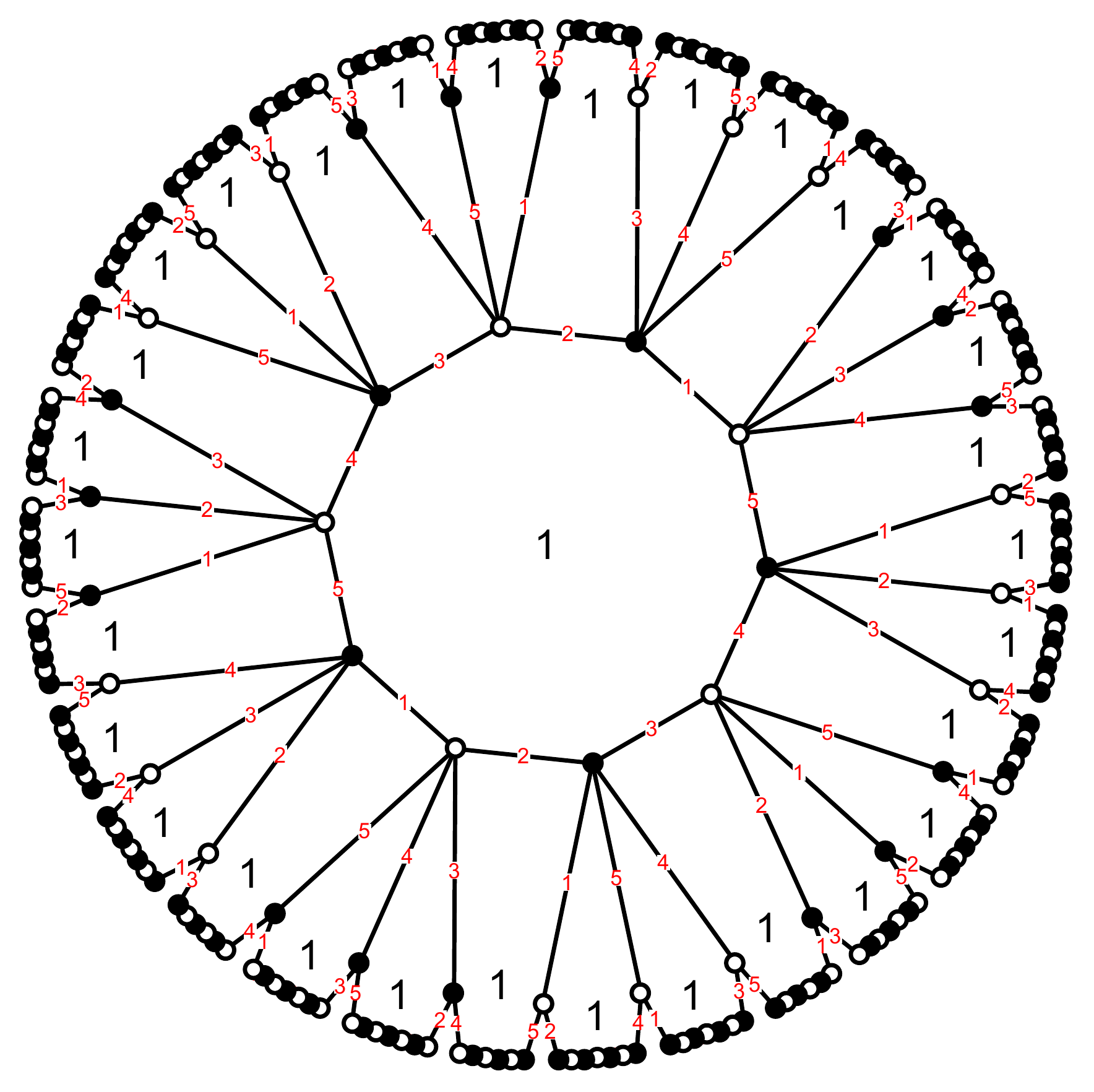}\\
\vspace{0.3cm}
\begin{tabular}{ccccc}
{\color[rgb]{1.000000,0.000000,0.000000} 1} &
{\color[rgb]{1.000000,0.000000,0.000000} 2} &
{\color[rgb]{1.000000,0.000000,0.000000} 3} &
{\color[rgb]{1.000000,0.000000,0.000000} 4} &
{\color[rgb]{1.000000,0.000000,0.000000} 5} 
\\
$X_{11}^{1}$ & $X_{11}^{2}$ & $X_{11}^{3}$ & $X_{11}^{4}$ & $X_{11}^{5}$
\end{tabular}
\caption{The Model 5.2 brane tiling on a $g=2$ Riemann surface with 5 fields and 2 superpotential terms.
\label{fm1n1t}}
 \end{center}
 \end{figure}

\begin{figure}[ht!!]
\begin{center}
\includegraphics[trim=0cm 0cm 0cm 0cm,totalheight=1 cm]{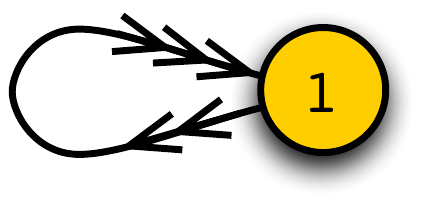}
\caption{The quiver diagram for Model 5.2, a brane tiling on a $g=2$ Riemann surface with 5 fields and 2 superpotential terms.
\label{fm1n1q}}
 \end{center}
 \end{figure}
 
The first $g=2$ brane tiling of our classification and the corresponding quiver diagram are shown in \fref{fm1n1t} and \fref{fm1n1q} respectively. The brane tiling is made of a single decagonal face which is the single gauge group with 5 adjoints in the quiver diagram. The superpotential is
\beal{e20i1}
W= 
+ X_{11}^{1} X_{11}^{2} X_{11}^{3} X_{11}^{4} X_{11}^{5}
- X_{11}^{5} X_{11}^{4} X_{11}^{3} X_{11}^{2} X_{11}^{1}
~~.
\eea

A single adjoint on its own forms a perfect matching of the brane tiling. Accordingly, the perfect matching matrix is the identity matrix
\beal{e20i2}
P=\left(
\begin{array}{c|ccccc}
\; & a_1 & a_2 & a_3 & a_4 & a_5\\
\hline
 X_{11}^{1} & 1 & 0 & 0 & 0 & 0 \\
 X_{11}^{2} & 0 & 1 & 0 & 0 & 0 \\
 X_{11}^{3} & 0 & 0 & 1 & 0 & 0 \\
 X_{11}^{4} & 0 & 0 & 0 & 1 & 0 \\
 X_{11}^{5} & 0 & 0 & 0 & 0 & 1
\end{array}
\right)~~.
\eea
The perfect matching matrix is always the identity matrix for models with just 2 superpotential terms. The zig-zag paths of the brane tiling are 
\beal{e20i2bb}
&\eta_1 = ( X_{11}^{1}, X_{11}^{2}) ~,~ 
\eta_2 = ( X_{11}^{2},  X_{11}^{3}) ~,~
\eta_3 = ( X_{11}^{3}, X_{11}^{4})~,~&
\nn\\
&
\eta_4 = ( X_{11}^{4}, X_{11}^{5})~,~
\eta_5 = ( X_{11}^{5},  X_{11}^{1})~.&
\eea

There are only trivial F- and D-terms. The mesonic moduli space is a toric Calabi-Yau 5-fold. More specifically, Model 5.2's mesonic moduli space is $\mathbb{C}^5$ with the refined Hilbert series being
\beal{e20i2}
g_1(\alpha_i;\mesonic) &=& 
\frac{1}{
\prod_{i=1}^{5}(1-\alpha_i)
}
~~,
\eea
where the fugacities $\alpha_i$ count the perfect matchings $a_i$ respectively. 

Given that the mesonic moduli space is $\mathbb{C}^5$, the global symmetry group is found as $SU(5)\times U(1)_R$, where the $U(1)_R$ is the R-symmetry. 
The global symmetry charges assigned to perfect matchings are shown below.

\begin{center}
\begin{tabular}{|c|cc|l|}
\hline
\; & $SU(5)_{x_i}$ & $U(1)_R$ & fugacity\\
\hline
\hline
$a_1$ & $(1,0,0,0)$ & 1 & $\alpha_1=x_1 t$ \\
$a_2$ & $(-1,1,0,0)$ & 1 & $\alpha_2=x_1^{-1} x_2 t$ \\
$a_3$ & $(0,-1,1,0)$ & 1 & $\alpha_3=x_2^{-1} x_3 t$ \\
$a_4$ & $(0,0,-1,1)$ & 1 & $\alpha_4=x_3^{-1} x_4 t$ \\
$a_5$ & $(0,0,0,-1)$ & 1 & $\alpha_5=x_4^{-1} t$ \\
\hline
\end{tabular}
\end{center}

Under the above global symmetry charge assignment, the Hilbert series can be expressed in terms of characters of irreducible representations of $SU(5)$,
\beal{e20i3}
g_1(x_i, t;\mesonic) = \sum_{n=0}^{\infty} [n,0,0,0]_{SU(5)} t^n~~.
\eea

The toric diagram of the mesonic moduli space is a 4 dimensional lattice polytope. The coordinates of the toric points are encoded in the matrix
\beal{e9}
G_t=
\left(
\begin{array}{ccccc}
a_1 & a_2 & a_3 & a_4 & a_5\\
\hline
 1 & 0 & 0 & 0 & 0 \\
 0 & 1 & 0 & 0 & 0 \\
 0 & 0 & 1 & 0 & 0 \\
 0 & 0 & 0 & 1 & 0 \\
 0 & 0 & 0 & 0 & 1 \\
\end{array}
\right)
~~.
\eea
The projected toric diagram is a unit lattice 4-simplex.
\\

\subsection{6 Fields, 2 Superpotential Terms, 2 Gauge Groups \label{s_m2}}

\subsubsection{Model 6.2a: $\mathbb{C}^5$ \label{s_m62a}}

\begin{figure}[ht!!]
\begin{center}
\includegraphics[trim=0cm 0cm 0cm 0cm,totalheight=8 cm]{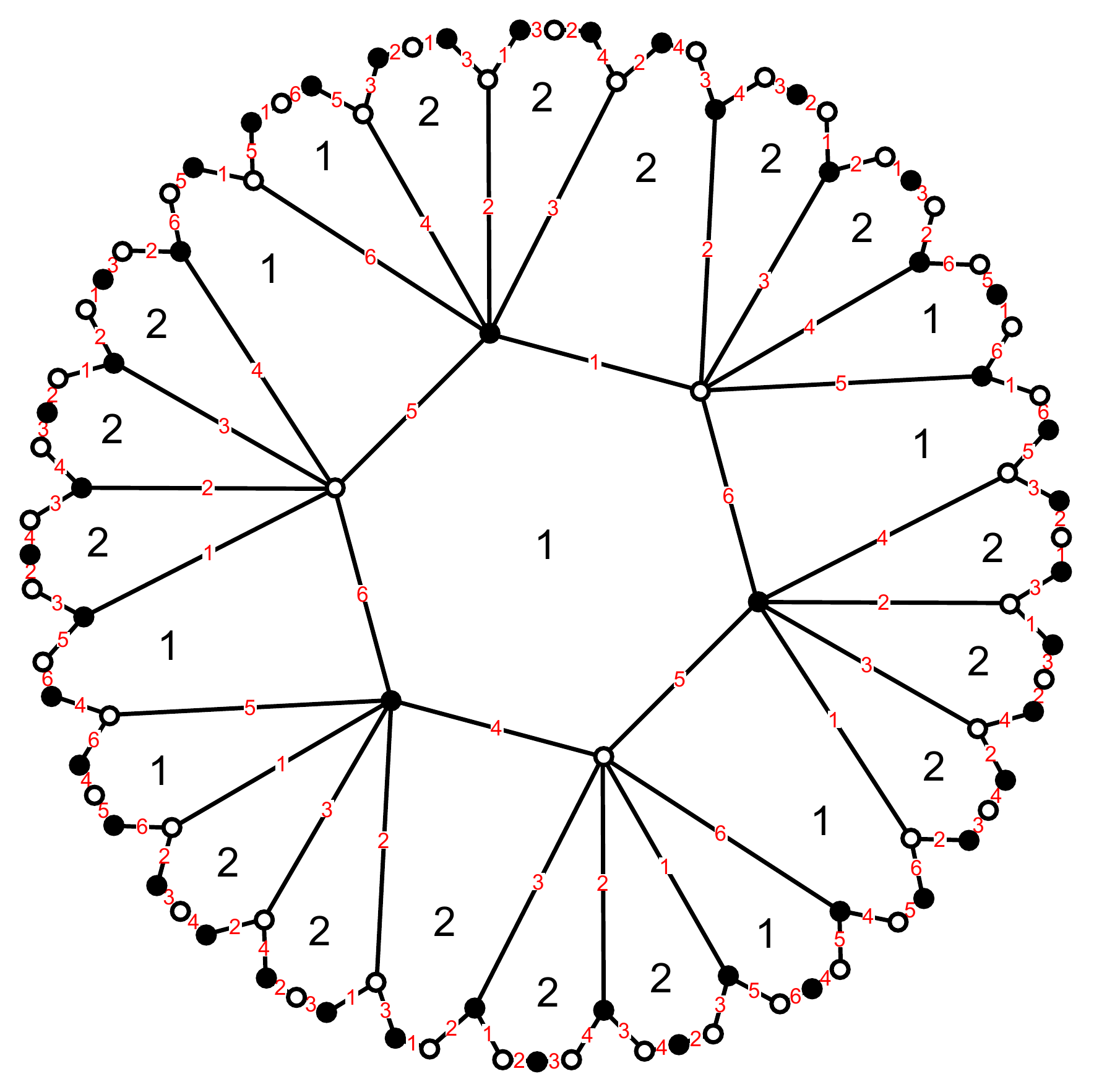}\\
\vspace{0.3cm}
\begin{tabular}{cccccc}
{\color[rgb]{1.000000,0.000000,0.000000} 1} &
{\color[rgb]{1.000000,0.000000,0.000000} 2} &
{\color[rgb]{1.000000,0.000000,0.000000} 3} &
{\color[rgb]{1.000000,0.000000,0.000000} 4} &
{\color[rgb]{1.000000,0.000000,0.000000} 5} &
{\color[rgb]{1.000000,0.000000,0.000000} 6} 
\\
 $X_{12}$ & $X_{22}^1$ & $X_{22}^2$ & $X_{21}$ & $X_{11}^1$ & $X_{11}^2$ 
\end{tabular}
\caption{The Model 6.2a brane tiling on a $g=2$ Riemann surface with 6 fields and 2 superpotential terms.
\label{fm2n1t}}
 \end{center}
 \end{figure}

\begin{figure}[ht!!]
\begin{center}
\includegraphics[trim=0cm 0cm 0cm 0cm,totalheight=1 cm]{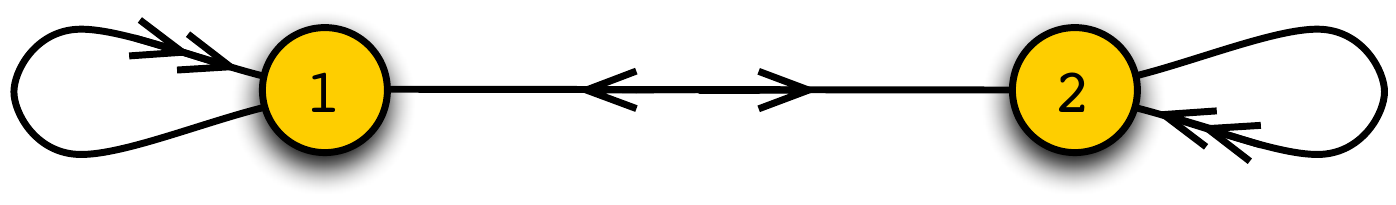}
\caption{The quiver diagram for Model 6.2a, a brane tiling on a $g=2$ Riemann surface with 6 fields and 2 superpotential terms.
\label{fm2n1q}}
 \end{center}
 \end{figure}
 
The brane tiling on a $g=2$ Riemann surface and the corresponding quiver diagram are shown in \fref{fm2n1t} and \fref{fm2n1q} respectively. The superpotential is
\beal{e1}
W= 
+ X_{12} X_{22}^{1} X_{22}^{2} X_{21} X_{11}^1 X_{11}^2 
- X_{12} X_{22}^{2} X_{22}^{1} X_{21} X_{11}^2 X_{11}^1
~~.
\eea
The quiver incidence matrix is
\beal{e2}
d=\left(
\begin{array}{cccccc}
 X_{11}^1 & X_{11}^2 & X_{22}^1 & X_{22}^2 & X_{12} & X_{21}  \\
 \hline
 0 & 0 & 0 & 0 & 1 &-1 \\
 0 & 0 & 0 & 0 &-1 & 1
\end{array}
\right)
~~.
\eea

The brane tiling has $6$ perfect matchings. Since there are only 2 superpotential terms, every field on its own represents a perfect matching. The perfect matching matrix is therefore the identity matrix,
\beal{e3}
P=\left(
\begin{array}{c|cccccc}
\; & a_1 & a_2 & a_3 & a_4 & p_1 & p_2 \\
\hline
X_{11}^1 & 1 & 0 & 0 & 0 & 0 & 0 \\
X_{11}^2 & 0 & 1 & 0 & 0 & 0 & 0 \\
X_{22}^1 & 0 & 0 & 1 & 0 & 0 & 0 \\
X_{22}^2 & 0 & 0 & 0 & 1 & 0 & 0 \\
X_{12} & 0 & 0 & 0 & 0 & 1 & 0 \\
X_{21} & 0 & 0 & 0 & 0 & 0 & 1
\end{array}
\right)
\eea
The zig-zag paths in the brane tiling of Model 6.2a are 
\beal{e3bbbb}
&
\eta_1 = (X_{11}^1,X_{11}^2) ~,~
\eta_2 = (X_{22}^1,X_{22}^2) ~,~
&
\nn\\
&
\eta_3 = (X_{12},X_{22}^2,X_{21} ,X_{11}^1)~,~
\eta_4 = (X_{12},X_{22}^2,X_{21},X_{11}^2) ~.
&
\eea

The superpotential for a theory with only $U(1)$ gauge groups vanishes $W=0$, and therefore the kernel of the perfect matching matrix is empty. There are no F-terms, and there are no F-term charges
\beal{e4}
Q_F =0 
~~.
\eea
The D-term charges are encoded in the quiver incidence matrix $d$ and are summarized in the following charge matrix,
\beal{e5}
Q_D = 
\left(
\begin{array}{cccccc}
a_1 & a_2 & a_3 & a_4 & p_1 & p_2\\
\hline
 0 & 0 & 0 & 0 & 1 & -1
\end{array}
\right)~~.
\eea
\comment{
\beal{e5}
Q_D = 
\left(
\begin{array}{cccccc}
b_1 & a_1 & a_2 & b_2 & a_3 & a_4 \\
p_1 & p_2 & p_3 & p_4 & p_5 & p_6 \\
\hline
 1 & 0 & 0 & -1 & 0 & 0 
\end{array}
\right)~~.
\eea
}
Accordingly, the total charge matrix $Q_t=Q_F$, and the mesonic moduli space is given by the symplectic quotient of the form
\beal{es6}
\mesonic = \mathbb{C}^6 // Q_t ~~.
\eea

By associating the fugacities $\alpha_i,t_i$ to the perfect matchings $a_i,p_i$ respectively, the fully refined Hilbert series of $\mesonic$ is given by the following Molien integral
\beal{e7}
g_1(\alpha_i,\beta_i;\mesonic) &=& 
\frac{1}{(2\pi i)}
\oint_{|z_1|=1} 
\frac{\ud z_1}{z_1} ~~
\frac{1}{
\prod_{i=1}^{4}(1-\alpha_i)
}
\times
\frac{1}{
(1-z_1 t_1)
(1-z_1^{-1} t_2)
}
\nn\\
&=&
\frac{1}{
\prod_{i=1}^{4}(1-\alpha_i)
}
\times
\frac{1}{
(1-t_1 t_2)
}
~~.
\eea
Accordingly, the mesonic moduli space is a freely generated space, $\mesonic=\mathbb{C}^5$.

The $Q_D$ charge matrix in \eref{e5} indicates a symmetry of $SU(4)\times U(1)\times U(1)_R$.

\begin{center}
\begin{tabular}{|c|ccc|l|}
\hline
\; & $SU(4)_{x_i}$ & $U(1)_b$ & $U(1)_R$ & fugacity\\
\hline
\hline
$a_1$ & $(1,0,0)$ & 0 & 1 &  $\alpha_1=x_1 t$\\
$a_2$ & $(-1,1,0)$ & 0 & 1 & $\alpha_2=x_1^{-1} x_2 t$\\
$a_3$ & $(0,-1,1)$ & 0 & 1 & $\alpha_3=x_2^{-1}x_3 t$\\
$a_4$ & $(0,0,-1)$ & 0 & 1 & $\alpha_4=x_3^{-1} t$\\
\hline
$p_1$ & $(0,0,0)$ &  1 & 1 & $t_1=b t$\\
$p_2$ & $(0,0,0)$ & -1 & 1 & $t_2=b^{-1}t$ \\
\hline
\end{tabular}
\end{center}

\noindent Under the above charge assignment, the Hilbert series of $\mesonic$ can be expressed as
\beal{e8}
g_1(x_i,t;\mesonic) &=& 
\frac{1}{1-t^2}\sum_{n=0}^{\infty} [n,0,0]_{SU(4)}t^n~~.
\eea

Since the moduli space space is $\mathbb{C}^5$, we expect a $SU(5)$ symmetry. The fully enhanced global symmetry is therefore $SU(5)\times U(1)_R$. This can be observed by modifying the global charges on the perfect matchings $p_1$ and $p_2$. A possible choice can be:

\begin{center}
\begin{tabular}{|c|ccc|l|}
\hline
\; & $SU(5)_{x_i}$  & $U(1)_R$ & fugacity\\
\hline
\hline
$a_1$ & $(1,0,0,0)$  & 1 &  $\alpha_1=x_1 t$\\
$a_2$ & $(-1,1,0,0)$  & 1 & $\alpha_2=x_1^{-1} x_2 t$\\
$a_3$ & $(0,-1,1,0)$  & 1 & $\alpha_3=x_2^{-1}x_3 t$\\
$a_4$ & $(0,0,-1,1)$  & 1 & $\alpha_4=x_4 x_3^{-1} t$\\
\hline
$p_1$ & $(0,0,0,-1/2)$  & 1/2 & $t_1=x_4^{-1/2} t^{-1/2}$\\
$p_2$ & $(0,0,0,-1/2)$  & 1/2 & $t_2=x_4^{-1/2} t^{-1/2}$ \\
\hline
\end{tabular}
\end{center}

Under the above charge assignment, the mesonic Hilbert series can be expressed as expected in terms of characters of $SU(5)$ irreducible representations,
\beal{e8}
g_1(x_i,t;\mesonic) &=& 
\sum_{n=0}^{\infty} [n,0,0,0]_{SU(5)}t^n~~.
\eea

The toric diagram of the mesonic moduli space is a 4 dimensional lattice polytope. The coordinates of the toric points are encoded in the matrix
\beal{e9}
G_t=
\left(
\begin{array}{cccccc}
a_1 & a_2 & a_3 & a_4 & p_1 & p_2 \\
\hline
 1 & 0 & 0 & 0 & 0 & 0 \\
 0 & 1 & 0 & 0 & 0 & 0 \\
 0 & 0 & 1 & 0 & 0 & 0 \\
 0 & 0 & 0 & 1 & 0 & 0 \\
 0 & 0 & 0 & 0 & 1 & 1
\end{array}
\right)
~~.
\eea
\comment{
\beal{e9}
G_t=
\left(
\begin{array}{cccccc}
p_1 & p_2 & p_3 & p_4 & p_5 & p_6 \\
\hline
 0 & 0 & 0 & 0 & 0 & 1 \\
 0 & 0 & 0 & 0 & 1 & 0 \\
 1 & 0 & 0 & 1 & 0 & 0 \\
 0 & 0 & 1 & 0 & 0 & 0 \\
0 & 1 & 0 & 0 & 0 & 0
\end{array}
\right)
~~.
\eea
}
Recall that perfect matchings correspond to toric points. We observe that the perfect matchings $p_1$ and $p_2$ correspond to the same toric point.
\\

\subsubsection{Model 6.2b: $NC1$ \label{s_m62b}} 

\begin{figure}[ht!!]
\begin{center}
\includegraphics[trim=0cm 0cm 0cm 0cm,totalheight=8 cm]{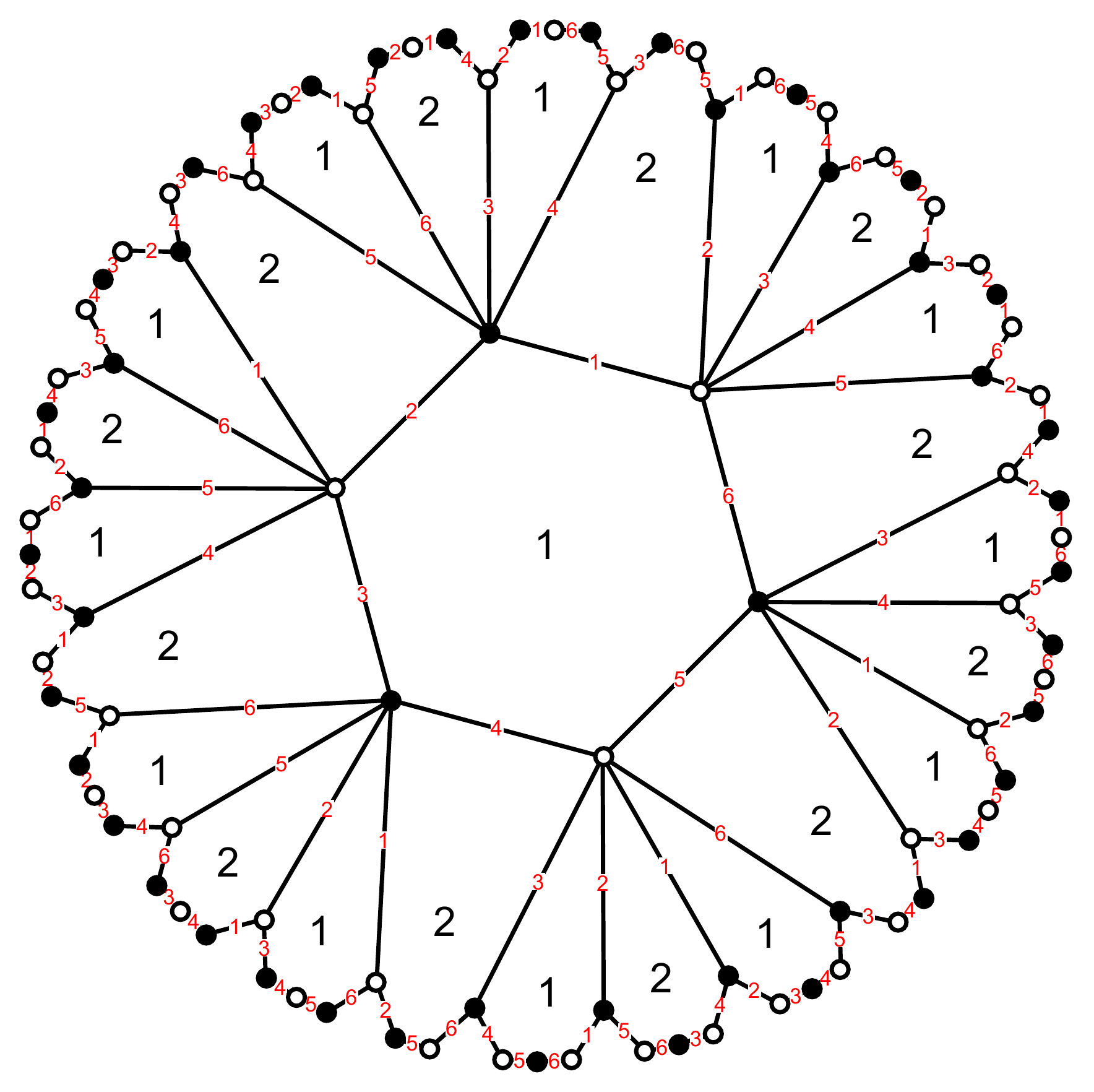}\\
\vspace{0.3cm}
\begin{tabular}{cccccc}
{\color[rgb]{1.000000,0.000000,0.000000} 1} &
{\color[rgb]{1.000000,0.000000,0.000000} 2} &
{\color[rgb]{1.000000,0.000000,0.000000} 3} &
{\color[rgb]{1.000000,0.000000,0.000000} 4} &
{\color[rgb]{1.000000,0.000000,0.000000} 5} &
{\color[rgb]{1.000000,0.000000,0.000000} 6} 
\\
$X_{12}^1$ & $X_{21}^{1}$ & $X_{12}^{2}$ & $X_{21}^2$ &
$X_{12}^3$ & $X_{21}^3$
\end{tabular}
\caption{The Model 6.2b brane tiling on a $g=2$ Riemann surface
with 6 fields and 2 superpotential terms.
\label{fm2n2t}}
 \end{center}
 \end{figure}

\begin{figure}[ht!!]
\begin{center}
\includegraphics[trim=0cm 0cm 0cm 0cm,totalheight=1
cm]{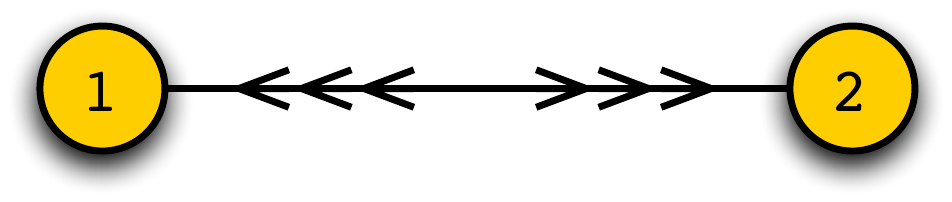}
\caption{The quiver diagram for Model 6.2b, a brane tiling on a
$g=2$ Riemann surface with 6 fields and 2 superpotential terms.
\label{fm2n2q}}
 \end{center}
 \end{figure}

The second brane tiling on a $g=2$ Riemann surface with 2
superpotential terms with 6 fields is shown with the
corresponding quiver diagram in \fref{fm2n2t} and \fref{fm2n2q}
respectively. The superpotential is
\beal{e21}
W= 
+ X_{12}^1 X_{21}^{1} X_{12}^{2} X_{21}^2 X_{12}^3 X_{21}^3 
- X_{12}^1 X_{21}^{2} X_{12}^{2} X_{21}^3 X_{12}^3 X_{21}^1 
~~.
\eea
The quiver incidence matrix is
\beal{e22}
d=\left(
\begin{array}{cccccc}
X_{12}^1 & X_{12}^{2} & X_{12}^{3} & X_{21}^1 & X_{21}^2 &
X_{21}^3 \\
 \hline
 1 & 1 & 1 & -1 & -1 & -1 \\
 -1 & -1 & -1 & 1 & 1 & 1
\end{array}
\right)
~~.
\eea

The brane tiling has $c=6$ perfect matchings, each of them given
by a bifundamental field. The perfect matching matrix is the
identity matrix,
\beal{e23}
P=\left(
\begin{array}{c|cccccc}
\; & a_1 & a_2 & a_3 & b_1 & b_2 & b_3 \\
\hline
X_{12}^1& 1 & 0 & 0 & 0 & 0 & 0 \\
X_{12}^2& 0 & 1 & 0 & 0 & 0 & 0 \\
X_{12}^3& 0 & 0 & 1 & 0 & 0 & 0 \\
X_{21}^1& 0 & 0 & 0 & 1 & 0 & 0 \\
X_{21}^2& 0 & 0 & 0 & 0 & 1 & 0 \\
X_{21}^3& 0 & 0 & 0 & 0 & 0 & 1
\end{array}
\right)
\eea
The zig-zag paths of the brane tiling are
\beal{e23bb}
&\eta_1 = (X_{12}^1,X_{21}^1) ~,~
\eta_2 = (X_{12}^2,X_{21}^2) ~,~
\eta_3 = (X_{12}^3,X_{21}^3) ~,~
&\nn\\
&
\eta_4 = (X_{12}^1,X_{21}^2,X_{12}^3,X_{21}^1,X_{12}^2,X_{21}^3)~.
&
\eea

The Abelian superpotential vanishes $W=0$, and the kernel of the
perfect matching matrix is empty. There are no F-terms,
therefore no F-term charges.
The D-term charges are encoded in the quiver incidence matrix
$d$:
\beal{e25}
Q_D = 
\left(
\begin{array}{cccccc}
a_1 & a_2 & a_3 & b_1 & b_2 & b_3 \\
\hline
 1 & 1 & 1 & -1 & -1 & -1 
\end{array}
\right)~~.
\eea
The total charge matrix $Q_t=Q_D$, and the mesonic moduli space
is the symplectic quotient
\beal{es26}
\mesonic = \mathbb{C}^6 // Q_t ~~.
\eea

By associating the fugacities $\alpha_i$ and $\beta_j$ to the
perfect matchings $a_i$ and $b_j$ respectively, the fully
refined Hilbert series of $\mesonic$ is given by the Molien
integral
\beal{e27}
g_1(\alpha_i,\beta_i;\mesonic) =
\oint_{|z|=1} 
\frac{\ud z}{2\pi i z} ~~
\frac{1}{
\prod\limits_{i=1}^3 (1-z \alpha_i)
(1-z^{-1} \beta_i)}
=
\frac{(\prod\limits_{i=1}^3 \alpha_i \beta_i)\,P(\alpha_i, \beta_i)}{\prod \limits_{i,j=1}^3 (1- \alpha_i \beta_j)} ~~,\nn\\\eea
where 
\be\label{es27b}
P(\alpha_i,\beta_i)=  
\prod\limits_{i=1}^3 \alpha_i^{-1} \beta_i^{-1} - \sum_{i,j=1}^3  \alpha_i^{-1}\beta_j^{-1}+\sum_{i,j=1}^3 (\alpha_i \alpha_j^{-1}+ \beta_i  \beta_j^{-1})-2 - \sum_{i,j=1}^3 \alpha_i \beta_j +  \prod\limits_{i=1}^3 \alpha_i \beta_i  \;.
\ee
Accordingly, the mesonic moduli space is a non-complete
intersection of dimension 5. By setting the fugacities $\alpha_i=\beta_i=t$,
the unrefined Hilbert series is
\beal{es27bb}
g_1(t;\mesonic) =
\frac{1 + 4 t^2 + t^4}{(1 - t^2)^5}~~.
\eea
The palindromic numerator of the Hilbert series indicates that
$\mesonic$ is a Calabi-Yau 5-fold. The plethystic logarithm of
the refined Hilbert series of $\mesonic$ is of the form
\beal{e27c}
PL[g_1(\alpha_i,\beta_i;\mesonic)] &=& \sum_{i,j=1}^3 \alpha_i\beta_j
-\sum_{i_1\neq i_2,j_1\neq j_2}^3 \alpha_{i_1}\beta_{j_1}\alpha_{i_2}\beta_{j_2}
+ \dots~~.
\eea
The generators of the mesonic moduli space in terms of perfect
matching variables are
\begin{center}
\begin{tabular}{|c|c|}
\hline
generator & perfect matchings\\
\hline\hline
$A_{ij}$ & $a_i b_j$\\
\hline
\end{tabular}
\end{center}
which are subject to the first order relations 
\be\label{e27d}
\epsilon^{i_1 i_2 i_3} \epsilon^{j_1 j_2 j_3} A_{i_2 j_2} A_{i_3 j_3} = 0\;.
\ee

One can assign the following enhanced $SU(3) \times SU(3) \times
U(1)_R$ global charges to the perfect matching variables
\begin{center}
\begin{tabular}{|c|ccc|l|}
\hline
\; & $SU(3)_x$ & $SU(3)_{y}$ & $U(1)_R$ & fugacity\\
\hline
\hline
$a_1$ & $(1,0)$ & 0 & 1 		& $\al_1=x_1 t$\\
$a_2$ & $(-1,1)$ & 0 & 1  		& $\al_2=x_1^{-1} x_2  t$\\
$a_3$ & $(0,-1)$ & 0 & 1 		& $\al_3=x_2^{-1} t$\\
\hline
$b_1$ & 0 & $(-1,0)$ & 1 		& $\beta_1=y_1^{-1}  t$\\
$b_2$ & 0 & $(1,-1)$ & 1  		& $\beta_2= y_1 y_2^{-1} t$ \\
$b_3$ & 0 & $(0, 1)$ & 1 		& $\beta_3= y_2 t$\\
\hline
\end{tabular}
\end{center}

\noindent Under the above charge assignment, the Hilbert series
of $\mesonic$ can be expressed as
\beal{e28}
g_1(x_i,y_i,t;\mesonic) &=& 
\sum_{n=0}^{\infty}
[n,0;0,n] t^{2n}~~,
\eea
where $[n,0;0,n]\equiv [n,0]_{SU(3)_x} [0,n]_{SU(3)_{y}}$. The
generators and the first order relations formed by them are
encoded in the plethystics logarithm, which now takes the form
\beal{e28b}
PL[g_1(x_i,y_i,t;\mesonic)]=
[1,0;0,1] t^2 - [0,1;1,0] t^4 + \dots ~~.
\eea

The toric diagram of the mesonic moduli space is a 4 dimensional
lattice polytope. The coordinates of the toric points are
encoded in the matrix 
\beal{e29}
G_t=
\left(
\begin{array}{cccccc}
p_1 & p_2 & p_3 & p_4 & p_5 & p_6 \\
\hline
 1 & 0 & 0 & 0 & 0 & 1 \\
 -1 & 0 & 0 & 0 & 1 & 0 \\
 1 & 0 & 0 & 1 & 0 & 0 \\
 -1 & 0 & 1 & 0 & 0 & 0 \\
 1 & 1 & 1 & 1 & 1 & 1
\end{array}
\right)
~~.
\eea
Note that the mesonic moduli space here is the same as the master space of $\mathbb{C}^3/\mathbb{Z}_3$ \cite{Butti:2007jv}.
\\

\subsubsection{Model 6.2c: $NC1$ \label{s_m62c}} 

\begin{figure}[ht!!]
\begin{center}
\includegraphics[trim=0cm 0cm 0cm 0cm,totalheight=8 cm]{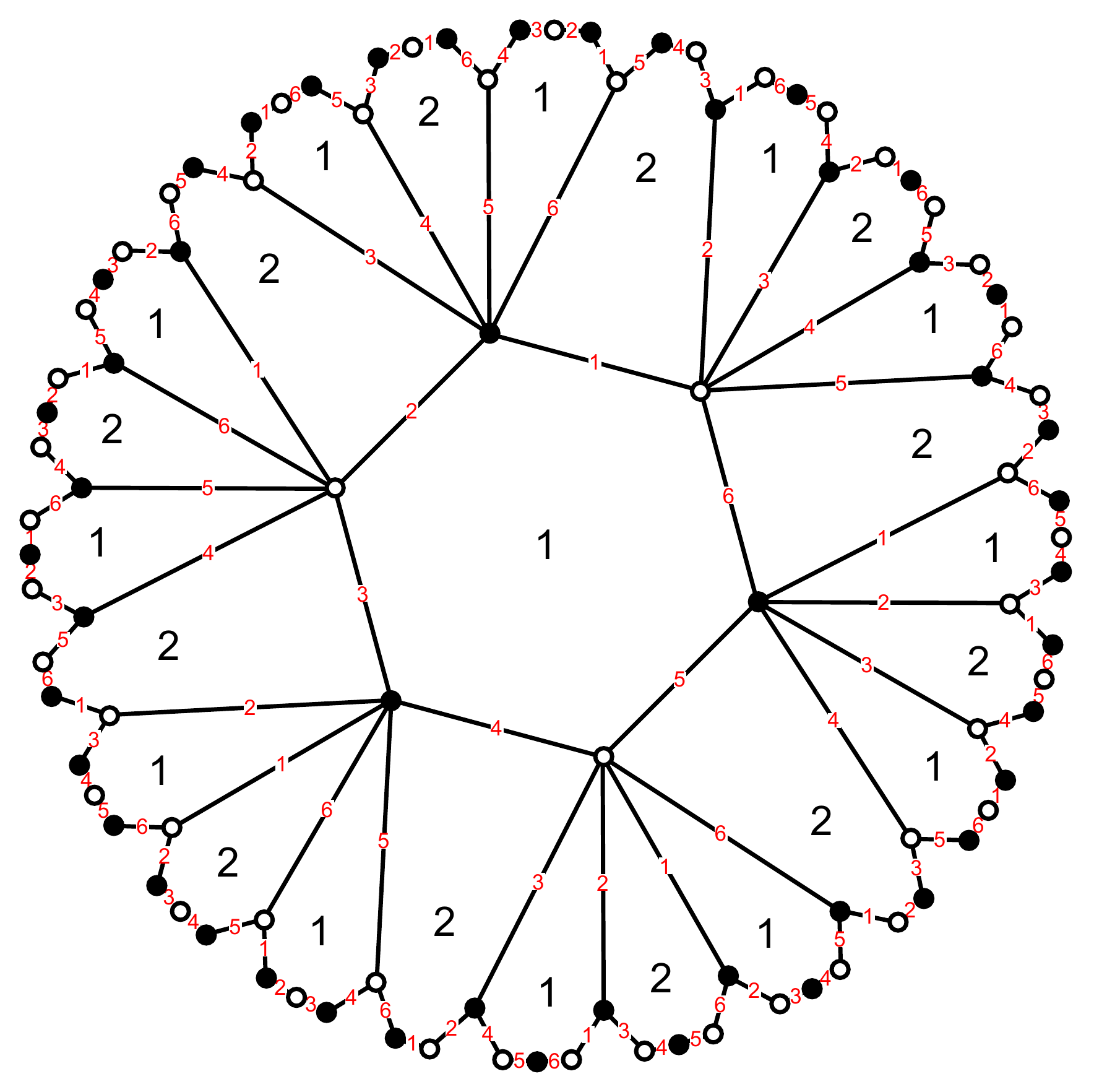}\\
\vspace{0.3cm}
\begin{tabular}{cccccc}
{\color[rgb]{1.000000,0.000000,0.000000} 1} &
{\color[rgb]{1.000000,0.000000,0.000000} 2} &
{\color[rgb]{1.000000,0.000000,0.000000} 3} &
{\color[rgb]{1.000000,0.000000,0.000000} 4} &
{\color[rgb]{1.000000,0.000000,0.000000} 5} &
{\color[rgb]{1.000000,0.000000,0.000000} 6} 
\\
$X_{12}^{1}$& $X_{21}^{1}$& $X_{12}^{2}$& $X_{21}^{2}$&
$X_{12}^{3}$& $X_{21}^{3}$
\end{tabular}
\caption{The Model 6.2c brane tiling on a $g=2$ Riemann surface
with 6 fields and 2 superpotential terms.
\label{fm2n3t}}
 \end{center}
 \end{figure}

\begin{figure}[ht!!]
\begin{center}
\includegraphics[trim=0cm 0cm 0cm 0cm,totalheight=1
cm]{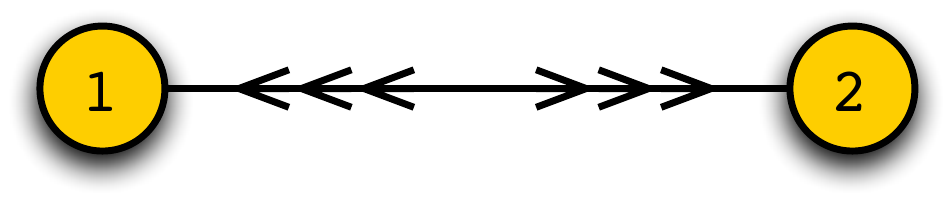}
\caption{The quiver diagram for Model 6.2c, a brane tiling on a
$g=2$ Riemann surface with 6 fields and 2 superpotential terms.
\label{fm2n3q}}
 \end{center}
 \end{figure}

The brane tiling and quiver for Model 6.2c are shown in
\fref{fm2n3t} and \fref{fm2n3q} respectively. The superpotential
is
\beal{e11}
W= 
+ X_{12}^{1} X_{21}^{1} X_{12}^{2} X_{21}^{2} X_{12}^{3}
X_{21}^{3}
- X_{21}^{3} X_{12}^{3} X_{21}^{2} X_{12}^{2}
X_{21}^{1} X_{12}^{1} 
~~.
\eea
In the Abelian gauge theory the superpotential vanishes, giving the same model as in the previous section. (The non-Abelian gauge theories differ by superpotential interactions.)
There is a difference in the zig-zag paths, which now are  
\beal{e13bb}
&
\eta_1 = (X_{21}^{3}, X_{12}^{3}) ~,~
\eta_2 = (X_{12}^{3}, X_{21}^{2}) ~,~
\eta_3 = (X_{21}^{2}, X_{12}^{2}) ~,~
&\nn\\
&
\eta_4 = (X_{12}^{2}, X_{21}^{1}) ~,~
\eta_5 = (X_{21}^{1}, X_{12}^{3}) ~,~
\eta_6 = (X_{12}^{3}, X_{21}^{3}) ~.~
&
\eea
\\

\subsection{7 Fields, 2 Superpotential Terms, 3 Gauge Groups \label{s_m3}}

\subsubsection{Model 7.2: $\mathbb{C}^2\times \mathcal{C}$ \label{s_m72a}}

\begin{figure}[ht!!]
\begin{center}
\includegraphics[trim=0cm 0cm 0cm 0cm,totalheight=8 cm]{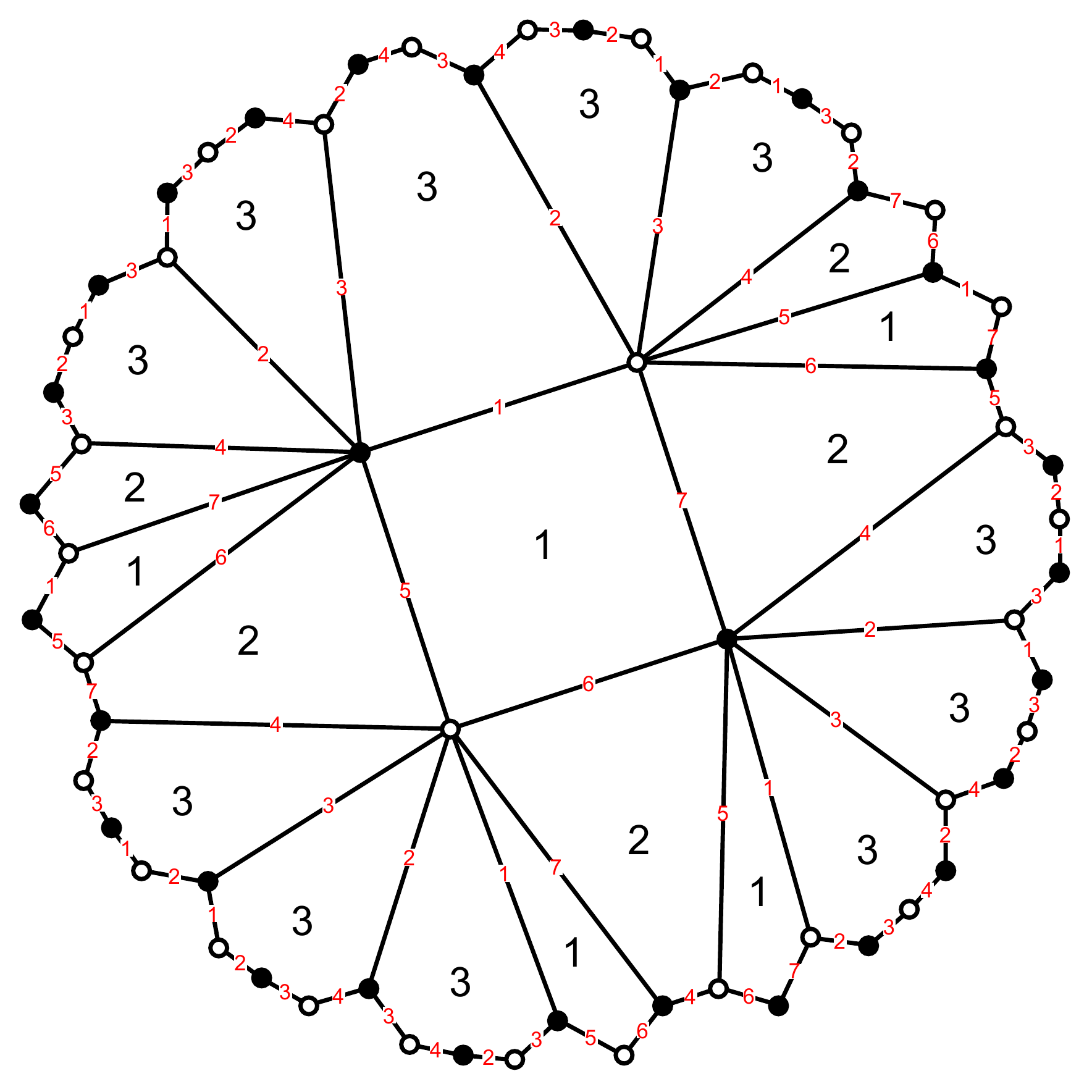}\\
\vspace{0.3cm}
\begin{tabular}{ccccccc}
{\color[rgb]{1.000000,0.000000,0.000000} 1} &
{\color[rgb]{1.000000,0.000000,0.000000} 2} &
{\color[rgb]{1.000000,0.000000,0.000000} 3} &
{\color[rgb]{1.000000,0.000000,0.000000} 4} &
{\color[rgb]{1.000000,0.000000,0.000000} 5} &
{\color[rgb]{1.000000,0.000000,0.000000} 6} &
{\color[rgb]{1.000000,0.000000,0.000000} 7} 
\\
$X_{13}$ & $X_{33}^{1}$ & $X_{33}^{2}$ & $X_{32}$ & $X_{21}^{1}$ & $X_{12}$ & $X_{21}^{2}$
\end{tabular}
\caption{The Model 7.2 brane tiling on a $g=2$ Riemann surface with 3 gauge groups, 7 fields and 2 superpotential terms.
\label{fm3n1t}}
 \end{center}
 \end{figure}

\begin{figure}[ht!!]
\begin{center}
\includegraphics[trim=0cm 0cm 0cm 0cm,totalheight=4 cm]{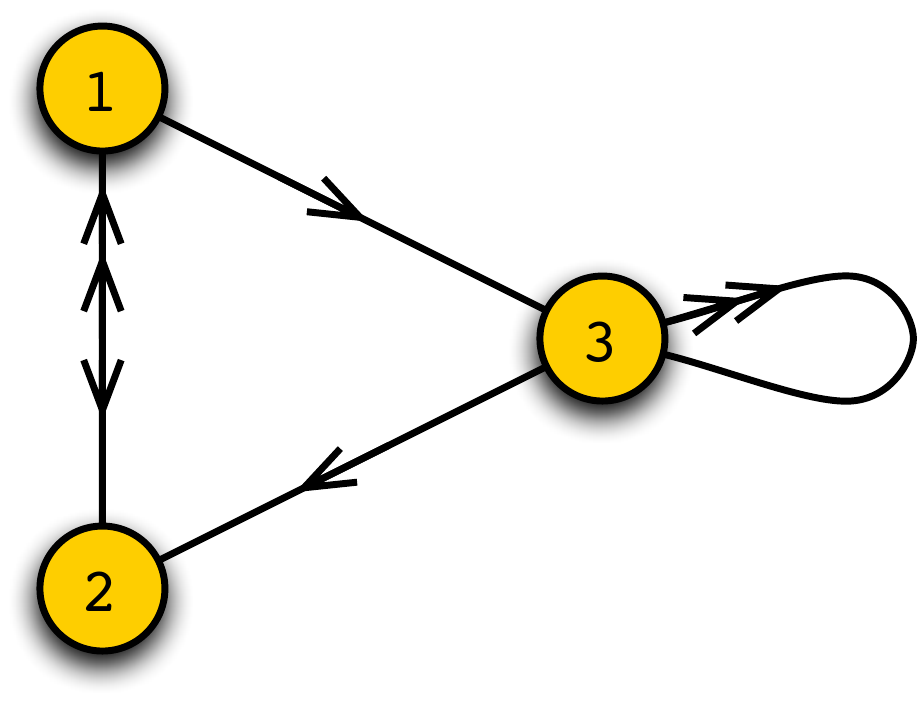}
\caption{The quiver diagram for Model 7.2, a brane tiling on a $g=2$ Riemann surface with 3 gauge groups, 7 fields and 2 superpotential terms.
\label{fm3n1q}}
 \end{center}
 \end{figure}
 
The brane tiling and corresponding quiver for Model 7.2 is shown in \fref{fm3n1t} and \fref{fm3n1q} respectively. The superpotential is
\beal{e12i0}
W= 
+ X_{13} X_{33}^{1} X_{33}^{2} X_{32} X_{21}^{1} X_{12} X_{21}^{2}
- X_{13} X_{33}^{2} X_{33}^{1} X_{32} X_{21}^{2} X_{12} X_{21}^{1}
~~.
\nn\\
\eea
The quiver incidence matrix is
\comment{\beal{e12i1}
d=\left(
\begin{array}{cccccccc}
X_{13} & X_{33}^{1} & X_{33}^{2} & X_{32} & X_{21}^{1} & X_{12} & X_{21}^{2}
\\
\hline
 -1 & 0 & 0 & 0 & 1 & -1 & 1 \\
 0 & 0 & 0 & 1 & -1 & 1 & -1 \\
 1 & 0 & 0 & -1 & 0 & 0 & 0
 \end{array}
\right)
~~.
\eea
}
\beal{e12i1}
d=\left(
\begin{array}{cccccccc}
X_{33}^{1} & X_{33}^{2} & X_{21}^{1} & X_{21}^{2} & X_{12} & X_{13} &  X_{32}  
\\
\hline
 0 & 0 & 1 & 1 &-1 &-1 & 0 \\
 0 & 0 &-1 &-1 & 1 & 0 & 1 \\
 0 & 0 & 0 & 0 & 0 & 1 &-1
 \end{array}
\right)
~~.
\eea

Model 7.2 has $c=7$ perfect matchings, each made out of a single field in the quiver. The perfect matching matrix is therefore the identity matrix,
\beal{e12i2}
P=\left(
\begin{array}{c|ccccccc}
\; & a_1 & a_2 & b_1 & b_2 & p_1 & p_2 & p_3 \\
\hline
X_{33}^{1} & 1 & 0 & 0 & 0 & 0 & 0 & 0 \\
X_{33}^{2} & 0 & 1 & 0 & 0 & 0 & 0 & 0 \\
X_{21}^{1} & 0 & 0 & 1 & 0 & 0 & 0 & 0 \\
X_{21}^{2} & 0 & 0 & 0 & 1 & 0 & 0 & 0 \\
X_{12}  & 0 & 0 & 0 & 0 & 1 & 0 & 0 \\
X_{13}  & 0 & 0 & 0 & 0 & 0 & 1 & 0 \\
X_{32}  & 0 & 0 & 0 & 0 & 0 & 0 & 1
\end{array}
\right)
~~.
\eea
\comment{\beal{e12i2}
P=\left(
\begin{array}{c|ccccccc}
\; & p_1 & p_2 & p_3 & p_4 & p_5 & p_6 & p_7 \\
\hline
X_{13} & 1 & 0 & 0 & 0 & 0 & 0 & 0 \\
X_{33}^{1} & 0 & 1 & 0 & 0 & 0 & 0 & 0 \\
X_{33}^{2} & 0 & 0 & 1 & 0 & 0 & 0 & 0 \\
X_{32} & 0 & 0 & 0 & 1 & 0 & 0 & 0 \\
X_{21}^{1} & 0 & 0 & 0 & 0 & 1 & 0 & 0 \\
X_{12} & 0 & 0 & 0 & 0 & 0 & 1 & 0 \\
X_{21}^{2} & 0 & 0 & 0 & 0 & 0 & 0 & 1
\end{array}
\right)
~~.
\eea}
The brane tiling has the following zig-zag paths,
\beal{e12i2bb}
&\eta_1 = (X_{33}^{1},X_{33}^{2})~,~
\eta_2 = (X_{21}^{1},X_{12})~,~
\eta_3 = (X_{12},X_{21}^{2})~,~
&\nn\\
&
\eta_4 = (X_{13},X_{33}^{1},X_{32},X_{21}^{1})~,~
\eta_5 = (X_{13},X_{33}^{2},X_{32},X_{21}^{2})~.~&
\eea

There are only trivial F-term constraints. The D-term constraints are encoded in 
the charge matrix
\beal{e12i4}
Q_D=\left(
\begin{array}{ccccccc}
a_1 & a_2 & b_1 & b_2 & p_1 & p_2 & p_3 \\
\hline
 0 & 0 &  1 &  1 & -1 & 0 & -1 \\
 0 & 0 &  0 &  0 &  0 & 1 & -1
\end{array}
\right)~~.
\eea
\comment{
\beal{e12i4}
Q_D=\left(
\begin{array}{ccccccc}
c_2 & a_1 & a_2 & c_3 & b_1 & c_1 & b_2 \\
p_1 & p_2 & p_3 & p_4 & p_5 & p_6 & p_7 \\
\hline
 0 & 0 & 0 & 1 & -1 & 1 & -1 \\
 1 & 0 & 0 & -1 & 0 & 0 & 0
\end{array}
\right)~~.
\eea}

Model 7.2's mesonic moduli space is expressed as the following symplectic quotient,
\beal{es12i6}
\mesonic = \mathbb{C}^{7} // Q_D ~~.
\eea

By associating the fugacities $\alpha_i,\beta_i,t_i$ to the perfect matchings $a_i,b_i,p_i$ respectively, the fully refined Hilbert series of $\mesonic$ is given by the following Molien integral
\beal{e12i7}
g_1(\alpha_i,\beta_i,t_i;\mesonic) &=& 
\frac{1}{(2\pi i)^3}
\oint_{|z_1|=1} 
\frac{\ud z_1}{z_1} 
\oint_{|z_2|=1} 
\frac{\ud z_2}{z_2}
\oint_{|z_3|=1} 
\frac{\ud z_3}{z_3}
~~
\nn\\
&& 
\hspace{1cm}
\times
\frac{1}{
\prod_{i=1}^{2} 
(1- \alpha_i)
(1- z_1 \beta_i)
}
\nn\\
&& 
\hspace{1cm}
\times
\frac{1}{
(1-z_1^{-1} t_1)
(1-z_2 t_2)
(1-z_1^{-1} z_2^{-1} t_3)
}
\nn\\
&=&
\frac{
1 - \beta_1 \beta_2 t_1 t_2 t_3
}{
\prod_{i=1}^{2}
(1-\alpha_i)
(1-\beta_i t_1)
(1-\beta_i t_2 t_3)
}
~~.
\nn\\
\eea

From the Hilbert series, we observe that the mesonic moduli space is a complete intersection. It is a 5-dimensional Calabi-Yau space. More specifically, the mesonic moduli space is $\mesonic=\mathbb{C}^2\times\mathcal{C}$ where the conifold generators are

\begin{center}
\begin{tabular}{|c|c|}
\hline
generator & perfect matchings\\
\hline\hline
$A_i$ & $b_i p_1$\\
$B_i$ & $b_i p_2 p_3$\\
\hline
\end{tabular}
\end{center}
The conifold relation is
\beal{es128}
\epsilon^{ij} A_i B_j = 0 ~~.
\eea

The global symmetry is enhanced to $SU(2)\times SU(2) \times U(1)^2 \times U(1)_R$ according to the charge matrix in \eref{e12i4}. One can assign the following global symmetry charges to the perfect matchings.

\begin{center}
\begin{tabular}{|c|ccccc|l|}
\hline
\; & $SU(2)_x$ & $SU(2)_y$ & $U(1)_{b_1}$ & $U(1)_{b_2}$ & $U(1)_R$ & fugacity \\
\hline \hline
$a_1$ & +1 &  0 & 0 & 0 & +1 & $\alpha_1=x t$\\
$a_2$ & -1 &  0 & 0 & 0 & +1 & $\alpha_2=x^{-1} t$\\
\hline
$b_1$ &  0 & +1 & 0 & -1 & +1 & $\beta_1=y b_2^{-1} t$\\
$b_2$ &  0 & -1 & 0 & -1 & +1 & $\beta_2=y^{-1} b_2^{-1} t$\\
\hline
$p_1$ & 0 & 0 & 0 &+1 & +1 & $t_1=b_2 t$\\
$p_2$ & 0 & 0 &+1 & 0 & +1 & $t_2=b_1 t$\\
$p_3$ & 0 & 0 &-1 &+1 & +1 & $t_3=b_1^{-1} b_2 t$\\
\hline
\end{tabular}
\end{center}
Under the above charge assignment, the Hilbert series of $\mesonic$ can be expressed in terms of characters of irreducible representations of the global symmetry,
\beal{es12i9}
g_1(x,y,t;\mesonic) &=&
\frac{1}{(1-x t)(1-x^{-1} t)} \sum_{n_1=0}^{\infty} \sum_{n_2=0}^{\infty} [n_1+n_2]_y t^{2n_1+3n_2}
\nn\\
&=&
\sum_{m=0}^{\infty} \sum_{n_1=0}^{\infty} \sum_{n_2=0}^{\infty} [m]_x [n_1+n_2]_y t^{m+2n_1+3n_2}
~~.
\eea
\\

We expect however from the conifold itself two $SU(2)$ symmetries and therefore a fully enhanced symmetry of $SU(2)^3 \times U(1) \times U(1)_R$. The full symmetry can be probed by modifying the above charge assignment on perfect matchings as follows.

\begin{center}
\begin{tabular}{|c|ccccc|l|}
\hline
\; & $SU(2)_x$ & $SU(2)_y$ & $SU(2)_z$ & $U(1)_{b}$ & $U(1)_R$ & fugacity \\
\hline \hline
$a_1$ & +1 &  0 & 0 & 0 & +1 & $\alpha_1=x t$\\
$a_2$ & -1 &  0 & 0 & 0 & +1 & $\alpha_2=x^{-1} t$\\
\hline
$b_1$ &  0 & +1 & 0 & -1 & +1 & $\beta_1=y b_2^{-1} t$\\
$b_2$ &  0 & -1 & 0 & -1 & +1 & $\beta_2=y^{-1} b_2^{-1} t$\\
\hline
$p_1$ & 0 & 0 &+1   &+1 & +1     & $t_1=z b t$\\
$p_2$ & 0 & 0 &-1/2 &+1/2 & +1/2 & $t_2=z^{-1/2} b^{1/2} t^{1/2}$\\
$p_3$ & 0 & 0 &-1/2 &+1/2 & +1/2 & $t_3=z^{-1/2} b^{1/2} t^{1/2}$\\
\hline
\end{tabular}
\end{center}
With the above refinement, the Hilbert series displays the full $SU(2)^3$ symmetry,
\beal{es12i9bb}
g_1(x,y,z,t;\mesonic) &=&
\sum_{n_1=0}^{\infty}
\sum_{n_2=0}^{\infty} 
[n_1]_x [n_2]_y [n_2]_z t^{n_1+2 n_2}
~~.
\eea
\\

The toric diagram of $\mesonic$ is given by
\beal{es12i10}
G_t=\left(
\begin{array}{ccccccc}
a_1 & a_2 & b_1 & b_2 & p_1 & p_2 & p_3 \\
\hline
 1 & 0 & 0 & 0 & 0 & 0 & 0 \\
 0 & 1 & 0 & 0 & 0 & 0 & 0 \\
 0 & 0 & 1 & 0 & 1 & 0 & 0 \\
 0 & 0 & 0 & 1 & 1 & 0 & 0 \\
 0 & 0 & 0 & 0 & 1 & -1 & -1
\end{array}
\right)~~,
\eea
where we notice that the perfect matchings $p_2$ and $p_3$ relate to the same toric point.
\\

\subsection{7 Fields, 4 Superpotential Terms, 1 Gauge Group \label{s_m4}}

\subsubsection{Model 7.4: $\mathbb{C}\times \mathcal{M}_{3,2}$ \label{s_m74a}}

\begin{figure}[ht!!]
\begin{center}
\includegraphics[trim=0cm 0cm 0cm 0cm,totalheight=8 cm]{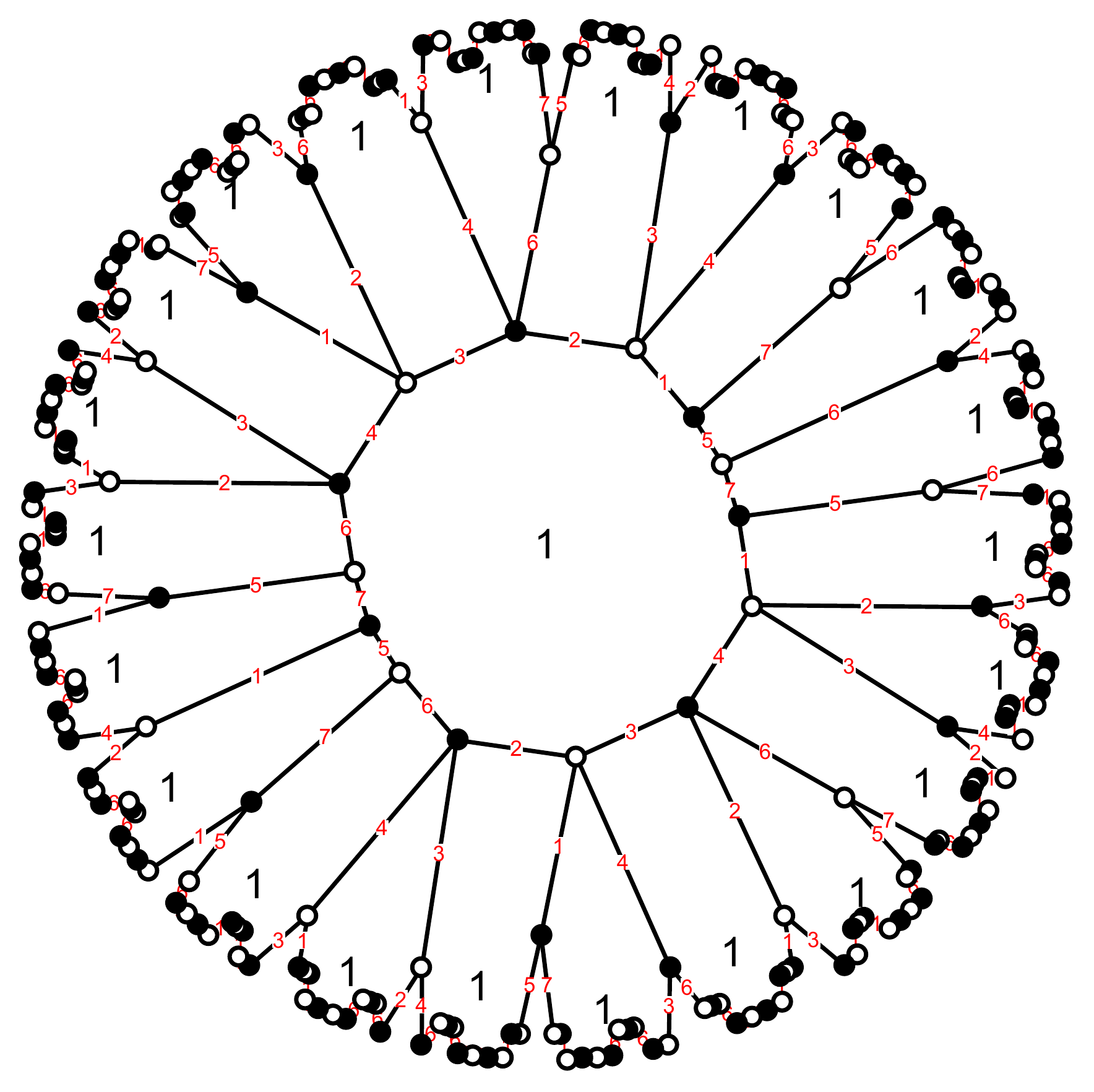}\\
\vspace{0.3cm}
\begin{tabular}{ccccccc}
{\color[rgb]{1.000000,0.000000,0.000000} 1} &
{\color[rgb]{1.000000,0.000000,0.000000} 2} &
{\color[rgb]{1.000000,0.000000,0.000000} 3} &
{\color[rgb]{1.000000,0.000000,0.000000} 4} &
{\color[rgb]{1.000000,0.000000,0.000000} 5} &
{\color[rgb]{1.000000,0.000000,0.000000} 6} &
{\color[rgb]{1.000000,0.000000,0.000000} 7} 
\\
$X_{11}^{1}$ & $X_{11}^{2}$ & $X_{11}^{3}$ & $X_{11}^{4}$ & $X_{11}^{5}$ & $X_{11}^{6}$ & $X_{11}^{7}$
\end{tabular}
\caption{The Model 7.4 brane tiling on a $g=2$ Riemann surface with 1 gauge group, 7 fields and 4 superpotential terms.
\label{fm4n1t}}
 \end{center}
 \end{figure}

\begin{figure}[ht!!]
\begin{center}
\includegraphics[trim=0cm 0cm 0cm 0cm,totalheight=1 cm]{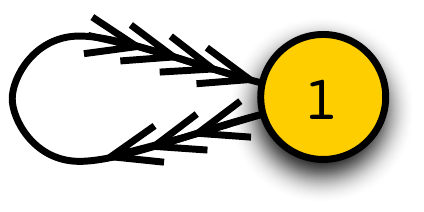}
\caption{The quiver diagram for Model 7.4, a brane tiling on a $g=2$ Riemann surface with 1 gauge group, 7 fields and 4 superpotential terms.
\label{fm4n1q}}
 \end{center}
 \end{figure}
 
The brane tiling and corresponding quiver for Model 7.4 is shown in \fref{fm4n1t} and \fref{fm4n1q} respectively. The superpotential is
\beal{e21i0}
W= 
+ X_{11}^{1} X_{11}^{2} X_{11}^{3} X_{11}^{4} 
+ X_{11}^{5} X_{11}^{6} X_{11}^{7}
- X_{11}^{2} X_{11}^{6} X_{11}^{4} X_{11}^{3} 
- X_{11}^{1} X_{11}^{5} X_{11}^{7}
~~.
\nn\\
\eea
The brane tiling is made of a single 14-sided face with the quiver having 7 adjoints. The brane tiling has overall $c=9$ perfect matchings,
\beal{e21i2}
P=\left(
\begin{array}{c|ccccccc}
\; & p_1 & p_2 & p_3 & p_4 & p_5 & p_6 & p_7\\
\hline
X_{11}^{1} & 0 & 0 & 0 & 0 & 0 & 0 & 1 \\
X_{11}^{2} & 1 & 0 & 0 & 1 & 0 & 0 & 0 \\
X_{11}^{3} & 0 & 1 & 0 & 0 & 1 & 0 & 0 \\
X_{11}^{4} & 0 & 0 & 1 & 0 & 0 & 1 & 0 \\
X_{11}^{5} & 1 & 1 & 1 & 0 & 0 & 0 & 0 \\
X_{11}^{6} & 0 & 0 & 0 & 0 & 0 & 0 & 1 \\
X_{11}^{7} & 0 & 0 & 0 & 1 & 1 & 1 & 0
\end{array}
\right)
~~.
\eea
The zig-zag paths of the brane tiling are,
\beal{e21i2}
&
\eta_1 = (X_{11}^{2},X_{11}^{3})~,~
\eta_2 = (X_{11}^{3},X_{11}^{4})~,~
\eta_3 = (X_{11}^{5},X_{11}^{7})~,~
&
\nn\\
&
\eta_4 = (X_{11}^{1},X_{11}^{2},X_{11}^{6},X_{11}^{7})~,~
\eta_5 = (X_{11}^{1},X_{11}^{5},X_{11}^{6},X_{11}^{4})~.~&
\eea

The F-term constraints are summarized by
\beal{e21i4}
Q_F=\left(
\begin{array}{ccccccc}
p_1 & p_2 & p_3 & p_4 & p_5 & p_6 & p_7\\
\hline
 1 & 0 & -1 & -1 &  0 & 1 & 0 \\
 0 & 1 & -1 &  0 & -1 & 1 & 0
\end{array}
\right)~~.
\eea
There are only trivial D-term constraints.

Overall, Model 7.4's mesonic moduli space is expressed as the following symplectic quotient,
\beal{es21i6}
\mesonic = \mathbb{C}^{7} // Q_F ~~.
\eea

By associating the fugacity $t_i$ to the perfect matching $p_i$, the fully refined Hilbert series of $\mesonic$ is given by the following Molien integral
\beal{e21i7}
g_1(t_i;\mesonic) &=& 
\frac{1}{(2\pi i)^2}
\oint_{|z_1|=1} 
\frac{\ud z_1}{z_1} 
\oint_{|z_2|=1} 
\frac{\ud z_2}{z_2}
~~
\nn\\
&& 
\hspace{1cm}
\times
\frac{1}{
(1-z_1 t_1)
(1-z_2 t_2)
(1-z_1^{-1} z_2^{-1} t_3)
}
\nn\\
&& 
\hspace{1cm}
\times
\frac{1}{
(1-z_1^{-1} t_4)
(1-z_2^{-1} t_5)
(1-z_1 z_2 t_6)
}
\times
\frac{1}{
(1-t_7)
}
\nn\\
&=&
\frac{1}{(1 - t_7)}
\times
\frac{
1 - t_1 t_2 t_3 t_4 t_5 t_6
}{
(1 - t_1 t_4) (1 - t_2 t_5) (1 - t_3 t_6)
(1 - t_1 t_2 t_3)  (1 - t_4 t_5 t_6) 
}
~~.
\nn\\
\eea

From the Hilbert series, we observe that the mesonic moduli space is a complete intersection. It is a 5-dimensional Calabi-Yau space. The generators of the moduli space are shown below.

\begin{center}
\begin{tabular}{|c|c|}
\hline
generator & perfect matchings\\
\hline\hline
$A_1$ & $p_1 p_4$\\
$A_2$ & $p_2 p_5$\\
$A_3$ & $p_3 p_6$\\
\hline
$B_1$ & $p_1 p_2 p_3$ \\
$B_2$ & $p_4 p_5 p_6$ \\
\hline
$C$ & $p_7$\\
\hline
\end{tabular}
\end{center}
The relation formed by the above generators is
\beal{es21i7b}
A_1 A_2 A_3= B_1 B_2~~.
\eea

The global symmetry is $U(1)^4 \times U(1)_R$. The toric diagram of $\mesonic$ is given by
\beal{es21i10}
G_t=\left(
\begin{array}{ccccccc}
p_1 & p_2 & p_3 & p_4 & p_5 & p_6 & p_7 \\
\hline
 1 & 0 & 0 & 1 & 0 & 0 & 0 \\
 0 & 1 & 0 & 0 & 1 & 0 & 0 \\
 0 & 0 & 1 & 0 & 0 & 1 & 0 \\
 0 & 0 & 0 & 1 & 1 & 1 & 0 \\
 0 & 0 & 0 & 0 & 0 & 0 & 1
\end{array}
\right)~~.
\eea
\\

\subsection{8 Fields, 2 Superpotential Terms, 4 Gauge Groups \label{s_m5}}

\subsubsection{Model 8.2a: $NC2$ \label{s_m82a}}

\begin{figure}[ht!!]
\begin{center}
\includegraphics[trim=0cm 0cm 0cm 0cm,totalheight=8 cm]{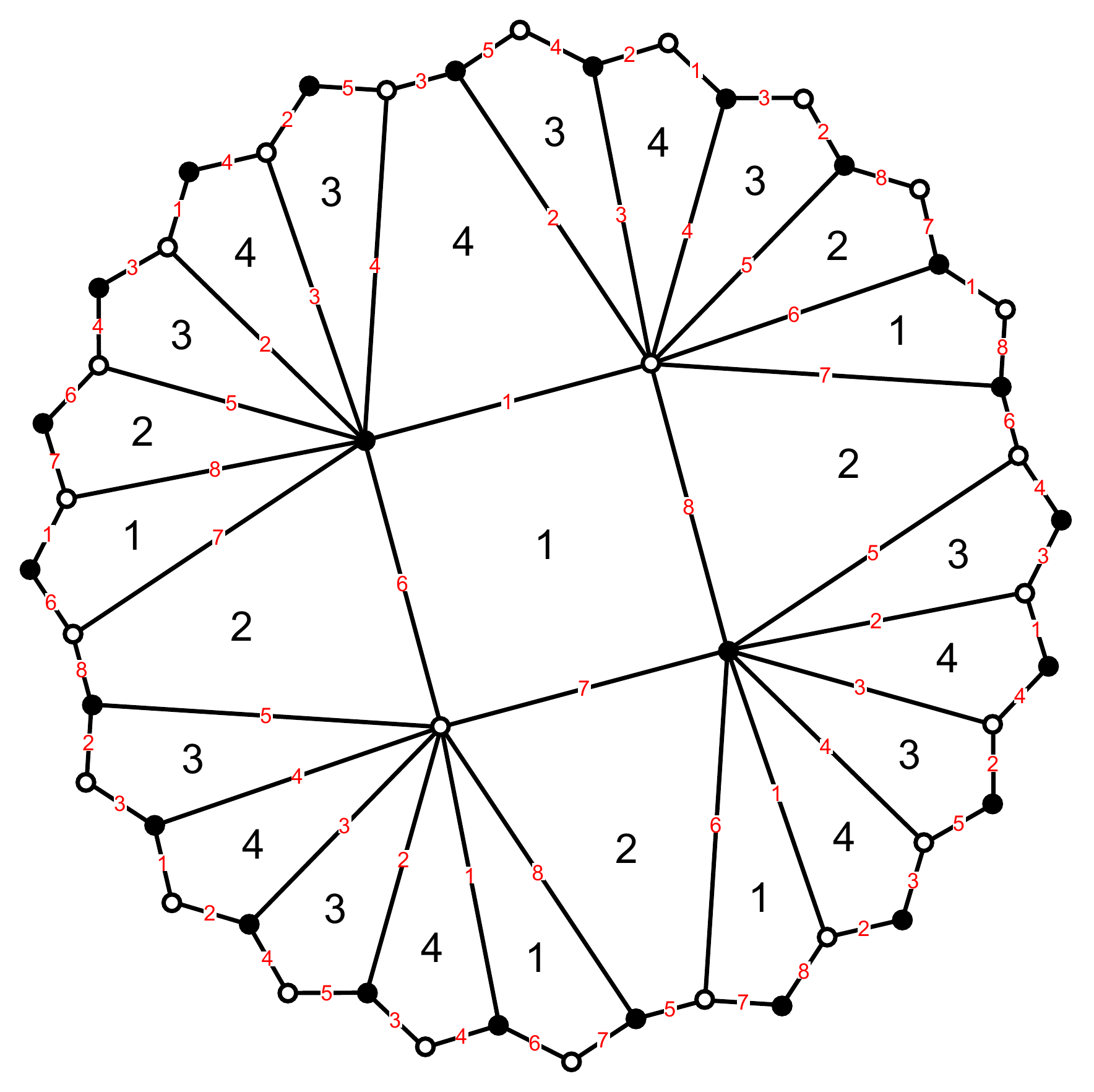}\\
\vspace{0.3cm}
\begin{tabular}{cccccccc}
{\color[rgb]{1.000000,0.000000,0.000000} 1} &
{\color[rgb]{1.000000,0.000000,0.000000} 2} &
{\color[rgb]{1.000000,0.000000,0.000000} 3} &
{\color[rgb]{1.000000,0.000000,0.000000} 4} &
{\color[rgb]{1.000000,0.000000,0.000000} 5} &
{\color[rgb]{1.000000,0.000000,0.000000} 6} &
{\color[rgb]{1.000000,0.000000,0.000000} 7} &
{\color[rgb]{1.000000,0.000000,0.000000} 8} 
\\
$X_{14}$ &$X_{43}^{1}$ &$X_{34}$ &$X_{43}^{2}$ &$X_{32}$ &$X_{21}^{1}$ &$X_{12}$ &$X_{21}^{2}$
\end{tabular}
\caption{The Model 8.2a brane tiling on a $g=2$ Riemann surface with 4 gauge groups, 8 fields and 2 superpotential terms.
\label{fm5n1t}}
 \end{center}
 \end{figure}

\begin{figure}[ht!!]
\begin{center}
\includegraphics[trim=0cm 0cm 0cm 0cm,totalheight=4 cm]{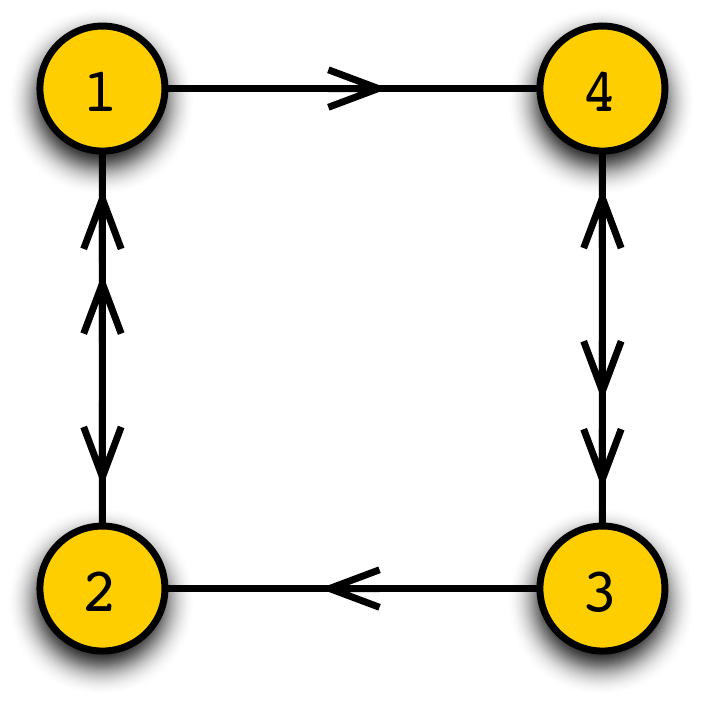}
\caption{The quiver diagram for Model 8.2a, a brane tiling on a $g=2$ Riemann surface with 4 gauge groups, 8 fields and 2 superpotential terms.
\label{fm5n1q}}
 \end{center}
 \end{figure}
 
The brane tiling and quiver of Model 8.2a are shown in \fref{fm5n1t} and \fref{fm5n1q} respectively. The superpotential is
\beal{e16i0}
W= 
+ X_{14} X_{43}^{1} X_{34} X_{43}^{2} X_{32} X_{21}^{1} X_{12} X_{21}^{2}
- X_{14} X_{43}^{1} X_{34} X_{43}^{2} X_{32} X_{21}^{1} X_{12} X_{21}^{2}
~~.
\nn\\
\eea
The quiver incidence matrix is
\comment{
\beal{e16i1}
d=\left(
\begin{array}{cccccccc}
X_{14} &X_{43}^{1} &X_{34} &X_{43}^{2} &X_{32} &X_{21}^{1} &X_{12} &X_{21}^{2}
\\
\hline
 -1 & 0 & 0 & 0 & 0 & 1 & -1 & 1 \\
 0 & 0 & 0 & 0 & 1 & -1 & 1 & -1 \\
 0 & 1 & -1 & 1 & -1 & 0 & 0 & 0 \\
 1 & -1 & 1 & -1 & 0 & 0 & 0 & 0
 \end{array}
\right)
~~.
\eea
}
\beal{e16i1}
d=\left(
\begin{array}{cccccccc}
X_{21}^{1} &X_{21}^{2} &X_{43}^{1} &X_{43}^{2} &X_{12} &X_{14} &X_{32} &X_{34}   
\\
\hline
 1 & 1 & 0 & 0 &-1 &-1 & 0 & 0 \\
-1 &-1 & 0 & 0 & 1 & 0 & 1 & 0 \\
 0 & 0 & 1 & 1 & 0 & 0 &-1 &-1 \\
 0 & 0 &-1 &-1 & 0 & 1 & 0 & 1
 \end{array}
\right)
~~.
\eea

The brane tiling has $c=8$ perfect matchings, each made out of a single field. The perfect matching matrix is therefore the identity matrix,
\beal{e16i2}
P=\left(
\begin{array}{c|cccccccc}
\; & a_1 & a_2 & b_1 & b_2 & p_1 & p_2 & p_3 & p_4\\
\hline
X_{21}^{1} & 1 & 0 & 0 & 0 & 0 & 0 & 0 & 0 \\
X_{21}^{2} & 0 & 1 & 0 & 0 & 0 & 0 & 0 & 0 \\
X_{43}^{1} & 0 & 0 & 1 & 0 & 0 & 0 & 0 & 0 \\
X_{43}^{2} & 0 & 0 & 0 & 1 & 0 & 0 & 0 & 0 \\
X_{12} & 0 & 0 & 0 & 0 & 1 & 0 & 0 & 0 \\
X_{14} & 0 & 0 & 0 & 0 & 0 & 1 & 0 & 0 \\
X_{32} & 0 & 0 & 0 & 0 & 0 & 0 & 1 & 0 \\
X_{34} & 0 & 0 & 0 & 0 & 0 & 0 & 0 & 1
\end{array}
\right)
~~.
\eea
The brane tiling of Model 8.2a has the following zig-zag paths
\beal{e16i2bb}
&
\eta_1 = (X_{43}^{1},X_{34})~,~
\eta_2 = (X_{34},X_{43}^{2})~,~
\eta_3 = (X_{21}^{1},X_{12})~,~
\eta_4 = (X_{12},X_{21}^{2})~,~
&
\nn\\
&
\eta_5 = (X_{14},X_{43}^{1},X_{32},X_{21}^{1})~,~
\eta_6 = (X_{14},X_{43}^{2},X_{32},X_{21}^{2})~.~
&
\eea

There are only trivial F-terms due to the identity perfect matching matrix. The D-term charge matrix is as follows
\beal{e16i4}
Q_D=\left(
\begin{array}{cccccccc}
a_1 & a_2 & b_1 & b_2 & p_1 & p_2 & p_3 & p_4\\
\hline
  1 &  1 &  0 &  0 & -1 & 0 & -1 &  0 \\
  0 &  0 &  1 &  1 &  0 & 0 & -1 & -1 \\
  0 &  0 &  0 &  0 &  0 & 1 & -1 & 0
\end{array}
\right)~~.
\eea
\comment{
\beal{e16i4}
Q_D=\left(
\begin{array}{cccccccc}
 p_1 & p_2 & p_3 & p_4 & p_5 & p_6 & p_7 & p_8\\
\hline
 0 & 0 & 0 & 0 & 1 & -1 & 1 & -1 \\
 0 & 1 & -1 & 1 & -1 & 0 & 0 & 0 \\
 1 & -1 & 1 & -1 & 0 & 0 & 0 & 0
\end{array}
\right)~~.
\eea
}

The symplectic quotient describing the mesonic moduli space is as follows,
\beal{es16i6}
\mesonic = \mathbb{C}^{8} // Q_D ~~.
\eea

By associating the fugacities $\alpha_i,\beta_i,t_i$ to the perfect matchings $a_i,b_i,p_i$ respectively, the fully refined Hilbert series of $\mesonic$ is given by the following Molien integral
\beal{e16i7}
g_1(\alpha_i,\beta_i,t_i;\mesonic) &=& 
\frac{1}{(2\pi i)^3}
\oint_{|z_1|=1} 
\frac{\ud z_1}{z_1} 
\oint_{|z_2|=1} 
\frac{\ud z_2}{z_2}
\oint_{|z_3|=1} 
\frac{\ud z_3}{z_3}
~~
\nn\\
&& 
\hspace{1cm}
\times
\frac{1}{
\prod_{i=1}^{2}
(1-z_1 \alpha_i)
(1-z_2 \beta_i)
}
\nn\\
&& 
\hspace{1cm}
\times
\frac{1}{
(1-z_1^{-1} t_1)
(1-z_3 t_2)
(1-z_1^{-1} z_2^{-1} z_3^{-1} t_3)
(1-z_2^{-1} t_4)
}
\nn\\
&=&
\frac{
(\alpha_1 \alpha_2 \beta_1 \beta_2 t_1 t_2 t_3 t_4)
P(\alpha_i,\beta_i,t_i)
}{
\prod_{i=1}^{2}
(1-\alpha_i t_1)
(1-\beta_i t_4)
\prod_{i,j=1}^{2}
(1-\alpha_i \beta_j t_2 t_3)
}
~~,
\nn\\
\eea
where the numerator is
\beal{e16i7b}
P(\alpha_i,\beta_i,\gamma_i)&=&
\alpha_1^{-1} \alpha_2^{-1} \beta_1^{-1} \beta_2^{-1} t_1^{-1} t_2^{-1} t_3^{-1} t_4^{-1}
-\sum_{i=1}^{2}\alpha_i^{-1}t_1^{-1}
-\sum_{i=1}^{2}\beta_i^{-1}t_4^{-1}
+1
\nn\\
&&
-t_1^{-1} t_2 t_3 t_4^{-1}
+\sum_{i=1}^{2}\alpha_i t_2 t_3 t_4^{-1}
+\sum_{i=1}^{2}\beta_i t_1^{-1} t_2 t_3 
-\alpha_1 \alpha_2 \beta_1 \beta_2 t_2^2 t_3^2
~~.\nn\\
\eea
By setting the fugacities $\alpha_i=\beta_i=t_i=t$, the unrefined Hilbert series is
\beal{e16i7bb}
g_1(t;\mesonic)&=&
\frac{
1 - 4 t^6 + 4 t^{10} - t^{16}
}{
(1 - t^2)^4 (1 - t^4)^4
}
~~.
\eea

The Hilbert series above indicates that the mesonic moduli space is not a complete intersection. The plethystic logarithm of the Hilbert series is,
\beal{e16i7c}
PL[g_1(\alpha_i,\beta_i,t_i;\mesonic)]
&=&
\sum_{i=1}^{2}(
\alpha_i t_1
+\beta_i t_4
)
+
\sum_{i,j=1}^{2}
\alpha_i \beta_j t_2 t_3
\nn\\
&&
-\sum_{i=1}^{2}(
\alpha_1 \alpha_2 \beta_i t_1 t_2 t_3
+ \alpha_i \beta_1 \beta_2 t_2 t_3 t_4
)
 + \dots ~~.
\eea
The first order generators are as follows.

\begin{center}
\begin{tabular}{|c|c|}
\hline
generator & perfect matchings\\
\hline\hline
$A_i$ & $a_i p_1$\\
$B_j$ & $b_j p_4$\\
$C_{ij}$ & $a_i b_j p_2 p_3$\\
\hline
\end{tabular}
\end{center}
The generators form the following first order relations
\beal{es16i8}
\epsilon^{i_1 i_2} A_{i_1} C_{i_2 j} =0~,~
\epsilon^{j_1 j_2} B_{j_1} C_{i j_2} =0
~~.
\eea

The global symmetry is enhanced to $SU(2)\times SU(2) \times U(1)^2 \times U(1)_R$. The perfect matchings carry the following global charges.

\begin{center}
\begin{tabular}{|c|ccccc|l|}
\hline
\; & $SU(2)_{x}$ & $SU(2)_{y}$ & $U(1)_{b_1}$ & $U(1)_{b_2}$ & $U(1)_R$ & fugacity\\
\hline
\hline
$a_1$ &  1 &  0 & 1 & 0 & 1 & $\alpha_1=x b_1 t$ \\
$a_2$ & -1 &  0 & 1 & 0 & 1 & $\alpha_2=x^{-1} b_1 t$ \\
\hline
$b_1$ &  0 &  1 & 0 & 1 & 1 & $\beta_1= y b_2 t$ \\
$b_2$ &  0 & -1 & 0 & 1 & 1 & $\beta_2= y^{-1} b_2 t$ \\
\hline
$p_1$ & 0 & 0 & -1 &  0 & 1 & $t_1=b_1^{-1} t$ \\
$p_2$ & 0 & 0 & -1 &  0 & 1 & $t_2=b_1^{-1} t$ \\
$p_3$ & 0 & 0 &  0 & -1 & 1 & $t_3=b_2^{-1} t$ \\
$p_4$ & 0 & 0 &  0 & -1 & 1 & $t_4=b_2^{-1} t$ \\
\hline
\end{tabular}
\end{center}

The Hilbert series of the mesonic moduli space can be expressed in terms of characters of irreducible representations of the global symmetry group. It is
\beal{es16i8b}
g_1(x,y,b_i,t;\mesonic) &=& 
\sum_{n_1=0}^{\infty}
\sum_{n_2=0}^{\infty}
\sum_{n_3=0}^{\infty}
[n_2+n_3;n_1+n_3]t^{2n_1+2n_2+4n_3}~~,
\eea
where $[n_2+n_3;n_1+n_3]\equiv [n_2+n_3]_{SU(2)_x} [n_1+n_2]_{SU(2)_y}$.

The toric diagram of $\mesonic$ is given by
\beal{es16i10}
G_t=\left(
\begin{array}{cccccccc}
a_1 & a_2 & b_1 & b_2 & p_1 & p_2 & p_3 & p_4\\
\hline
 1 & 0 & 0 & 0 & 1 & 0 & 0 & 0 \\
 0 & 1 & 0 & 0 & 1 & 0 & 0 & 0 \\
 0 & 0 & 1 & 0 & 0 & 0 & 0 & 1 \\
 0 & 0 & 0 & 1 & 0 & 0 & 0 & 1 \\
 1 & 1 & 1 & 1 & 1 & 1 & 1 & 1
 \end{array}
\right)~~.
\eea
\\

\subsubsection{Model 8.2b: $NC3$ \label{s_m82b}}

\begin{figure}[ht!!]
\begin{center}
\includegraphics[trim=0cm 0cm 0cm 0cm,totalheight=8 cm]{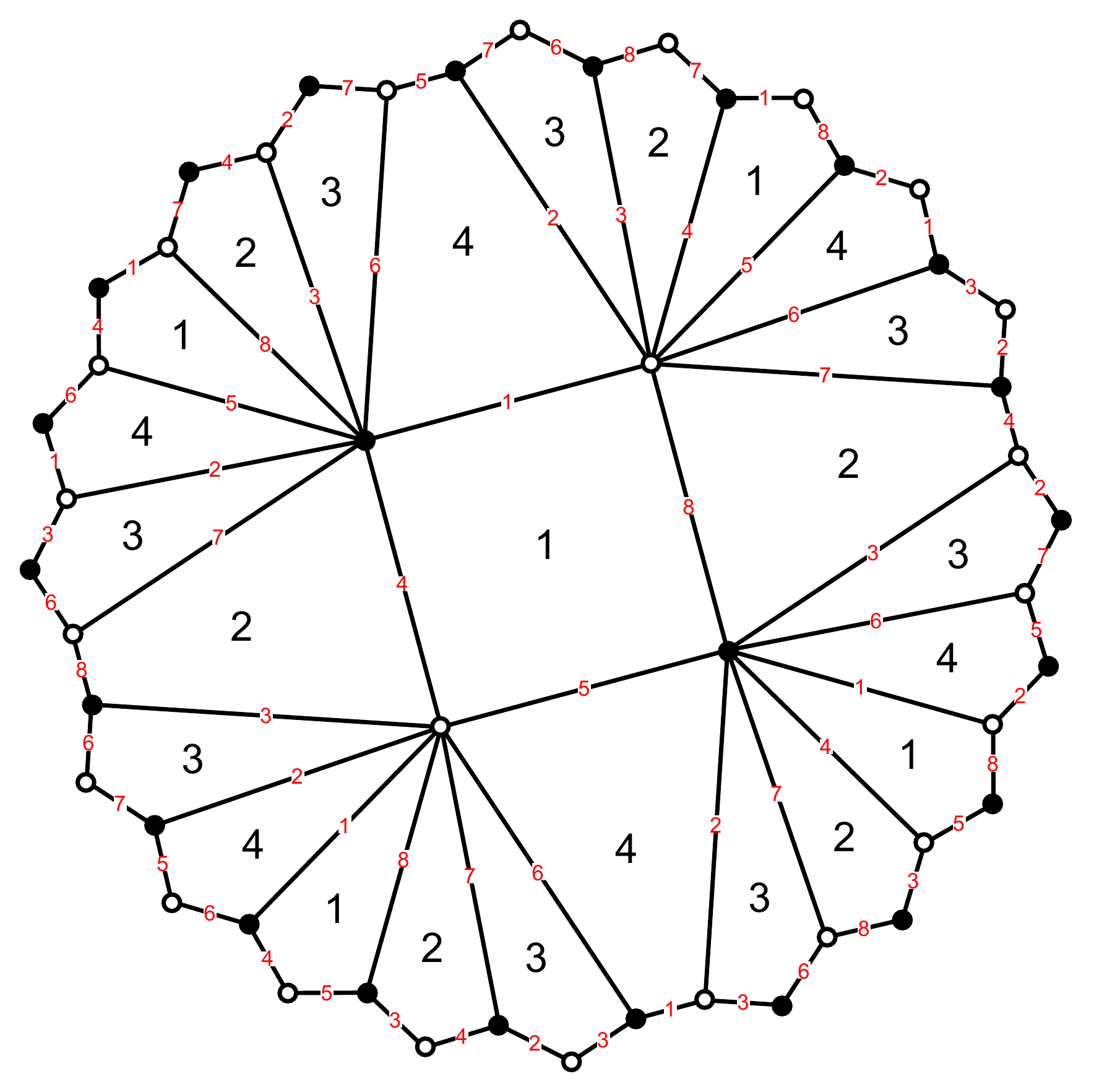}\\
\vspace{0.3cm}
\begin{tabular}{cccccccc}
{\color[rgb]{1.000000,0.000000,0.000000} 1} &
{\color[rgb]{1.000000,0.000000,0.000000} 2} &
{\color[rgb]{1.000000,0.000000,0.000000} 3} &
{\color[rgb]{1.000000,0.000000,0.000000} 4} &
{\color[rgb]{1.000000,0.000000,0.000000} 5} &
{\color[rgb]{1.000000,0.000000,0.000000} 6} &
{\color[rgb]{1.000000,0.000000,0.000000} 7} &
{\color[rgb]{1.000000,0.000000,0.000000} 8} 
\\
$X_{14}^{1}$ & $X_{43}^{1}$ & $X_{32}^{1}$ & $X_{21}^{1}$ & $X_{14}^{2}$ & $X_{43}^{2}$ & $X_{32}^{2}$ & $X_{21}^{2}$
\end{tabular}
\caption{The Model 8.2b brane tiling on a $g=2$ Riemann surface with 4 gauge groups, 8 fields and 2 superpotential terms.
\label{fm5n4t}}
 \end{center}
 \end{figure}

\begin{figure}[ht!!]
\begin{center}
\includegraphics[trim=0cm 0cm 0cm 0cm,totalheight=4 cm]{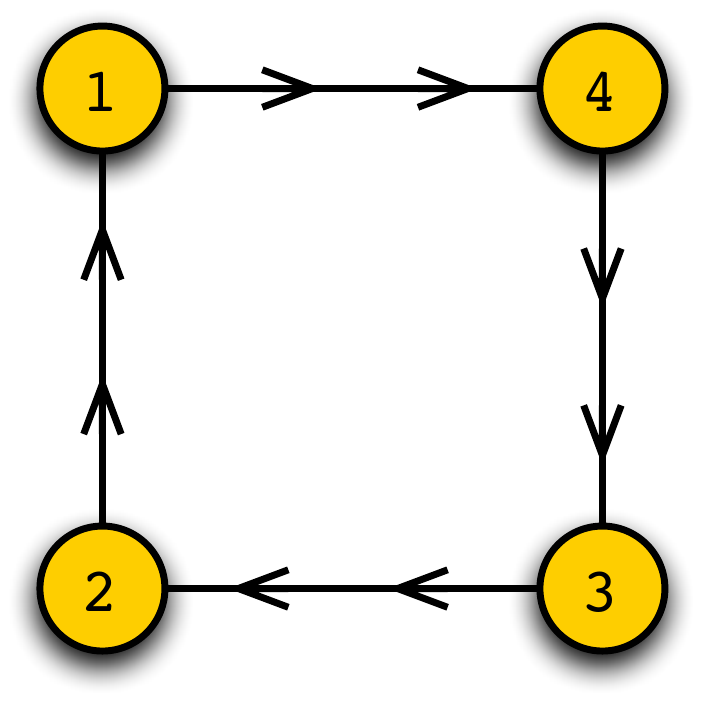}
\caption{The quiver diagram for Model 8.2b, a brane tiling on a $g=2$ Riemann surface with 4 gauge groups, 8 fields and 2 superpotential terms.
\label{fm5n4q}}
 \end{center}
 \end{figure}
 
The brane tiling and quiver of Model 8.2b are shown in \fref{fm5n4t} and \fref{fm5n4q} respectively. The superpotential is
\beal{e19i0}
W= 
+ X_{14}^{1} X_{43}^{1} X_{32}^{1} X_{21}^{1} X_{14}^{2} X_{43}^{2} X_{32}^{2} X_{21}^{2}
- X_{14}^{1} X_{43}^{2} X_{32}^{1} X_{21}^{2} X_{14}^{2} X_{43}^{1} X_{32}^{2} X_{21}^{1}
~~.
\nn\\
\eea
The quiver incidence matrix is
\beal{e19i1}
d=\left(
\begin{array}{cccccccc}
X_{14}^{1} & X_{43}^{1} & X_{32}^{1} & X_{21}^{1} & X_{14}^{2} & X_{43}^{2} & X_{32}^{2} & X_{21}^{2}
\\
\hline
 -1 & 0 & 0 & 1 & -1 & 0 & 0 & 1 \\
 0 & 0 & 1 & -1 & 0 & 0 & 1 & -1 \\
 0 & 1 & -1 & 0 & 0 & 1 & -1 & 0 \\
 1 & -1 & 0 & 0 & 1 & -1 & 0 & 0
 \end{array}
\right)
~~.
\eea

The brane tiling has $c=8$ perfect matchings, each made of a single quiver field. The perfect matching matrix is the identity matrix,
\beal{e19i2}
P=\left(
\begin{array}{c|cccccccc}
\; & a_1 & a_2 & b_1 & b_2 & c_1 & c_2 & d_1 & d_2\\
\hline
X_{32}^{1} & 1 & 0 & 0 & 0 & 0 & 0 & 0 & 0 \\
X_{32}^{2} & 0 & 1 & 0 & 0 & 0 & 0 & 0 & 0 \\
X_{43}^{1} & 0 & 0 & 1 & 0 & 0 & 0 & 0 & 0 \\
X_{43}^{2} & 0 & 0 & 0 & 1 & 0 & 0 & 0 & 0 \\
X_{14}^{1} & 0 & 0 & 0 & 0 & 1 & 0 & 0 & 0 \\
X_{14}^{2} & 0 & 0 & 0 & 0 & 0 & 1 & 0 & 0 \\
X_{21}^{1} & 0 & 0 & 0 & 0 & 0 & 0 & 1 & 0 \\
X_{21}^{2} & 0 & 0 & 0 & 0 & 0 & 0 & 0 & 1
\end{array}
\right)
~~.
\eea
The zig-zag paths of the brane tiling are
\beal{e19i1bb}
&
\eta_1 = (X_{14}^{1},X_{43}^{1},X_{32}^{2},X_{21}^{2},X_{14}^{2},X_{43}^{2},X_{32}^{1},X_{21}^{1})~,~&\nn\\
&
\eta_2 = (X_{14}^{1},X_{43}^{2},X_{32}^{2},X_{21}^{1},X_{14}^{2},X_{43}^{1},X_{32}^{1},X_{21}^{2})~.~&
\eea

There is no F-term charge matrix. The D-term charge matrix is as follows
\beal{e19i4}
Q_D=\left(
\begin{array}{cccccccc}
a_1 & a_2 & b_1 & b_2 & c_1 & c_2 & d_1 & d_2\\
\hline
 1 &  1 &  0 &  0 &  0 &  0 & -1 & -1 \\
 0 &  0 &  1 &  1 &  0 &  0 & -1 & -1 \\
 0 &  0 &  0 &  0 &  1 &  1 & -1 & -1
\end{array}
\right)~~.
\eea
\comment{
\beal{e19i4}
Q_D=\left(
\begin{array}{cccccccc}
p_1 & p_2 & p_3 & p_4 & p_5 & p_6 & p_7 & p_8\\
\hline
 0 & 0 & 1 & -1 & 0 & 0 & 1 & -1 \\
 0 & 1 & -1 & 0 & 0 & 1 & -1 & 0 \\
 1 & -1 & 0 & 0 & 1 & -1 & 0 & 0
\end{array}
\right)~~.
\eea
}

The symplectic quotient describing the mesonic moduli space is
\beal{es19i6}
\mesonic = \mathbb{C}^{8} // Q_D ~~.
\eea

By associating the fugacity $t_i$ to the perfect matching $p_i$, the fully refined Hilbert series of $\mesonic$ is given by the following Molien integral
\beal{e19i7}
g_1(\alpha_i,\beta_i,\gamma_i,\delta_i;\mesonic) &=& 
\frac{1}{(2\pi i)^3}
\oint_{|z_1|=1} 
\frac{\ud z_1}{z_1} 
\oint_{|z_2|=1} 
\frac{\ud z_2}{z_2}
\oint_{|z_3|=1} 
\frac{\ud z_3}{z_3}
~~
\nn\\
&& 
\hspace{1cm}
\times
\frac{1}{
\prod_{i=1}^{2}
(1-z_1 \alpha_i)
(1-z_2 \beta_i)
(1-z_3 \gamma_i)
(1-z_1^{-1} z_2^{-1} z_3^{-1} \delta_i)
}
\nn\\
&=&
\frac{
P(\alpha_i,\beta_i,\gamma_i,\delta_i)
}{
\prod_{i,j,k,l=1}^{2}
(1-\alpha_i \beta_j \gamma_k \delta_l)
}
~~.
\nn\\
\eea
The numerator $P(\alpha_i,\beta_i,\gamma_i,\delta_i)$ is too large to be presented here. We unrefine the above Hilbert series by setting the fugacities $\alpha_i=\beta_i=\gamma_i=\delta_i=t$. The unrefined Hilbert series is
\beal{e19i7b}
g_1(t;\mesonic) &=& 
\frac{1 + 11 t^4 + 11 t^8 + t^{12}}{
(1 - t^4)^5}
~~.
\eea
The Hilbert series above indicates that the mesonic moduli space is not a complete intersection. It is a Calabi-Yau 5-fold.

The plethystic logarithm of the refined Hilbert series is,
\beal{e19i7c}
&&
PL[g_1(\alpha_i,\beta_i,\gamma_i,\delta_i;\mesonic)]
=
\sum_{i,j,k,l=1}^{2} \alpha_i \beta_j \gamma_k \delta_l
\nn\\
&& \hspace{1cm}
- \prod_{m=1}^{2} \alpha_m \beta_m \gamma_m \delta_m
\Big(  7 
+ 3 \sum_{i\neq j}^{2} (\alpha_i \alpha_j^{-1} + \beta_i \beta_j^{-1} +\gamma_i \gamma_j^{-1} +\delta_i \delta_j^{-1})
\nn\\
&& \hspace{4cm}
+ \mathop{\sum_{i\neq j}^{2}}_{k \neq l}
(
\alpha_i \alpha_j^{-1} \beta_k \beta_l^{-1}
+ \alpha_i \alpha_j^{-1} \gamma_k \gamma_l^{-1}
+ \alpha_i \alpha_j^{-1} \delta_k \delta_l^{-1}
\nn\\
&& \hspace{5cm}
+ \beta_i \beta_j^{-1} \gamma_k \gamma_l^{-1}
+ \beta_i \beta_j^{-1} \delta_k \delta_l^{-1}
+ \gamma_i \gamma_j^{-1} \delta_k \delta_l^{-1}
)\Big)
 + \dots ~~.\nn\\
\eea
The first order generators are shown below.

\begin{center}
\begin{tabular}{|c|c|}
\hline
generator & perfect matchings\\
\hline\hline
$A_{ijkl}$ & $a_i b_j c_k d_l$\\
\hline
\end{tabular}
\end{center}
The generators form the following first order simplified relations
\beal{es19i8}
&
\epsilon^{i_1 i_2} A_{i_1 j_1 k_1 l_1} A_{i_2 j_2 k_2 l_2} = 0 ~,~
\epsilon^{j_1 j_2} A_{i_1 j_1 k_1 l_1} A_{i_2 j_2 k_2 l_2} = 0 ~,~
&
\nn\\
&
\epsilon^{k_1 k_2} A_{i_1 j_1 k_1 l_1} A_{i_2 j_2 k_2 l_2} = 0 ~,~
\epsilon^{l_1 l_2} A_{i_1 j_1 k_1 l_1} A_{i_2 j_2 k_2 l_2} = 0 ~.~
&
\eea
The above are 112 relations which reduce to 55 independent ones in the representations $[2;2;0;0]$ with permutations and $[0;0;0;0]$.

The global symmetry is enhanced to $SU(2)^4 \times U(1)_R$. The perfect matchings carry the following global charges.

\begin{center}
\begin{tabular}{|c|ccccc|l|}
\hline
\; & $SU(2)_{x}$ & $SU(2)_{y}$ & $SU(2)_{z}$ & $SU(2)_{w}$ & $U(1)_R$ & fugacity\\
\hline
\hline
$a_1$ &  1 &  0 &  0 &  0 & 1 & $\alpha_1= x t$ \\
$a_2$ & -1 &  0 &  0 &  0 & 1 & $\alpha_2= x^{-1} t$ \\
\hline
$b_1$ &  0 &  1 &  0 &  0 & 1 & $\beta_1= y t$ \\
$b_2$ &  0 & -1 &  0 &  0 & 1 & $\beta_2= y^{-1} t$ \\
\hline
$c_1$ &  0 &  0 &  1 &  0 & 1 & $\gamma_1= z t$ \\
$c_2$ &  0 &  0 & -1 &  0 & 1 & $\gamma_2= z^{-1} t$ \\
\hline
$d_1$ &  0 &  0 &  0 &  1 & 1 & $\delta_1= w t$ \\
$d_2$ &  0 &  0 &  0 & -1 & 1 & $\delta_2= w^{-1} t$ \\
\hline
\end{tabular}
\end{center}

The Hilbert series of the mesonic moduli space can be expressed in terms of characters of irreducible representations of the global symmetry group. It is
\beal{es18i8b}
&&
g_1(x,y,z,w,t;\mesonic) = 
\sum_{n=0}^{\infty}
[n;n;n;n]t^{4n}
~~,
\eea
where $[n;n;n;n]\equiv [n]_{SU(2)_x}[n]_{SU(2)_y}[n]_{SU(2)_z}[n]_{SU(2)_w}$ is the character of the irreducible representation of $SU(2)^4$.

The toric diagram of $\mesonic$ is given by
\beal{es19i10}
G_t=\left(
\begin{array}{cccccccc}
a_1 & a_2 & b_1 & b_2 & c_1 & c_2 & d_1 & d_2\\
\hline
 1 & 0 & 1 & 0 & 1 & 0 & 1 & 0 \\
 1 &-1 & 0 & 0 & 0 & 0 & 0 & 0 \\
 0 & 0 & 1 & -1 & 0 & 0 & 0 & 0 \\
 0 & 0 & 0 & 0 & 1 & -1 & 0 & 0 \\
 0 & 0 & 0 & 0 & 0 & 0 & 1 & -1
 \end{array}
\right)~~.
\eea
\\

\subsection{8 Fields, 4 Superpotential Terms, 2 Gauge Groups \label{s_m6}}

\subsubsection{Model 8.4a: $\mathcal{M}_{3,3}$ \label{s_m6n1}}

\begin{figure}[ht!!]
\begin{center}
\includegraphics[trim=0cm 0cm 0cm 0cm,totalheight=8 cm]{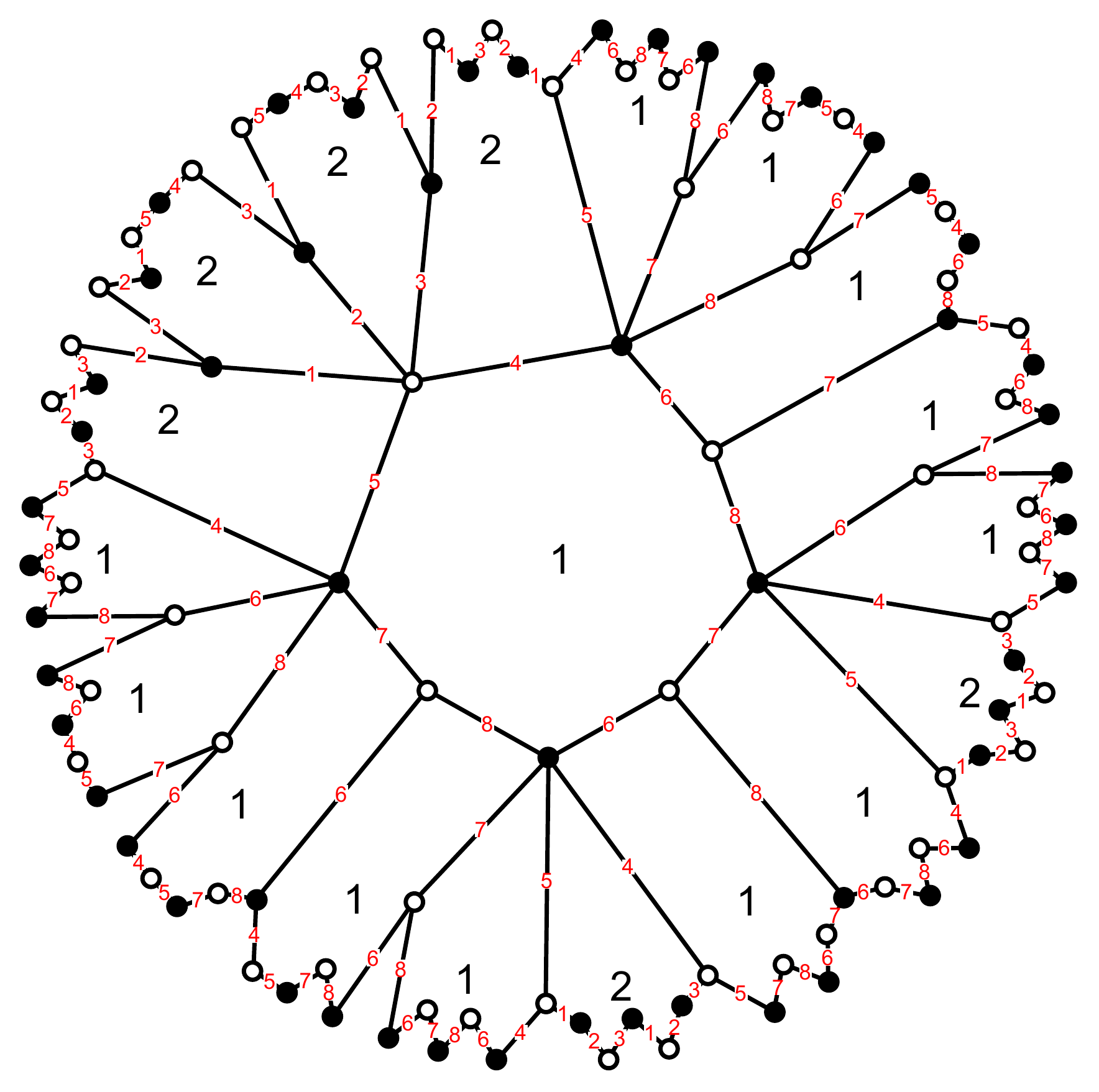}\\
\vspace{0.3cm}
\begin{tabular}{cccccccc}
{\color[rgb]{1.000000,0.000000,0.000000} 1} &
{\color[rgb]{1.000000,0.000000,0.000000} 2} &
{\color[rgb]{1.000000,0.000000,0.000000} 3} &
{\color[rgb]{1.000000,0.000000,0.000000} 4} &
{\color[rgb]{1.000000,0.000000,0.000000} 5} &
{\color[rgb]{1.000000,0.000000,0.000000} 6} &
{\color[rgb]{1.000000,0.000000,0.000000} 7} &
{\color[rgb]{1.000000,0.000000,0.000000} 8} 
\\
$X_{22}^{1}$ & $X_{22}^{2}$ & $X_{22}^{3}$ & $X_{21}$ & $X_{12}$ & $X_{11}^{1}$ & $X_{11}^{2}$ & $X_{11}^{3}$
\end{tabular}
\caption{The Model 8.4a brane tiling on a $g=2$ Riemann surface with 2 gauge groups, 8 fields and 4 superpotential terms.
\label{fm6n1t}}
 \end{center}
 \end{figure}

\begin{figure}[ht!!]
\begin{center}
\includegraphics[trim=0cm 0cm 0cm 0cm,totalheight=1 cm]{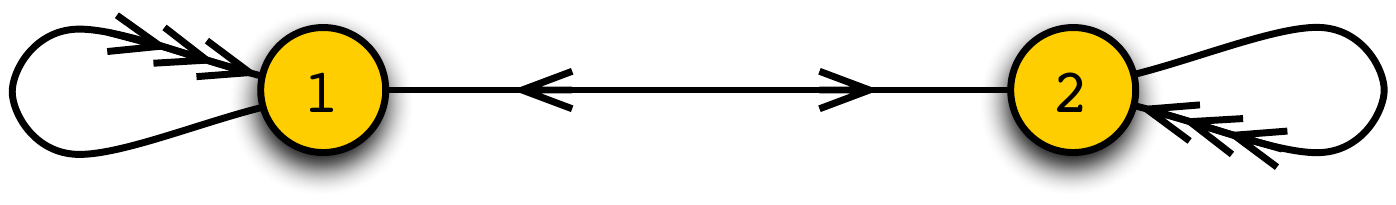}
\caption{The quiver diagram for Model 8.4a, a brane tiling on a $g=2$ Riemann surface with 2 gauge groups, 8 fields and 4 superpotential terms.
\label{fm6n1q}}
 \end{center}
 \end{figure}
 
The brane tiling on a $g=2$ Riemann surface and the corresponding quiver are shown in \fref{fm6n1t} and \fref{fm6n1q} respectively. The quartic superpotential is
\beal{e40}
W= 
+ X_{22}^{1} X_{22}^{2} X_{22}^{3} X_{21} X_{12}
+ X_{11}^{1} X_{11}^{2} X_{11}^{3}
- X_{21} X_{11}^{1} X_{11}^{3} X_{11}^{2} X_{12}
- X_{22}^{1} X_{22}^{3} X_{22}^{2}
~~.
\nn\\
\eea
The quiver incidence matrix is
\beal{e41}
d=\left(
\begin{array}{cccccccc}
X_{22}^{1} & X_{22}^{2} & X_{22}^{3} & X_{21} & X_{12} & X_{11}^{1} & X_{11}^{2} & X_{11}^{3} \\
\hline
 0 & 0 & 0 & 1 & -1 & 0 & 0 & 0 \\
 0 & 0 & 0 & -1 & 1 & 0 & 0 & 0
\end{array}
\right)
~~.
\eea

The brane tiling has $c=9$ perfect matchings. The perfect matchings are encoded in the matrix
\beal{e42}
P=\left(
\begin{array}{c|ccccccccc}
\; & p_1 & p_2 & p_3 & p_4 & p_5 & p_6 & p_7 & p_8 & p_9 \\
\hline
X_{22}^{1} &  1 & 0 & 0 & 1 & 0 & 0 & 1 & 0 & 0 \\
X_{22}^{2} &  0 & 1 & 0 & 0 & 1 & 0 & 0 & 1 & 0 \\
X_{22}^{3} &  0 & 0 & 1 & 0 & 0 & 1 & 0 & 0 & 1 \\
X_{21} &  0 & 0 & 0 & 0 & 0 & 0 & 0 & 0 & 0 \\
X_{12} &  0 & 0 & 0 & 0 & 0 & 0 & 0 & 0 & 0 \\
X_{11}^{1} &  1 & 1 & 1 & 0 & 0 & 0 & 0 & 0 & 0 \\
X_{11}^{2} &  0 & 0 & 0 & 1 & 1 & 1 & 0 & 0 & 0 \\
X_{11}^{3} &  0 & 0 & 0 & 0 & 0 & 0 & 1 & 1 & 1
\end{array}
\right)
~~.
\eea
The brane tiling has the following zig-zag paths,
\beal{e42bb}
&
\eta_1= (X_{22}^{1},X_{22}^{2})~,~
\eta_2= (X_{22}^{2},X_{22}^{2})~,~
\eta_3= (X_{21},X_{12})~,~
\eta_4= (X_{11}^{1},X_{11}^{3})~,~
&\nn\\
&
\eta_5= (X_{11}^{2},X_{11}^{3})~,~
\eta_6= (X_{22}^{1},X_{22}^{3},X_{21},X_{11}^{1},X_{11}^{2},X_{12})~.~
&
\eea

The F-term constraints can be expressed as charges carried by the perfect matchings. The charges are given by
\beal{e43}
Q_F =
\left(
\begin{array}{ccccccccc}
p_1 & p_2 & p_3 & p_4 & p_5 & p_6 & p_7 & p_8 & p_9\\
\hline
 1 & 0 & -1 & 0 & 0 & 0 & -1 & 0 & 1 \\
 0 & 1 & -1 & 0 & 0 & 0 & 0 & -1 & 1 \\
 0 & 0 & 0 & 1 & 0 & -1 & -1 & 0 & 1 \\
 0 & 0 & 0 & 0 & 1 & -1 & 0 & -1 & 1
\end{array}
\right)
~~.
\eea
There are no D-term constraints.
The mesonic moduli space can be expressed as the symplectic quotient
\beal{es46}
\mesonic = \mathbb{C}^{9} // Q_F ~~.
\eea

By associating the fugacity $t_i$ to the perfect matching $p_i$, the fully refined Hilbert series of $\mesonic$ is given by the following Molien integral
\beal{e47}
g_1(t_i;\mesonic) &=& 
\frac{1}{(2\pi i)^5}
\oint_{|z_1|=1} 
\frac{\ud z_1}{z_1} 
\oint_{|z_2|=1} 
\frac{\ud z_2}{z_2}
\oint_{|z_3|=1} 
\frac{\ud z_3}{z_3}
\oint_{|z_4|=1} 
\frac{\ud z_4}{z_4}
~~
\nn\\
&& 
\hspace{1cm}
\times
\frac{1}{
(1-z_1 t_1)
(1-z_2 t_2)
(1-z_1^{-1} z_2^{-1} t_3)
}
\nn\\
&& 
\hspace{1cm}
\times
\frac{1}{
(1-z_3 t_4)
(1-z_4 t_5)
(1-z_3^{-1} z_4^{-1} t_6)
}
\nn\\
&& 
\hspace{1cm}
\times
\frac{1}{
(1-z_1^{-1} z_3^{-1} t_7)
(1-z_2^{-1} z_4^{-1} t_8)
(1-z_1 z_2 z_3 z_4 t_9)
}
\nn\\
&=&
\frac{
1 - t_1 t_2 t_3 t_4 t_5 t_6 t_7 t_8 t_9
}{
(1 - t_1 t_2 t_3) (1 - t_4 t_5 t_6) (1 - t_7 t_8 t_9)
(1 - t_1 t_4 t_7) (1 - t_2 t_5 t_8) (1 - t_3 t_6 t_9) 
}
~~.
\nn\\
\eea
Accordingly, the mesonic moduli space is a complete intersection of dimension $5$. It is a Calabi-Yau 5-fold and its generators can be written in terms of perfect matching variables as follows:

\begin{center}
\begin{tabular}{|c|c|}
\hline
generator & perfect matchings\\
\hline\hline
$A_1$ & $p_1 p_2 p_3$\\
$A_2$ & $p_4 p_5 p_6$\\
$A_3$ & $p_7 p_8 p_9$\\
$B_1$ & $p_1 p_4 p_7$ \\
$B_2$ & $p_2 p_5 p_8$ \\
$B_3$ & $p_3 p_6 p_9$ \\
\hline
\end{tabular}
\end{center}
The generators form a single relation of the form
\beal{es48}
A_1 A_2 A_3 = B_1 B_2 B_3~~.
\eea

The global symmetry is $U(1)^4 \times U(1)_R$ and experiences no enhancement. The toric diagram of the Calabi-Yau 5-fold is given by
\beal{e47bb}
G_t = 
\left(
\begin{array}{ccccccccc}
p_1 & p_2 & p_3 & p_4 & p_5 & p_6 & p_7 & p_8 & p_9 \\
\hline
 1 & 0 & 0 & 1 & 0 & 0 & 1 & 0 & 0 \\
 0 & 1 & 0 & 0 & 1 & 0 & 0 & 1 & 0 \\
 0 & 0 & 1 & 0 & 0 & 1 & 0 & 0 & 1 \\
 1 & 1 & 1 & 0 & 0 & 0 & 0 & 0 & 0 \\
 0 & 0 & 0 & 1 & 1 & 1 & 0 & 0 & 0 
\end{array}
\right)~~.
\eea
\\

\subsubsection{Model 8.4b: $\mathbb{C}^3\times\mathbb{C}^2/\mathbb{Z}_2$ \label{s_m84b}}

\begin{figure}[ht!!]
\begin{center}
\includegraphics[trim=0cm 0cm 0cm 0cm,totalheight=8 cm]{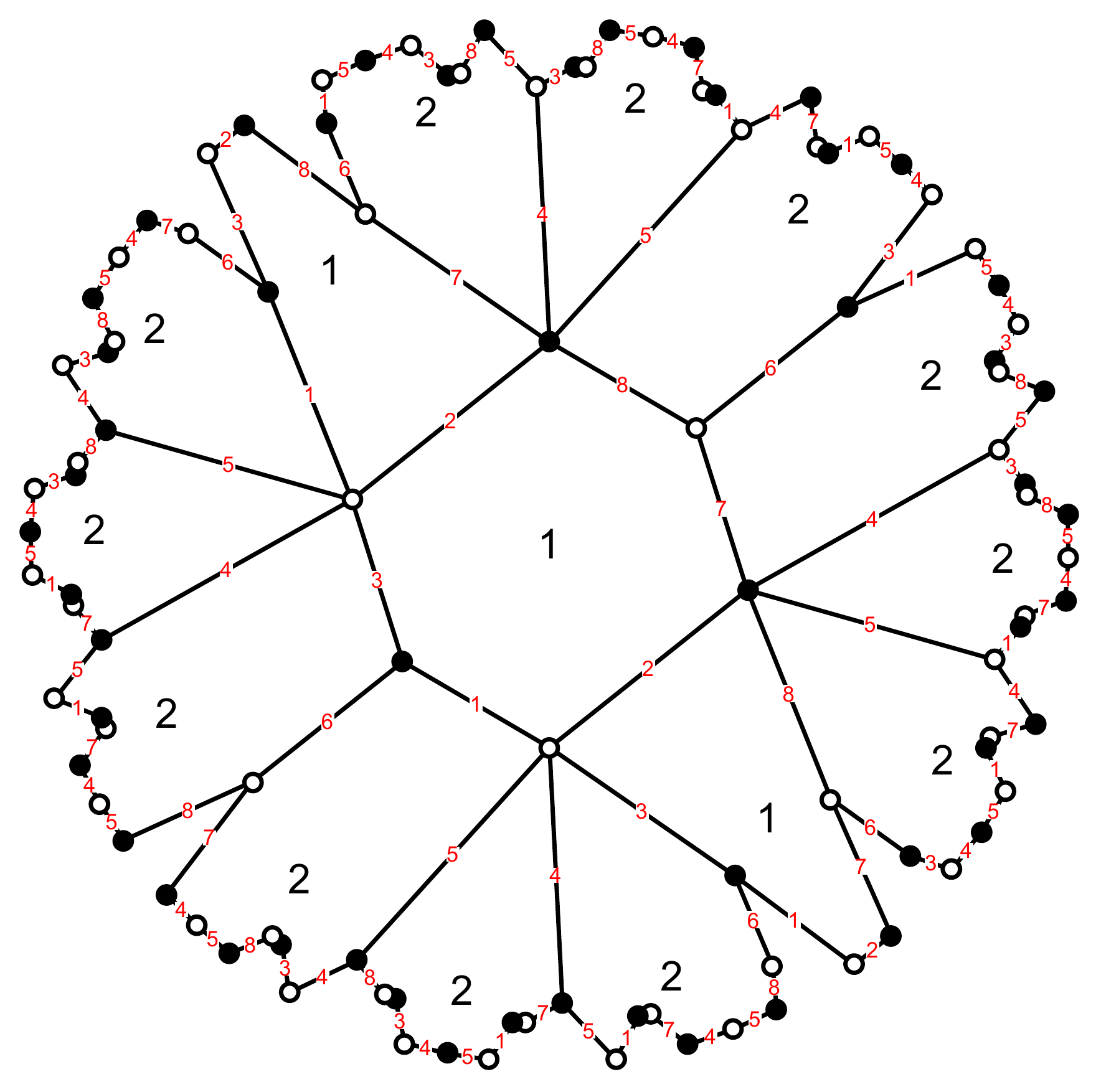}\\
\vspace{0.3cm}
\begin{tabular}{cccccccc}
{\color[rgb]{1.000000,0.000000,0.000000} 1} &
{\color[rgb]{1.000000,0.000000,0.000000} 2} &
{\color[rgb]{1.000000,0.000000,0.000000} 3} &
{\color[rgb]{1.000000,0.000000,0.000000} 4} &
{\color[rgb]{1.000000,0.000000,0.000000} 5} &
{\color[rgb]{1.000000,0.000000,0.000000} 6} &
{\color[rgb]{1.000000,0.000000,0.000000} 7} &
{\color[rgb]{1.000000,0.000000,0.000000} 8} 
\\
$X_{21}^{1}$ & $X_{11}$ & $X_{12}^{1}$ & $X_{22}^{1}$ & $X_{22}^{2}$ & $X_{22}^{3}$ & $X_{21}^{2}$ & $X_{12}^{2}$
\end{tabular}
\caption{The Model 8.4b brane tiling on a $g=2$ Riemann surface with 2 gauge groups, 8 fields and 4 superpotential terms.
\label{fm6n2t}}
 \end{center}
 \end{figure}

\begin{figure}[ht!!]
\begin{center}
\includegraphics[trim=0cm 0cm 0cm 0cm,totalheight=1 cm]{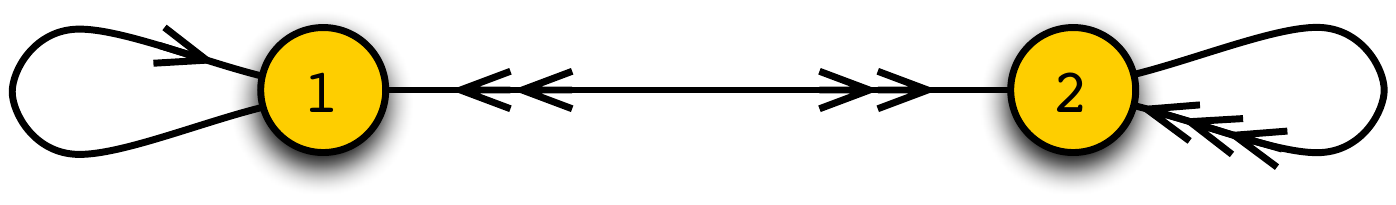}
\caption{The quiver diagram for Model 8.4b, a brane tiling on a $g=2$ Riemann surface with 2 gauge groups, 8 fields and 4 superpotential terms.
\label{fm6n2q}}
 \end{center}
 \end{figure}
 
For Model 8.4b, the brane tiling and corresponding quiver is shown in \fref{fm6n2t} and \fref{fm6n2q} respectively. The quartic superpotential is
\beal{e50}
W= 
+ X_{21}^{1} X_{11} X_{12}^{1} X_{22}^{1} X_{22}^{2}
+ X_{22}^{3} X_{21}^{2} X_{12}^{2}
- X_{11} X_{12}^{2} X_{22}^{2} X_{22}^{1} X_{21}^{2}
- X_{21}^{1} X_{12}^{1} X_{22}^{3}
~~.
\nn\\
\eea
The quiver incidence matrix is
\beal{e51}
d=\left(
\begin{array}{cccccccc}
X_{21}^{1} & X_{11} & X_{12}^{1} & X_{22}^{1} & X_{22}^{2} &
X_{22}^{3} & X_{21}^{2} & X_{12}^{2}
\\
\hline
 1 & 0 & -1 & 0 & 0 & 0 & 1 & -1 \\
 -1 & 0 & 1 & 0 & 0 & 0 & -1 & 1
\end{array}
\right)
~~.
\eea

The brane tiling has $c=7$ perfect matchings. The perfect matchings are encoded in the matrix
\beal{e52}
P=\left(
\begin{array}{c|ccccccc}
\; & a_1 & a_2 & a_3 & b_1 & b_2 & p_1 & p_2 \\
\hline
X_{11} & 1 & 0 & 0 & 0 & 0 & 0 & 0 \\
X_{22}^{1} & 0 & 1 & 0 & 0 & 0 & 0 & 0 \\
X_{22}^{2} & 0 & 0 & 1 & 0 & 0 & 0 & 0 \\
X_{22}^{3} & 1 & 1 & 1 & 0 & 0 & 0 & 0 \\
X_{12}^{1} & 0 & 0 & 0 & 0 & 1 & 1 & 0 \\
X_{12}^{2} & 0 & 0 & 0 & 1 & 0 & 1 & 0 \\
X_{21}^{1} & 0 & 0 & 0 & 1 & 0 & 0 & 1 \\
X_{21}^{2} & 0 & 0 & 0 & 0 & 1 & 0 & 1
\end{array}
\right)
~~.
\eea
The brane tiling has the zig-zag paths,
\beal{e52bb}
&
\eta_1 = (X_{22}^{1},X_{22}^{2})~,~
\eta_2 = (X_{11},X_{12}^{1},X_{22}^{3},X_{21}^{2})~,~
\eta_3 = (X_{21}^{1},X_{11},X_{12}^{2},X_{22}^{3})~,~
&\nn\\
&
\eta_4 = (X_{21}^{1},X_{12}^{1},X_{22}^{1},X_{21}^{2},X_{12}^{2},X_{22}^{2})~.~
&
\eea

The F-term constraints can be expressed as charges carried by the perfect matchings. The charges are given by
\beal{e53}
Q_F =
\left(
\begin{array}{ccccccc}
a_1 & a_2 & a_3 & b_1 & b_2 & p_1 & p_2 \\
\hline
 0 & 0 & 0 & 1 & 1 & -1 & -1
\end{array}
\right)
~~.
\eea
The D-term charges are encoded in the quiver incidence matrix $d$ and are
\beal{e54}
Q_D = 
\left(
\begin{array}{ccccccc}
a_1 & a_2 & a_3 & b_1 & b_2 & p_1 & p_2 \\
\hline
 0 & 0 & 0 & 1 & 1 & 0 & -2
\end{array}
\right)~~.
\eea
The combined charges can be written as
\beal{e54}
Q_t = 
\left(
\begin{array}{ccccccc}
a_1 & a_2 & a_3 & b_1 & b_2 & p_1 & p_2 \\
\hline
 0 & 0 & 0 & 1 & 1 & -1 & -1 \\
 0 & 0 & 0 & 0 & 0 & 1 & -1
\end{array}
\right)~~,
\eea
where the mesonic moduli space can be expressed as the symplectic quotient
\beal{es56}
\mesonic = \mathbb{C}^{7} // Q_t ~~.
\eea

By associating to perfect matchings $a_i,b_i,p_i$ the fugacities $\alpha_i,\beta_i,t_i$, the fully refined Hilbert series of $\mesonic$ is given by the following Molien integral
\beal{e57}
g_1(\alpha_i,\beta_i,t_i;\mesonic) &=& 
\frac{1}{(2\pi i)^5}
\oint_{|z_1|=1} 
\frac{\ud z_1}{z_1} 
\oint_{|z_2|=1} 
\frac{\ud z_2}{z_2}
~~
\frac{1}{
\prod_{i=1}^{3} (1-\alpha_i) 
\prod_{i=1}^{2} (1-z_1 \beta_i)
}
\nn\\
&& 
\hspace{1cm}
\times
\frac{1}{
(1-z_1^{-1} z_2  t_1)
(1-z_1^{-1} z_2^{-1} t_2)
}
\nn\\
&=&
\frac{1}{\prod_{i=1}^{3} (1-\alpha_i)}\times
\frac{
1-\prod_{i=1}^{2} \beta_i^2 t_i^2
}{
(1-\beta_1\beta_2 t_1 t_2)
\prod_{i=1}^{2} (1-\beta_i^2 t_1 t_2)
}
~~.
\nn\\
\eea
Accordingly, the mesonic moduli space is a complete intersection of dimension $5$. It is a Calabi-Yau 5-fold and its generators can be found in terms of perfect matching variables as follows:

\begin{center}
\begin{tabular}{|c|c|}
\hline
generator & perfect matchings\\
\hline\hline
$A_i$ & $a_i$ \\
\hline
$B_{ij}$ & $b_i b_j p_1 p_2$\\
\hline
\end{tabular}
\end{center}
$A_i$ generate $\mathbb{C}^3$ and $B_{ij}$ form a single relation of $\mathbb{C}^2/\mathbb{Z}_2$ which can be expressed as 
\beal{es48}
\det B = 0 ~~.
\eea

The global symmetry is $SU(3) \times SU(2) \times U(1) \times U(1)_R$. The perfect matchings carry the global symmetry charges as follows.
\begin{center}
\begin{tabular}{|c|cccc|l|}
\hline
\; & $SU(3)_x$ & $SU(2)_y$ & $U(1)_{h}$ & $U(1)_R$ & fugacity \\
\hline \hline
$a_1$ & $(1,0)$ & 0 & 0 & 1 & $\alpha_1=x_1 t$\\
$a_2$ & $(-1,1)$ & 0 & 0 & 1 & $\alpha_2=x_1^{-1} x_2 t$\\
$a_3$ & $(0,-1)$ & 0 & 0 & 1 & $\alpha_3=x_2^{-1} t$\\
\hline
$b_1$ & $(0,0)$ &  1 & 0 & 1 & $\beta_1=y t$\\
$b_2$ & $(0,0)$ & -1 & 0 & 1 & $\beta_2=y^{-1} t$\\
\hline
$p_1$ & $(0,0)$ & 0 &  1 & 1 & $t_1=h t$\\
$p_2$ & $(0,0)$ & 0 & -1 & 1 & $t_2=h^{-1} t$\\
\hline
\end{tabular}
\end{center}
Under the above assignment of global charges the refined Hilbert series of the mesonic moduli space can be written as
\beal{es48bb}
g_1(x_i,y,t;\mesonic) &=& 
\sum_{n_1=0}^{\infty}
\sum_{n_2=0}^{\infty}
[n_1,0;n_2] t^{n_1+4n_2}~~,
\eea
where $[n_1,0;n_2]\equiv [n_1,0]_{SU(3)_x} [n_2]_{SU(2)_y}$ are characters of irreducible representations of $SU(3)_{x} \times SU(2)_{y}$.

The toric diagram is given by
\beal{e54nn}
G_t = 
\left(
\begin{array}{ccccccc}
a_1 & a_2 & a_3 & b_1 & b_2 & p_1 & p_2 \\
\hline
 1 & 0 & 0 & 0 & 0 & 0 & 0 \\
 0 & 1 & 0 & 0 & 0 & 0 & 0 \\
 0 & 0 & 1 & 0 & 0 & 0 & 0 \\
 0 & 0 & 0 & 1 & -1 & 0 & 0 \\
 0 & 0 & 0 & 0 & 2 & 1 & 1
\end{array}
\right)~~.
\eea
\\

\subsubsection{Model 8.4c: $\mathbb{C}^3\times\mathbb{C}^2/\mathbb{Z}_2$ \label{s_m84c}}

\begin{figure}[ht!!]
\begin{center}
\includegraphics[trim=0cm 0cm 0cm 0cm,totalheight=8 cm]{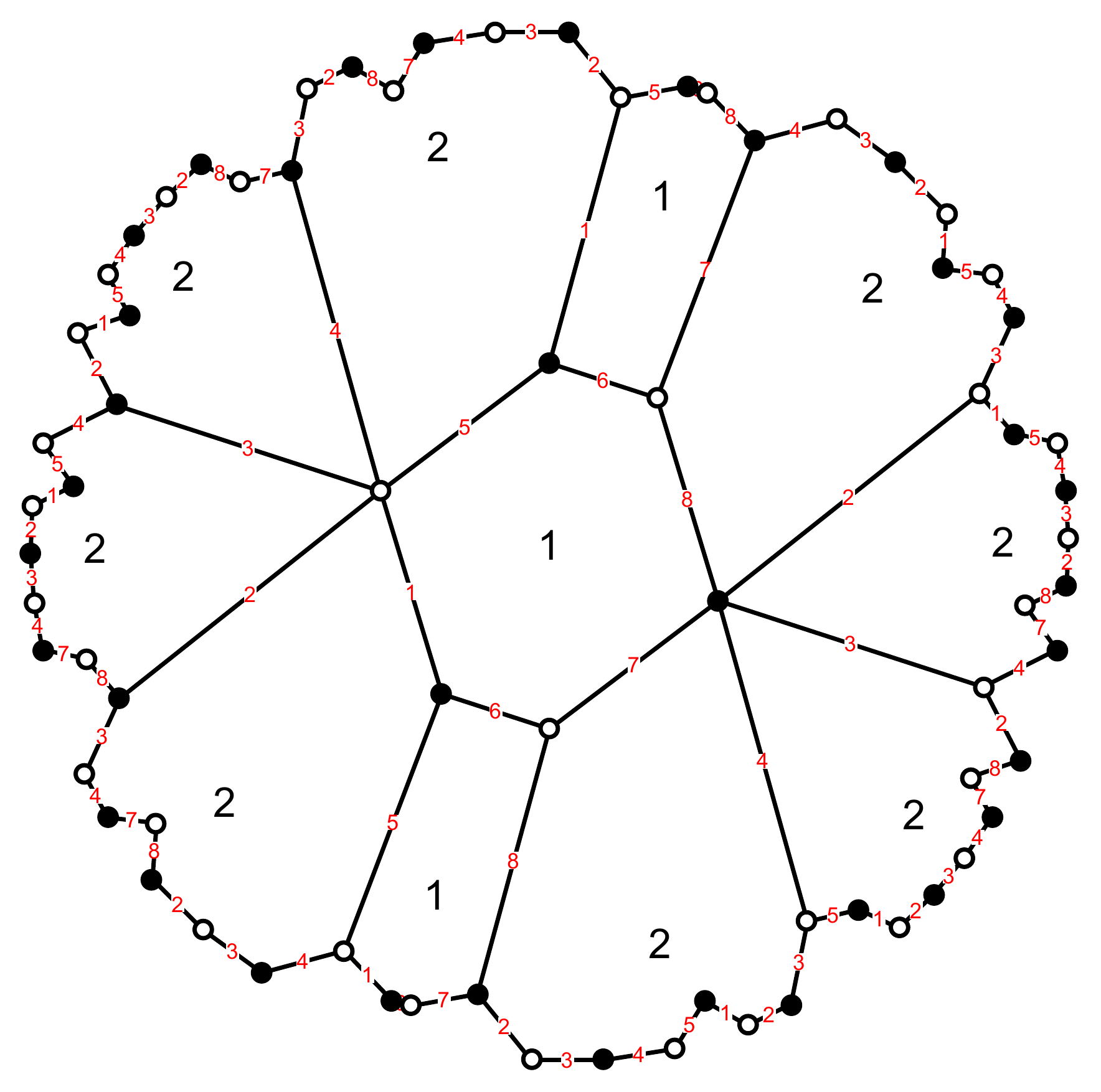}\\
\vspace{0.3cm}
\begin{tabular}{cccccccc}
{\color[rgb]{1.000000,0.000000,0.000000} 1} &
{\color[rgb]{1.000000,0.000000,0.000000} 2} &
{\color[rgb]{1.000000,0.000000,0.000000} 3} &
{\color[rgb]{1.000000,0.000000,0.000000} 4} &
{\color[rgb]{1.000000,0.000000,0.000000} 5} &
{\color[rgb]{1.000000,0.000000,0.000000} 6} &
{\color[rgb]{1.000000,0.000000,0.000000} 7} &
{\color[rgb]{1.000000,0.000000,0.000000} 8} 
\\
$X_{12}^{1}$ & $X_{22}^{1}$ & $X_{22}^{2}$ & $X_{22}^{3}$ & $X_{21}^{1}$ & $X_{11}$ & $X_{12}^{2}$ & $X_{21}^{2}$
\end{tabular}
\caption{The Model 8.4c brane tiling on a $g=2$ Riemann surface with 2 gauge groups, 8 fields and 4 superpotential terms.
\label{fm6n3t}}
 \end{center}
 \end{figure}

\begin{figure}[ht!!]
\begin{center}
\includegraphics[trim=0cm 0cm 0cm 0cm,totalheight=1 cm]{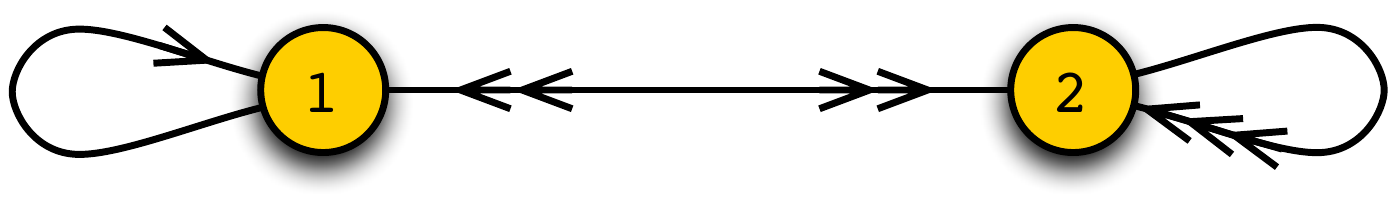}
\caption{The quiver diagram for Model 8.4c, a brane tiling on a $g=2$ Riemann surface with 2 gauge groups, 8 fields and 4 superpotential terms.
\label{fm6n3q}}
 \end{center}
 \end{figure}
 
For Model 8.4c, the brane tiling and corresponding quiver is shown in \fref{fm6n3t} and \fref{fm6n3q} respectively. The quartic superpotential is
\beal{e60}
W= 
+ X_{12}^{1} X_{22}^{1} X_{22}^{2} X_{22}^{3} X_{21}^{1}
+ X_{11} X_{12}^{2} X_{21}^{2}
- X_{22}^{1} X_{21}^{2} X_{12}^{2} X_{22}^{3} X_{22}^{2}
- X_{11} X_{12}^{1} X_{21}^{1}
~~.
\nn\\
\eea
The quiver incidence matrix is
\beal{e61}
d=\left(
\begin{array}{cccccccc}
X_{12}^{1} & X_{22}^{1} & X_{22}^{2} & X_{22}^{3} & X_{21}^{1} & X_{11} & X_{12}^{2} & X_{21}^{2}
\\
\hline
 -1 & 0 & 0 & 0 & 1 & 0 & -1 & 1 \\
 1 & 0 & 0 & 0 & -1 & 0 & 1 & -1
\end{array}
\right)
~~.
\eea

The brane tiling has $c=7$ perfect matchings. The perfect matchings are encoded in the matrix
\beal{e62}
P=\left(
\begin{array}{c|ccccccc}
\; & a_1 & a_2 & a_3 & b_1 & b_2 & p_1 & p_2\\
\hline
X_{12}^{1} & 0 & 0 & 0 & 1 & 0 & 0 & 1 \\
X_{12}^{2} & 0 & 0 & 0 & 0 & 1 & 0 & 1 \\
X_{21}^{1} & 0 & 0 & 0 & 0 & 1 & 1 & 0 \\
X_{21}^{2} & 0 & 0 & 0 & 1 & 0 & 1 & 0\\
X_{11} & 1 & 1 & 1 & 0 & 0 & 0 & 0 \\
X_{22}^{1} & 1 & 0 & 0 & 0 & 0 & 0 & 0 \\
X_{22}^{2} & 0 & 1 & 0 & 0 & 0 & 0 & 0 \\
X_{22}^{3} & 0 & 0 & 1 & 0 & 0 & 0 & 0 
\end{array}
\right)
~~.
\eea
The zig-zag paths of the brane tiling of Model 8.4c are
\beal{e62bb}
&
\eta_1=(X_{12}^{1},X_{21}^{1})~,~
\eta_2=(X_{22}^{1},X_{22}^{3})~,~
\eta_3=(X_{22}^{2},X_{22}^{3})~,~
\eta_4=(X_{12}^{2},X_{21}^{2})~,~
&\nn\\
&
\eta_5=(X_{22}^{3},X_{21}^{1},X_{11},X_{12}^{2})~,~
\eta_6=(X_{12}^{1},X_{22}^{1},X_{21}^{2},X_{11})~.~
&
\eea
As we will see below, and seen above with the quiver diagram, Model 8.4c has many similar properties as Model 8.4b in section \sref{s_m84b}. The zig-zag paths of Model 8.4c in \eref{e62bb} are however distinct from the ones for Model 8.4b in \eref{e52bb}.

The F-term constraints can be expressed as charges carried by the perfect matchings. The charges are given by
\beal{e63}
Q_F =
\left(
\begin{array}{ccccccc}
a_1 & a_2 & a_3 & b_1 & b_2 & p_1 & p_2\\
\hline
 0 & 0 & 0 & 1 & 1 & -1 & -1
\end{array}
\right)
~~.
\eea
The D-term charges are encoded in the quiver incidence matrix $d$ and are
\beal{e64}
Q_D = 
\left(
\begin{array}{ccccccc}
a_1 & a_2 & a_3 & b_1 & b_2 & p_1 & p_2\\
\hline
 0 & 0 & 0 & 1 & 1 & -2 & 0
\end{array}
\right)~~.
\eea
The charges can be combined to give
\beal{e64b}
Q_t = 
\left(
\begin{array}{ccccccc}
a_1 & a_2 & a_3 & b_1 & b_2 & p_1 & p_2\\
\hline
 0 & 0 & 0 & 1 & 1 & -1 & -1 \\
 0 & 0 & 0 & 0 & 0 & 1 & -1
\end{array}
\right)~~,
\eea
which is precisely the total charge matrix for Model 8.4b in \sref{s_m84b}.

Accordingly, the mesonic moduli space as the following symplectic quotient
\beal{es66}
\mesonic = \mathbb{C}^{7} // Q_t ~~,
\eea
is identical to the one in Model 8.4b. The mesonic moduli space is $\mathbb{C}^3\times\mathbb{C}^2/\mathbb{Z}_2$ which is a toric Calabi-Yau 5-fold.
\\

\subsubsection{Model 8.4d: $\mathbb{C}\times\mathcal{M}_{3,2}$ \label{s_m84d}}

\begin{figure}[ht!!]
\begin{center}
\includegraphics[trim=0cm 0cm 0cm 0cm,totalheight=8 cm]{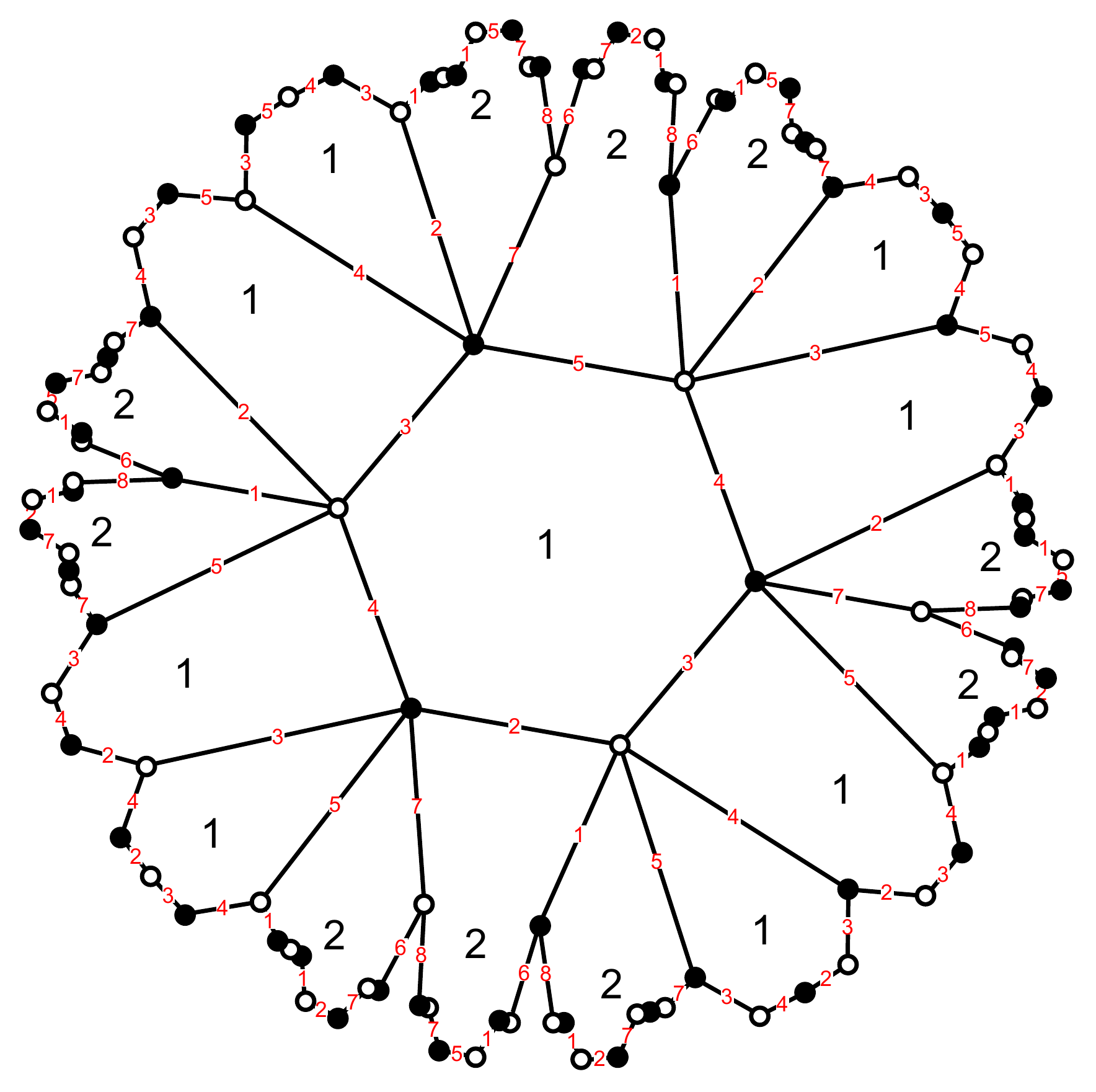}\\
\vspace{0.3cm}
\begin{tabular}{cccccccc}
{\color[rgb]{1.000000,0.000000,0.000000} 1} &
{\color[rgb]{1.000000,0.000000,0.000000} 2} &
{\color[rgb]{1.000000,0.000000,0.000000} 3} &
{\color[rgb]{1.000000,0.000000,0.000000} 4} &
{\color[rgb]{1.000000,0.000000,0.000000} 5} &
{\color[rgb]{1.000000,0.000000,0.000000} 6} &
{\color[rgb]{1.000000,0.000000,0.000000} 7} &
{\color[rgb]{1.000000,0.000000,0.000000} 8} 
\\
$X_{22}^{1}$ & $X_{21}$ & $X_{11}^{1}$ & $X_{11}^{2}$ & $X_{12}$ & $X_{22}^{2}$ & $X_{22}^{3}$ & $X_{22}^{4}$
\end{tabular}
\caption{The Model 8.4d brane tiling on a $g=2$ Riemann surface with 2 gauge groups, 8 fields and 4 superpotential terms.
\label{fm6n4t}}
 \end{center}
 \end{figure}

\begin{figure}[ht!!]
\begin{center}
\includegraphics[trim=0cm 0cm 0cm 0cm,totalheight=1 cm]{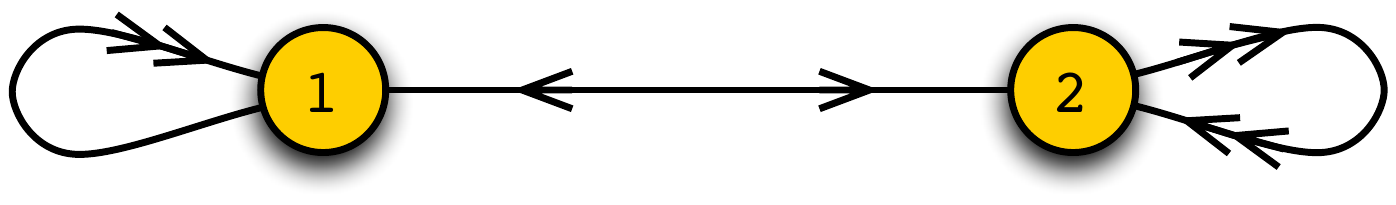}
\caption{The quiver diagram for Model 8.4d, a brane tiling on a $g=2$ Riemann surface with 2 gauge groups, 8 fields and 4 superpotential terms.
\label{fm6n4q}}
 \end{center}
 \end{figure}
 
The brane tiling and corresponding quiver for Model 8.4d is shown in \fref{fm6n4t} and \fref{fm6n4q} respectively. The superpotential is
\beal{e80}
W= 
+ X_{22}^{1} X_{21} X_{11}^{1} X_{11}^{2} X_{12}
+ X_{22}^{2} X_{22}^{3} X_{22}^{4}
- X_{21} X_{11}^{2} X_{11}^{1} X_{12} X_{22}^{2}
- X_{22}^{1} X_{22}^{2} X_{22}^{4}
~~.
\nn\\
\eea
The quiver incidence matrix is
\beal{e81}
d=\left(
\begin{array}{cccccccc}
X_{22}^{1} & X_{21} & X_{11}^{1} & X_{11}^{2} & X_{12} & X_{22}^{2} & X_{22}^{3} & X_{22}^{4}
\\
\hline
 0 & 1 & 0 & 0 & -1 & 0 & 0 & 0 \\
 0 & -1 & 0 & 0 & 1 & 0 & 0 & 0
\end{array}
\right)
~~.
\eea

Model 8.4d's brane tiling has $c=9$ perfect matchings. The perfect matchings are encoded in the matrix
\beal{e82}
P=\left(
\begin{array}{c|ccccccccc}
\; & p_1 & p_2 & p_3 & p_4 & p_5 & p_6 & p_7 & p_8 & p_9\\
\hline
X_{22}^{1} &  0 & 0 & 0 & 0 & 1 & 0 & 0 & 0 & 0 \\
X_{21} &  1 & 0 & 0 & 0 & 0 & 1 & 0 & 0 & 0 \\
X_{11}^{1} &  0 & 1 & 0 & 0 & 0 & 0 & 1 & 0 & 0 \\
X_{11}^{2} &  0 & 0 & 1 & 0 & 0 & 0 & 0 & 1 & 0 \\
X_{12} &  0 & 0 & 0 & 1 & 0 & 0 & 0 & 0 & 1 \\
X_{22}^{2} &  1 & 1 & 1 & 1 & 0 & 0 & 0 & 0 & 0 \\
X_{22}^{3} &  0 & 0 & 0 & 0 & 1 & 0 & 0 & 0 & 0 \\
X_{22}^{4} &  0 & 0 & 0 & 0 & 0 & 1 & 1 & 1 & 1
\end{array}
\right)
~~.
\eea
The brane tiling has the following zig-zag paths,
\beal{e82bb}
&
\eta_1=(X_{11}^{1},X_{11}^{2})~,~
\eta_2=(X_{22}^{2},X_{22}^{4})~,~
&\nn\\
&
\eta_3=(X_{22}^{1},X_{21},X_{11}^{2},X_{12},X_{22}^{3},X_{22}^{4})~,~
\eta_4=(X_{22}^{1},X_{22}^{2},X_{22}^{3},X_{21},X_{11}^{1},X_{12})~.~
&
\eea

The F-term charge matrix is
\beal{e83}
Q_F =
\left(
\begin{array}{ccccccccc}
p_1 & p_2 & p_3 & p_4 & p_5 & p_6 & p_7 & p_8 & p_9 \\
\hline
 1 & 0 & 0 & -1 & 0 & -1 & 0 & 0 & 1 \\
 0 & 1 & 0 & -1 & 0 & 0 & -1 & 0 & 1 \\
 0 & 0 & 1 & -1 & 0 & 0 & 0 & -1 & 1
\end{array}
\right)
~~.
\eea
The D-term charge matrix is
\beal{e84}
Q_D =
\left(
\begin{array}{ccccccccc}
p_1 & p_2 & p_3 & p_4 & p_5 & p_6 & p_7 & p_8 & p_9 \\
\hline
 1 & 0 & 0 & -1 & 0 & 0 & 0 & 0 & 0
\end{array}
\right)
~~.
\eea

The mesonic moduli space of Model 8.4d in terms of a symplectic quotient is
\beal{es86}
\mesonic = \mathbb{C}^{9} // Q_t ~~.
\eea

By associating the fugacity $t_i$ to the perfect matching $p_i$, the fully refined Hilbert series of $\mesonic$ is given by the following Molien integral
\beal{e87}
g_1(t_i;\mesonic) &=& 
\frac{1}{(2\pi i)^4}
\oint_{|z_1|=1} 
\frac{\ud z_1}{z_1} 
\oint_{|z_2|=1} 
\frac{\ud z_2}{z_2}
\oint_{|z_3|=1} 
\frac{\ud z_3}{z_3}
\oint_{|z_4|=1} 
\frac{\ud z_4}{z_4}
~~
\nn\\
&& 
\hspace{1cm}
\times
\frac{1}{
(1-z_1 z_4 t_1)
(1-z_2 t_2)
(1-z_3 t_3)
(1-z_1^{-1} z_2^{-1} z_3^{-1} z_4^{-1} t_4)
}
\nn\\
&& 
\hspace{1cm}
\times
\frac{1}{
(1-t_5)
(1-z_1^{-1}t_6)
(1-z_2^{-1} t_7)
(1-z_3^{-1} t_8)
(1-z_1 z_2 z_3 t_9)
}
\nn\\
&=&
\frac{
1 - t_1 t_2 t_3 t_4 t_6 t_7 t_8 t_9
}{
(1-  t_5) 
(1-  t_1 t_4 t_6 t_9) (1-  t_2 t_7) (1-  t_3 t_8)  
(1-  t_1 t_2 t_3 t_4) (1-  t_6 t_7 t_8 t_9)
}
~~.
\nn\\
\eea

From the Hilbert series, we observe that the mesonic moduli space is a complete intersection. It is a 5-dimensional Calabi-Yau space. The generators of the mesonic moduli space are:

\begin{center}
\begin{tabular}{|c|c|}
\hline
generator & perfect matchings\\
\hline\hline
$A_1$ & $p_1 p_4 p_6 p_9$\\
$A_2$ & $p_2 p_7$\\
$A_3$ & $p_3 p_8$\\
$B_1$ & $p_1 p_2 p_3 p_4$ \\
$B_2$ & $p_6 p_7 p_8 p_9$ \\
\hline
$C$ & $p_5$ \\
\hline
\end{tabular}
\end{center}
The $A_i,B_i$ generators form a single relation,
\beal{es88}
A_1 A_2 A_3 = B_1 B_2~~.
\eea

The global symmetry is $U(1)^4 \times U(1)_R$ and has no enhancement. The toric diagram of the mesonic moduli space is given by
\beal{e83nn}
G_t =
\left(
\begin{array}{ccccccccc}
p_1 & p_2 & p_3 & p_4 & p_5 & p_6 & p_7 & p_8 & p_9 \\
\hline
 0 & 1 & 0 & 0 & 0 & 0 & 1 & 0 & 0 \\
 0 & 0 & 1 & 0 & 0 & 0 & 0 & 1 & 0 \\
 1 & 0 & 0 & 1 & 0 & 1 & 0 & 0 & 1 \\
 0 & 0 & 0 & 0 & 1 & 0 & 0 & 0 & 0 \\
 1 & 1 & 1 & 1 & 0 & 0 & 0 & 0 & 0
\end{array}
\right)
~~.
\eea
\\

\subsubsection{Model 8.4e: $NC4$ \label{s_m84e}}

\begin{figure}[ht!!]
\begin{center}
\includegraphics[trim=0cm 0cm 0cm 0cm,totalheight=8 cm]{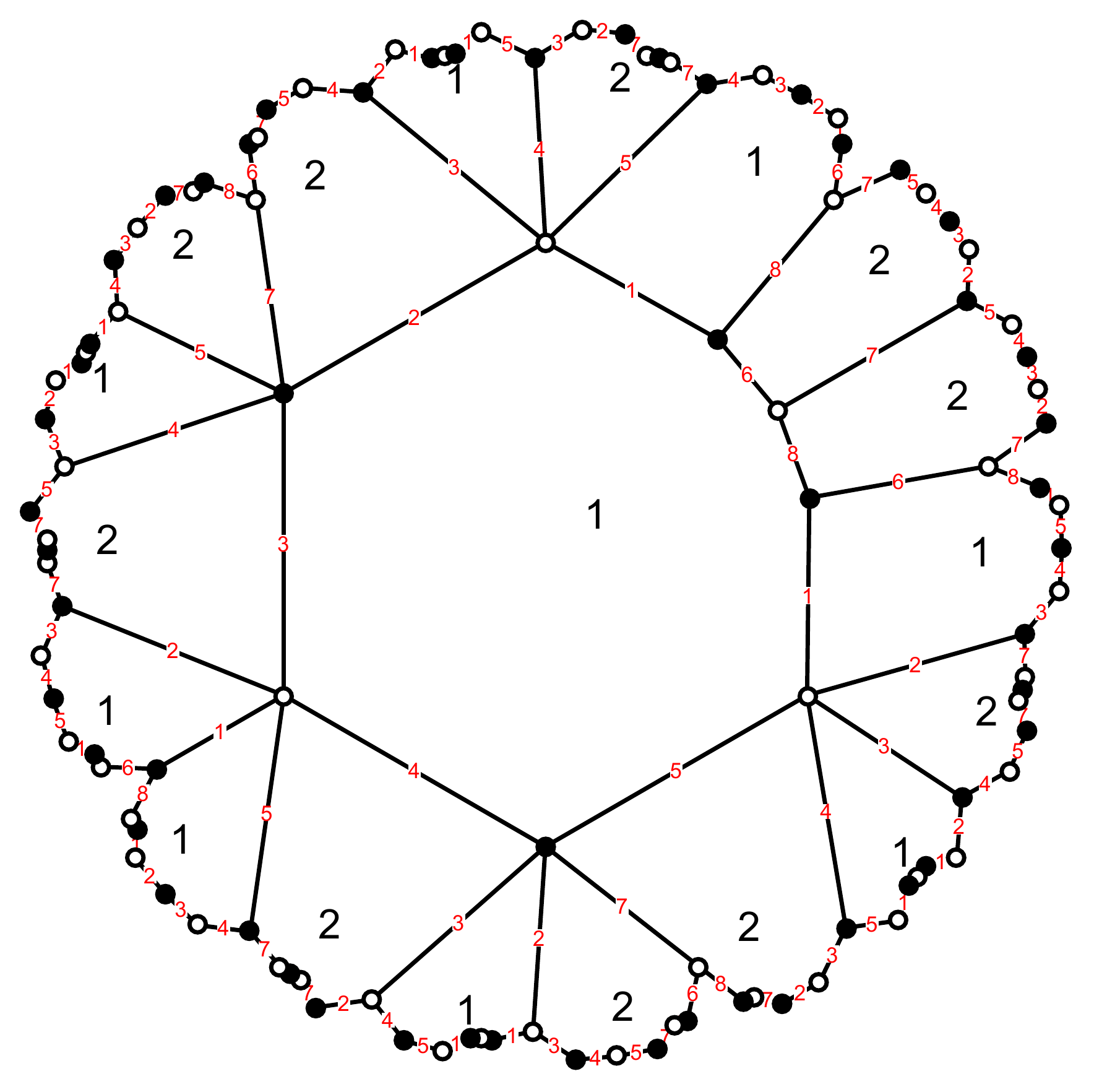}\\
\vspace{0.3cm}
\begin{tabular}{cccccccc}
{\color[rgb]{1.000000,0.000000,0.000000} 1} &
{\color[rgb]{1.000000,0.000000,0.000000} 2} &
{\color[rgb]{1.000000,0.000000,0.000000} 3} &
{\color[rgb]{1.000000,0.000000,0.000000} 4} &
{\color[rgb]{1.000000,0.000000,0.000000} 5} &
{\color[rgb]{1.000000,0.000000,0.000000} 6} &
{\color[rgb]{1.000000,0.000000,0.000000} 7} &
{\color[rgb]{1.000000,0.000000,0.000000} 8} 
\\
$X_{11}$ & $X_{12}^{1}$ & $X_{21}^{1}$ & $X_{12}^{2}$ & $X_{21}^{2}$ &  $X_{12}^{3}$ & $X_{22}$ & $X_{21}^{3}$
\end{tabular}
\caption{The Model 8.4e brane tiling on a $g=2$ Riemann surface with 2 gauge groups, 8 fields and 4 superpotential terms.
\label{fm6n5t}}
 \end{center}
 \end{figure}

\begin{figure}[ht!!]
\begin{center}
\includegraphics[trim=0cm 0cm 0cm 0cm,totalheight=1 cm]{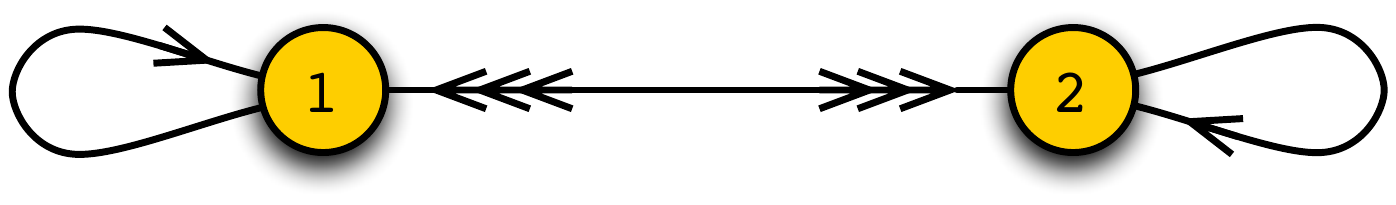}
\caption{The quiver diagram for Model 8.4e, a brane tiling on a $g=2$ Riemann surface with 2 gauge groups, 8 fields and 4 superpotential terms.
\label{fm6n5q}}
 \end{center}
 \end{figure}
 
Model 8.4e's brane tiling and quiver are shown in \fref{fm6n5t} and \fref{fm6n5q} respectively. The superpotential is
\beal{e90}
W= 
+ X_{11} X_{12}^{1} X_{21}^{1} X_{12}^{2} X_{21}^{2} 
+ X_{12}^{3} X_{22} X_{21}^{3}
- X_{12}^{1} X_{22} X_{21}^{2} X_{12}^{2} X_{21}^{1}
- X_{11} X_{12}^{3} X_{21}^{3} 
~~.
\nn\\
\eea
The quiver incidence matrix is
\beal{e91}
d=\left(
\begin{array}{cccccccc}
X_{11} & X_{12}^{1} & X_{21}^{1} & X_{12}^{2} & X_{21}^{2} &  X_{12}^{3} & X_{22} & X_{21}^{3}
\\
\hline
 0 & -1 & 1 & -1 & 1 & -1 & 0 & 1 \\
 0 & 1 & -1 & 1 & -1 & 1 & 0 & -1
\end{array}
\right)
~~.
\eea

Model 8.4e has $c=9$ perfect matchings which are
\beal{e92}
P=\left(
\begin{array}{c|ccccccccc}
\; & p_1 & p_2 & p_3 & p_4 & p_5 & p_6 & p_7 & p_8 & p_9\\
\hline
X_{11} &  0 & 0 & 0 & 0 & 1 & 0 & 0 & 0 & 0 \\
X_{12}^{1} &  1 & 0 & 0 & 0 & 0 & 1 & 0 & 0 & 0 \\
X_{21}^{1} &  0 & 1 & 0 & 0 & 0 & 0 & 1 & 0 & 0 \\
X_{12}^{2} &  0 & 0 & 1 & 0 & 0 & 0 & 0 & 1 & 0 \\
X_{21}^{2} &  0 & 0 & 0 & 1 & 0 & 0 & 0 & 0 & 1 \\
X_{12}^{3} &  1 & 1 & 1 & 1 & 0 & 0 & 0 & 0 & 0 \\
X_{22} &  0 & 0 & 0 & 0 & 1 & 0 & 0 & 0 & 0 \\
X_{21}^{3} &  0 & 0 & 0 & 0 & 0 & 1 & 1 & 1 & 1
\end{array}
\right)
~~.
\eea
The brane tiling has 6 zig-zag paths, which are
\beal{e92bbb}
&
\eta_1 = (X_{12}^{1},X_{21}^{1})~,~
\eta_2 = (X_{12}^{2},X_{21}^{1})~,~
\eta_3 = (X_{12}^{2},X_{21}^{2})~,~
\eta_4 = (X_{12}^{3},X_{21}^{3})~,~
&
\nn\\
&
\eta_5 = (X_{11},X_{12}^{1},X_{22},X_{21}^{3})~,~
\eta_6 = (X_{11},X_{12}^{3},X_{22},X_{21}^{2})~.~
&
\eea

The F-terms are encoded in the charge matrix
\beal{e93}
Q_F =
\left(
\begin{array}{ccccccccc}
p_1 & p_2 & p_3 & p_4 & p_5 & p_6 & p_7 & p_8 & p_9 \\
\hline
 1 & 0 & 0 & -1 & 0 & -1 & 0 & 0 & 1 \\
 0 & 1 & 0 & -1 & 0 & 0 & -1 & 0 & 1 \\
 0 & 0 & 1 & -1 & 0 & 0 & 0 & -1 & 1
\end{array}
\right)
~~.
\eea
The D-terms are given by the matrix
\beal{e94}
Q_D =
\left(
\begin{array}{ccccccccc}
p_1 & p_2 & p_3 & p_4 & p_5 & p_6 & p_7 & p_8 & p_9 \\
\hline
 2 & -1 & 1 & -1 & 0 & -1 & 0 & 0 & 0
\end{array}
\right)
~~.
\eea

As a symplectic quotient the mesonic moduli space is
\beal{es96}
\mesonic = \mathbb{C}^{9} // Q_t ~~.
\eea

By associating the fugacity $t_i$ to the perfect matching $p_i$, the fully refined Hilbert series of $\mesonic$ is given by the following Molien integral
\beal{es96bb}
g_1(t_i;\mesonic) &=& 
\frac{1}{(2\pi i)^4}
\oint_{|z_1|=1} 
\frac{\ud z_1}{z_1} 
\oint_{|z_2|=1} 
\frac{\ud z_2}{z_2}
\oint_{|z_3|=1} 
\frac{\ud z_3}{z_3}
\oint_{|z_4|=1} 
\frac{\ud z_4}{z_4}
~~
\nn\\
&& 
\hspace{1cm}
\times
\frac{1}{
(1-z_1 z_4^2 t_1)
(1-z_2 z_4^{-1} t_2)
(1-z_3 z_4 t_3)
(1-z_1^{-1} z_2^{-1} z_3^{-1} z_4^{-1} t_4)
}
\nn\\
&& 
\hspace{1cm}
\times
\frac{1}{
(1-t_5)
(1-z_1^{-1} z_4^{-1}t_6)
(1-z_2^{-1} t_7)
(1-z_3^{-1} t_8)
(1-z_1 z_2 z_3 t_9)
}
\nn\\
&=&
\frac{
P(t_i)
}{
(1 - t_5) 
(1 - t_1 t_2 t_6 t_7) (1 - t_2 t_3 t_7 t_8) (1 - t_1 t_4 t_6 t_9) (1 - t_3 t_4 t_8 t_9) 
}\nn\\
&&
\times
\frac{1}{
(1 - t_1 t_2^2 t_3 t_4 t_7)  (1 - t_1 t_2 t_3 t_4^2 t_9) (1 - t_1 t_6^2 t_7 t_8 t_9) (1 - t_3 t_6 t_7 t_8^2 t_9)
}
~~,
\nn\\
\eea
where the numerator is
{\tiny
\beal{es96bbb}
P(t_i)&=&
1 - t_1^2 t_2^2 t_3 t_4^2 t_6 t_7 t_9 - t_1 t_2^2 t_3^2 t_4^2 t_7 t_8 t_9 - 
 t_1 t_2 t_3 t_4 t_6 t_7 t_8 t_9 + t_1^2 t_2^3 t_3^2 t_4^2 t_6 t_7^2 t_8 t_9 - 
 t_1^2 t_2^2 t_3 t_4 t_6^2 t_7^2 t_8 t_9 - t_1 t_2^2 t_3^2 t_4 t_6 t_7^2 t_8^2 t_9 
 \nn\\
 &&
 - 
 t_1 t_2 t_3 t_6^2 t_7^2 t_8^2 t_9 + t_1^2 t_2^3 t_3^2 t_4 t_6^2 t_7^3 t_8^2 t_9 + 
 t_1^2 t_2^2 t_3^2 t_4^3 t_6 t_7 t_8 t_9^2 - t_1^2 t_2 t_3 t_4^2 t_6^2 t_7 t_8 t_9^2 +
  t_1^3 t_2^3 t_3^2 t_4^3 t_6^2 t_7^2 t_8 t_9^2 + 
 t_1^3 t_2^2 t_3 t_4^2 t_6^3 t_7^2 t_8 t_9^2 
 \nn\\
 &&
 - 
 t_1 t_2 t_3^2 t_4^2 t_6 t_7 t_8^2 t_9^2 - t_1 t_3 t_4 t_6^2 t_7 t_8^2 t_9^2 + 
 t_1^2 t_2^3 t_3^3 t_4^3 t_6 t_7^2 t_8^2 t_9^2 + 
 4 t_1^2 t_2^2 t_3^2 t_4^2 t_6^2 t_7^2 t_8^2 t_9^2 + 
 t_1^2 t_2 t_3 t_4 t_6^3 t_7^2 t_8^2 t_9^2 - 
 t_1^3 t_2^4 t_3^3 t_4^3 t_6^2 t_7^3 t_8^2 t_9^2 
 \nn\\
 &&
 - 
 t_1^3 t_2^3 t_3^2 t_4^2 t_6^3 t_7^3 t_8^2 t_9^2 + 
 t_1 t_2^2 t_3^3 t_4^2 t_6 t_7^2 t_8^3 t_9^2 + 
 t_1 t_2 t_3^2 t_4 t_6^2 t_7^2 t_8^3 t_9^2 - 
 t_1^2 t_2^3 t_3^3 t_4^2 t_6^2 t_7^3 t_8^3 t_9^2 + 
 t_1^2 t_2^2 t_3^2 t_4 t_6^3 t_7^3 t_8^3 t_9^2 + 
 t_1^2 t_2 t_3^2 t_4^3 t_6^2 t_7 t_8^2 t_9^3 
 \nn\\
 &&
 - 
 t_1^3 t_2^3 t_3^3 t_4^4 t_6^2 t_7^2 t_8^2 t_9^3 - 
 t_1^3 t_2^2 t_3^2 t_4^3 t_6^3 t_7^2 t_8^2 t_9^3 - 
 t_1^2 t_2^2 t_3^3 t_4^3 t_6^2 t_7^2 t_8^3 t_9^3 + 
 t_1^2 t_2 t_3^2 t_4^2 t_6^3 t_7^2 t_8^3 t_9^3 - 
 t_1^3 t_2^3 t_3^3 t_4^3 t_6^3 t_7^3 t_8^3 t_9^3 - 
 t_1^3 t_2^2 t_3^2 t_4^2 t_6^4 t_7^3 t_8^3 t_9^3 
 \nn\\
 &&
 - 
 t_1^2 t_2^2 t_3^3 t_4^2 t_6^3 t_7^3 t_8^4 t_9^3 + 
 t_1^4 t_2^4 t_3^4 t_4^4 t_6^4 t_7^4 t_8^4 t_9^4
 ~~.
\eea
\tiny}

By setting all perfect matching fugacities to $t_i=t$, the Hilbert series takes the form
\beal{es96cc}
g_1(t;\mesonic) &=& 
\frac{1}{(1 - t)} \times
\frac{1 + 2 t^4 + 2 t^6 + 2 t^8 + t^{12}
}{ 
(1 - t^4)^2 (1 - t^6)^2
}~~.
\eea
It can be seen that the mesonic moduli space is a Calabi-Yau 5-fold. It is not a complete intersection. The plethystic logarithm of the refined Hilbert series in \eref{es96bb} is
\beal{es96dd}
PL[g_1(t_i;\mesonic)] &=& 
t_5  + (t_1 t_2 t_6 t_7 + t_2 t_3 t_7 t_8 + t_1 t_4 t_6 t_9 + t_3 t_4 t_8 t_9)  + (t_1 t_2^2 t_3 t_4 t_7 \nn\\
&&
+ t_1 t_2 t_3 t_4^2 t_9 + t_1 t_6^2 t_7 t_8 t_9 + t_3 t_6 t_7 t_8^2 t_9)  
-t_1 t_2 t_3 t_4 t_6 t_7 t_8 t_9 \nn\\
&&
- (t_1^2 t_2^2 t_3 t_4^2 t_6 t_7 t_9 + t_1 t_2^2 t_3^2 t_4^2 t_7 t_8 t_9 + t_1 t_2 t_3 t_6^2 t_7^2 t_8^2 t_9 + t_1 t_3 t_4 t_6^2 t_7 t_8^2 t_9^2) 
\nn\\
&&
-(t_1^2 t_2^2 t_3 t_4 t_6^2 t_7^2 t_8 t_9 + t_1 t_2^2 t_3^2 t_4 t_6 t_7^2 t_8^2 t_9 + t_1^2 t_2 t_3 t_4^2 t_6^2 t_7 t_8 t_9^2 
\nn\\
&&
+ t_1 t_2 t_3^2 t_4^2 t_6 t_7 t_8^2 t_9^2) + \dots
\eea

The first order generators of the mesonic moduli space can be found from the above plethystic logarithm and are shown below.

\begin{center}
\begin{tabular}{|c|c|}
\hline
generator & perfect matchings\\
\hline\hline
$A_1$ & $p_1 p_2 p_6 p_7$\\
$A_2$ & $p_2 p_3 p_7 p_8$\\
$A_3$ & $p_1 p_4 p_6 p_9$\\
$A_4$ & $p_3 p_4 p_8 p_9$\\
$B_1$ & $p_1 p_2^2 p_3 p_4 p_7$ \\
$B_2$ & $p_1 p_2 p_3 p_4^2 p_9$ \\
$B_3$ & $p_1 p_6^2 p_7 p_8 p_9$ \\
$B_4$ & $p_3 p_6 p_7 p_8^2 p_9$ \\
\hline
$C$ & $p_5$ \\
\hline
\end{tabular}
\end{center}
The generators above form the following first order relations,
\beal{es98}
&\{&
A_2 A_3 = A_1 A_4 ~,~
A_3 B_1 = A_1 B_2 ~,~
A_2 B_3 = A_1 B_4 ~,~
A_4 B_3 = A_3 B_4 ~,~
A_4 B_1 = A_2 B_2 ~,~
\nn\\
&&
A_1 A_2 A_3 = B_1 B_3 ~,~
A_2 A_3 A_4 = B_2 B_4 ~,~
A_2 A_3^2 = B_2 B_3 ~,~
A_2^2 A_3 = B_1 B_4 
\}~~.
\eea

The global symmetry is not enhanced and remains $U(1)^4 \times U(1)_R$. The toric diagram is given by
\beal{e93nn}
G_t =
\left(
\begin{array}{ccccccccc}
p_1 & p_2 & p_3 & p_4 & p_5 & p_6 & p_7 & p_8 & p_9 \\
\hline
 1 & 0 & 0 & 1 & 0 & 0 & 1 & 0 & 0 \\
 0 & 1 & 0 & 0 & 1 & 0 & 0 & 1 & 0 \\
 0 & 0 & 1 & 0 & 0 & 1 & 0 & 0 & 1 \\
 1 & 1 & 1 & 0 & 0 & 0 & 0 & 0 & 0 \\ 
 0 & 0 & 0 & 1 & 1 & 1 & 0 & 0 & 0 
\end{array}
\right)
~~.
\eea
\\

\subsubsection{Model 8.4f: $\mathcal{M}_{4,2}$ \label{s_m84f}}

\begin{figure}[ht!!]
\begin{center}
\includegraphics[trim=0cm 0cm 0cm 0cm,totalheight=8 cm]{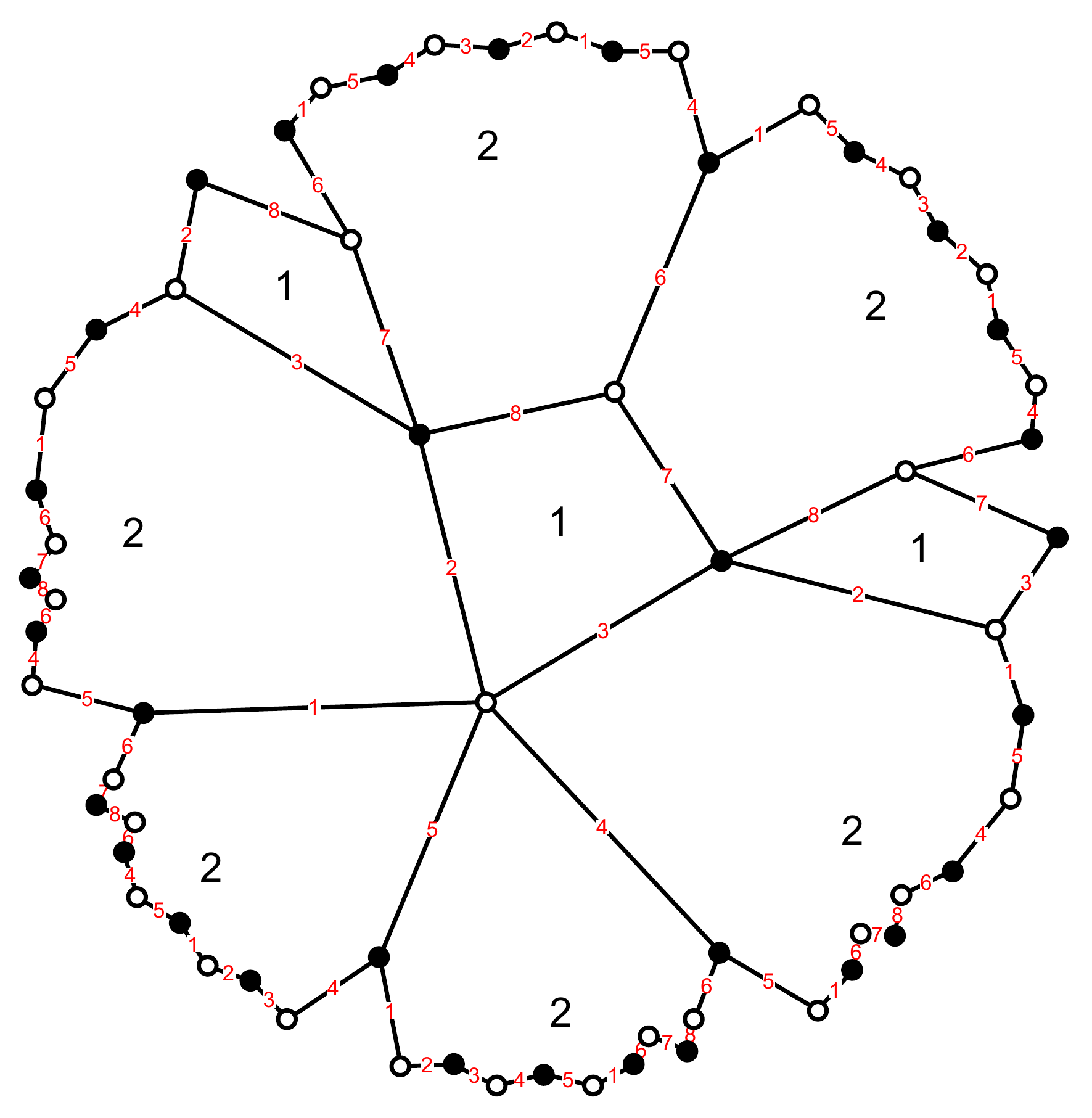}\\
\vspace{0.3cm}
\begin{tabular}{cccccccc}
{\color[rgb]{1.000000,0.000000,0.000000} 1} &
{\color[rgb]{1.000000,0.000000,0.000000} 2} &
{\color[rgb]{1.000000,0.000000,0.000000} 3} &
{\color[rgb]{1.000000,0.000000,0.000000} 4} &
{\color[rgb]{1.000000,0.000000,0.000000} 5} &
{\color[rgb]{1.000000,0.000000,0.000000} 6} &
{\color[rgb]{1.000000,0.000000,0.000000} 7} &
{\color[rgb]{1.000000,0.000000,0.000000} 8} 
\\
$X_{22}^{1}$ & $X_{21}^{1}$ & $X_{12}^{1}$ & $X_{22}^{2}$ & $X_{22}^{3}$ & $X_{22}^{4}$ & $X_{21}^{2}$ & $X_{12}^{2}$
\end{tabular}
\caption{The Model 8.5f brane tiling on a $g=2$ Riemann surface with 2 gauge groups, 8 fields and 4 superpotential terms.
\label{fm6n6t}}
 \end{center}
 \end{figure}

\begin{figure}[ht!!]
\begin{center}
\includegraphics[trim=0cm 0cm 0cm 0cm,totalheight=1 cm]{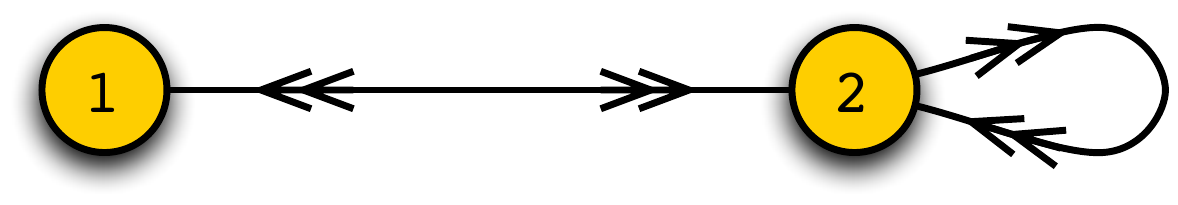}
\caption{The quiver diagram for Model 8.5f, a brane tiling on a $g=2$ Riemann surface with 2 gauge groups, 8 fields and 4 superpotential terms.
\label{fm6n6q}}
 \end{center}
 \end{figure}
 
The brane tiling and corresponding quiver for Model 8.4f is shown in \fref{fm6n6t} and \fref{fm6n6q} respectively. The superpotential is
\beal{e100}
W
=
+ X_{22}^{1} X_{21}^{1} X_{12}^{1} X_{22}^{2} X_{22}^{3} 
+ X_{22}^{4} X_{21}^{2} X_{12}^{2}
- X_{22}^{1} X_{22}^{3} X_{22}^{2} X_{22}^{4}
- X_{21}^{1} X_{12}^{2} X_{21}^{2} X_{12}^{1}
~~.
\nn\\
\eea
The quiver incidence matrix is
\beal{e101}
d=\left(
\begin{array}{cccccccc}
X_{22}^{1} & X_{21}^{1} & X_{12}^{1} & X_{22}^{2} & X_{22}^{3} & X_{22}^{4} & X_{21}^{2} & X_{12}^{2}
\\
\hline
 0 & 1 & -1 & 0 & 0 & 0 & 1 & -1 \\
 0 & -1 & 1 & 0 & 0 & 0 & -1 & 1
\end{array}
\right)
~~.
\eea

Model 8.4f's brane tiling has $c=8$ perfect matchings. The perfect matchings are encoded in the matrix
\beal{e102}
P=\left(
\begin{array}{c|cccccccccccc}
\; & p_1 & p_2 & p_3 & p_4 & p_5 & p_6 & p_7 & p_8\\
\hline
X_{12}^{1} & 0 & 1 & 0 & 0 & 0 & 0 & 0 & 0 \\
X_{12}^{2} & 0 & 0 & 1 & 0 & 1 & 0 & 1 & 0 \\
X_{21}^{1} & 1 & 0 & 0 & 0 & 0 & 0 & 0 & 0 \\
X_{21}^{2} & 0 & 0 & 0 & 1 & 0 & 1 & 0 & 1 \\
X_{22}^{1} & 0 & 0 & 1 & 1 & 0 & 0 & 0 & 0 \\
X_{22}^{2} & 0 & 0 & 0 & 0 & 1 & 1 & 0 & 0 \\
X_{22}^{3} & 0 & 0 & 0 & 0 & 0 & 0 & 1 & 1 \\
X_{22}^{4} & 1 & 1 & 0 & 0 & 0 & 0 & 0 & 0 
\end{array}
\right)
~~.
\eea
The brane tiling has 6 zig-zag paths, which are
\beal{e102bbb}
&
\eta_1 = (X_{12}^{1},X_{21}^{1})~,~
\eta_2 = (X_{12}^{2},X_{21}^{2})~,~
\eta_3 = (X_{22}^{1},X_{22}^{3})~,~
\eta_4 = (X_{22}^{2},X_{22}^{3})~,~
&\nn\\
&
\eta_5 = (X_{21}^{1},X_{12}^{2},X_{22}^{4},X_{22}^{1})~,~
\eta_6 = (X_{21}^{2},X_{12}^{1},X_{22}^{2},X_{22}^{4})~.~
&
\eea

The F-term charge matrix is
\beal{e103}
Q_F =
\left(
\begin{array}{cccccccc}
p_1 & p_2 & p_3 & p_4 & p_5 & p_6 & p_7 & p_8\\
\hline
 0 & 0 & 1 & -1 & 0 & 0 & -1 & 1 \\
 0 & 0 & 0 & 0 & 1 & -1 & -1 & 1
\end{array}
\right)
~~.
\eea
The D-terms are encoded in the matrix
\beal{e104}
Q_D =
\left(
\begin{array}{cccccccc}
p_1 & p_2 & p_3 & p_4 & p_5 & p_6 & p_7 & p_8\\
\hline
 1 & -1 & -1 & 1 & 0 & 0 & 0 & 0
\end{array}
\right)
~~.
\eea

The symplectic quotient description of the mesonic moduli space of Model 8.4f is given in terms of the total charge matrix $Q_t$, 
\beal{es106}
\mesonic = \mathbb{C}^{8} // Q_t ~~.
\eea

By associating the fugacity $t_i$ to the perfect matching $p_i$, the fully refined Hilbert series of $\mesonic$ is given by the following Molien integral
\beal{e107}
g_1(t_i;\mesonic) &=& 
\frac{1}{(2\pi i)^3}
\oint_{|z_1|=1} 
\frac{\ud z_1}{z_1} 
\oint_{|z_2|=1} 
\frac{\ud z_2}{z_2} 
\oint_{|z_3|=1} 
\frac{\ud z_3}{z_3} 
~~
\nn\\
&& 
\hspace{1cm}
\times
\frac{1}{
(1- z_3 t_1)
(1- z_3^{-1} t_2)
(1- z_1 z_3^{-1} t_3)
(1- z_1^{-1} z_3 t_4)
}
\nn\\
&&
\hspace{1cm}
\times
\frac{1}{
(1- z_2 t_5)
(1- z_2^{-1} t_6)
(1- z_1^{-1} z_2^{-1} t_7)
(1- z_1 z_2 t_8)
}
\nn\\
&=&
\frac{
1 - t_1 t_2 t_3 t_4 t_5 t_6 t_7 t_8
}{
(1 - t_1 t_2) (1 - t_3 t_4) (1 - t_5 t_6) (1 - t_7 t_8) 
(1 - t_1 t_3 t_5 t_7) (1 - t_2 t_4 t_6 t_8) 
}
~~.
\nn\\
\eea

From the Hilbert series, we observe that the mesonic moduli space is a complete intersection. As expected for a $g=2$ tiling, it is a 5-dimensional Calabi-Yau space. The generators of the mesonic moduli space are:

\begin{center}
\begin{tabular}{|c|c|}
\hline
generator & perfect matchings\\
\hline\hline
$A_1$ & $p_1 p_2$\\
$A_2$ & $p_3 p_4$\\
$A_3$ & $p_5 p_6$\\
$A_4$ & $p_7 p_8$ \\
$B_1$ & $p_1 p_3 p_5 p_7$ \\
$B_2$ & $p_2 p_4 p_6 p_8$ \\
\hline
\end{tabular}
\end{center}
The generators form a single relation
\beal{es88}
A_1 A_2 A_3 A_4 = B_1 B_2 ~~.
\eea
The global symmetry is $U(1)^4 \times U(1)_R$ and has no enhancement. The toric diagram of the Calabi-Yau 5-fold is given by
\beal{e104nn}
G_t =
\left(
\begin{array}{cccccccc}
p_1 & p_2 & p_3 & p_4 & p_5 & p_6 & p_7 & p_8\\
\hline
 1 & 1 & 0 & 0 & 0 & 0 & 0 & 0 \\
 0 & 0 & 1 & 1 & 0 & 0 & 0 & 0 \\
 0 & 0 & 0 & 0 & 1 & 1 & 0 & 0 \\
 0 & 0 & 0 & 0 & 0 & 0 & 1 & 1 \\
 0 & 1 & 0 & 1 & 0 & 1 & 0 & 1 
\end{array}
\right)
~~.
\eea

We further note that one can apply the urban renewal move on face 1 of the brane tiling. It can be shown that Model 8.4f is self-dual under toric duality on face 1 up to a sign of the superpotential.
\\

\subsubsection{Model 8.4g: $NC5$ \label{s_m84g}}
 
\begin{figure}[ht!!]
\begin{center}
\includegraphics[trim=0cm 0cm 0cm 0cm,totalheight=8 cm]{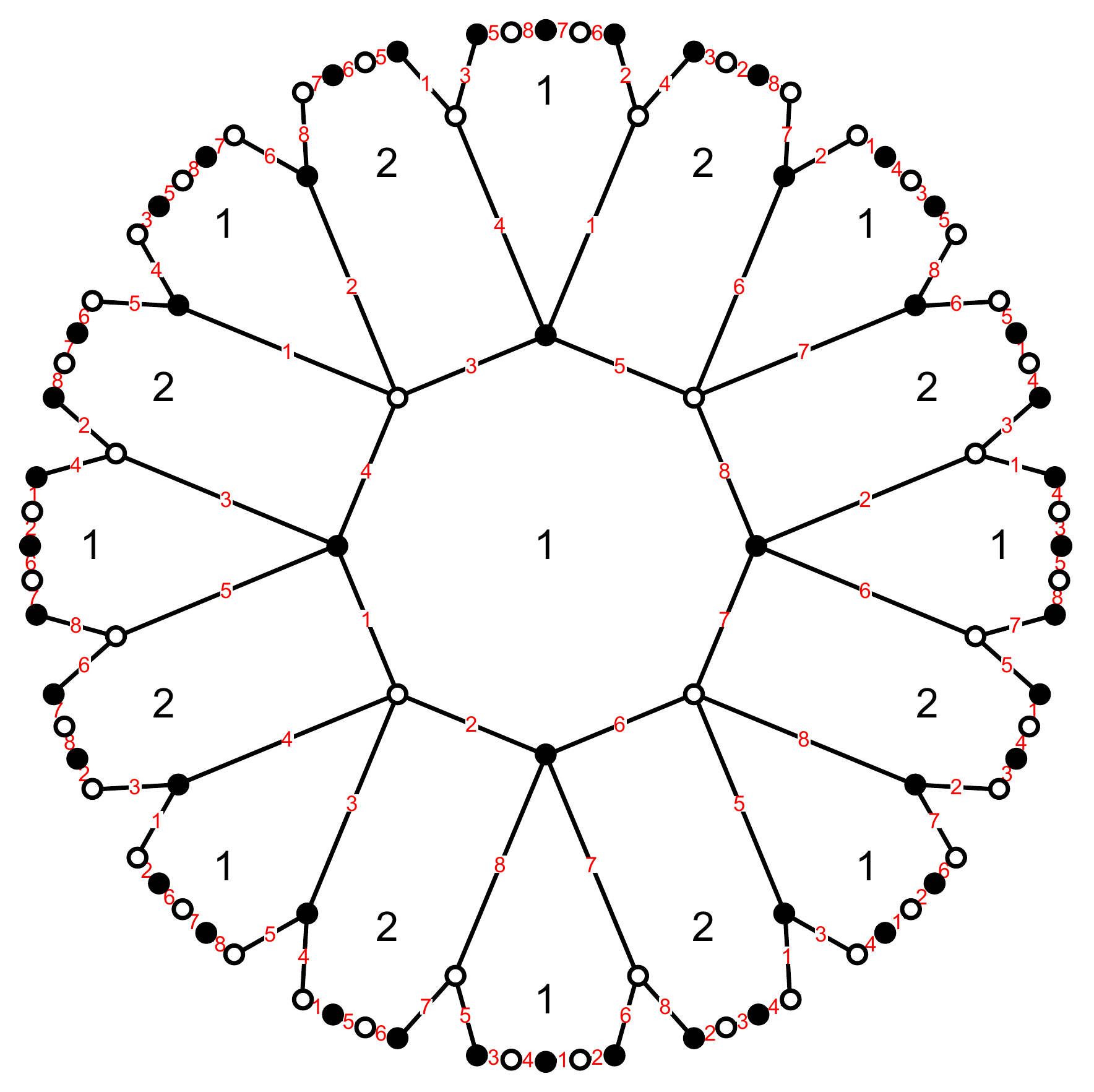}\\
\vspace{0.3cm}
\begin{tabular}{cccccccc}
{\color[rgb]{1.000000,0.000000,0.000000} 1} &
{\color[rgb]{1.000000,0.000000,0.000000} 2} &
{\color[rgb]{1.000000,0.000000,0.000000} 3} &
{\color[rgb]{1.000000,0.000000,0.000000} 4} &
{\color[rgb]{1.000000,0.000000,0.000000} 5} &
{\color[rgb]{1.000000,0.000000,0.000000} 6} &
{\color[rgb]{1.000000,0.000000,0.000000} 7} &
{\color[rgb]{1.000000,0.000000,0.000000} 8} 
\\
$X_{21}^{1}$ & $X_{12}^{1}$ & $X_{21}^{2}$ & $X_{12}^{2}$
& $X_{21}^{3}$ & $X_{12}^{3}$ & $X_{21}^{4}$ & $X_{12}^{4}$
\end{tabular}
\caption{The Model 8.4g brane tiling on a $g=2$ Riemann surface with 2 gauge groups, 8 fields and 4 superpotential terms.
\label{fm6n3t}}
 \end{center}
 \end{figure}

\begin{figure}[ht!!]
\begin{center}
\includegraphics[trim=0cm 0cm 0cm 0cm,totalheight=1 cm]{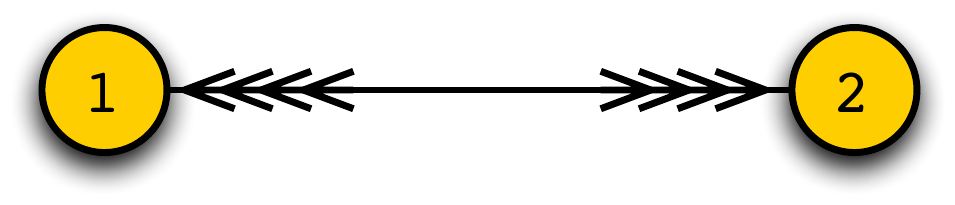}
\caption{The quiver diagram for Model 8.4g, a brane tiling on a $g=2$ Riemann surface with 2 gauge groups, 8 fields and 4 superpotential terms.
\label{fm6n3q}}
 \end{center}
 \end{figure}
 
For Model 8.4g, the brane tiling and corresponding quiver is shown in \fref{fm6n3t} and \fref{fm6n3q} respectively. The quartic superpotential is
\beal{e60}
W= 
+ X_{21}^{1} X_{12}^{1} X_{21}^{2} X_{12}^{2}
+ X_{21}^{3} X_{12}^{3} X_{21}^{4} X_{12}^{4}
- X_{21}^{1} X_{12}^{2} X_{21}^{2} X_{12}^{3}
- X_{21}^{1} X_{12}^{4} X_{21}^{4} X_{12}^{3}
~~.
\nn\\
\eea
The quiver incidence matrix is
\beal{e61}
d=\left(
\begin{array}{cccccccc}
X_{21}^{1} & X_{12}^{1} & X_{21}^{2} & X_{12}^{2}
& X_{21}^{3} & X_{12}^{3} & X_{21}^{4} & X_{12}^{4}
\\
\hline
 1 & -1 & 1 & -1 & -1 & 1 & -1 & 1 \\
 -1 & 1 & -1 & 1 & 1 & -1 & 1 & -1
\end{array}
\right)
~~.
\eea

The brane tiling has $c=10$ perfect matchings. The perfect matchings are encoded in the matrix
\beal{e62}
P=\left(
\begin{array}{c|cccccccccc}
\; & p_1 & p_2 & p_3 & p_4 & p_5 & p_6 & p_7 & p_8 & p_9 & p_{10}\\
\hline
X_{21}^{1} & 0 & 1 & 0 & 0 & 1 & 0 & 0 & 1 & 0 & 0 \\
X_{12}^{1} & 1 & 0 & 0 & 0 & 0 & 0 & 0 & 0 & 0 & 0 \\
X_{21}^{2} & 0 & 0 & 1 & 0 & 0 & 1 & 0 & 0 & 1 & 0 \\
X_{12}^{2} & 0 & 0 & 0 & 1 & 0 & 0 & 1 & 0 & 0 & 1 \\
X_{21}^{3} & 1 & 0 & 0 & 0 & 0 & 0 & 0 & 0 & 0 & 0 \\
X_{12}^{3} & 0 & 1 & 1 & 1 & 0 & 0 & 0 & 0 & 0 & 0 \\
X_{21}^{4} & 0 & 0 & 0 & 0 & 1 & 1 & 1 & 0 & 0 & 0 \\
X_{12}^{4} & 0 & 0 & 0 & 0 & 0 & 0 & 0 & 1 & 1 & 1
\end{array}
\right)
~~.
\eea
The brane tiling has 6 zig-zag paths, which are
\beal{e62bb}
&
\eta_1 = (X_{21}^{1},X_{12}^{2})~,~
\eta_2 = (X_{21}^{2},X_{12}^{2})~,~
\eta_3 = (X_{21}^{3},X_{12}^{4})~,~
\eta_4 = (X_{21}^{4},X_{12}^{4})~,~
&\nn\\
&
\eta_5 = (X_{21}^{1},X_{12}^{1},X_{21}^{4},X_{12}^{3})~,~
\eta_6 = (X_{12}^{1},X_{21}^{2},X_{12}^{3},X_{21}^{3})~.~
&
\eea

The F-term constraints can be expressed as charges carried by the perfect matchings. The charges are given by
\beal{e63}
Q_F =
\left(
\begin{array}{cccccccccc}
p_1 & p_2 & p_3 & p_4 & p_5 & p_6 & p_7 & p_8 & p_9 & p_{10}\\
\hline
 0 & 1 & 0 & -1 & 0 & 0 & 0 & -1 & 0 & 1 \\
 0 & 0 & 1 & -1 & 0 & 0 & 0 & 0 & -1 & 1 \\
 0 & 0 & 0 & 0 & 1 & 0 & -1 & -1 & 0 & 1 \\
 0 & 0 & 0 & 0 & 0 & 1 & -1 & 0 & -1 & 1
\end{array}
\right)
~~.
\eea
The D-term charges are encoded in the quiver incidence matrix $d$ and are
\beal{e64}
Q_D = 
\left(
\begin{array}{cccccccccc}
p_1 & p_2 & p_3 & p_4 & p_5 & p_6 & p_7 & p_8 & p_9 & p_{10}\\
\hline
 1 & -1 & -1 & 1 & 1 & 0 & 0 & -1 & 0 & 0
\end{array}
\right)~~.
\eea

Using the total charge matrix, the mesonic moduli space can be expressed as the symplectic quotient
\beal{es66}
\mesonic = \mathbb{C}^{10} // Q_t ~~.
\eea

By associating the fugacity $t_i$ to the perfect matching $p_i$, the fully refined Hilbert series of $\mesonic$ is given by the following Molien integral
\beal{e67}
g_1(t_i;\mesonic) &=& 
\frac{1}{(2\pi i)^5}
\oint_{|z_1|=1} 
\frac{\ud z_1}{z_1} 
\oint_{|z_2|=1} 
\frac{\ud z_2}{z_2}
\oint_{|z_3|=1} 
\frac{\ud z_3}{z_3}
\oint_{|z_4|=1} 
\frac{\ud z_4}{z_4}
\oint_{|z_5|=1} 
\frac{\ud z_5}{z_5}
~~
\nn\\
&& 
\hspace{1cm}
\times
\frac{1}{
(1-z_5 t_1)
(1-z_1 z_5^{-1} t_2)
(1-z_2 z_5^{-1} t_3)
(1-z_1^{-1} z_2^{-1} z_5 t_4)
}
\nn\\
&& 
\hspace{1cm}
\times
\frac{1}{
(1-z_3 z_5 t_5)
(1-z_4 t_6)
(1-z_3^{-1} z_4^{-1} t_7)
(1-z_1^{-1} z_3^{-1} z_5^{-1} t_8)
}
\nn\\
&& 
\hspace{1cm}
\times
\frac{1}{
(1-z_2^{-1} z_4^{-1} t_9)
(1-z_1 z_2 z_3 z_4 t_{10})
}
\nn\\
&=&
\frac{
P(t_i)
}{
(1 - t_1 t_2 t_3 t_4) 
(1 - t_2 t_3 t_4^2 t_7 t_{10}) 
(1 - t_2 t_3 t_4 t_5 t_6 t_7) 
(1 - t_1 t_2 t_5 t_8) 
}
\nn\\
&&
\hspace{0.5cm}
\times
\frac{1}{
(1 - t_2 t_4 t_5 t_7 t_8 t_{10}) 
(1 - t_2 t_5^2 t_6 t_7 t_8) 
(1 - t_1 t_3 t_6 t_9) 
(1 - t_3 t_4 t_6 t_7 t_9 t_{10}) 
}
\nn\\
&&
\hspace{0.5cm}
\times
\frac{1}{
(1 - t_3 t_5 t_6^2 t_7 t_9) 
(1 - t_1 t_8 t_9 t_{10}) 
(1 - t_4 t_7 t_8 t_9 t_{10}^2) 
(1 - t_5 t_6 t_7 t_8 t_9 t_{10})
}
~~,
\nn\\
\eea
where the numerator is 
{\tiny
\beal{e67b}
&&
P(t_i) =
1 - t_1 t_2^2 t_3 t_4 t_5^2 t_6 t_7 t_8 - t_1 t_2 t_3^2 t_4 t_5 t_6^2 t_7 t_9 - 
 t_1 t_2 t_3 t_5^2 t_6^2 t_7 t_8 t_9 + t_1^2 t_2^2 t_3^2 t_4 t_5^2 t_6^2 t_7 t_8 t_9 + 
 t_1 t_2^2 t_3^2 t_4 t_5^3 t_6^3 t_7^2 t_8 t_9 - t_1 t_2^2 t_3 t_4^2 t_5 t_7 t_8 t_{10} \nn\\&& \hspace{0.5cm}
 -
  t_2^2 t_3 t_4^2 t_5^2 t_6 t_7^2 t_8 t_{10} + 
 t_1 t_2^3 t_3^2 t_4^3 t_5^2 t_6 t_7^2 t_8 t_{10}  
 + 
 t_1 t_2^3 t_3 t_4^2 t_5^3 t_6 t_7^2 t_8^2 t_{10} - 
 t_1 t_2 t_3^2 t_4^2 t_6 t_7 t_9 t_{10} - t_2 t_3^2 t_4^2 t_5 t_6^2 t_7^2 t_9 t_{10} \nn\\&& \hspace{0.5cm}
 + 
 t_1 t_2^2 t_3^3 t_4^3 t_5 t_6^2 t_7^2 t_9 t_{10} - 
 3 t_1 t_2 t_3 t_4 t_5 t_6 t_7 t_8 t_9 t_{10} + 
 t_1^2 t_2^2 t_3^2 t_4^2 t_5 t_6 t_7 t_8 t_9 t_{10} - 
 2 t_2 t_3 t_4 t_5^2 t_6^2 t_7^2 t_8 t_9 t_{10} + 
 5 t_1 t_2^2 t_3^2 t_4^2 t_5^2 t_6^2 t_7^2 t_8 t_9 t_{10} 
\nn\\&& \hspace{0.5cm}
 - 
 t_1^2 t_2^3 t_3^3 t_4^3 t_5^2 t_6^2 t_7^2 t_8 t_9 t_{10} + 
 t_2^2 t_3^2 t_4^2 t_5^3 t_6^3 t_7^3 t_8 t_9 t_{10} - 
 t_1 t_2^3 t_3^3 t_4^3 t_5^3 t_6^3 t_7^3 t_8 t_9 t_{10} - 
 t_1 t_2 t_5^2 t_6 t_7 t_8^2 t_9 t_{10} + 
 t_1^2 t_2^2 t_3 t_4 t_5^2 t_6 t_7 t_8^2 t_9 t_{10}
\nn\\&& \hspace{0.5cm} + 
 3 t_1 t_2^2 t_3 t_4 t_5^3 t_6^2 t_7^2 t_8^2 t_9 t_{10} - 
 t_1^2 t_2^3 t_3^2 t_4^2 t_5^3 t_6^2 t_7^2 t_8^2 t_9 t_{10} - 
 t_1 t_2^3 t_3^2 t_4^2 t_5^4 t_6^3 t_7^3 t_8^2 t_9 t_{10} + 
 t_1 t_2 t_3^3 t_4^2 t_5 t_6^3 t_7^2 t_9^2 t_{10} - 
 t_1 t_3 t_5 t_6^2 t_7 t_8 t_9^2 t_{10}
\nn\\&& \hspace{0.5cm} + 
 t_1^2 t_2 t_3^2 t_4 t_5 t_6^2 t_7 t_8 t_9^2 t_{10} + 
 3 t_1 t_2 t_3^2 t_4 t_5^2 t_6^3 t_7^2 t_8 t_9^2 t_{10} - 
 t_1^2 t_2^2 t_3^3 t_4^2 t_5^2 t_6^3 t_7^2 t_8 t_9^2 t_{10} - 
 t_1 t_2^2 t_3^3 t_4^2 t_5^3 t_6^4 t_7^3 t_8 t_9^2 t_{10} + 
 t_1^2 t_2 t_3 t_5^2 t_6^2 t_7 t_8^2 t_9^2 t_{10}
\nn\\&& \hspace{0.5cm} - 
 t_1^3 t_2^2 t_3^2 t_4 t_5^2 t_6^2 t_7 t_8^2 t_9^2 t_{10} + 
 t_1 t_2 t_3 t_5^3 t_6^3 t_7^2 t_8^2 t_9^2 t_{10} - 
 t_1^2 t_2^2 t_3^2 t_4 t_5^3 t_6^3 t_7^2 t_8^2 t_9^2 t_{10} - 
 t_1 t_2^2 t_3^2 t_4 t_5^4 t_6^4 t_7^3 t_8^2 t_9^2 t_{10} - 
 t_1^2 t_2^3 t_3^3 t_4^2 t_5^4 t_6^4 t_7^3 t_8^2 t_9^2 t_{10}
 \nn\\&& \hspace{0.5cm}
  - 
 t_1 t_2 t_3 t_4^2 t_7 t_8 t_9 t_{10}^2 - 2 t_2 t_3 t_4^2 t_5 t_6 t_7^2 t_8 t_9 t_{10}^2 + 
 3 t_1 t_2^2 t_3^2 t_4^3 t_5 t_6 t_7^2 t_8 t_9 t_{10}^2 + 
 2 t_2^2 t_3^2 t_4^3 t_5^2 t_6^2 t_7^3 t_8 t_9 t_{10}^2 - 
 2 t_1 t_2^3 t_3^3 t_4^4 t_5^2 t_6^2 t_7^3 t_8 t_9 t_{10}^2
 \nn\\&& \hspace{0.5cm} - 
 t_1 t_2 t_4 t_5 t_7 t_8^2 t_9 t_{10}^2 + 
 t_1^2 t_2^2 t_3 t_4^2 t_5 t_7 t_8^2 t_9 t_{10}^2 - 
 t_2 t_4 t_5^2 t_6 t_7^2 t_8^2 t_9 t_{10}^2 + 
 5 t_1 t_2^2 t_3 t_4^2 t_5^2 t_6 t_7^2 t_8^2 t_9 t_{10}^2 - 
 t_1^2 t_2^3 t_3^2 t_4^3 t_5^2 t_6 t_7^2 t_8^2 t_9 t_{10}^2
 \nn\\&& \hspace{0.5cm} + 
 2 t_2^2 t_3 t_4^2 t_5^3 t_6^2 t_7^3 t_8^2 t_9 t_{10}^2 - 
 3 t_1 t_2^3 t_3^2 t_4^3 t_5^3 t_6^2 t_7^3 t_8^2 t_9 t_{10}^2 + 
 t_1 t_2^2 t_4 t_5^3 t_6 t_7^2 t_8^3 t_9 t_{10}^2 - 
 t_1^2 t_2^3 t_3 t_4^2 t_5^3 t_6 t_7^2 t_8^3 t_9 t_{10}^2 - 
 2 t_1 t_2^3 t_3 t_4^2 t_5^4 t_6^2 t_7^3 t_8^3 t_9 t_{10}^2
 \nn\\&& \hspace{0.5cm} - 
 t_1 t_3 t_4 t_6 t_7 t_8 t_9^2 t_{10}^2 + 
 t_1^2 t_2 t_3^2 t_4^2 t_6 t_7 t_8 t_9^2 t_{10}^2 - 
 t_3 t_4 t_5 t_6^2 t_7^2 t_8 t_9^2 t_{10}^2 + 
 5 t_1 t_2 t_3^2 t_4^2 t_5 t_6^2 t_7^2 t_8 t_9^2 t_{10}^2 - 
 t_1^2 t_2^2 t_3^3 t_4^3 t_5 t_6^2 t_7^2 t_8 t_9^2 t_{10}^2
 \nn\\&& \hspace{0.5cm} + 
 2 t_2 t_3^2 t_4^2 t_5^2 t_6^3 t_7^3 t_8 t_9^2 t_{10}^2 - 
 3 t_1 t_2^2 t_3^3 t_4^3 t_5^2 t_6^3 t_7^3 t_8 t_9^2 t_{10}^2 + 
 t_1^2 t_2 t_3 t_4 t_5 t_6 t_7 t_8^2 t_9^2 t_{10}^2 - 
 t_1^3 t_2^2 t_3^2 t_4^2 t_5 t_6 t_7 t_8^2 t_9^2 t_{10}^2 + 
 5 t_1 t_2 t_3 t_4 t_5^2 t_6^2 t_7^2 t_8^2 t_9^2 t_{10}^2 
 \nn\\&& \hspace{0.5cm}- 
 5 t_1^2 t_2^2 t_3^2 t_4^2 t_5^2 t_6^2 t_7^2 t_8^2 t_9^2 t_{10}^2 + 
 t_1^3 t_2^3 t_3^3 t_4^3 t_5^2 t_6^2 t_7^2 t_8^2 t_9^2 t_{10}^2 + 
 t_2 t_3 t_4 t_5^3 t_6^3 t_7^3 t_8^2 t_9^2 t_{10}^2 - 
 6 t_1 t_2^2 t_3^2 t_4^2 t_5^3 t_6^3 t_7^3 t_8^2 t_9^2 t_{10}^2 - 
 t_2^2 t_3^2 t_4^2 t_5^4 t_6^4 t_7^4 t_8^2 t_9^2 t_{10}^2
 \nn\\&& \hspace{0.5cm} + 
 2 t_1^2 t_2^4 t_3^4 t_4^4 t_5^4 t_6^4 t_7^4 t_8^2 t_9^2 t_{10}^2 - 
 t_1^2 t_2^2 t_3 t_4 t_5^3 t_6^2 t_7^2 t_8^3 t_9^2 t_{10}^2 + 
 t_1^3 t_2^3 t_3^2 t_4^2 t_5^3 t_6^2 t_7^2 t_8^3 t_9^2 t_{10}^2 - 
 t_1 t_2^2 t_3 t_4 t_5^4 t_6^3 t_7^3 t_8^3 t_9^2 t_{10}^2 + 
 t_1 t_2^3 t_3^2 t_4^2 t_5^5 t_6^4 t_7^4 t_8^3 t_9^2 t_{10}^2 
 \nn\\&& \hspace{0.5cm} + 
 t_1^2 t_2^4 t_3^3 t_4^3 t_5^5 t_6^4 t_7^4 t_8^3 t_9^2 t_{10}^2 + 
 t_1 t_3^2 t_4 t_5 t_6^3 t_7^2 t_8 t_9^3 t_{10}^2 - 
 t_1^2 t_2 t_3^3 t_4^2 t_5 t_6^3 t_7^2 t_8 t_9^3 t_{10}^2 - 
 2 t_1 t_2 t_3^3 t_4^2 t_5^2 t_6^4 t_7^3 t_8 t_9^3 t_{10}^2 - 
 t_1^2 t_2 t_3^2 t_4 t_5^2 t_6^3 t_7^2 t_8^2 t_9^3 t_{10}^2 
 \nn\\&& \hspace{0.5cm} + 
 t_1^3 t_2^2 t_3^3 t_4^2 t_5^2 t_6^3 t_7^2 t_8^2 t_9^3 t_{10}^2 - 
 t_1 t_2 t_3^2 t_4 t_5^3 t_6^4 t_7^3 t_8^2 t_9^3 t_{10}^2 + 
 t_1 t_2^2 t_3^3 t_4^2 t_5^4 t_6^5 t_7^4 t_8^2 t_9^3 t_{10}^2 + 
 t_1^2 t_2^3 t_3^4 t_4^3 t_5^4 t_6^5 t_7^4 t_8^2 t_9^3 t_{10}^2 - 
 t_1^2 t_2^2 t_3^2 t_4 t_5^4 t_6^4 t_7^3 t_8^3 t_9^3 t_{10}^2
 \nn\\&& \hspace{0.5cm} + 
 t_1^3 t_2^3 t_3^3 t_4^2 t_5^4 t_6^4 t_7^3 t_8^3 t_9^3 t_{10}^2 + 
 t_1^2 t_2^3 t_3^3 t_4^2 t_5^5 t_6^5 t_7^4 t_8^3 t_9^3 t_{10}^2 + 
 t_1 t_2^2 t_3 t_4^3 t_5 t_7^2 t_8^2 t_9 t_{10}^3 + 
 t_2^2 t_3 t_4^3 t_5^2 t_6 t_7^3 t_8^2 t_9 t_{10}^3 - 
 t_1 t_2^3 t_3^2 t_4^4 t_5^2 t_6 t_7^3 t_8^2 t_9 t_{10}^3
 \nn
 \eea
 }
 
{\tiny
\beal{e67bII} 
 && \hspace{0.5cm} - 
 t_1 t_2^3 t_3 t_4^3 t_5^3 t_6 t_7^3 t_8^3 t_9 t_{10}^3 + 
 t_1 t_2 t_3^2 t_4^3 t_6 t_7^2 t_8 t_9^2 t_{10}^3 + 
 t_2 t_3^2 t_4^3 t_5 t_6^2 t_7^3 t_8 t_9^2 t_{10}^3 - 
 t_1 t_2^2 t_3^3 t_4^4 t_5 t_6^2 t_7^3 t_8 t_9^2 t_{10}^3 + 
 3 t_1 t_2 t_3 t_4^2 t_5 t_6 t_7^2 t_8^2 t_9^2 t_{10}^3 
 \nn\\&& \hspace{0.5cm} - 
 t_1^2 t_2^2 t_3^2 t_4^3 t_5 t_6 t_7^2 t_8^2 t_9^2 t_{10}^3 + 
 2 t_2 t_3 t_4^2 t_5^2 t_6^2 t_7^3 t_8^2 t_9^2 t_{10}^3 - 
 6 t_1 t_2^2 t_3^2 t_4^3 t_5^2 t_6^2 t_7^3 t_8^2 t_9^2 t_{10}^3 - 
 t_2^2 t_3^2 t_4^3 t_5^3 t_6^3 t_7^4 t_8^2 t_9^2 t_{10}^3 + 
 t_1^2 t_2^4 t_3^4 t_4^5 t_5^3 t_6^3 t_7^4 t_8^2 t_9^2 t_{10}^3 \nn\\&& \hspace{0.5cm} - 
 t_2^3 t_3^3 t_4^4 t_5^4 t_6^4 t_7^5 t_8^2 t_9^2 t_{10}^3 + 
 t_1 t_2^4 t_3^4 t_4^5 t_5^4 t_6^4 t_7^5 t_8^2 t_9^2 t_{10}^3 + 
 t_1 t_2 t_4 t_5^2 t_6 t_7^2 t_8^3 t_9^2 t_{10}^3 - 
 t_1^2 t_2^2 t_3 t_4^2 t_5^2 t_6 t_7^2 t_8^3 t_9^2 t_{10}^3 - 
 3 t_1 t_2^2 t_3 t_4^2 t_5^3 t_6^2 t_7^3 t_8^3 t_9^2 t_{10}^3 
 \nn\\&& \hspace{0.5cm} + 
 3 t_1^2 t_2^4 t_3^3 t_4^4 t_5^4 t_6^3 t_7^4 t_8^3 t_9^2 t_{10}^3 + 
 t_1 t_2^4 t_3^3 t_4^4 t_5^5 t_6^4 t_7^5 t_8^3 t_9^2 t_{10}^3 - 
 t_1^2 t_2^5 t_3^4 t_4^5 t_5^5 t_6^4 t_7^5 t_8^3 t_9^2 t_{10}^3 + 
 t_1^2 t_2^4 t_3^2 t_4^3 t_5^5 t_6^3 t_7^4 t_8^4 t_9^2 t_{10}^3 - 
 t_1 t_2 t_3^3 t_4^3 t_5 t_6^3 t_7^3 t_8 t_9^3 t_{10}^3 
 \nn\\&& \hspace{0.5cm} + 
 t_1 t_3 t_4 t_5 t_6^2 t_7^2 t_8^2 t_9^3 t_{10}^3 - 
 t_1^2 t_2 t_3^2 t_4^2 t_5 t_6^2 t_7^2 t_8^2 t_9^3 t_{10}^3 - 
 3 t_1 t_2 t_3^2 t_4^2 t_5^2 t_6^3 t_7^3 t_8^2 t_9^3 t_{10}^3 + 
 3 t_1^2 t_2^3 t_3^4 t_4^4 t_5^3 t_6^4 t_7^4 t_8^2 t_9^3 t_{10}^3 + 
 t_1 t_2^3 t_3^4 t_4^4 t_5^4 t_6^5 t_7^5 t_8^2 t_9^3 t_{10}^3 
 \nn\\&& \hspace{0.5cm} - 
 t_1^2 t_2^4 t_3^5 t_4^5 t_5^4 t_6^5 t_7^5 t_8^2 t_9^3 t_{10}^3 - 
 t_1^2 t_2 t_3 t_4 t_5^2 t_6^2 t_7^2 t_8^3 t_9^3 t_{10}^3 + 
 t_1^3 t_2^2 t_3^2 t_4^2 t_5^2 t_6^2 t_7^2 t_8^3 t_9^3 t_{10}^3 - 
 t_1 t_2 t_3 t_4 t_5^3 t_6^3 t_7^3 t_8^3 t_9^3 t_{10}^3 + 
 t_1^3 t_2^3 t_3^3 t_4^3 t_5^3 t_6^3 t_7^3 t_8^3 t_9^3 t_{10}^3 
 \nn\\&& \hspace{0.5cm} + 
 6 t_1^2 t_2^3 t_3^3 t_4^3 t_5^4 t_6^4 t_7^4 t_8^3 t_9^3 t_{10}^3 - 
 2 t_1^3 t_2^4 t_3^4 t_4^4 t_5^4 t_6^4 t_7^4 t_8^3 t_9^3 t_{10}^3 + 
 t_1 t_2^3 t_3^3 t_4^3 t_5^5 t_6^5 t_7^5 t_8^3 t_9^3 t_{10}^3 - 
 3 t_1^2 t_2^4 t_3^4 t_4^4 t_5^5 t_6^5 t_7^5 t_8^3 t_9^3 t_{10}^3 + 
 t_1^2 t_2^3 t_3^2 t_4^2 t_5^5 t_6^4 t_7^4 t_8^4 t_9^3 t_{10}^3 \nn\\&& \hspace{0.5cm} 
 - 
 t_1^3 t_2^4 t_3^3 t_4^3 t_5^5 t_6^4 t_7^4 t_8^4 t_9^3 t_{10}^3 - 
 t_1^2 t_2^4 t_3^3 t_4^3 t_5^6 t_6^5 t_7^5 t_8^4 t_9^3 t_{10}^3 + 
 t_1^2 t_2^2 t_3^4 t_4^3 t_5^3 t_6^5 t_7^4 t_8^2 t_9^4 t_{10}^3 + 
 t_1^2 t_2^2 t_3^3 t_4^2 t_5^4 t_6^5 t_7^4 t_8^3 t_9^4 t_{10}^3 - 
 t_1^3 t_2^3 t_3^4 t_4^3 t_5^4 t_6^5 t_7^4 t_8^3 t_9^4 t_{10}^3 \nn\\&& \hspace{0.5cm} 
 - 
 t_1^2 t_2^3 t_3^4 t_4^3 t_5^5 t_6^6 t_7^5 t_8^3 t_9^4 t_{10}^3 - 
 t_1 t_2^2 t_3^2 t_4^4 t_5 t_6 t_7^3 t_8^2 t_9^2 t_{10}^4 - 
 t_2^2 t_3^2 t_4^4 t_5^2 t_6^2 t_7^4 t_8^2 t_9^2 t_{10}^4 + 
 t_1 t_2^3 t_3^3 t_4^5 t_5^2 t_6^2 t_7^4 t_8^2 t_9^2 t_{10}^4 - 
 t_1 t_2^2 t_3 t_4^3 t_5^2 t_6 t_7^3 t_8^3 t_9^2 t_{10}^4 
 \nn\\&& \hspace{0.5cm}
 - 
 t_1^2 t_2^3 t_3^2 t_4^4 t_5^2 t_6 t_7^3 t_8^3 t_9^2 t_{10}^4 + 
 t_1^2 t_2^4 t_3^3 t_4^5 t_5^3 t_6^2 t_7^4 t_8^3 t_9^2 t_{10}^4 - 
 t_2^3 t_3^2 t_4^4 t_5^4 t_6^3 t_7^5 t_8^3 t_9^2 t_{10}^4 + 
 t_1 t_2^4 t_3^3 t_4^5 t_5^4 t_6^3 t_7^5 t_8^3 t_9^2 t_{10}^4 + 
 2 t_1^2 t_2^4 t_3^2 t_4^4 t_5^4 t_6^2 t_7^4 t_8^4 t_9^2 t_{10}^4 
 \nn\\&& \hspace{0.5cm}
 + 
 t_1 t_2^4 t_3^2 t_4^4 t_5^5 t_6^3 t_7^5 t_8^4 t_9^2 t_{10}^4 - 
 t_1^2 t_2^5 t_3^3 t_4^5 t_5^5 t_6^3 t_7^5 t_8^4 t_9^2 t_{10}^4 - 
 t_1 t_2 t_3^2 t_4^3 t_5 t_6^2 t_7^3 t_8^2 t_9^3 t_{10}^4 - 
 t_1^2 t_2^2 t_3^3 t_4^4 t_5 t_6^2 t_7^3 t_8^2 t_9^3 t_{10}^4 + 
 t_1^2 t_2^3 t_3^4 t_4^5 t_5^2 t_6^3 t_7^4 t_8^2 t_9^3 t_{10}^4 
 \nn\\&& \hspace{0.5cm}- 
 t_2^2 t_3^3 t_4^4 t_5^3 t_6^4 t_7^5 t_8^2 t_9^3 t_{10}^4 + 
 t_1 t_2^3 t_3^4 t_4^5 t_5^3 t_6^4 t_7^5 t_8^2 t_9^3 t_{10}^4 - 
 2 t_1 t_2 t_3 t_4^2 t_5^2 t_6^2 t_7^3 t_8^3 t_9^3 t_{10}^4 + 
 t_1^3 t_2^3 t_3^3 t_4^4 t_5^2 t_6^2 t_7^3 t_8^3 t_9^3 t_{10}^4 + 
 6 t_1^2 t_2^3 t_3^3 t_4^4 t_5^3 t_6^3 t_7^4 t_8^3 t_9^3 t_{10}^4 
 \nn\\&& \hspace{0.5cm}
 - 
 t_1^3 t_2^4 t_3^4 t_4^5 t_5^3 t_6^3 t_7^4 t_8^3 t_9^3 t_{10}^4 - 
 t_2^2 t_3^2 t_4^3 t_5^4 t_6^4 t_7^5 t_8^3 t_9^3 t_{10}^4 + 
 5 t_1 t_2^3 t_3^3 t_4^4 t_5^4 t_6^4 t_7^5 t_8^3 t_9^3 t_{10}^4 - 
 5 t_1^2 t_2^4 t_3^4 t_4^5 t_5^4 t_6^4 t_7^5 t_8^3 t_9^3 t_{10}^4 + 
 t_2^3 t_3^3 t_4^4 t_5^5 t_6^5 t_7^6 t_8^3 t_9^3 t_{10}^4 
 \nn\\&& \hspace{0.5cm} - 
 t_1 t_2^4 t_3^4 t_4^5 t_5^5 t_6^5 t_7^6 t_8^3 t_9^3 t_{10}^4 + 
 3 t_1^2 t_2^3 t_3^2 t_4^3 t_5^4 t_6^3 t_7^4 t_8^4 t_9^3 t_{10}^4 - 
 2 t_1^3 t_2^4 t_3^3 t_4^4 t_5^4 t_6^3 t_7^4 t_8^4 t_9^3 t_{10}^4 + 
 t_1 t_2^3 t_3^2 t_4^3 t_5^5 t_6^4 t_7^5 t_8^4 t_9^3 t_{10}^4 - 
 5 t_1^2 t_2^4 t_3^3 t_4^4 t_5^5 t_6^4 t_7^5 t_8^4 t_9^3 t_{10}^4 
 \nn\\&& \hspace{0.5cm} + 
 t_1^3 t_2^5 t_3^4 t_4^5 t_5^5 t_6^4 t_7^5 t_8^4 t_9^3 t_{10}^4 - 
 t_1 t_2^4 t_3^3 t_4^4 t_5^6 t_6^5 t_7^6 t_8^4 t_9^3 t_{10}^4 + 
 t_1^2 t_2^5 t_3^4 t_4^5 t_5^6 t_6^5 t_7^6 t_8^4 t_9^3 t_{10}^4 + 
 2 t_1^2 t_2^2 t_3^4 t_4^4 t_5^2 t_6^4 t_7^4 t_8^2 t_9^4 t_{10}^4 + 
 t_1 t_2^2 t_3^4 t_4^4 t_5^3 t_6^5 t_7^5 t_8^2 t_9^4 t_{10}^4 
 \nn\\&& \hspace{0.5cm}
 - 
 t_1^2 t_2^3 t_3^5 t_4^5 t_5^3 t_6^5 t_7^5 t_8^2 t_9^4 t_{10}^4 + 
 3 t_1^2 t_2^2 t_3^3 t_4^3 t_5^3 t_6^4 t_7^4 t_8^3 t_9^4 t_{10}^4 - 
 2 t_1^3 t_2^3 t_3^4 t_4^4 t_5^3 t_6^4 t_7^4 t_8^3 t_9^4 t_{10}^4 + 
 t_1 t_2^2 t_3^3 t_4^3 t_5^4 t_6^5 t_7^5 t_8^3 t_9^4 t_{10}^4 - 
 5 t_1^2 t_2^3 t_3^4 t_4^4 t_5^4 t_6^5 t_7^5 t_8^3 t_9^4 t_{10}^4 
 \nn\\&& \hspace{0.5cm}
 + 
 t_1^3 t_2^4 t_3^5 t_4^5 t_5^4 t_6^5 t_7^5 t_8^3 t_9^4 t_{10}^4 - 
 t_1 t_2^3 t_3^4 t_4^4 t_5^5 t_6^6 t_7^6 t_8^3 t_9^4 t_{10}^4 + 
 t_1^2 t_2^4 t_3^5 t_4^5 t_5^5 t_6^6 t_7^6 t_8^3 t_9^4 t_{10}^4 + 
 2 t_1^2 t_2^2 t_3^2 t_4^2 t_5^4 t_6^4 t_7^4 t_8^4 t_9^4 t_{10}^4 - 
 2 t_1^3 t_2^3 t_3^3 t_4^3 t_5^4 t_6^4 t_7^4 t_8^4 t_9^4 t_{10}^4 
 \nn\\&& \hspace{0.5cm}
 - 
 3 t_1^2 t_2^3 t_3^3 t_4^3 t_5^5 t_6^5 t_7^5 t_8^4 t_9^4 t_{10}^4 + 
 2 t_1^3 t_2^4 t_3^4 t_4^4 t_5^5 t_6^5 t_7^5 t_8^4 t_9^4 t_{10}^4 + 
 t_1^2 t_2^4 t_3^4 t_4^4 t_5^6 t_6^6 t_7^6 t_8^4 t_9^4 t_{10}^4 + 
 t_1 t_2^2 t_3^2 t_4^4 t_5^2 t_6^2 t_7^4 t_8^3 t_9^3 t_{10}^5 + 
 t_1^2 t_2^3 t_3^3 t_4^5 t_5^2 t_6^2 t_7^4 t_8^3 t_9^3 t_{10}^5 
 \nn\\&& \hspace{0.5cm}
 + 
 t_1 t_2^3 t_3^3 t_4^5 t_5^3 t_6^3 t_7^5 t_8^3 t_9^3 t_{10}^5 - 
 t_1^2 t_2^4 t_3^4 t_4^6 t_5^3 t_6^3 t_7^5 t_8^3 t_9^3 t_{10}^5 + 
 t_2^3 t_3^3 t_4^5 t_5^4 t_6^4 t_7^6 t_8^3 t_9^3 t_{10}^5 - 
 t_1 t_2^4 t_3^4 t_4^6 t_5^4 t_6^4 t_7^6 t_8^3 t_9^3 t_{10}^5 + 
 t_1^2 t_2^3 t_3^2 t_4^4 t_5^3 t_6^2 t_7^4 t_8^4 t_9^3 t_{10}^5 
 \nn\\&& \hspace{0.5cm} 
 + 
 t_1 t_2^3 t_3^2 t_4^4 t_5^4 t_6^3 t_7^5 t_8^4 t_9^3 t_{10}^5 - 
 3 t_1^2 t_2^4 t_3^3 t_4^5 t_5^4 t_6^3 t_7^5 t_8^4 t_9^3 t_{10}^5 - 
 t_1 t_2^4 t_3^3 t_4^5 t_5^5 t_6^4 t_7^6 t_8^4 t_9^3 t_{10}^5 + 
 t_1^2 t_2^5 t_3^4 t_4^6 t_5^5 t_6^4 t_7^6 t_8^4 t_9^3 t_{10}^5 - 
 t_1^2 t_2^4 t_3^2 t_4^4 t_5^5 t_6^3 t_7^5 t_8^5 t_9^3 t_{10}^5 
 \nn\\&& \hspace{0.5cm}
 + 
 t_1^2 t_2^2 t_3^3 t_4^4 t_5^2 t_6^3 t_7^4 t_8^3 t_9^4 t_{10}^5 + 
 t_1 t_2^2 t_3^3 t_4^4 t_5^3 t_6^4 t_7^5 t_8^3 t_9^4 t_{10}^5 - 
 3 t_1^2 t_2^3 t_3^4 t_4^5 t_5^3 t_6^4 t_7^5 t_8^3 t_9^4 t_{10}^5 - 
 t_1 t_2^3 t_3^4 t_4^5 t_5^4 t_6^5 t_7^6 t_8^3 t_9^4 t_{10}^5 + 
 t_1^2 t_2^4 t_3^5 t_4^6 t_5^4 t_6^5 t_7^6 t_8^3 t_9^4 t_{10}^5 
 \nn\\&& \hspace{0.5cm}+ 
 t_1^2 t_2^2 t_3^2 t_4^3 t_5^3 t_6^3 t_7^4 t_8^4 t_9^4 t_{10}^5 - 
 t_1^3 t_2^3 t_3^3 t_4^4 t_5^3 t_6^3 t_7^4 t_8^4 t_9^4 t_{10}^5 + 
 t_1 t_2^2 t_3^2 t_4^3 t_5^4 t_6^4 t_7^5 t_8^4 t_9^4 t_{10}^5 - 
 5 t_1^2 t_2^3 t_3^3 t_4^4 t_5^4 t_6^4 t_7^5 t_8^4 t_9^4 t_{10}^5 + 
 2 t_1^3 t_2^4 t_3^4 t_4^5 t_5^4 t_6^4 t_7^5 t_8^4 t_9^4 t_{10}^5 
 \nn\\&& \hspace{0.5cm}- 
 t_1 t_2^3 t_3^3 t_4^4 t_5^5 t_6^5 t_7^6 t_8^4 t_9^4 t_{10}^5 + 
 3 t_1^2 t_2^4 t_3^4 t_4^5 t_5^5 t_6^5 t_7^6 t_8^4 t_9^4 t_{10}^5 - 
 t_1^2 t_2^3 t_3^2 t_4^3 t_5^5 t_6^4 t_7^5 t_8^5 t_9^4 t_{10}^5 + 
 t_1^3 t_2^4 t_3^3 t_4^4 t_5^5 t_6^4 t_7^5 t_8^5 t_9^4 t_{10}^5 + 
 t_1^2 t_2^4 t_3^3 t_4^4 t_5^6 t_6^5 t_7^6 t_8^5 t_9^4 t_{10}^5 
 \nn\\&& \hspace{0.5cm} 
 - 
 t_1^2 t_2^2 t_3^4 t_4^4 t_5^3 t_6^5 t_7^5 t_8^3 t_9^5 t_{10}^5 - 
 t_1^2 t_2^2 t_3^3 t_4^3 t_5^4 t_6^5 t_7^5 t_8^4 t_9^5 t_{10}^5 + 
 t_1^3 t_2^3 t_3^4 t_4^4 t_5^4 t_6^5 t_7^5 t_8^4 t_9^5 t_{10}^5 + 
 t_1^2 t_2^3 t_3^4 t_4^4 t_5^5 t_6^6 t_7^6 t_8^4 t_9^5 t_{10}^5 - 
 t_1^2 t_2^3 t_3^3 t_4^5 t_5^3 t_6^3 t_7^5 t_8^4 t_9^4 t_{10}^6 
 \nn\\&& \hspace{0.5cm} - 
 t_1 t_2^3 t_3^3 t_4^5 t_5^4 t_6^4 t_7^6 t_8^4 t_9^4 t_{10}^6 + 
 t_1^2 t_2^4 t_3^4 t_4^6 t_5^4 t_6^4 t_7^6 t_8^4 t_9^4 t_{10}^6 + 
 t_1^2 t_2^4 t_3^3 t_4^5 t_5^5 t_6^4 t_7^6 t_8^5 t_9^4 t_{10}^6 + 
 t_1^2 t_2^3 t_3^4 t_4^5 t_5^4 t_6^5 t_7^6 t_8^4 t_9^5 t_{10}^6 - 
 t_1^3 t_2^5 t_3^5 t_4^6 t_5^6 t_6^6 t_7^7 t_8^5 t_9^5 t_{10}^6
 ~~.\nn\\
\eea
\footnotesize
}
The mesonic moduli space is a non-complete intersection. The unrefined Hilbert series is
\beal{e67bbb}
g_1(t;\mesonic)
&=&
\frac{(1-t^2)^3}{
(1-t^4)^4 (1-t^6)^4
 }
  \times
 (1 + 3 t^2 + 6 t^4 + 14 t^6 + 27 t^8 + 32 t^{10} + 31 t^{12}
 \nn\\
 &&
 + 32 t^{14} + 
 27 t^{16} + 14 t^{18} + 6 t^{20} + 3 t^{22} + t^{24})
\eea
 It is a 5-dimensional Calabi-Yau space. The plethystic logarithm of the refined Hilbert series is
\beal{e67c}
&&
PL[g_1(t_i;\mesonic)]=
(t_1 t_2 t_3 t_4 + t_1 t_2 t_5 t_8 + t_1 t_3 t_6 t_9 + t_1 t_8 t_9 t_{10})
+
(t_2 t_3 t_4 t_5 t_6 t_7 + t_2 t_5^2 t_6 t_7 t_8 
\nn\\
&&
\hspace{0.5cm}
 + t_3 t_5 t_6^2 t_7 t_9 + t_2 t_3 t_4^2 t_7 t_{10} 
 + t_2 t_4 t_5 t_7 t_8 t_{10} 
 + t_3 t_4 t_6 t_7 t_9 t_{10} 
 + t_5 t_6 t_7 t_8 t_9 t_{10} 
 + t_4 t_7 t_8 t_9 t_{10}^2)
 \nn\\
 &&
 \hspace{0.5cm}
- (
t_1 t_2^2 t_3 t_4 t_5^2 t_6 t_7 t_8 
+ t_1 t_2 t_3^2 t_4 t_5 t_6^2 t_7 t_9 
+ t_1 t_2 t_3 t_5^2 t_6^2 t_7 t_8 t_9 
+ t_1 t_2^2 t_3 t_4^2 t_5 t_7 t_8 t_{10} 
\nn\\
&&
\hspace{0.5cm}
+ t_1 t_2 t_3^2 t_4^2 t_6 t_7 t_9 t_{10} 
+ 3 t_1 t_2 t_3 t_4 t_5 t_6 t_7 t_8 t_9 t_{10} 
+ t_1 t_2 t_5^2 t_6 t_7 t_8^2 t_9 t_{10} 
+ t_1 t_3 t_5 t_6^2 t_7 t_8 t_9^2 t_{10} 
\nn\\
&&
\hspace{0.5cm}
+ t_1 t_2 t_3 t_4^2 t_7 t_8 t_9 t_{10}^2 
+ t_1 t_2 t_4 t_5 t_7 t_8^2 t_9 t_{10}^2 
+ t_1 t_3 t_4 t_6 t_7 t_8 t_9^2 t_{10}^2
) 
- (
t_2^2 t_3 t_4^2 t_5^2 t_6 t_7^2 t_8 t_{10} 
\nn\\
&&
\hspace{0.5cm}
+ t_2 t_3^2 t_4^2 t_5 t_6^2 t_7^2 t_9 t_{10} 
+ 2 t_2 t_3 t_4 t_5^2 t_6^2 t_7^2 t_8 t_9 t_{10} 
+  2 t_2 t_3 t_4^2 t_5 t_6 t_7^2 t_8 t_9 t_{10}^2 
+ t_2 t_4 t_5^2 t_6 t_7^2 t_8^2 t_9 t_{10}^2 
\nn\\
&&
\hspace{0.5cm}
+ t_3 t_4 t_5 t_6^2 t_7^2 t_8 t_9^2 t_{10}^2
)
+ \dots ~~.
\eea
We can read from the plethystic logarithm the lowest order generators of the mesonic moduli space and are

\begin{center}
\begin{tabular}{|c|c|}
\hline
generator & perfect matchings\\
\hline\hline
$A_1$ & $p_1 p_2 p_3 p_4$\\
$A_2$ & $p_1 p_2 p_5 p_8$\\
$A_3$ & $p_1 p_3 p_6 p_9$\\
$A_4$ & $p_1 p_8 p_9 p_{10}$\\
$B_1$ & $p_2 p_3 p_4 p_5 p_6 p_7$ \\
$B_2$ & $p_2 p_5^2 p_6 p_7 p_8$ \\
$B_3$ & $p_3 p_5 p_6^2 p_7 p_9$ \\
$B_4$ & $p_2 p_4 p_5 p_7 p_8 p_{10}$ \\
$B_5$ & $p_3 p_4 p_6 p_7 p_9 p_{10}$ \\
$B_6$ & $p_5 p_6 p_7 p_8 p_9 p_{10}$ \\
$B_7$ & $p_4 p_7 p_8 p_7 p_{10}^2$ \\
\hline
\end{tabular}
\end{center}
The generators form the following first order relations amongst them which correspond to the presented negative terms in the expansion of the plethystic logarithm in \eref{e67c},
\beal{es48}
&&
\{B_{6} B_{7}-B_{5} B_{8},~B_{4} B_{7}-B_{3} B_{8},~B_{2} B_{7}-B_{1} B_{8},~B_{3}
B_{6}-B_{2} B_{8},~A_{4} B_{6}-A_{3} B_{8}, \nn\\
&&
~B_{4} B_{5}-B_{2} B_{8},~B_{3} B_{5}-B_{1}
B_{8},~B_{2} B_{5}-B_{1} B_{6},~A_{4} B_{5}-A_{3} B_{7},~A_{2} B_{5}-A_{1} B_{8},
\nn\\
&&
~A_{4}
B_{4}-A_{2} B_{8},~A_{3} B_{4}-A_{2} B_{6},~B_{2} B_{3}-B_{1} B_{4},~A_{4} B_{3}-A_{2}
B_{7},~A_{3} B_{3}-A_{1} B_{8},\nn\\
&&
~A_{4} B_{2}-A_{1} B_{8},~A_{3} B_{2}-A_{1} B_{6},~A_{2}
B_{2}-A_{1} B_{4},~A_{4} B_{1}-A_{1} B_{7},~A_{3} B_{1}-A_{1} B_{5},\nn\\
&&
~A_{2} B_{1}-A_{1}
B_{3}\}
~~.
\eea

The global symmetry is $U(1)^4 \times U(1)_R$ and has no enhancement. The toric diagram of the Calabi-Yau 5-fold is given by
\beal{e64nn}
G_t = 
\left(
\begin{array}{cccccccccc}
p_1 & p_2 & p_3 & p_4 & p_5 & p_6 & p_7 & p_8 & p_9 & p_{10}\\
\hline
 1 & 1 & 0 & 0 & 1 & 0 & 0 & 1 & 0 & 0 \\
 1 & 0 & 1 & 0 & 0 & 1 & 0 & 0 & 1 & 0 \\
-1 & 0 & 0 & 1 & 0 & 0 & 1 & 0 & 0 & 1 \\
 -1 & 0 & 0 & 0 & 1 & 1 & 1 & 0 & 0 & 0 \\
 1 & 1 & 1 & 1 & 0 & 0 & 0 & 0 & 0 & 0
\end{array}
\right)~~.
\eea
\\

\subsubsection{Model 8.4h: $NC3$ \label{s_m84h}}

\begin{figure}[ht!!]
\begin{center}
\includegraphics[trim=0cm 0cm 0cm 0cm,totalheight=8 cm]{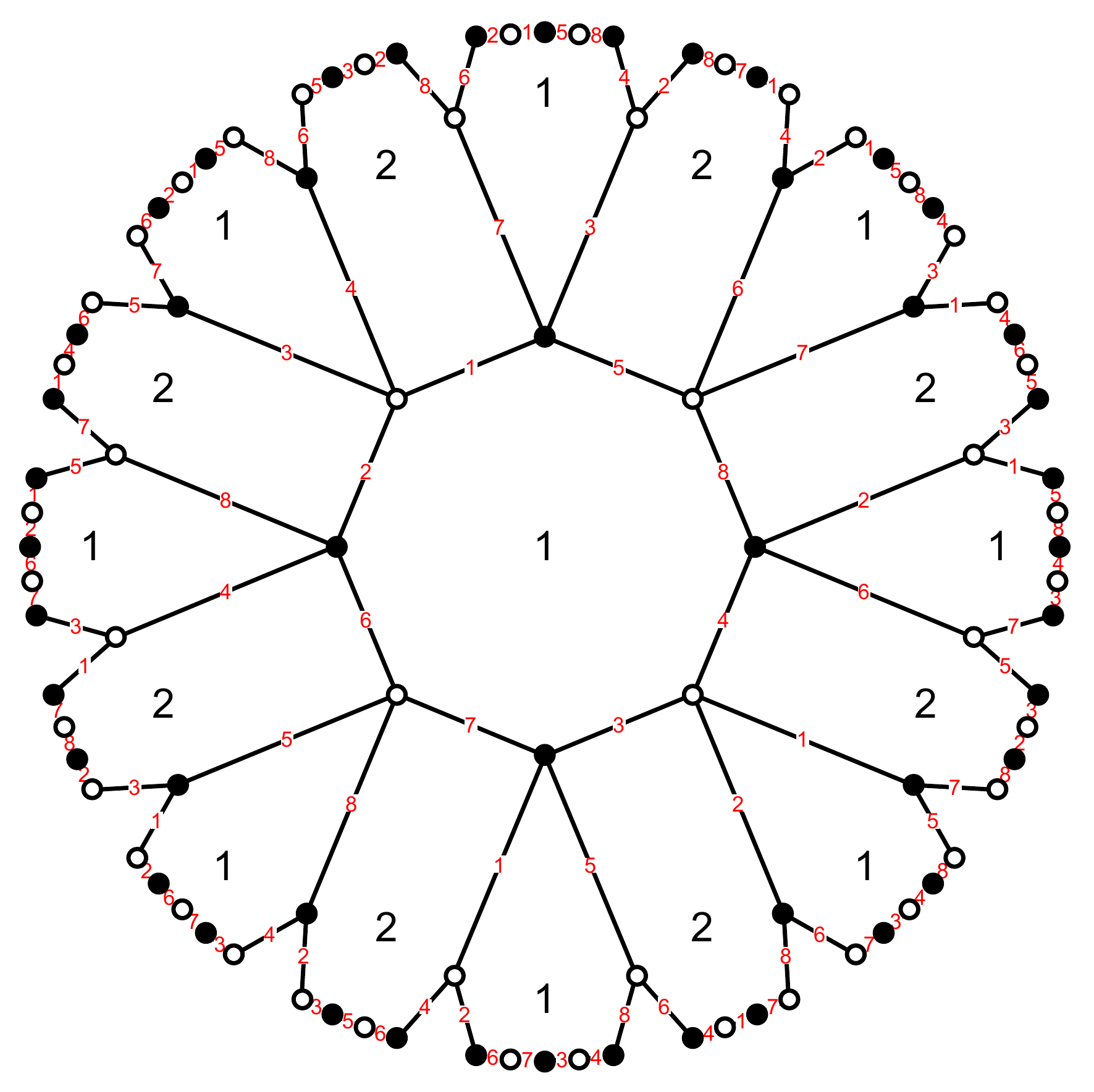}\\
\vspace{0.3cm}
\begin{tabular}{cccccccc}
{\color[rgb]{1.000000,0.000000,0.000000} 1} &
{\color[rgb]{1.000000,0.000000,0.000000} 2} &
{\color[rgb]{1.000000,0.000000,0.000000} 3} &
{\color[rgb]{1.000000,0.000000,0.000000} 4} &
{\color[rgb]{1.000000,0.000000,0.000000} 5} &
{\color[rgb]{1.000000,0.000000,0.000000} 6} &
{\color[rgb]{1.000000,0.000000,0.000000} 7} &
{\color[rgb]{1.000000,0.000000,0.000000} 8} 
\\
$X_{21}^{1}$ & $X_{12}^{1}$ & $X_{21}^{2}$ & $X_{12}^{2}$ & $X_{12}^{3}$ & $X_{21}^{3}$ & $X_{12}^{4}$ & $X_{21}^{4}$
\end{tabular}
\caption{The Model 8.4h brane tiling on a $g=2$ Riemann surface with 2 gauge groups, 8 fields and 4 superpotential terms.
\label{fm84ht}}
 \end{center}
 \end{figure}

\begin{figure}[ht!!]
\begin{center}
\includegraphics[trim=0cm 0cm 0cm 0cm,totalheight=1 cm]{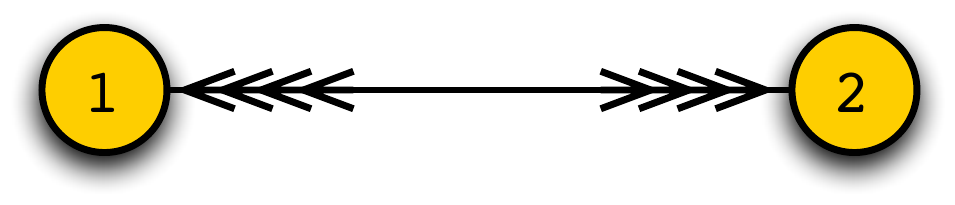}
\caption{The quiver diagram for Model 8.4h, a brane tiling on a $g=2$ Riemann surface with 2 gauge groups, 8 fields and 4 superpotential terms.
\label{fm84hq}}
 \end{center}
 \end{figure}
 
For Model 8.4h, the brane tiling and corresponding quiver is shown in \fref{fm84ht} and \fref{fm84hq} respectively. The quartic superpotential is
\beal{e8450}
W= 
+ X_{21}^{1} X_{12}^{1} X_{21}^{2} X_{12}^{2}
+ X_{12}^{3} X_{21}^{3} X_{12}^{4} X_{21}^{4}
- X_{21}^{1} X_{12}^{3} X_{21}^{2} X_{12}^{4}
- X_{12}^{1} X_{21}^{4} X_{12}^{2} X_{21}^{3}
~~.
\nn\\
\eea
The quiver incidence matrix is
\beal{e8451}
d=\left(
\begin{array}{cccccccc}
X_{21}^{1} & X_{12}^{1} & X_{21}^{2} & X_{12}^{2} & X_{12}^{3} & X_{21}^{3} & X_{12}^{4} & X_{21}^{4}
\\
\hline
 1 & -1 & 1 & -1 & -1 & 1 & -1 & 1 \\
 -1 & 1 & -1 & 1 & 1 & -1 & 1 & -1
\end{array}
\right)
~~.
\eea

The brane tiling has $c=8$ perfect matchings. The perfect matchings are encoded in the matrix
\beal{e8452}
P=\left(
\begin{array}{c|cccccccc}
\; & a_1 & a_2 & b_1 & b_2 & c_1 & c_2 & d_1 & d_2\\
\hline
X_{12}^{1} & 0 & 0 & 0 & 0 & 1 & 0 & 0 & 1 \\
X_{12}^{2} & 0 & 0 & 0 & 0 & 0 & 1 & 1 & 0 \\
X_{12}^{3} & 0 & 0 & 0 & 0 & 1 & 0 & 1 & 0 \\
X_{12}^{4} & 0 & 0 & 0 & 0 & 0 & 1 & 0 & 1 \\
X_{21}^{1} & 1 & 0 & 0 & 1 & 0 & 0 & 0 & 0 \\
X_{21}^{2} & 0 & 1 & 1 & 0 & 0 & 0 & 0 & 0 \\
X_{21}^{3} & 1 & 0 & 1 & 0 & 0 & 0 & 0 & 0 \\
X_{21}^{4} & 0 & 1 & 0 & 1 & 0 & 0 & 0 & 0
\end{array}
\right)
~~.
\eea
The brane tiling has the zig-zag paths,
\beal{e8452bb}
&
\eta_1 = (X_{21}^{1},X_{12}^{1},X_{21}^{4},X_{12}^{3},X_{21}^{2},X_{12}^{2},X_{21}^{3},X_{12}^{4})~,~
&\nn\\
&
\eta_2 = (X_{21}^{1},X_{12}^{3},X_{21}^{3},X_{12}^{1},X_{21}^{2},X_{12}^{4},X_{21}^{4},X_{12}^{2})~.~
&
\eea

The F-term constraints can be expressed as charges carried by the perfect matchings. The charges are given by
\beal{e8453}
Q_F =
\left(
\begin{array}{cccccccc}
a_1 & a_2 & b_1 & b_2 & c_1 & c_2 & d_1 & d_2 \\
\hline
 1 	&  1	  &  0  &  0  &  0  & 0   & -1  & -1 \\
 0 	&  0  & -1  & -1  &  1  & 1   &  0  &  0
\end{array}
\right)
~~.
\eea
The D-term charges are encoded in the quiver incidence matrix $d$ and are
\beal{e8454}
Q_D = 
\left(
\begin{array}{cccccccc}
a_1 & a_2 & b_1 & b_2 & c_1 & c_2 & d_1 & d_2 \\
\hline
 1 &  1 &  -1 &  -1 &   0 &   0 &   0 &   0
\end{array}
\right)~~.
\eea
\comment{\beal{e8454}
Q_D = 
\left(
\begin{array}{cccccccc}
c_1 & b_1 & d_1 & a_1 & b_2 & c_2 & a_2 & d_2 \\
p_1 & p_2 & p_3 & p_4 & p_5 & p_6 & p_7 & p_8\\
\hline
  0 &   1 &   0 &  -1 &   1 &   0 &  -1 &   0
\end{array}
\right)~~.
\eea}
When reduced, the total charge matrix
\beal{e8454bb}
Q_t = 
\left(
\begin{array}{cccccccc}
a_1 & a_2 & b_1 & b_2 & c_1 & c_2 & d_1 & d_2 \\
\hline
 1 	&  1	  &  0  &  0  &  0  & 0   & -1  & -1 \\
 0  &  0  &  1  &  1  &   0 &   0 & -1  & -1 \\
 0 	&  0  &  0  &  0  &  1  & 1   & -1  & -1
\end{array}
\right)~~.
\eea
is identical to the total charge matrix of Model 8.2b in section \sref{s_m82b}.
The mesonic moduli space of Model 8.4h which can be expressed as a symplectic quotient,
\beal{es8456}
\mesonic = \mathbb{C}^{8} // Q_t ~~,
\eea
is the same as Model 8.2b. It is a toric Calabi-Yau 5-fold and is a non-complete intersection.
\\

\section{Conclusions and Future Directions}

We have discovered a new set of field theories with the classification of the first few brane tilings on a $g=2$ Riemann surface. The classification identifies 16 of what we call \textit{restricted} $g=2$ brane tilings with up to 8 fields and 4 superpotential terms. Their mesonic moduli spaces are specified by calculating the refined Hilbert series and are shown to be toric Calabi-Yau 5-folds. 

A feature that has not been highlighted so far is that although the $g=2$ brane tilings in the classification have no self-intersecting zig-zag paths and no multi-bonded edges, some of them have multiple perfect matchings associated to extremal points in the toric diagram. This is one of a series of new observations which requires further studies in the near future. In summary, the new observations are as follows:
\begin{itemize}
\item For the following $g=2$ brane tilings in the classification, more than one perfect matching is assigned to extremal toric points:
 \begin{center}
 6.2a~,~7.2~,~8.4d~.
 \end{center}
 These are however restricted brane tilings with no self-intersecting zig-zag paths and no multi-bonded edges. We expect that the brane tilings on a $g=2$ Riemann surface feature graphical properties beyond zig-zag paths and multi-bonded edges that indicate the assignment of multiple GLSM fields to extremal toric points.

\item Zig-zag paths that play a pivotal role in relating geometry and field theory for torus brane tilings appear to play a lesser role in $g=2$ brane tilings. In fact, for all models in the classification, we observe that the number of zig-zag paths is less than the number of facets of the corresponding 4-dimensional toric diagram. The only exception is Model 5.2 where the numbers are equal.
 
\item For torus brane tilings with Calabi-Yau 3-fold mesonic moduli spaces, the area of the toric diagram corresponds to the number of gauge groups in the corresponding quiver gauge theory. The analogue of the area for the Calabi-Yau 5-fold mesonic moduli spaces for $g=2$ brane tilings is the 4-dimensional volume of the toric diagram. For the brane tilings in our classification, the volumes of their toric diagrams are as follows:
\begin{center}
\begin{tabular}{|c|cc|}
\hline
\# & Volume & Gauge Groups\\
\hline \hline
5.2 	 &1 & 1\\
6.2a &1 & 2\\
6.2b &3 & 2\\
6.2c &3 & 2\\
7.2  &2 & 3\\
7.4  &3 & 1\\
8.2a &4 & 4\\
8.2b &8 & 4\\
\hline
\end{tabular}~~~~
\begin{tabular}{|c|cc|}
\hline
\# & Volume & Gauge Groups\\
\hline \hline
8.4a &6 & 2\\
8.4b &2 & 2\\
8.4c &2 & 2\\
8.4d &3 & 2\\
8.4e &7 & 2\\
8.4f &4 & 2\\
8.4g &12& 2\\
8.4h &8 & 2\\
\hline
\end{tabular}
\end{center}
We observe that only Models 5.2, 8.2a, 8.4b and 8.4c have matching values for the number of gauge groups and toric diagram volumes. It is an interesting question to investigate when and why these two values match for $g=2$ brane tilings.
\end{itemize}

On the field theory side, we observe another array of open questions from our classification of $g=2$ brane tilings. As noted in the introduction, we have a limited understanding of the IR behaviour of these brane tilings. We hope to obtain more answers by doing the following in future studies:
\begin{itemize}
\item The ranks of the gauge groups can be varied, and one needs to study the IR behaviour for non-Abelian theories as well as their vacuum moduli spaces.
\item Boundaries, which represent flavor groups, can be added to a brane tiling. The IR behaviour of these theories with their vacuum moduli spaces needs to be studied.
\end{itemize}

As a final note of our work, we would like to point out that the mesonic moduli spaces of brane tilings on any Riemann surface are always odd dimensional toric Calabi-Yau. The natural question given this property is to ask whether even dimensional toric Calabi-Yau spaces can be related to brane tilings on Riemann surfaces via a modification of the bipartite graphs. 

As in the studies of Chern-Simons theories and brane tilings for M2-branes at toric Calabi-Yau fourfolds \cite{Hanany:2008cd,Hanany:2008fj,Ueda:2008hx,Davey:2009qx,Davey:2011mz,Martelli:2008si}, one may assign integer weights to edges in the tiling such that the alternated sum of these integers along the boundary of a face gives the Chern-Simons level of the associated gauge group. Solving the classical moduli space for $3d$ Chern-Simons theories introduces a symplectic quotient by a further $U(1)$ action, increasing the complex dimension by 1. Therefore the dimension of the mesonic moduli space of a $3d$ Abelian Chern-Simons quiver theory associated to a weighted tiling on a genus $g$ Riemann surface is $2(g+1)$.

\begin{figure}[ht!!]
\begin{center}

$\vcenter{\hbox{
\includegraphics[trim=0cm 0cm 0cm 0cm,totalheight=7
cm]{m62bt.pdf}
}}$
$\vcenter{\hbox{
\begin{tabular}{c}
\includegraphics[trim=0cm 0cm 0cm 0cm,totalheight=2
cm]{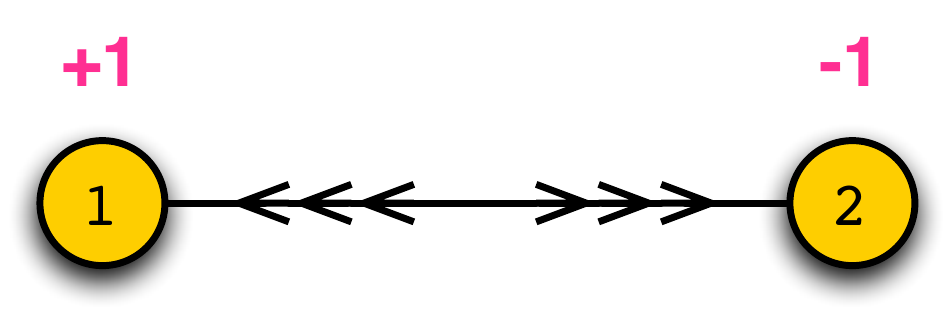}\\
\begin{tabular}{cccccc}
{\color[rgb]{1.000000,0.000000,0.000000} 1} &
{\color[rgb]{1.000000,0.000000,0.000000} 2} &
{\color[rgb]{1.000000,0.000000,0.000000} 3} &
{\color[rgb]{1.000000,0.000000,0.000000} 4} &
{\color[rgb]{1.000000,0.000000,0.000000} 5} &
{\color[rgb]{1.000000,0.000000,0.000000} 6} 
\\
$X_{12}^1$ & $X_{21}^{1}$ & $X_{12}^{2}$ & $X_{21}^2$ &
$X_{12}^3$ & $X_{21}^3$
\\
{\color{strawberry} +1} &
{\color{strawberry} 0} &
{\color{strawberry} 0} &
{\color{strawberry} 0} &
{\color{strawberry} 0} &
{\color{strawberry} 0} 
\end{tabular}
\end{tabular}
}}$

\caption{The Model 6.2b brane tiling with level assignment on the quiver and bifundamental fields.
\label{fm62bCS}}
 \end{center}
 \end{figure}

Let us consider as a quick example Model 6.2b in section \sref{s_m62b} with the mesonic moduli space being a non-complete intersection Calabi-Yau 5-fold. This model is a generalised conifold and we can assign levels $\pm 1$ to the two gauge groups of the theory as illustrated in \fref{fm62bCS}. This for instance can be achieved by assigning the level $+1$ to the bifundamental $X_{12}^{1}$ and by assigning level $0$ to all other bifundamental fields. By adopting the forward algorithm for Chern-Simons brane tilings \cite{Hanany:2008cd,Hanany:2008fj,Ueda:2008hx,Davey:2009qx,Davey:2011mz}\footnote{\textit{cf.} forward algorithm for $4d$ quiver gauge theories in appendix \sref{app_mesonic}.}, the level matrix $C$ then is
\beal{econc10}
C=\left(
\ba{cc}
U(1)_1 & U(1)_2\\
\hline
~1 & ~1 \\
~1 & -1
\ea
\right)~~,
\eea
and 
\beal{econc11}
&
d = \tilde{Q} \cdot P^t~,~
Q_F = \ker(P)~,~
Q_D = \ker(C)\cdot \tilde{Q}~,~
&
\nn\\
&
Q_t = (Q_F~ Q_D) \rightarrow G_t = \ker(Q_t) ~.
&
\eea 
Accordingly, with the above level assignment $C$, the $g=2$ brane tiling of Model 6.2b gives the charge matrices
\beal{esconc12}
Q_F = 0 ~,~
\tilde{Q}=
\left(
\begin{array}{cccccc}
 1 & 1 & 1 & -1 & -1 & -1 \\
 -1 & -1 & -1 & 1 & 1 & 1
\end{array}
\right)
~,~
Q_D = 0~,~
\eea
and hence the toric diagram
\beal{esconc13}
G_t=
\left(
\begin{array}{cccccc}
1 & 0 & 0 & 0 & 0 & 0 \\
0 & 1 & 0 & 0 & 0 & 0 \\
0 & 0 & 1 & 0 & 0 & 0 \\
0 & 0 & 0 & 1 & 0 & 0 \\
0 & 0 & 0 & 0 & 1 & 0 \\
0 & 0 & 0 & 0 & 0 & 1 \\
\end{array}
\right)~~.
\eea
This is the toric diagram for $\mathbb{C}^6$, the unit 5-simplex. We see here the precise analogue of obtaining the $\mathbb{C}^4$ mesonic moduli space by assigning Chern-Simons levels to the conifold theory.

With our classification of the first few $g=2$ brane tilings we have paved the path for new exciting problems. Most importantly, we have obtained a new class of quiver gauge theories which exhibit interesting moduli spaces. We plan to report on more progress in the near future.
\\


\section*{Acknowledgements}
We thank Vishnu Jejjala and Sanjaye Ramgoolam for discussions on a related project, and especially Massimo Bianchi for numerous fruitful discussions.
R.-K. S. likes to thank the Korea Institute for Advanced Study, l'Universit\'e libre de Bruxelles, the Princeton Center for Theoretical Science at Princeton University and the Perimeter Institute for Theoretical Physics for kind hospitality during various stages of this work.
The work of S. C. was supported in part by the STFC Consolidated Grant ST/J000353/1.

\appendix
\section{Summary of restricted $g=2$ Brane Tilings \label{app_con}}

 \begin{table}[H]
\begin{center}
\begin{tabular}{p{1cm}|p{5cm}|p{8cm}}
\hline \hline
E.T\# & Brane Tiling & Quiver \& Superpotential \& $\mesonic$ \\
\hline  
\vspace{-5cm}
5.2
& 
\includegraphics[trim=0cm 0cm 0cm 0cm,totalheight=5 cm]{m52at.pdf}
&
\vspace{-4.5cm}
\begin{tabular}[b]{l}
\vspace{0.25cm}
\includegraphics[trim=0cm 0cm 0cm 0cm,totalheight=1 cm]{m52aq.pdf}
\\
\vspace{0.25cm}
$W = (1~2~3~4~5) - (5~4~3~2~1)$
\\
$\mesonic=\mathbb{C}^5$
\\
$\eta_i=((5~ 4), (4~ 3), (3~ 2), (2~ 1), (1~ 5))$
\vspace{0.2cm}
\end{tabular}
\\
\hline  
\vspace{-5cm}
6.2a
& 
\includegraphics[trim=0cm 0cm 0cm 0cm,totalheight=5 cm]{m62at.pdf}
&
\vspace{-4.5cm}
\begin{tabular}[b]{l}
\vspace{0.25cm}
\includegraphics[trim=0cm 0cm 0cm 0cm,totalheight=1 cm]{m62aq.pdf}
\\
\vspace{0.25cm}
$W = (1~ 2~ 3~ 4~ 5~ 6) - (2~ 4~ 6~ 5~ 1~ 3)$
\\
$\mesonic=\mathbb{C}^5$
\\
$\eta_i=(
(6~ 5), (3~ 2), (5~ 1~ 2~ 4), (4~ 6~ 1~ 3)
)$
\vspace{0.2cm}
\end{tabular}
\\
\hline  
\vspace{-5cm}
6.2b
& 
\includegraphics[trim=0cm 0cm 0cm 0cm,totalheight=5 cm]{m62bt.pdf}
&
\vspace{-4.5cm}
\begin{tabular}[b]{l}
\vspace{0.25cm}
\includegraphics[trim=0cm 0cm 0cm 0cm,totalheight=1 cm]{m62bq.pdf}
\\
\vspace{0.25cm}
$W = (1~ 2~ 3~ 4~ 5~ 6) - (2~ 1~ 4~ 3~ 6~ 5)$
\\
$\mesonic=NC1$
\\
$\eta_i=(
(6~ 5), (4~ 3), (2~ 1), (5~ 2~ 3~ 6~ 1~ 4)
)$
\vspace{0.2cm}
\end{tabular}
\\
\hline
\end{tabular}
\end{center}
\caption{Restricted $g=2$ brane tilings \textbf{(1/6)}. \label{tdata1}}
\end{table}


 \begin{table}[H]
\begin{center}
\begin{tabular}{p{1cm}|p{5cm}|p{8cm}}
\hline \hline
E.T\# & Brane Tiling & Quiver \& Superpotential \& $\mesonic$ \\
\hline  
\vspace{-5cm}
6.2c
& 
\includegraphics[trim=0cm 0cm 0cm 0cm,totalheight=5 cm]{m62ct.pdf}
&
\vspace{-4.5cm}
\begin{tabular}[b]{l}
\vspace{0.25cm}
\includegraphics[trim=0cm 0cm 0cm 0cm,totalheight=1 cm]{m62cq.pdf}
\\
\vspace{0.25cm}
$W = (1~2~3~4~5~6) - (6~ 5~ 4~ 3~ 2~ 1)$
\\
$\mesonic=NC1$
\\
$\eta_i=((6~ 5), (5~ 4), (4~ 3), (3~ 2), (2~ 1), (1~ 6))$
\vspace{0.2cm}
\end{tabular}
\\
\hline  
\vspace{-5cm}
7.2
& 
\includegraphics[trim=0cm 0cm 0cm 0cm,totalheight=5 cm]{m72at.pdf}
&
\vspace{-4.5cm}
\begin{tabular}[b]{l}
\vspace{0.25cm}
\includegraphics[trim=0cm 0cm 0cm 0cm,totalheight=2.75 cm]{m72aq.pdf}
\\
\vspace{0.25cm}
$W = (1~ 2~ 3~ 4~ 5~ 6~ 7) - (1~ 3~ 2~ 4~ 7~ 6~ 5)$
\\
$\mesonic=\mathbb{C}^2\times\mathcal{C}$
\\
$\eta_i=((7~ 6), (6~ 5), (3~ 2), (5~ 1~ 2~ 4), (4~ 7~ 1~ 3))$
\vspace{0.2cm}
\end{tabular}
\\
\hline  
\vspace{-5cm}
7.4
& 
\includegraphics[trim=0cm 0cm 0cm 0cm,totalheight=5 cm]{m74at.pdf}
&
\vspace{-4.5cm}
\begin{tabular}[b]{l}
\vspace{0.25cm}
\includegraphics[trim=0cm 0cm 0cm 0cm,totalheight=1 cm]{m74aq.pdf}
\\
\vspace{0.25cm}
$W = (1~ 2~ 3~ 4) + (5~ 6~ 7)$
\\
$
\hspace{1cm}
-(1~ 5~ 7) - (2~ 6~ 4~ 3)$
\\
$\mesonic=\mathbb{C}\times\mathcal{M}_{3,2}$
\\
$\eta_i=((5~ 7), (4~ 3), (3~ 2), (7~ 1~ 2~ 6), (6~ 4~ 1~ 5))$
\vspace{0.2cm}
\end{tabular}
\\
\hline
\end{tabular}
\end{center}
\caption{Restricted $g=2$ brane tilings \textbf{(2/6)}. \label{tdata2}}
\end{table}


 \begin{table}[H]
\begin{center}
\begin{tabular}{p{1cm}|p{5cm}|p{8cm}}
\hline \hline
E.T\# & Brane Tiling & Quiver \& Superpotential \& $\mesonic$ \\
\hline  
\vspace{-5cm}
8.2a
& 
\includegraphics[trim=0cm 0cm 0cm 0cm,totalheight=5 cm]{m82at.pdf}
&
\vspace{-4.5cm}
\begin{tabular}[b]{l}
\vspace{0.25cm}
\includegraphics[trim=0cm 0cm 0cm 0cm,totalheight=2.75 cm]{m82aq.pdf}
\\
\vspace{0.25cm}
$W = (1~2~3~4~5~6~7~8) - (1~ 4~ 3~ 2~ 5~ 8~ 7~ 6)$
\\
$\mesonic=NC2$
\\
$\eta_i=((8~ 7), (7~ 6), (4~ 3), (3~ 2),$
\\
$\hspace{1cm}(6~ 1~ 2~ 5), (5~ 8~ 1~ 4))$
\vspace{0.2cm}
\end{tabular}
\\
\hline  
\vspace{-5cm}
8.2b
& 
\includegraphics[trim=0cm 0cm 0cm 0cm,totalheight=5 cm]{m82bt.pdf}
&
\vspace{-4.5cm}
\begin{tabular}[b]{l}
\vspace{0.25cm}
\includegraphics[trim=0cm 0cm 0cm 0cm,totalheight=2.75 cm]{m82bq.pdf}
\\
\vspace{0.25cm}
$W = (1~ 2~ 3~ 4~ 5~ 6~ 7~ 8) - (1~ 6~ 3~ 8~ 5~ 2~ 7~ 4)$
\\
$\mesonic=NC3$
\\
$\eta_i=((8~ 5~ 6~ 3~ 4~ 1~ 2~ 7), (7~ 4~ 5~ 2~ 3~ 8~ 1~ 6))$
\vspace{0.2cm}
\end{tabular}
\\
\hline  
\vspace{-5cm}
8.4a
& 
\includegraphics[trim=0cm 0cm 0cm 0cm,totalheight=5 cm]{m84at.pdf}
&
\vspace{-4.5cm}
\begin{tabular}[b]{l}
\vspace{0.25cm}
\includegraphics[trim=0cm 0cm 0cm 0cm,totalheight=1 cm]{m84aq.pdf}
\\
\vspace{0.25cm}
$W = (1~ 2~ 3~ 4~ 5)+ (6~ 7~ 8)$\\
$\hspace{1cm}-(1~ 3~ 2)- (4~ 6~ 8~ 7~ 5)$
\\
$\mesonic=\mathcal{M}_{3,3}$
\\
$\eta_i=((8, 7), (6~ 8), (5~ 4), (3~ 2), (2~ 1),$
\\
$\hspace{1cm} (7~ 5~ 1~ 3~ 4~ 6))$
\vspace{0.2cm}
\end{tabular}
\\
\hline
\end{tabular}
\end{center}
\caption{Restricted $g=2$ brane tilings \textbf{(3/6)}. \label{tdata3}}
\end{table}


 \begin{table}[H]
\begin{center}
\begin{tabular}{p{1cm}|p{5cm}|p{8cm}}
\hline \hline
E.T\# & Brane Tiling & Quiver \& Superpotential \& $\mesonic$ \\
\hline  
\vspace{-5cm}
8.4b
& 
\includegraphics[trim=0cm 0cm 0cm 0cm,totalheight=5 cm]{m84bt.pdf}
&
\vspace{-4.5cm}
\begin{tabular}[b]{l}
\vspace{0.25cm}
\includegraphics[trim=0cm 0cm 0cm 0cm,totalheight=1 cm]{m84bq.pdf}
\\
\vspace{0.25cm}
$W = (1~ 2~ 3~ 4~ 5)+ (6~ 7~ 8)$\\
$\hspace{1cm}-(1~ 3~ 6) - (2~ 8~ 5~ 4~ 7)$
\\
$\mesonic=\mathbb{C}^3\times\mathbb{C}^2/\mathbb{Z}_2$
\\
$\eta_i=(
(5~ 4), (7~ 2~ 3~ 6), (6~ 1~ 2~ 8),$
\\
$\hspace{1cm}(8~ 5~ 1~ 3~ 4~ 7)
)$
\vspace{0.2cm}
\end{tabular}
\\
\hline  
\vspace{-5cm}
8.4c
& 
\includegraphics[trim=0cm 0cm 0cm 0cm,totalheight=5 cm]{m84ct.pdf}
&
\vspace{-4.5cm}
\begin{tabular}[b]{l}
\vspace{0.25cm}
\includegraphics[trim=0cm 0cm 0cm 0cm,totalheight=1 cm]{m84cq.pdf}
\\
\vspace{0.25cm}
$W = (1~ 2~ 3~ 4~ 5)+ (6~ 7~ 8)$\\
$\hspace{1cm} -(1~ 5~ 6)- (2~ 8~ 7~ 4~ 3)$
\\
$\mesonic=\mathbb{C}^3\times\mathbb{C}^2/\mathbb{Z}_2$
\\
$\eta_i=(
(8~ 7), (4~ 3), (3~ 2), (1~ 5),$
\\
$\hspace{1cm} (7~ 4~ 5~ 6), (6~ 1~ 2~ 8)
)$
\vspace{0.2cm}
\end{tabular}
\\
\hline  
\vspace{-5cm}
8.4d
& 
\includegraphics[trim=0cm 0cm 0cm 0cm,totalheight=5 cm]{m84dt.pdf}
&
\vspace{-4.5cm}
\begin{tabular}[b]{l}
\vspace{0.25cm}
\includegraphics[trim=0cm 0cm 0cm 0cm,totalheight=1 cm]{m84dq.pdf}
\\
\vspace{0.25cm}
$W = (1~ 2~ 3~ 4~ 5)+ (6~ 7~ 8)$\\
$\hspace{1cm}-(1~ 6~ 8)- (2~ 4~ 3~ 5~ 7)$
\\
$\mesonic=\mathbb{C}\times \mathcal{M}_{3,2}$
\\
$\eta_i=(
(6~ 8), (4~ 3), (8~ 1~ 2~ 4~ 5~ 7),$
\\
$\hspace{1cm}(7~ 2~ 3~ 5~ 1~ 6)
)$
\vspace{0.2cm}
\end{tabular}
\\
\hline
\end{tabular}
\end{center}
\caption{Restricted $g=2$ brane tilings \textbf{(4/6)}. \label{tdata4}}
\end{table}


 \begin{table}[H]
\begin{center}
\begin{tabular}{p{1cm}|p{5cm}|p{8cm}}
\hline \hline
E.T\# & Brane Tiling & Quiver \& Superpotential \& $\mesonic$ \\
\hline  
\vspace{-5cm}
8.4e
& 
\includegraphics[trim=0cm 0cm 0cm 0cm,totalheight=5 cm]{m84et.pdf}
&
\vspace{-4.5cm}
\begin{tabular}[b]{l}
\vspace{0.25cm}
\includegraphics[trim=0cm 0cm 0cm 0cm,totalheight=1 cm]{m84eq.pdf}
\\
\vspace{0.25cm}
$W = (1~ 2~ 3~ 4~ 5)+ (6~ 7~ 8)$\\
$\hspace{1cm}-(1~ 6~ 8)- (2~ 7~ 5~ 4~ 3)$
\\
$\mesonic=NC4$
\\
$\eta_i=(
(8~ 1~ 5~ 4~ 3~ 2~ 1~ 6), (7~ 5~ 4~ 3~ 2~ 7~ 6~ 8)
)$
\vspace{0.2cm}
\end{tabular}
\\
\hline  
\vspace{-5cm}
8.4f
& 
\includegraphics[trim=0cm 0cm 0cm 0cm,totalheight=5 cm]{m84ft.pdf}
&
\vspace{-4.5cm}
\begin{tabular}[b]{l}
\vspace{0.25cm}
\includegraphics[trim=0cm 0cm 0cm 0cm,totalheight=1 cm]{m84fq.pdf}
\\
\vspace{0.25cm}
$W = (1~ 2~ 3~ 4~ 5)+ (6~ 7~ 8)$\\
$\hspace{1cm} -(1~ 5~ 4~ 6)- (2~ 8~ 7~ 3)$
\\
$\mesonic=\mathcal{M}_{4,2}$
\\
$\eta_i=(
(8~ 7), (5~ 4), (3~ 2), (1~ 5),$
\\
$\hspace{1cm}(7~ 3~ 4~ 6), (6~ 1~ 2~ 8)
)$
\vspace{0.2cm}
\end{tabular}
\\
\hline  
\vspace{-5cm}
8.4g
& 
\includegraphics[trim=0cm 0cm 0cm 0cm,totalheight=5 cm]{m84gt.pdf}
&
\vspace{-4.5cm}
\begin{tabular}[b]{l}
\vspace{0.25cm}
\includegraphics[trim=0cm 0cm 0cm 0cm,totalheight=1 cm]{m84gq.pdf}
\\
\vspace{0.25cm}
$W = (1~ 2~ 3~ 4)+ (5~ 6~ 7~ 8)$\\
$\hspace{1cm}-(1~ 4~ 3~ 5)- (2~ 8~ 7~ 6)$
\\
$\mesonic=NC5$
\\
$\eta_i=(
(8~ 7), (7~ 6), (4~ 3), (1~ 4),$
\\
$\hspace{1cm} (6~ 2~ 3~ 5), (5~ 1~ 2~ 8)
)$
\vspace{0.2cm}
\end{tabular}
\\
\hline
\end{tabular}
\end{center}
\caption{Restricted $g=2$ brane tilings \textbf{(5/6)}. \label{tdata5}}
\end{table}


 \begin{table}[H]
\begin{center}
\begin{tabular}{p{1cm}|p{5cm}|p{8cm}}
\hline \hline
E.T\# & Brane Tiling & Quiver \& Superpotential \& $\mesonic$ \\
\hline  
\vspace{-5cm}
8.4h
& 
\includegraphics[trim=0cm 0cm 0cm 0cm,totalheight=5 cm]{m84ht.pdf}
&
\vspace{-4.5cm}
\begin{tabular}[b]{l}
\vspace{0.25cm}
\includegraphics[trim=0cm 0cm 0cm 0cm,totalheight=1 cm]{m84hq.pdf}
\\
\vspace{0.25cm}
$W = (1~ 2~ 3~ 4)+ (5~ 6~ 7~ 8)$\\
$\hspace{1cm}-(1~ 5~ 3~ 7)- (2~ 8~ 4~ 6)$
\\
$\mesonic=NC4$
\\
$\eta_i=(
(8~ 4~ 1~ 5~ 6~ 2~ 3~ 7), (7~ 1~ 2~ 8~ 5~ 3~ 4~ 6)
)$
\vspace{0.2cm}
\end{tabular}
\\
\hline
\end{tabular}
\end{center}
\caption{Restricted $g=2$ brane tilings \textbf{(6/6)}. \label{tdata6}}
\end{table}
\vspace{1cm}

\section{Unrestricted Brane Tilings from Higgsing\label{app_incon}}


 \begin{table}[H]
\begin{center}
\begin{tabular}{p{1cm}|p{5cm}|p{8cm}}
\hline \hline
E.T\# & Brane Tiling & Quiver \& Superpotential \& $\mesonic$ \\
\hline  
\vspace{-5cm}
5.2b*
& 
\includegraphics[trim=0cm 0cm 0cm 0cm,totalheight=5 cm]{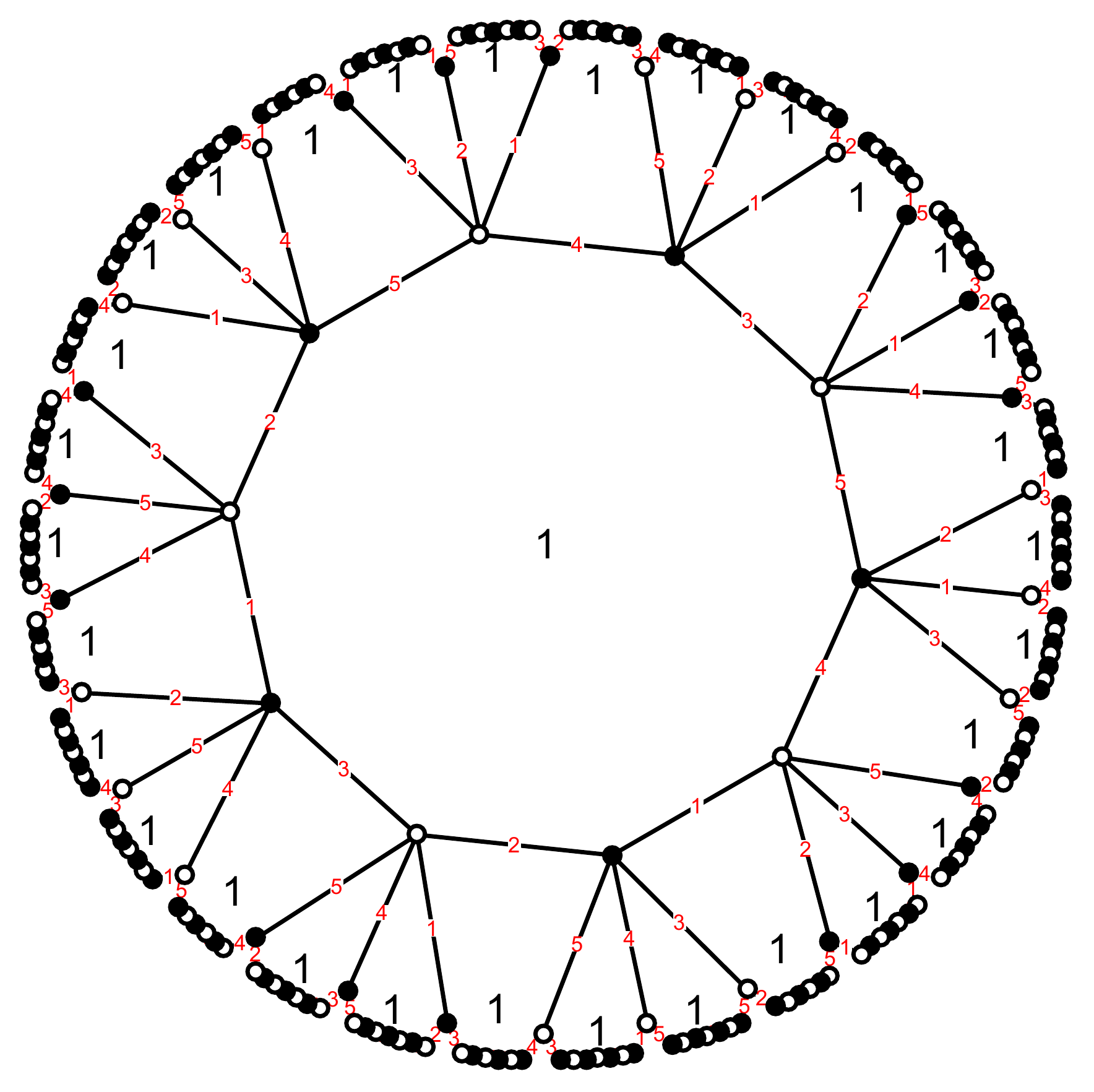}
&
\vspace{-4.5cm}
\begin{tabular}[b]{l}
\vspace{0.25cm}
\includegraphics[trim=0cm 0cm 0cm 0cm,totalheight=1 cm]{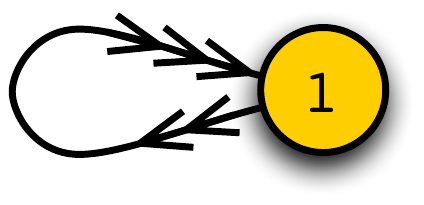}
\\
\vspace{0.25cm}
$W = (1~ 4~ 5~ 3~ 2)-(1~2~5~4~3)$
\\
$\mesonic=\mathbb{C}^5$
\\
$\eta_i=(
(5~ 4), (1~ 2), (4~ 3~ 2~ 5~ 3~ 1)
)$
\vspace{0.2cm}
\end{tabular}
\\
\hline
\end{tabular}
\end{center}
\caption{Unrestricted $g=2$ brane tilings from Higgsing \textbf{(1/4)}. \label{tdata6}}
\end{table}


 \begin{table}[H]
\begin{center}
\begin{tabular}{p{1cm}|p{5cm}|p{8cm}}
\hline \hline
E.T\# & Brane Tiling & Quiver \& Superpotential \& $\mesonic$ \\
\hline  
\vspace{-5cm}
5.2c*
& 
\includegraphics[trim=0cm 0cm 0cm 0cm,totalheight=5 cm]{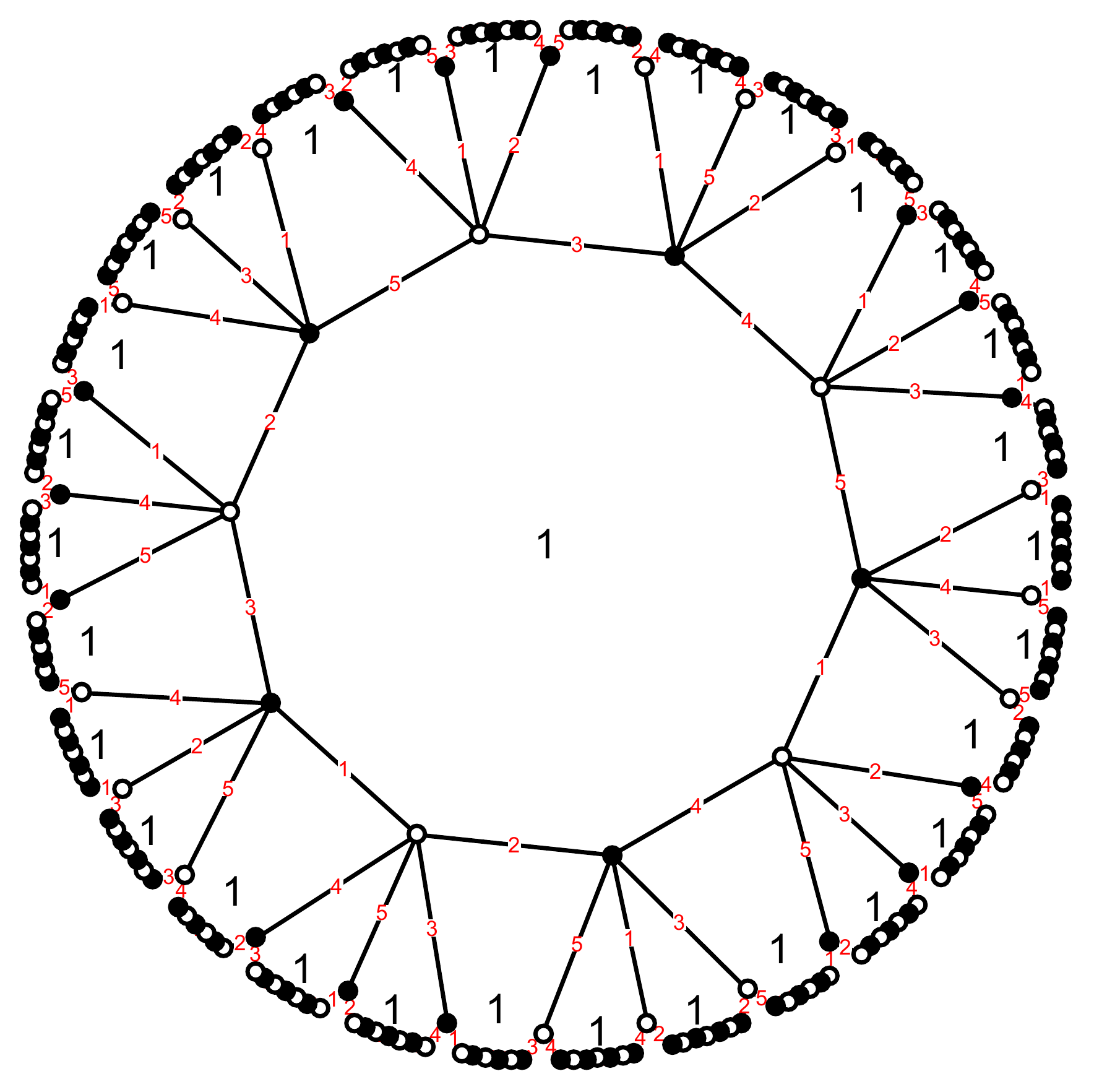}
&
\vspace{-4.5cm}
\begin{tabular}[b]{l}
\vspace{0.25cm}
\includegraphics[trim=0cm 0cm 0cm 0cm,totalheight=1 cm]{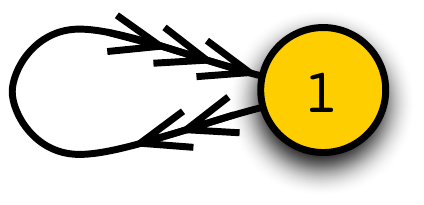}
\\
\vspace{0.25cm}
$W = (1~ 2~ 3~ 5~ 4)-(1~3~4~2~5)$
\\
$\mesonic=\mathbb{C}^5$
\\
$\eta_i=(
(5~ 1~ 2~ 5~ 4~ 2~ 3~ 4~ 1~ 3)
)$
\vspace{0.2cm}
\end{tabular}
\\
\hline  
\vspace{-5cm}
6.2d*
& 
\includegraphics[trim=0cm 0cm 0cm 0cm,totalheight=5 cm]{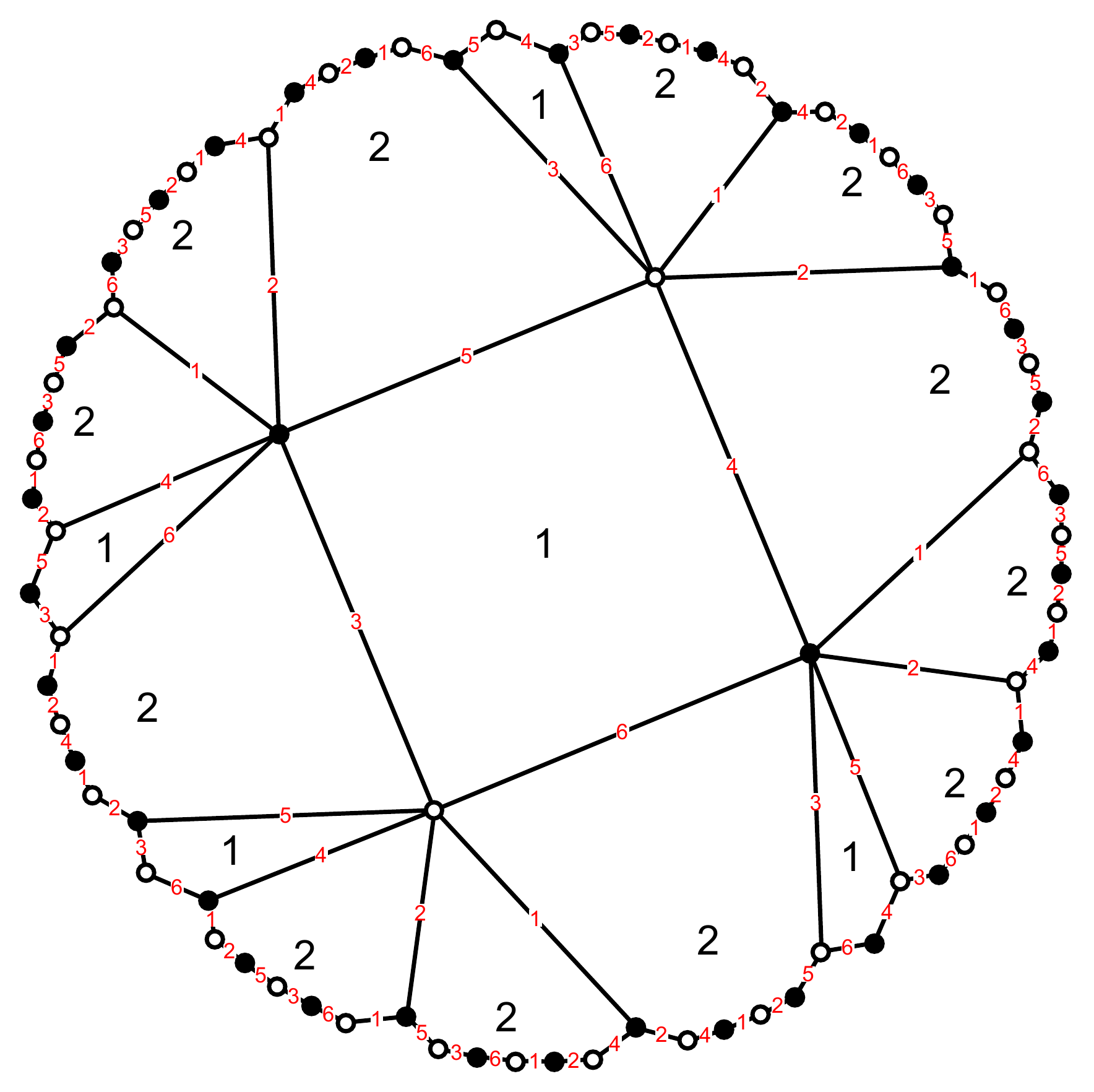}
&
\vspace{-4.5cm}
\begin{tabular}[b]{l}
\vspace{0.25cm}
\includegraphics[trim=0cm 0cm 0cm 0cm,totalheight=1 cm]{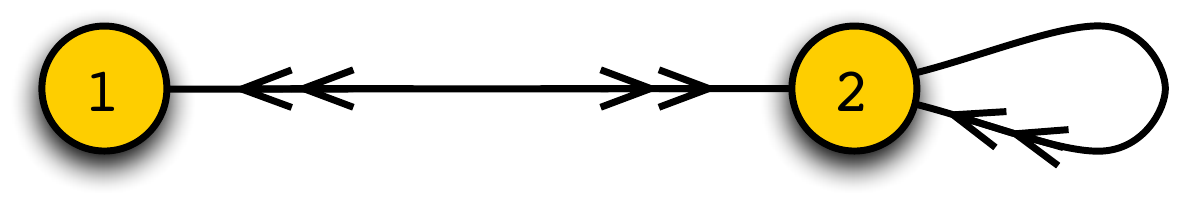}
\\
\vspace{0.25cm}
$W = (1~ 2~ 4~ 5~ 3~ 6)- (1~4~6~3~5~2)$
\\
$\mesonic=\mathcal{C}\times\mathbb{C}^2$
\\
$\eta_i=(
(6~ 3), (3~ 5), (2~ 1), (5~ 2~ 4~ 6~ 1~ 4)
)$
\vspace{0.2cm}
\end{tabular}
\\
\hline  
\vspace{-5cm}
6.2e*
& 
\includegraphics[trim=0cm 0cm 0cm 0cm,totalheight=5 cm]{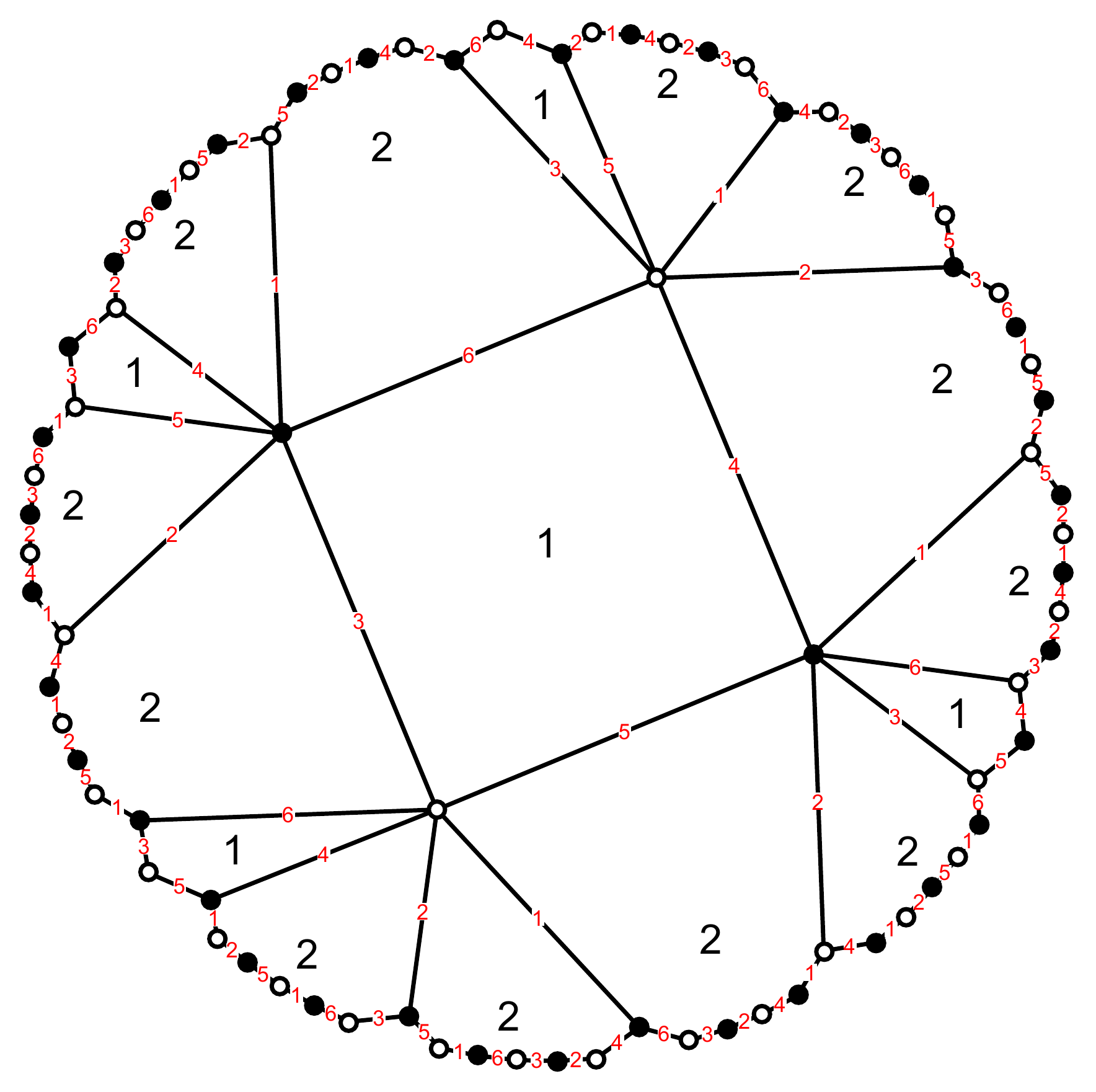}
&
\vspace{-4.5cm}
\begin{tabular}[b]{l}
\vspace{0.25cm}
\includegraphics[trim=0cm 0cm 0cm 0cm,totalheight=1 cm]{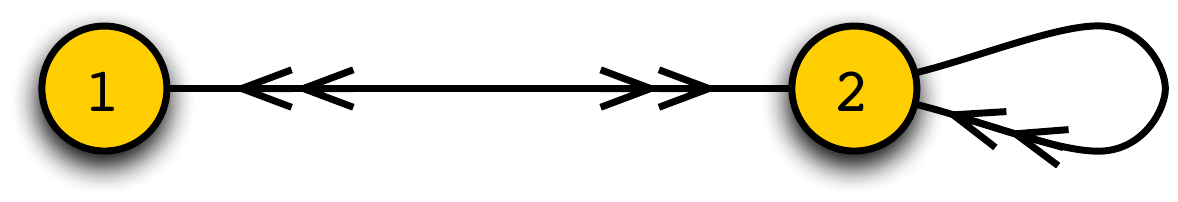}
\\
\vspace{0.25cm}
$W = (1~2~4~6~3~5) - (1~4~5~2~3~6)$
\\
$\mesonic=\mathcal{C}\times \mathbb{C}^2$
\\
$\eta_i=(
(3~ 6), (6~ 1~ 2~ 3~ 5~ 2~ 4~ 5~ 1~ 4)
)$
\vspace{0.2cm}
\end{tabular}
\\
\hline
\end{tabular}
\end{center}
\caption{Unrestricted $g=2$ brane tilings from Higgsing \textbf{(2/4)}. \label{tdata5}}
\end{table}


 \begin{table}[H]
\begin{center}
\begin{tabular}{p{1cm}|p{5cm}|p{8cm}}
\hline \hline
E.T\# & Brane Tiling & Quiver \& Superpotential \& $\mesonic$ \\
\hline  
\vspace{-5cm}
6.2f*
& 
\includegraphics[trim=0cm 0cm 0cm 0cm,totalheight=5 cm]{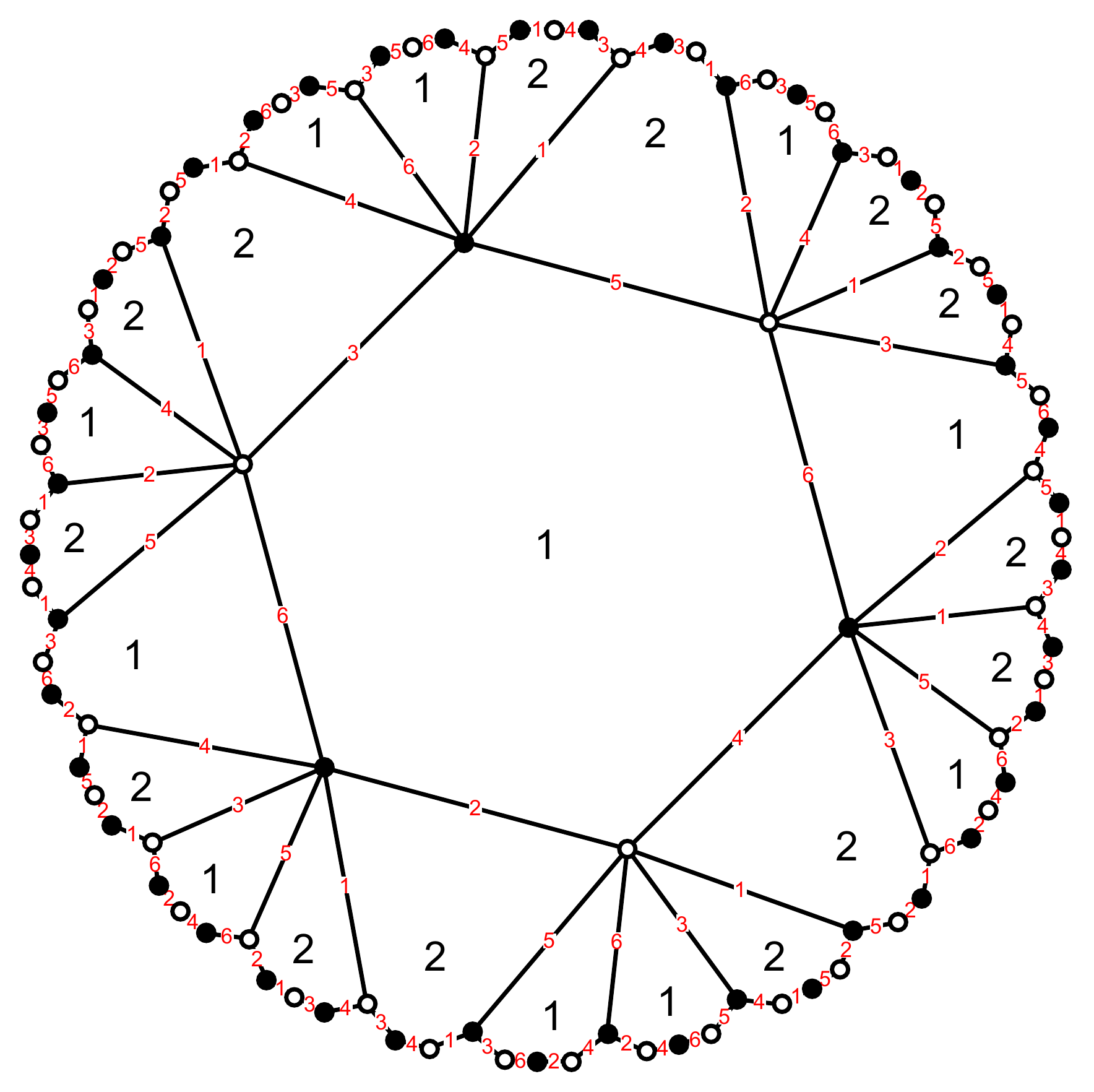}
&
\vspace{-4.5cm}
\begin{tabular}[b]{l}
\vspace{0.25cm}
\includegraphics[trim=0cm 0cm 0cm 0cm,totalheight=1 cm]{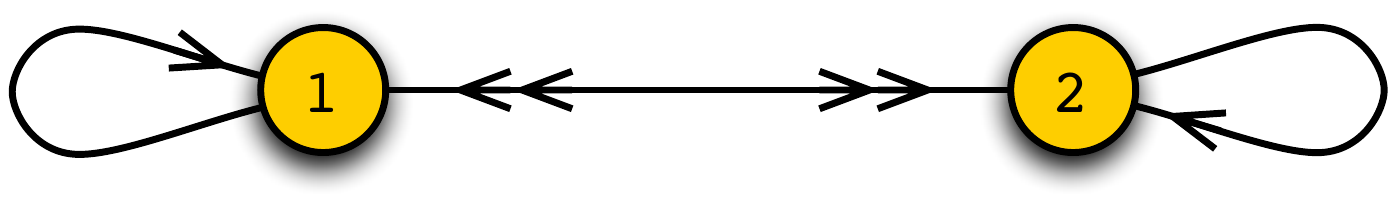}
\\
\vspace{0.25cm}
$W = (1~3~6~5~2~4) - (1~2~6~4~3~5)$
\\
$\mesonic=\mathcal{C}\times\mathbb{C}^2$
\\
$\eta_i=(
(6~ 4~ 1~ 2~ 4~ 3), (5~ 1~ 3~ 5~ 2~ 6)
)$
\vspace{0.2cm}
\end{tabular}
\\
\hline  
\vspace{-5cm}
7.2b*
& 
\includegraphics[trim=0cm 0cm 0cm 0cm,totalheight=5 cm]{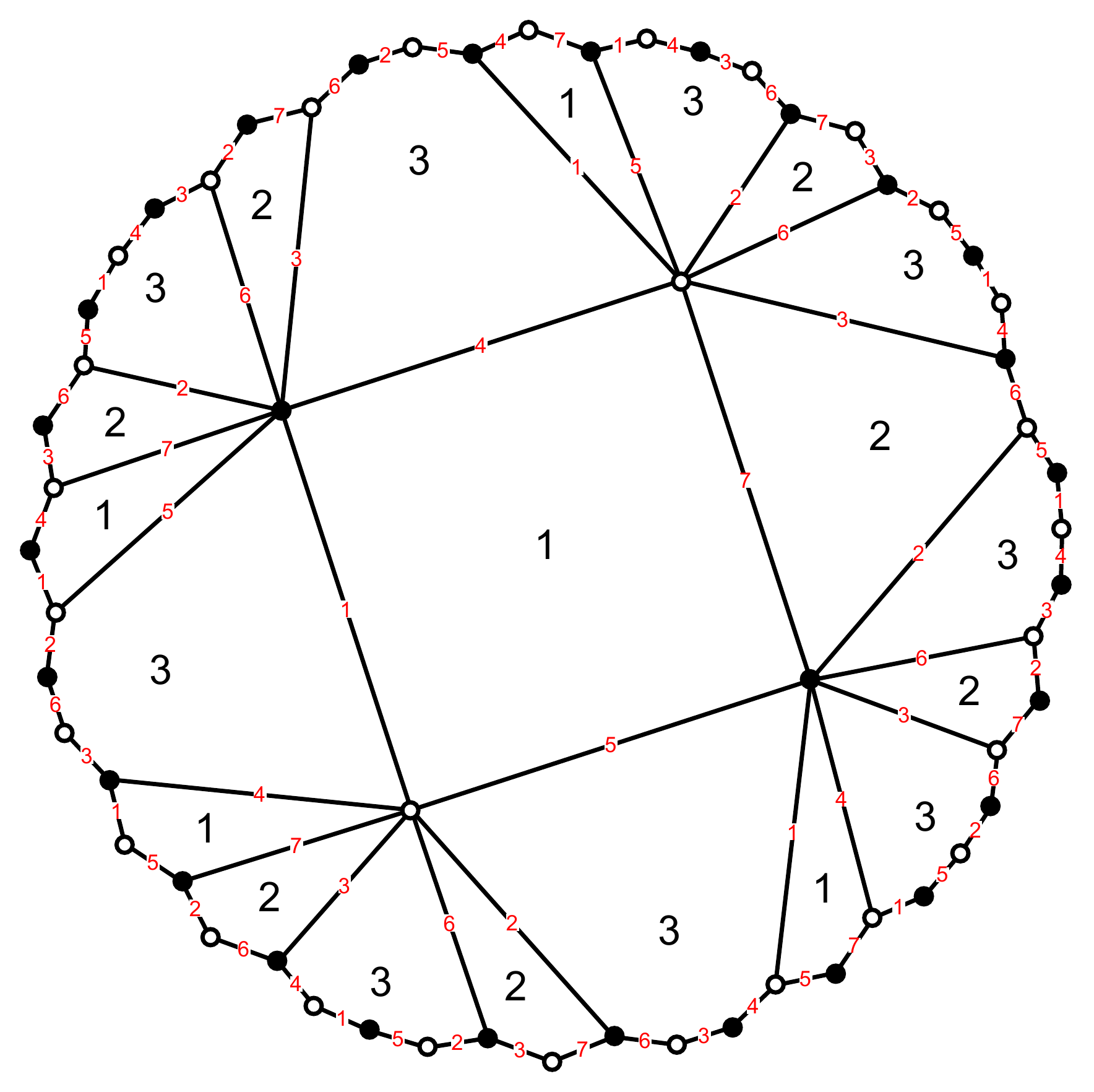}
&
\vspace{-4.5cm}
\begin{tabular}[b]{l}
\vspace{0.25cm}
\includegraphics[trim=0cm 0cm 0cm 0cm,totalheight=2.75 cm]{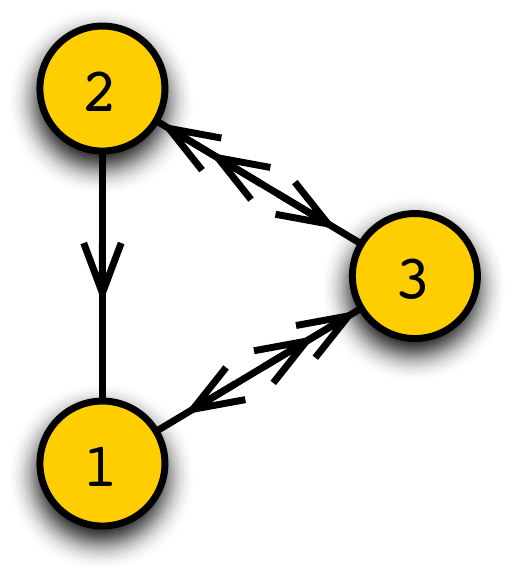}
\\
\vspace{0.25cm}
$W = (1~5~2~6~3~7~4) - (1~4~3~6~2~7~5)$
\\
$\mesonic=NC6$
\\
$\eta_i=(
(6~ 2), (5~ 1), (3~ 6), (1~ 4), (7~ 5~ 2~ 7~ 4~ 3)
)$
\vspace{0.2cm}
\end{tabular}
\\
\hline  
\vspace{-5cm}
7.2c*
& 
\includegraphics[trim=0cm 0cm 0cm 0cm,totalheight=5 cm]{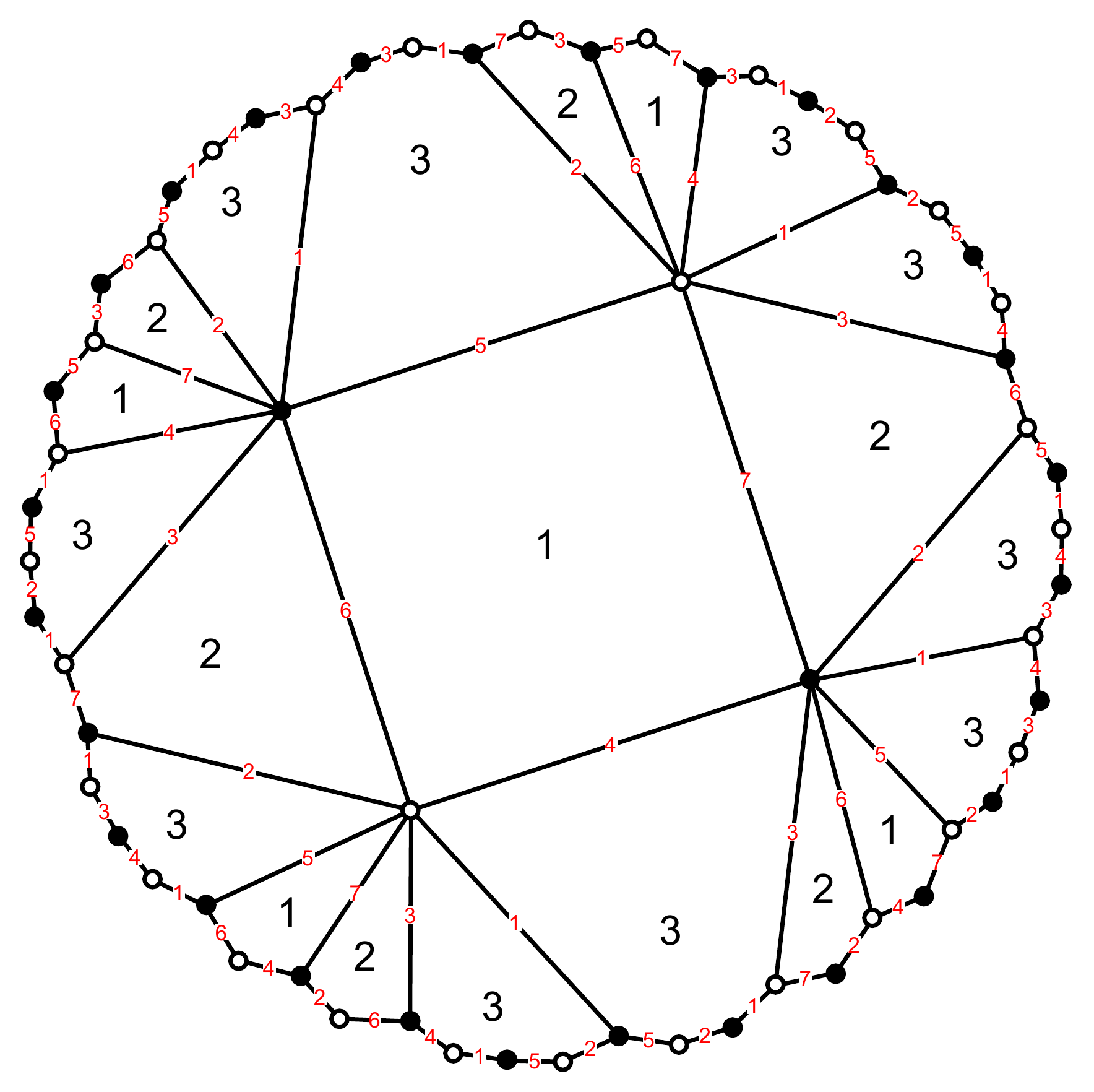}
&
\vspace{-4.5cm}
\begin{tabular}[b]{l}
\vspace{0.25cm}
\includegraphics[trim=0cm 0cm 0cm 0cm,totalheight=2.75 cm]{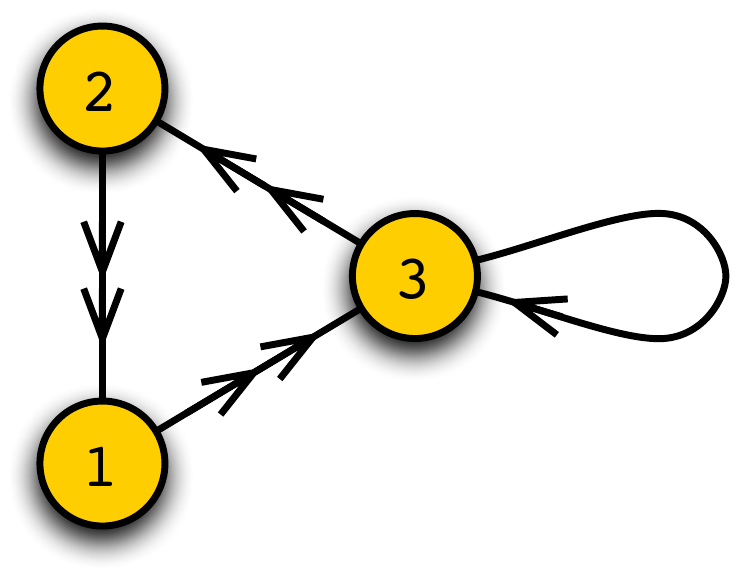}
\\
\vspace{0.25cm}
$W = (1~3~7~5~2~6~4) - (1~2~7~4~3~6~5)$
\\
$\mesonic=NC1$
\\
$\eta_i=(
(7~ 4~ 1~ 2~ 6~ 5~ 2~ 7~ 5~ 1~ 3~ 6~ 4~ 3)
)$
\vspace{0.2cm}
\end{tabular}
\\
\hline
\end{tabular}
\end{center}
\caption{Unrestricted $g=2$ brane tilings from Higgsing \textbf{(3/4)}. \label{tdata9}}
\end{table}


 \begin{table}[H]
\begin{center}
\begin{tabular}{p{1cm}|p{5cm}|p{8cm}}
\hline \hline
E.T\# & Brane Tiling & Quiver \& Superpotential \& $\mesonic$ \\
\hline  
\vspace{-5cm}
7.4b*
& 
\includegraphics[trim=0cm 0cm 0cm 0cm,totalheight=5 cm]{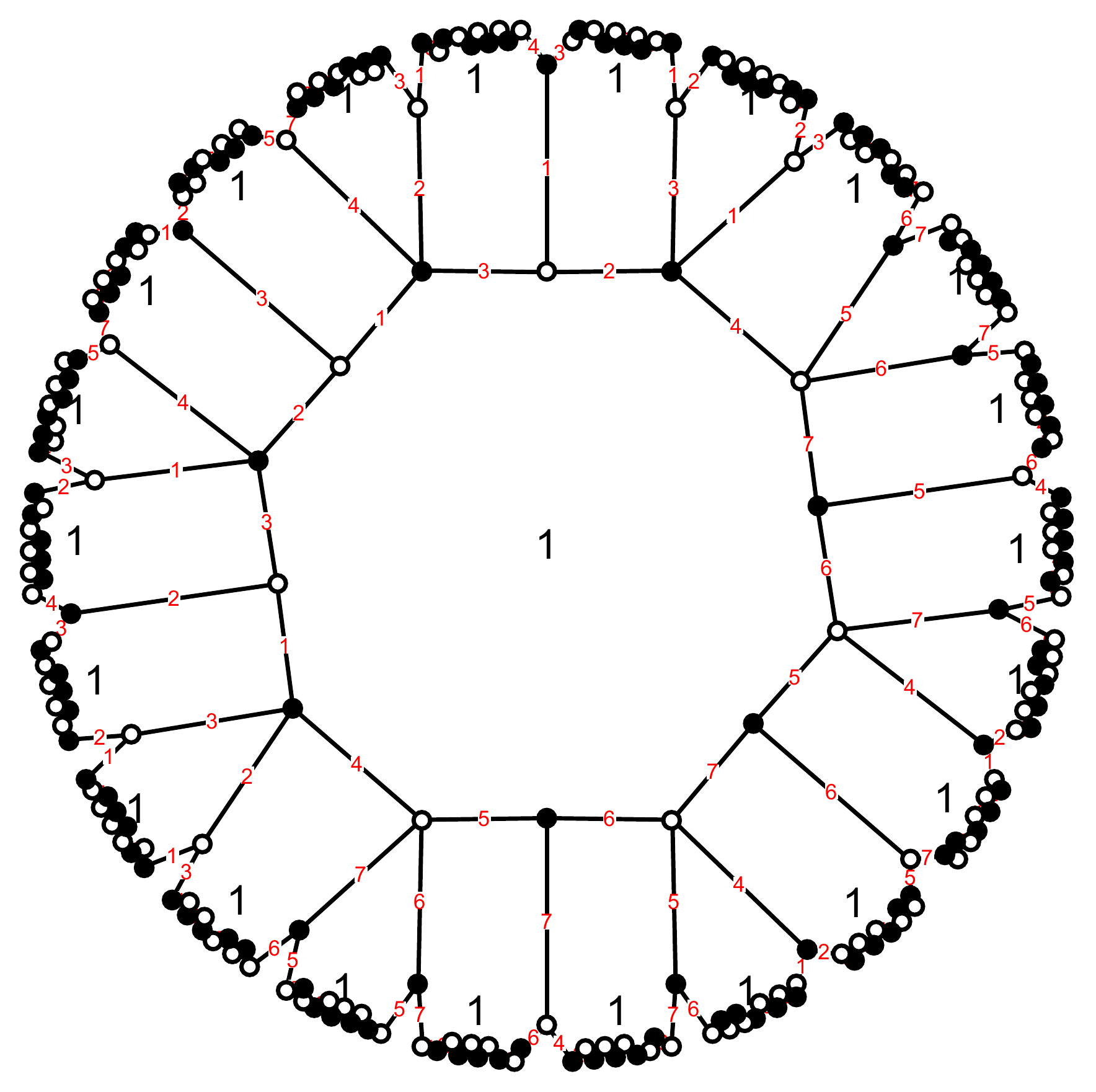}
&
\vspace{-4.5cm}
\begin{tabular}[b]{l}
\vspace{0.25cm}
\includegraphics[trim=0cm 0cm 0cm 0cm,totalheight=1 cm]{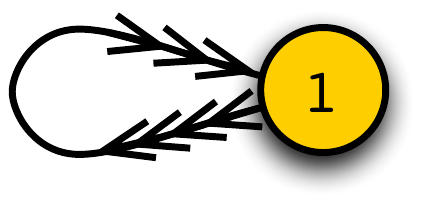}
\\
\vspace{0.25cm}
$W = (1~2~3) + (5~6~7~4)$\\
$\hspace{1cm}- (5~7~6) - (1~3~2~4)$
\\
$\mesonic=\mathcal{M}_{3,3}$
\\
$\eta_i=(
(7~ 6), (6~ 5), (3~ 2), (1~ 3), (5~ 7~ 4~ 1~ 2~ 4)
)$
\vspace{0.2cm}
\end{tabular}
\\
\hline  
\vspace{-5cm}
7.4c*
& 
\includegraphics[trim=0cm 0cm 0cm 0cm,totalheight=5 cm]{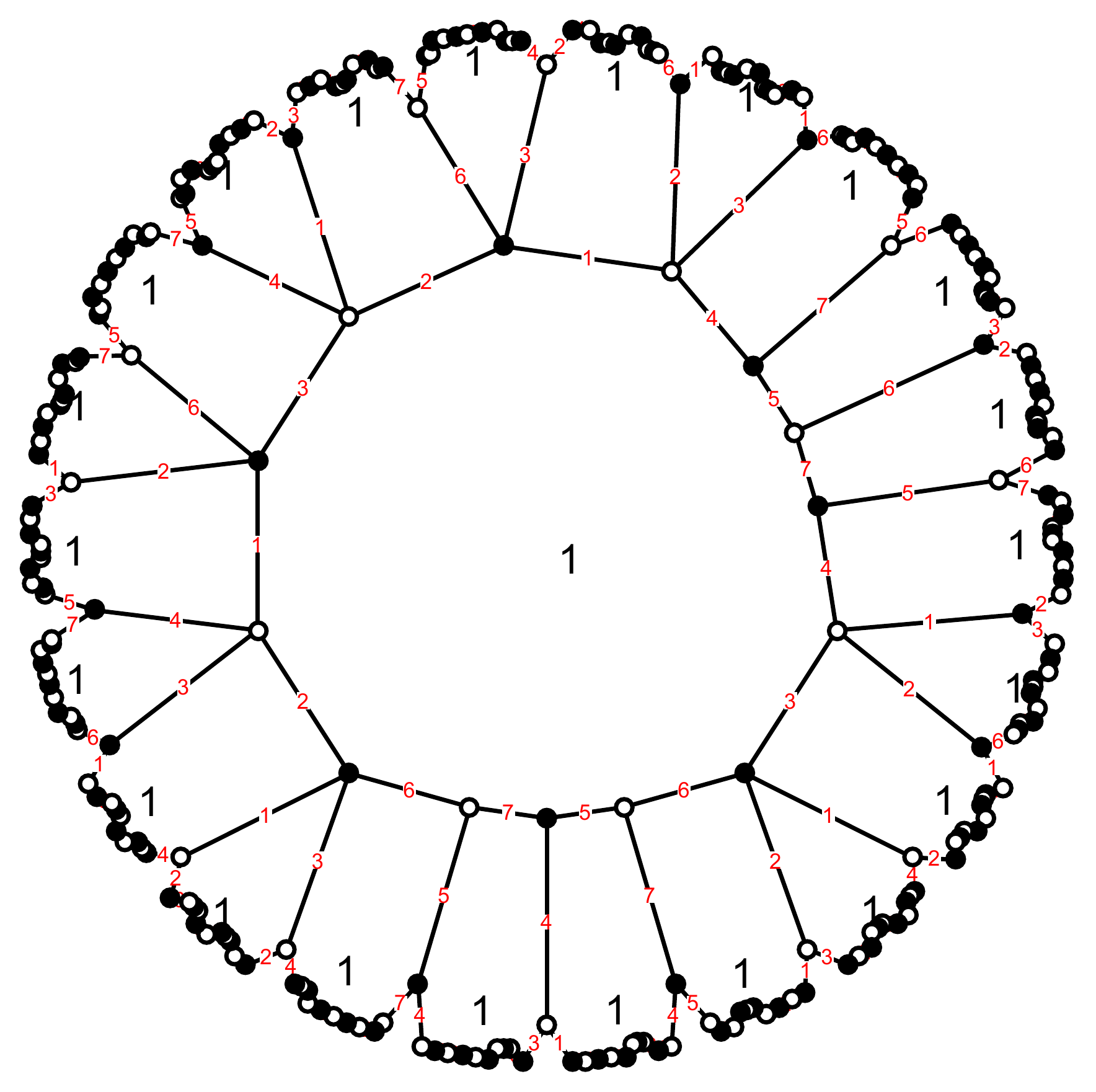}
&
\vspace{-4.5cm}
\begin{tabular}[b]{l}
\vspace{0.25cm}
\includegraphics[trim=0cm 0cm 0cm 0cm,totalheight=1 cm]{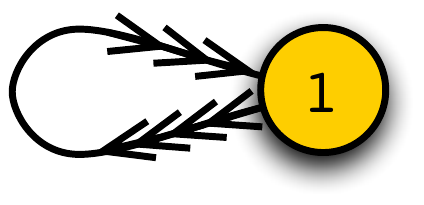}
\\
\vspace{0.25cm}
$W = (1~2~3~4) + (5~6~7)$\\
$\hspace{1cm}- (1~3~6~2) - (4~5~7)$
\\
$\mesonic=\mathcal{M}_{3,2}\times\mathbb{C}$
\\
$\eta_i=(
(5~ 7), (2~ 1), (7~ 4~ 1~ 3~ 4~ 5~ 6~ 2~ 3~ 6)
)$
\end{tabular}
\\
\hline  
\vspace{-5cm}
7.4d*
& 
\includegraphics[trim=0cm 0cm 0cm 0cm,totalheight=5 cm]{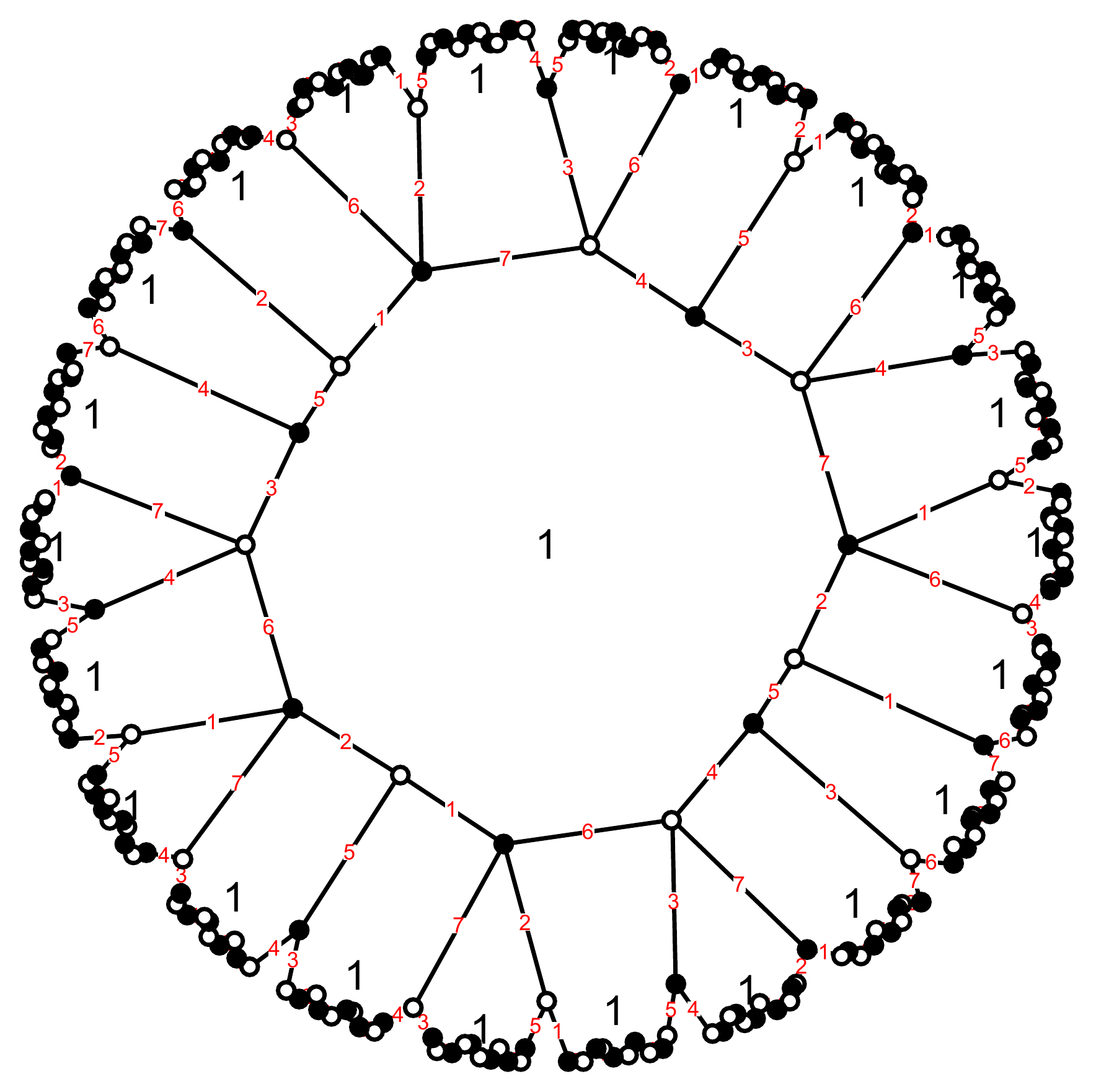}
&
\vspace{-4.5cm}
\begin{tabular}[b]{l}
\vspace{0.25cm}
\includegraphics[trim=0cm 0cm 0cm 0cm,totalheight=1 cm]{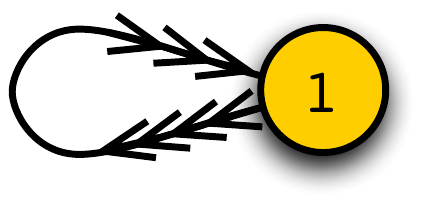}
\\
\vspace{0.25cm}
$W = (1~5~2) + (3~6~4~7)$\\
$\hspace{1cm}- (3~5~4) - (1~7~2~6)$
\\
$\mesonic=\mathcal{C}\times\mathbb{C}^2$
\\
$\eta_i=(
(7~ 2~ 1~ 7~ 3~ 5~ 2~ 6~ 4~ 3~ 6~ 1~ 5~ 4)
)$
\vspace{0.2cm}
\end{tabular}
\\
\hline
\end{tabular}
\end{center}
\caption{Unrestricted $g=2$ brane tilings from Higgsing \textbf{(4/4)}. \label{tdata11}}
\end{table}
\vspace{2cm}

\section{Forward Algorithm and the Mesonic Moduli Space \label{app_mesonic}}

\subsection{Perfect Matchings and the GLSM \label{app_pm}}

\noindent\textbf{Perfect Matchings/GLSM fields and F-and D-term charges.} A new basis of fields can be defined from the known set of bifundamental matter fields. The purpose of the new basis of fields is to describe both F-term and D-terms constraints of the supersymmetric gauge theory with a common setting. The new fields are known as gauge linear sigma model fields (GLSM) and are represented as perfect matchings in the brane tiling (see \cite{Feng:2000mi,Franco:2005rj} for a detailed explanation). They have the following properties:
\begin{itemize}
	\item A \textit{perfect matching} $p_i$ is a set of bifundamental fields which connect each node in the brane tiling precisely once. All points on the perimeter are called external, including extremal (corner) ones. They can be summarized in a matrix $P_{E\times c}$ where $E$ is the number of matter fields and $c$ the number of perfect matchings.
	\item The solutions to \textit{F-terms} are encoded in the perfect matching matrix $P_{E\times c}$ and can be translated into charges under an additional Abelian gauge symmetry. The associated charge matrix is
	\beal{es00_1c0}
	Q_{F~(c-G-2g)\times c} =  \ker{(P_{E \times c})}~~,
	\eea
	where $g$ is the genus of the Riemann surface.
	\item \textit{D-terms} are of the form \cite{Witten:1993yc},
\beal{es00_2}
D_i = - e^2 \left(\sum_{a} d_{ia}|X_a|^2 - \zeta_i\right)~~,
\eea
where $X_a$ is the matter field corresponding to the $a$-th column of the incidence matrix $d_{G\times E}$, $i$ runs over the $U(1)$ gauge groups in the quiver, $e$ is the gauge coupling, and $\zeta_i$ is the Fayet-Iliopoulos (FI) parameter. The D-terms are encoded via the reduced quiver matrix $\Delta_{(G-1)\times E}$\footnote{Since the sum of rows in $d_{G\times E}$ vanishes, there are $G-1$ independent rows giving the reduced matrix $\Delta_{(G-1)\times E}$.} and are related to the perfect matching matrix as follows,
\beal{es00_3}
\Delta_{(G-1)\times E} = Q_{D~(G-1)\times c}.P^{t}_{c \times E}~~,
\eea
where the $Q_{D~(G-1) \times c}$ matrix is the charge matrix under D-term constraints.
Equivalently, in terms of an interim matrix $\tilde{Q}_{G\times c}$, which maps perfect matchings into their quiver charges, one has the relation
\beal{es00_1c2}
d_{G\times E} = \tilde{Q}_{G \times c}.P^{t}_{c \times E}~~.
\eea
\end{itemize}

Overall, the charge matrices $Q_F$ and $Q_D$ can be concatenated to form a $(c-2g-1) \times c$ matrix,
\beal{es00_5}
Q_t = \left( \ba{c} Q_F \\ Q_D \ea \right)~~.
\eea
The kernel of the charge matrix,
\beal{es00_6}
G_t = \ker{(Q_t)}~~,
\eea
precisely encodes the coordinates of the \textbf{toric diagram} vertices with columns and hence perfect matchings and GLSM fields corresponding to vertices of the toric diagram.
\\

\noindent\textbf{Kasteleyn Matrix \cite{Kasteleyn19611209,kasteleyn1967graph,Kenyon:2003uj,Hanany:2005ve}.} The Kasteleyn matrix $K$ is the adjacency matrix of all unique edges in a given fundamental cell of a brane tiling on a torus. The matrix is a $N_w \times N_b$ matrix where $N_w$ and $N_b$ are the numbers of white and black nodes respectively in a given fundamental cell of the tiling. By the bipartite condition on the superpotential, $N_w=N_b$ and the Kasteleyn matrix is a square matrix. With the indices $i=1,\dots,N_w$ and $j=1,\dots,N_b$, the elements of the matrix are 
\beal{INe200}
K_{ij} = \sum_{X(i,j)} x^{h_{a}(X(i,j))} y^{h_b(X(i,j))}~~, 
\eea
where $X(i,j)$ is an edge between white node $w_i$ and black node $b_j$ in the brane tiling's fundamental cell. $(h_a(X(i,j)),h_b(X(i,j))\in \mathbb{Z}^2$ is the winding number of $X(i,j)$. The fugacities $x$ and $y$ count the winding number along the $a$- and $b$-cycles of the torus respectively.

The important property of the Kasteleyn matrix is that its permanent\footnote{The permanent of a matrix is the determinant of the matrix with all signs being positive.} satisfies the following identity,
\beal{INe201}
\text{perm}(K) = \sum_{p_\alpha} x^{h_a(p_\alpha)} y^{h_b^{p_\alpha}}~~,
\eea
which is a sum over all perfect matchings of the brane tiling weighted by their corresponding winding numbers $(h_a,h_b)$ for a given fundamental cell. As such, given that the winding numbers of perfect matchings correspond to the lattice coordinates of toric points, the permanent of the Kasteleyn matrix gives the toric diagram of the brane tiling.
\\

We can now generalise the above definition of the Kasteleyn matrix for brane tilings on Riemann surfaces of arbitrary genus. The genus $g$ Riemann surface has a $4g$-sided fundamental cell with fundamental cycles $a_k$ where $i=k,\dots,2g$. The generalisation takes the following form,
\beal{INe200b}
K_{ij} = \sum_{X(i,j)} \prod_{k=1}^{2g} x_k^{h_{a_k}(X(i,j))} 
~~, 
\eea
where $(h_{a_1}(X(i,j)),\dots,h_{a_{2g}}(X(i,j))$ is the winding number of $X(i,j)$ and $x_k$ is the fugacity for the winding number.
\\

\subsection{Mesonic Hilbert Series \label{app_hs}}

The mesonic moduli space is the space of invariants under F- and D-term charges introduced as $Q_F$ and $Q_D$ above. The $c$ GLSM fields corresponding to perfect matchings of the brane tiling form the space $\mathbb{C}^c$ known as the space of perfect matchings. 

\begin{itemize}
\item
The \textit{Symplectic Quotient}
\beal{es12_1}
\mathcal{M}^{mes}= (\mathbb{C}^c// Q_F) // Q_D ~~.
\eea
is the \textit{mesonic moduli space} of the quiver gauge theory.\footnote{The symplectic quotient $\mathcal{F}^{\flat}=\mathbb{C}^c // Q_F$ is known as the \textit{master space} \cite{Forcella:2008bb,Forcella:2008eh,Forcella:2008ng,Zaffaroni:2008zz,Forcella:2009bv,Hanany:2010zz}, and it is the space of invariants under F-term constraints.}  The invariants under the symplectic quotient are mesonic GIOs of the quiver gauge theory.

\item 
The mesonic Hilbert series is obtained via the Molien Integral formula,
\beal{es12_2}
g_1(y_i; \mathcal{M}^{mes}) =
\prod_{i=1}^{c-2g-1}  \oint_{|z_i|=1} \frac{\ud z_i}{2\pi i z_i} 
\prod_{\alpha=1}^{c} 
\frac{
1
}{
(1-t_i \prod_{j=1}^{c-2g-1} z_j^{(Q_t)_{j\alpha}})
}~~,
\eea
where $c$ is the number of perfect matchings labelled by $\alpha=1,\dots,c$ and $Q_t$ is the concatenated form of the F- and D-term charge matrices as shown in \eref{es00_5}. $g$ is the genus of the Riemann surface on which the brane tiling is drawn. The fugacity $t_i$ counts perfect matchings.  

\item The \textit{plethystic logarithm} of the Hilbert series encodes information about the generators of the moduli space and the relations formed by them. It is defined as
\beal{es12_3}
PL[g_1(y_i; \mathcal{M})] = 
\sum_{k=1}^{\infty} \frac{\mu(k)}{k} \log\left[
g_1(y_i^k; \mathcal{M})
\right]~~,
\eea
where $\mu(k)$ is the M\"obius function.
If the expansion of the plethystic logarithm is finite, the moduli space is a \textit{complete intersection} generated by a finite number of generators subject to a finite number of relations. If the expansion is infinite, the first positive terms refer to basic generators\footnote{The Gr\"obner basis of the set of gauge invariant operators forms the basic generators of the moduli space.} and all higher order terms refer to relations between generators as well as relations among relations which are known as syzygies. 
\end{itemize}

\bibliographystyle{JHEP}
\bibliography{mybib}


\end{document}

%% file: pref.tex
\newcommand{\be}{\begin{equation}}
\newcommand{\ee}{\end{equation}}
\newcommand{\beq}{\begin{equation}}
\newcommand{\beql}[1]{\begin{equation}\label{#1}}
\newcommand{\eeq}{\end{equation}}
\newcommand{\ba}{\begin{array}}
\newcommand{\ea}{\end{array}}
\newcommand{\bea}{\begin{eqnarray}}
\newcommand{\beal}[1]{\begin{eqnarray}\label{#1}}
\newcommand{\eea}{\end{eqnarray}}
\newcommand{\ben}{\begin{enumerate}}
\newcommand{\een}{\end{enumerate}}
\newcommand{\bean}{\begin{eqnarray*}}
\newcommand{\eean}{\end{eqnarray*}}
\newcommand{\eref}[1]{(\ref{#1})}
\newcommand{\sref}[1]{\S\ref{#1}}
\newcommand{\tref}[1]{Table~\ref{#1}}
\newcommand{\nn}{\nonumber}

\newcommand{\fref}[1]{Figure \ref{#1}}
\newcommand{\btab}[1]{\begin{tabular}{#1}}
\newcommand{\etab}{\end{tabular}}

\newcommand{\mesonic}{\mathcal{M}^{mes}}

\newcommand{\comment}[1]{}

\newcommand{\ud}{\mathrm{d}}

\newcommand{\qed}{\nobreak \ifvmode \relax \else
      \ifdim\lastskip<1.5em \hskip-\lastskip
      \hskip1.5em plus0em minus0.5em \fi \nobreak
      \vrule height0.75em width0.5em depth0.25em\fi}

\newcommand{\al}{\alpha}
